\documentclass[12pt,titlepage,twoside]{report}
\usepackage{german}
\usepackage[german,english]{babel}
\usepackage{amssymb}
\usepackage{a4}
\usepackage{fancyhdr}
\usepackage{color}
\usepackage{graphicx}
\newcommand{\dcauthorpre}{Dipl.-Phys. } 
\newcommand{\dcauthorsurname}{Kurth } 
\newcommand{\dcauthorname}{Stefan } 
\newcommand{\dcauthoradd}{geboren am 10.04.1971 in Berlin } 
\newcommand{\dctitle}{The Renormalised Quark Mass in the Schr\"{o}dinger Functional
of Lattice QCD} 
\newcommand{\dcsubtitle}{A One--Loop Calculation with a Non--Vanishing Background
Field} 

\newcommand{\dcapprovala}{Prof. Dr. Ulrich Wolff} 
\newcommand{\dcapprovalb}{Prof. Dr. Michael M\"{u}ller-Preu{\ss}ker} 
\newcommand{\dcapprovalc}{Dr. Peter Weisz} 

\newcommand{\dcdegree}{doctor rerum naturalium\\ (Dr. rer. nat.)} 
\newcommand{\dcsubject}{Physik} 
\newcommand{\dcfaculty}{Mathematisch-Naturwissenschaftlichen Fakult\"at I}
\newcommand{\dcuniversity}{der Humboldt-Universit\"at zu Berlin}
\newcommand{\dcdean}{Prof. Dr. M. Linscheid}
\newcommand{\dcpresident}{Prof. Dr. J. Mlynek}
\newcommand{\dcdatesubmitted}{06.06.2002} 
\newcommand{\dcdateexam}{04.09.2002} 

\newcommand{\dckeydea}{Gitter-QCD}
\newcommand{\dckeydeb}{renormierte Quarkmasse}
\newcommand{\dckeydec}{St\"orungstheorie}
\newcommand{\dckeyded}{O(a)-Ver\-bes\-se\-rung}

\newcommand{\dckeywordsde}{\vspace*{2cm}\\ {\bf{Schlagw\"orter:}}\\ \dckeydea, \dckeydeb, \dckeydec, \dckeyded}

\newcommand{\dckeyena}{lattice QCD}
\newcommand{\dckeyenb}{renormalised quark mass}
\newcommand{\dckeyenc}{perturbation theory}
\newcommand{\dckeyend}{O(a) improvement}
\newcommand{\dckeywordsen}{\vspace*{2cm}\\ {\bf{Keywords:}}\\ \dckeyena, \dckeyenb, \dckeyenc, \dckeyend}

\author{von \\ \dcauthorpre  \dcauthorname  \dcauthorsurname  \\ \dcauthoradd}

\title{ \vspace{-3cm}\dctitle \\
\vspace{0.5cm}
\large{\dcsubtitle} \\ 
\vspace{0.7cm} {\Large{D I S S E R T A T I O N }}\\ \vspace{0.5cm} \large{zur Erlangung des akademischen Grades \\ \dcdegree\\ im Fach \dcsubject \\ \vspace{0.5cm} eingereicht an der \\ \dcfaculty \\ \dcuniversity \\}}
\date{\vspace{0.7cm}
\raggedright{
Pr\"asident der Humboldt-Universit\"at zu Berlin:\\
\dcpresident \vspace{-0.3cm}
}\vspace{0.7cm}\\
\raggedright{
Dekan der \dcfaculty:\\
\dcdean \vspace{-0.3cm}
}\vspace{0.7cm}\\
\raggedright{
Gutachter:
\begin{enumerate} 
\item{\dcapprovala} \vspace{-0.3cm}
\item{\dcapprovalb} \vspace{-0.3cm}
\item{\dcapprovalc} \vspace{-0.3cm}
\end{enumerate}} \vspace{0.5cm}
\raggedright{
\begin{tabular}{lll}
eingereicht am: &  &\dcdatesubmitted\\
Tag der m\"undlichen Pr\"ufung: & & \dcdateexam
\end{tabular}}\\ 
}
\begin{document}
\setlength{\headheight}{16.5pt}
\selectlanguage{german}
\maketitle
\pagestyle{empty}
\cleardoublepage
\selectlanguage{english}
\abstract
The renormalised quark mass in the Schr\"{o}dinger functional is studied
perturbatively with a non--vanishing background field.

The framework in which the calculations are done is the Schr\"{o}dinger
functional. Its definition and basic properties are reviewed and it is
shown how to make the theory converge faster towards its
continuum limit by $\mathrm{O}(a)$ improvement.
It is explained how the Schr\"{o}dinger functional scheme avoids the
implications of treating a large energy range on a single lattice in
order to determine the scale dependence of renormalised quantities.
The description of the scale dependence by the step scaling function is
introduced both for the renormalised coupling and the renormalised quark masses.
The definition of the renormalised coupling in the Schr\"{o}dinger functional
is reviewed, and
the concept of the renormalised mass being defined by the axial current and
density via the PCAC--relation
is explained.  The running of the renormalised mass described by its step
scaling function is presented as a consequence of the fact that the
renormalisation constant of the axial density is scale dependent.

The central part of the thesis is the expansion of several correlation functions
up to 1--loop order. The expansion coefficients are used to compute the
critical quark mass at which the renormalised mass vanishes, as well as the
1--loop coefficient of the renormalisation constant of the axial 
density. Using the result
for this renormalisation constant, the 2--loop anomalous dimension 
is obtained by conversion from
the $\overline{\mathrm{MS}}$--scheme.

Another important application of perturbation theory carried
out in this thesis is the determination of discretisation errors. The critical
quark mass at 1--loop order is used to compute the deviation of the coupling's
step scaling function from its continuum limit at 2--loop order. Several
lattice artefacts of the current quark mass, defined by the PCAC relation with
the unrenormalised axial current and density, are computed at 1--loop order.
An essential property of the renormalised quark mass being computed in this
thesis at 1--loop order is the deviation of its step scaling function 
from the continuum limit, which was so far only known for the zero background
field case.
\dckeywordsen
\cleardoublepage
\selectlanguage{german}
\abstract
Diese Arbeit befasst sich mit st\"{o}rungstheoretischen Rechnungen zur 
renormierten Quarkmasse im Schr\"{o}dinger-Funktional mit nicht verschwindendem
Hintergrundfeld.

Als Grundlage der Rechnungen werden das Schr\"{o}dinger-Funktional und seine
grundlegenden Eigenschaften erl\"{a}utert. Auch die 
$\mathrm{O}(a)$-Verbesserung, die zu einem schnelleren Erreichen des 
Kontinuumslimes f\"{u}hren soll, wird in diesem Zusammenhang dargestellt.
Des weiteren wird erkl\"{a}rt, auf
welche Weise das Schr\"{o}dinger-Funktional dazu dient, das Skalenverhalten
renormierter Gr\"{o}\-{\ss}en \"{u}ber einen gro\-{\ss}en Energiebereich zu
untersuchen. Das Skalenverhalten sowohl der renormierten Kopplung als auch
der renormierten Quarkmassen wird in diesem Schema durch Step-Scaling-Funktionen
beschrieben. Die Definition der renormierten Kopplung wird dargestellt,
ebenso die Definition der renormierten Masse, die mit Hilfe der
PCAC-Relation \"{u}ber den 
Axialvektorstrom und die Pseudoskalardichte erfolgt. 
Die Skalenabh\"{a}ngigkeit der renormierten Masse
wird auf die Skalenabh\"{a}ngigkeit der Renormierungskonstanten der
Pseudoskalardichte zur\"{u}ckgef\"{u}hrt.

Breiten Raum nimmt die Berechnung verschiedener
Korrelationsfunktionen bis zur Ein-Loop-Ordnung in St\"{o}rungstheorie ein. 
Mit Hilfe der so
ermittelten Koeffizienten wird die kritische Quarkmasse, bei der die 
renormierte Masse verschwindet, in Ein-Loop-N\"{a}herung berechnet,
ebenso der Ein-Loop-Koeffizent der Renormierungskonstanten der 
Pseudoskalardichte. Mit Hilfe dieses Koeffizienten wird aus der bekannten
anomalen Dimension in Zwei-Loop-Ordnung im $\overline{\mathrm{MS}}$-Schema
die anomale Dimension im Schr\"{o}dinger-Funktional berechnet.

Als weitere Anwendung der St\"{o}rungstheorie werden verschiedene 
Diskretisierungsfehler bestimmt. Die kritische Quarkmasse in Ein-Loop-Ordnung
geht in den Zwei-Loop-Koeffizienten des Diskretisierungfehlers 
der Step-Scaling-Funktion der renormierten Kopplung ein, der durch
die Abweichung dieser Funktion von ihrem Kontinuumslimes definiert ist.
Verschiedene Diskretisierungsfehler der Strommasse, die durch die PCAC-Relation
mit unrenormiertem Axialvektorstrom und Pseudoskalardichte definiert ist, werden
in Ein-Loop-Ordnung berechnet. Ein wichtiger Diskretisierungsfehler der
renormierten Quarkmasse ist die Abweichung ihrer Step-Scaling-Funktion vom
Kontinuumslimes. Dieser Fehler ist in Ein-Loop-Ordnung bislang nur mit
verschwindendem Hintergrundfeld bekannt und wird in dieser Arbeit mit
nicht verschwindendem Hintergrundfeld berechnet.
\dckeywordsde
\cleardoublepage
\thispagestyle{empty}
\chapter*{Acknowledgements}

In one way or another, many people have contributed to this thesis and
deserve some words of thanks.

First of all, I would like to thank my supervisor Ulli Wolff, both for
taking me as a PhD student and for the guidance afterwards. Writing this
thesis would not have been possible without his advice, which I should 
have asked for more often.

Very important contributions came from Peter Weisz, 
whose numerical checks were
essential for the results in this thesis. During our correspondence, I
recognised that the title ``Mister Perturbation Theory'', given to him by a
colleague, is completely justified.

Furthermore, I would like to thank Rainer Sommer, who was so nice to have a
critical look at my results even when being on holiday. It were some remarks
of his that gave me the important idea that results looking wrong at a first
glance are sometimes correct. 

A colleague worth mentioning is Juri Rolf, not only for his ability to create
a humorous atmosphere in the office we shared, but also for essential checks
on the results. 

Without Burkhard Bunk and his skills in solving computer problems, the
project would have failed right from the beginning. In this context, I would
also like to thank Martin Hasenbusch for useful hints on using the right
Fortran compiler, and Bernd Gehrmann, both for his help in debugging some
programs and for critical reading of the manuscript of this thesis.

Concerning critical looks, I am also grateful to Francesco Knechtli and 
to my brother Martin for participating in a thorough--going discussion of 
the results. 

Furthermore, I would like to thank all members of the computational physics
group at Humboldt University not mentioned so far for contributing to the
pleasant atmosphere making my stay here an agreeable time. Gratefully, I
also have to mention 
the Graduiertenkolleg 271 for ensuring my survival on the financial front.

Last but not least, I would like to thank Alice Rolf and Paul Hasenbusch
for making my time as a PhD student more entertaining than it would have been
without them.

\thispagestyle{empty}

\selectlanguage{english}

\cleardoublepage
\pagenumbering{roman}
\pagestyle{plain}
\tableofcontents
\cleardoublepage
\listoffigures
\cleardoublepage
\listoftables
\cleardoublepage
\pagenumbering{arabic}
\pagestyle{fancy}
\renewcommand{\chaptermark}[1]{\markboth{#1}{}}
\renewcommand{\sectionmark}[1]{\markright{\thesection\ #1}}
\lhead[\fancyplain{}{\bfseries\thepage}]%
      {\fancyplain{}{\bfseries\rightmark}}
\rhead[\fancyplain{}{\bfseries\leftmark}]%
      {\fancyplain{}{\bfseries\thepage}}
\cfoot{}

\newcommand{\pgp}{P_{+}\Gamma P_{-}}
\newcommand{\ppf}{P_{+}\gamma_5}
\newcommand{\pfp}{\gamma_5 P_{+}}
\newcommand{\vp}{\mathrm{\bf p}}
\newcommand{\vq}{\mathrm{\bf q}}
\newcommand{\vr}{\mathrm{\bf r}}
\newcommand{\vo}{\mathrm{\bf 0}}
\newcommand{\vs}{\mathrm{\bf s}}
\newcommand{\vu}{\mathrm{\bf u}}
\newcommand{\vv}{\mathrm{\bf v}}
\newcommand{\vx}{\mathrm{\bf x}}
\newcommand{\vy}{\mathrm{\bf y}}
\newcommand{\vz}{\mathrm{\bf z}}
\newcommand{\rmg}{\mathrm{G}}
\newcommand{\rmf}{\mathrm{F}}
\newcommand{\rmfp}{\mathrm{FP}}
\newcommand{\rmt}{\mathrm{tot}}
\newcommand{\tr}{\mbox{tr}}
\newcommand{\slg}{\mathcal{L}_{\mathcal{G}}}
\newcommand{\slh}{\mathcal{L}_{\mathcal{H}}}

\newcommand{\beq}{\begin{equation}}
\newcommand{\eeq}{\end{equation}}
\newcommand{\beqn}{\begin{eqnarray}}
\newcommand{\eeqn}{\end{eqnarray}}
\newcommand{\bequ}{\begin{displaymath}}
\newcommand{\eequ}{\end{displaymath}}
\newcommand{\beqnu}{\begin{eqnarray*}}
\newcommand{\eeqnu}{\end{eqnarray*}}

\newcommand{\muhat}{\hat\mu}
\newcommand{\nuhat}{\hat\nu}
\newcommand{\sigmamunu}{\sigma_{\mu\nu}}
\newcommand{\khat}{\hat k}
\newcommand{\zerohat}{\hat 0} 
\newcommand{\gbar}{\bar{g}} 
\newcommand{\mbar}{\overline{m}}
\newcommand{\csw}{c_{\rm sw}}
\newcommand{\cttilde}{\tilde{c}_{\mathrm{t}}}
\newcommand{\ctt}{\tilde{c}_\mathrm{t}}
\newcommand{\cst}{\tilde{c}_\mathrm{s}}
\newcommand{\ct}{c_\mathrm{t}}
\newcommand{\cs}{c_\mathrm{s}}
\newcommand{\ca}{c_\mathrm{A}}
\newcommand{\pls}{p_{\mathrm{s}}}
\newcommand{\plt}{p_{\mathrm{t}}}
\newcommand{\fa}{f_\mathrm{A}}
\newcommand{\fp}{f_\mathrm{P}}
\newcommand{\mcrit}{m_{\mathrm{c}}}
\newcommand{\gren}{g_{\mathrm{R}}}
\newcommand{\mren}{m_{\mathrm{R}}}
\newcommand{\chig}{\chi_{\mathrm{g}}}
\newcommand{\chim}{\chi_{\mathrm{m}}}
\newcommand{\mq}{m_{\mathrm{q}}}

\newcommand{\cf}{C_{\mathrm{F}}}
\newcommand{\vbar}{\bar v}
\newcommand{\rmO}{{\rm O}}
\newcommand{\Nf}{N_{\rm f}}
\newcommand{\gsf}{\bar{g}_{\mathrm{SF}}}
\newcommand{\msbar}{\overline{\mathrm{MS}}}
\newcommand{\amsbar}{\alpha_{\overline{\mathrm{MS}}}}
\newcommand{\aqq}{\alpha_{q\bar{q}}}
\newcommand{\st}{S_{mathrm{tot}}}
\newcommand{\acal}{\mathcal{A}}
\newcommand{\bcal}{\mathcal{B}}
\newcommand{\vcal}{\mathcal{V}}
\newcommand{\wcal}{\mathcal{W}}

\newcommand{\Tr}{\mbox{Tr}}
\newcommand{\diag}{\mbox{diag}}
\renewcommand{\Re}{\mbox{Re}}
\renewcommand{\Im}{\mbox{Im}}

\chapter{Introduction\label{chapt:intro}}
\chaptermark{Introduction}

Since the idea of matter consisting of particles was invented by the Greek
philosopher Democrit (460--371 B.C.), there has been considerable progress
in particle physics. Especially the invention first of quantum mechanics
and later of quantum field theories have improved our basic understanding
of elementary particles and their interactions. At the same time, large
accelerators have made experiments in the high energy regime possible, giving
large amounts of results to compare to theoretical predictions.

The modern theoretical framework of particle physics is the Standard Model,
which is a gauge theory based on the 
invariance of the Lagrangian under the gauge group
$SU(3)_{\mathrm{c}}\times SU(2)_{\mathrm{I}}\times U(1)_{\mathrm{Y}}$.
The $SU(2)\times U(1)$ symmetry associated with the weak 
isospin I and the weak hypercharge Y describes the unified electroweak 
interaction~\cite{Glashow:1961tr,Goldstone:1962es,Weinberg:1967tq}.
In this context, particles acquire their masses via the Higgs 
mechanism~\cite{Higgs:1964ia}.

The $SU(3)$ is the gauge group of the strong interaction affecting the 
hadrons. The idea of hadrons consisting of several point--like constituents
was raised by Bjorken's analysis of the scaling properties of the structure
functions in deep inelastic lepton--nucleon scattering~\cite{Bjorken:1969dy}. 
This Bjorken scaling
could be explained by the assumption that the hadrons
consist of point-like particles called partons~\cite{Bjorken:1969ja,Feynman:1969ej}.
These partons could then be identified with the so called quarks proposed
earlier by Gell--Mann and Zweig in order to explain why hadrons can be
classified into multiplets~\cite{Gell-Mann:1964nj,Zweig:1964jf}. As a theory 
for the strong interaction,
Fritzsch, Gell--Mann, and Leutwyler proposed a non--Abelian gauge theory
containing quarks obeying an $SU(3)$ gauge symmetry, where the associated
quantum number is called colour~\cite{Fritzsch:1973pi}.

This theory, called quantum chromodynamics (QCD) and today widely accepted as a 
model for the strong interaction, does however pose some problems. First of
all, the theory is mathematically demanding, making it difficult to get 
predictions for experimental results. Second, the basic particles of the
theory, the quarks, cannot be observed as free particles in nature, due to
the phenomenon of confinement, binding them together in colour neutral
hadrons.

However, despite of these problems, QCD together with the Feynman path
integral approach~\cite{Feynman:PathInt1,Feynman:1948aa} 
turned out to be successful in
the high energy regime. This is due to a property of non--Abelian gauge
theories called asymptotic 
freedom~\cite{tHooft:NonAb1,'tHooft:1985ir,Politzer:1973fx}. 
Asymptotic freedom means that the 
coupling becomes small at high energies, allowing us to treat the quarks as
free particles and to expand expectation values in powers of the 
gauge coupling.
The coefficients in the expansion may be formally infinite, but can be made
finite by regularisation and renormalisation. Regularisation means introducing
an artificial momentum cutoff, and renormalisation means removing the 
divergences before removing the cutoff again.
This method gave, for example, good results for the violation of the
Bjorken scaling~\cite{Altarelli:1977zs,Peterman:1979tb,Altarelli:1982ax}
and for QCD corrections to cross sections in 
electron--positron 
annihilation~\cite{Dine:1979qh,Chetyrkin:1979bj,Celmaster:1980xr}.

At low energies, perturbation theory cannot be applied, since the gauge 
coupling is large at these scales. For this case, Wilson introduced the 
concept of lattice QCD~\cite{Wilson:Lattice}. 
In this approach, space--time is treated as a
four dimensional Euclidean lattice, the inverse lattice spacing $a^{-1}$
serving as an ultraviolet cutoff. Removing the cutoff then corresponds to
taking the continuum limit. On the lattice, QCD can be treated like a
classical statistical system, where expectation values of observables can
be obtained by Monte Carlo simulations. However, due to the limitations of
computer power, the simulations are restricted to small lattices, making
safe extrapolations to the continuum limit difficult. In order to improve
the situation, Symanzik has introduced the concept of systematically 
removing terms of order $a$ both from the action and from the observables,
thus making the theory converge at a rate proportional to $a^2$ rather than
$a$~\cite{Symanzik:1983dc,Symanzik:1983gh}.

In any case, both at high or low energies, the free parameters of the
theory like the coupling and the quark masses have to be fixed by a
set of observables. Since these observables have to be calculated in the
theory, one has to take observables computable in the respective framework,
i.e. lattice QCD or perturbation theory. This means that, at low energies,
the parameters have to be fixed by low energy quantities like hadron masses,
while for high energy perturbative calculations, one has to take high 
energy quantities like jet cross sections. It is not a priori clear if one
set of fixed parameters describes the effects of QCD both at low and at high
energy scales.

To study this question is one aim of the ALPHA collaboration. To this
end, one has to define the renormalised parameters at low energies on
the lattice and then evolve them to high energies, where they can be
compared to the quantities defined perturbatively. The technique 
developed for this purpose is the Schr\"{o}dinger functional 
scheme~\cite{Luscher:1991wu}, which
mainly amounts to defining the renormalised coupling and masses in a
four dimensional box of box size $L$ with special boundary conditions,
imposing a constant colour electric background field.
One thus gets a running coupling and running quark masses depending
on the scale $1/L$. Having computed these quantities at some low energy
scale, they may be evolved to higher energies step by step, using the so
called step scaling function.

In the Schr\"{o}dinger functional scheme, the renormalised strong coupling
is constructed by choosing the boundary conditions to be dependent on a parameter
$\eta$. The renormalised coupling is then defined as the derivative of 
an effective action with respect to $\eta$. For this reason, a non--vanishing
background field is required for the renormalisation of the coupling.

The renormalised quark mass is defined using the PCAC relation, which connects
the axial current $A^a_{\mu}(x)$ and the axial density $P^a(x)$ via
\beq
\partial_{\mu}A^a_{\mu}(x) = 2mP^a(x).
\eeq
A renormalised quark mass can be defined using appropriate correlation
functions containing the renormalised axial current and density. In
contrast to the continuum case, where the axial current does not need
to be renormalised, it does get a finite renormalisation in
lattice QCD. This is a
special consequence of the lattice regularisation, which explicitly
breaks the chiral symmetry of the massless continuum theory. The 
renormalisation needed on the lattice is, however, finite and scale 
independent. The only divergent and scale dependent renormalisation
constant needed for mass renormalisation 
is the one of the axial density $P^a(x)$. Thus, the running
of the mass is completely described by this renormalisation constant.
In contrast to the coupling, the mass renormalisation does not refer
to the boundary gauge fields and may thus be done both with a vanishing or
a non--vanishing background field.

In this scheme, the renormalised coupling and masses can be computed 
non--perturbatively by Monte Carlo simulations. Because of the high
computational costs of simulations in full QCD, first results have been
obtained in the quenched approximation~\cite{Luscher:1994gh}, 
where the fermion determinant is
constant, which means that the flavour number is set to $\Nf=0$. There
have, however, been recent results in full QCD with two 
flavours~\cite{Bode:2001jv}. A less
expensive toy model are so called bermions, which is a theory with
$\Nf=-2$~\cite{Anthony:1982fe,deDivitiis:1995au,Rolf:1999ih,Gehrmann:2001yn}. 
Although this is of course no realistic model of the physical
world, bermions may be used to study structural properties of the theory.
The original idea of extrapolating from $\Nf=-2$ and $\Nf=0$ to
$\Nf=2$ did, however, not turn out to be practicable.

In the high energy regime, one wants to compare the renormalised parameters
defined in the Schr\"{o}dinger functional scheme to those commonly used in
this energy region, i.e. to renormalised parameters in some perturbative
scheme. For this purpose, it is necessary to expand the renormalised coupling
and masses in powers of the bare coupling. For the coupling, this has been
done both at 1-- and 2--loop order~\cite{Luscher:1994gh,Sint:1996ch,Bode:1998hd,
Bode:1999sm}. 
For the mass, however, there has only been
a 1--loop calculation with a vanishing background 
field~\cite{Sint:1998iq}. One of the aims
of this thesis is to do this calculation with a non--vanishing background
field. The computation will give the finite part of the renormalisation
constant $Z_{\mathrm{P}}$ at 1--loop order as well as the 2--loop anomalous 
dimension.

Another useful application of perturbation theory on the lattice is the
estimation of discretisation errors. For the coupling, the discretisation
errors of the step scaling function have been calculated in~\cite{Bode:1999sm} up
to 2--loop order. The 2--loop coefficient does, however, contain the critical
quark mass, at which the renormalised mass vanishes, at 1--loop order.
In~\cite{Bode:1999sm}, the known continuum limit of the critical mass was used. In
order to get a precise result for the discretisation error, it is, however,
necessary to use the critical mass at the finite lattice spacing at which
the step scaling function is computed. The calculation of the critical quark
mass and the resulting discretisation errors of the step scaling function of
the coupling will be done in this theses.

For the quark mass, several lattice artefacts may be considered. One way 
to estimate the size of the lattice effects is to construct several 
different unrenormalised masses using different correlation functions
containing the axial current and density. The difference of these masses
should then be a lattice artefact of order $a$ (or $a^2$ in the 
improved theory). For the renormalised quark mass, the deviation of the
step scaling function from its continuum limit is of interest. This
deviation has already been calculated with a vanishing background field.
If one wishes to compute the quark masses in the same runs as the coupling,
it is, however, desirable to have an estimate for the discretisation errors
of the step scaling function of the running quark mass with a non--vanishing
background field. These discretisation errors will be computed in this
thesis at 1--loop order.

The thesis is structured as follows:
Chapter~\ref{chapt:latt} is an introduction to the basic ideas of lattice
QCD. Also Symanzik's $\rmO(a)$--improvement is presented here. 
In chapter~\ref{chap:schrodinger}, the Schr\"{o}dinger functional is defined,
and the renormalisation of the coupling constant in this scheme is outlined.
The Schr\"{o}dinger functional is then treated perturbatively in
chapter~\ref{chapt:schroedpert}, including the details of the gauge fixing
and the expansion of the coupling.
Chapter~\ref{chapt:curr_mass} describes how to calculate the PCAC mass, serving
as a mass later renormalised multiplicatively, on the lattice. The critical
bare mass, at which the lattice PCAC mass vanishes, is calculated at 1--loop
order by expanding correlation functions containing the axial current and
density.
In chapter~\ref{chapt:renmass}, the renormalisation of the mass is studied.
The renormalisation constant of the axial density and its step scaling function
are calculated at 1--loop order as well as the discretisation errors.
The results of all calculations can be found in chapter~\ref{chap:results}.
The thesis closes with a summary in 
chapter~\ref{chapt:summary}.
Finally, the appendices~\ref{app:group}--\ref{app:tables} contain notational
conventions, some computational techniques used in the calculations, and 
some numerical results.

\cleardoublepage
\chapter{QCD on the lattice \label{chapt:latt}}

\chaptermark{QCD on the lattice}

\section{Lattice gauge theory}
\sectionmark{Lattice gauge theory}
In this chapter, the basic concepts of lattice gauge theory will be 
introduced. Of course, only a short overview can be given here, with an
emphasis on ideas needed later on in this thesis. More details on quantum
fields on the lattice can be found in the literature~\cite{Montvay:Book,
Creutz:Book,Rothe:Book,Luscher:Advanced,Gupta:1997nd}.

Lattice gauge theory is set up on a 4--dimensional hyper-cubic lattice 
with lattice spacing $a$. In order to bring quantum field theories onto
the lattice, one has to discretise the action of the corresponding 
Euclidean continuum theory.

In the continuum, the gauge fields are represented by the vector potential
$A_{\mu}(x)$, lying in the Lie algebra of the gauge group.\footnote{Here and 
in the following, Greek letters denote Lorentz indices from 0 to 3, while
Latin indices run from 1 to 3.}
The calculations in this thesis will be done for the case of QCD, where one
has the gauge group $SU(3)$, but the construction of the theory is
applicable for a general gauge group $SU(N)$.
On the lattice, 
the gauge fields are expressed as parallel transporters between the lattice
points. Let $\muhat$ be a unit vector in direction $\mu$. Then the gauge field
on the link between the lattice site $x$ and the lattice site $x+a\muhat$ is
represented by a \emph{link variable} $U(x,\mu)$, which is related to the
continuum gauge field $A_{\mu}(x)$ by
\beq
U(x,\mu) = e^{aA_{\mu}(x)}.
\eeq
Local gauge transformations are represented by
gauge functions $\Omega(x)$ living on the lattice sites $x$ and acting on the
link variables according to
\beq
U(x,\mu)\rightarrow 
U^{\Omega}(x,\mu) = \Omega(x)U(x,\mu)\Omega(x+a\hat{\mu})^{-1}.
\eeq
Clearly, $U(x,\mu)$ is an element of the gauge group, allowing us to take
products of link variables on a curve consisting of several links. A special
curve giving a gauge invariant combination of link variables
is a closed loop consisting of four links in the
$\mu$-$\nu$ plane, which is called
a \emph{plaquette} $p$. The product of the link variables around this 
plaquette is then denoted by
\beq
  U(p) = U(x,\mu)U(x+a\muhat,\nu)U(x+a\nuhat,\mu)^{-1}U(x,\nu)^{-1}.
\eeq
According to Wilson~\cite{Wilson:Lattice}, an appropriate discretisation of the
gauge field action is given by
\beq
S_{\mathrm{G}}[U] = \frac{1}{g_0^2}\sum_{p}\tr\left\{1-U(p)\right\},
\eeq
where the sum is to be taken over all \emph{oriented} plaquettes. The Wilson
action can be shown to coincide with the continuum action at leading order in
the small $a$ expansion.

\section{Fermions on the lattice}
\sectionmark{Fermions on the lattice}

\subsection{The na\"{\i}ve fermion action}
In contrast to the gauge fields, the lattice fermions are not situated on the
links but on the lattice sites themselves. They are Grassmann valued fields
and carry Dirac, colour, and flavour indices. Thus they differ from the 
continuum case merely by the fact that they are only defined on discrete lattice
sites. 

To set up the theory, one has to define an action for the fermions
on the lattice. For this purpose, the covariant derivative has to be carried
over to the lattice. This may be done using the forward derivative,
\beq
\nabla_{\mu}\psi(x) = \frac{1}{a}\left[U(x,\mu)\psi(x+a\hat{\mu})-\psi(x)\right]
\eeq
or the backward derivative,
\beq
\nabla^{\ast}_{\mu}\psi(x) = \frac{1}{a}
\left[\psi(x)-U(x-a\hat{\mu},\mu)^{-1}\psi(x-a\hat{\mu})\right].
\eeq
Using the average of these two derivatives as a discretised version of the
covariant derivative, one gets the na\"{\i}ve fermion action
\beq
S_{\rmf,\mbox{\scriptsize na\"{\i}ve}}[U,\bar{\psi},\psi]
= a^4\sum_x\bar{\psi}(x)(D+m_0)\psi(x),
\eeq
where $m_0$ is the bare quark mass. Strictly speaking, $m_0$ is a diagonal
matrix containing the bare masses of the different quark flavours. In this
thesis, however, only degenerate quark masses will be considered. $D$ is the
Dirac operator,
\beq
D = \frac{1}{2}\gamma_{\mu}\left(\nabla^{\ast}_{\mu}+\nabla_{\mu}\right),
\label{eq:naivefermact}
\eeq
where $\gamma_{\mu}$ are the Dirac matrices, which can be found in
appendix~\ref{app:group}. 
Of course, one may add terms that vanish in the continuum limit. 
While~(\ref{eq:naivefermact}) seems to be the simplest formulation, it
turns out to cause some severe problems outlined in the next subsection.
For later use, a more complicated expression for the fermion action will
be needed due to these difficulties.
 
\subsection{Fermion doubling and chiral symmetry}

The main problem of the na\"{\i}ve fermion action~(\ref{eq:naivefermact})
is the phenomenon of \emph{fermion doubling}. This property is easily seen
when writing down the propagator one gets from the na\"{\i}ve action in
momentum space,
\beq
S^{-1}(p) = m_0 + \frac{i}{a}\sum_{\mu}\gamma_{\mu}\sin(p_{\mu}a).
\eeq
This expression has not only one but 16 zeros in the Brillouin cell in the
chiral limit $m_0\rightarrow 0$, and 
these spurious poles of the propagator cannot be ignored, since they survive
in the continuum limit.
To make things even worse, they come in pairs with opposite axial charge,
thus spoiling the axial anomaly. As a consequence of these problems, the 
na\"{\i}ve action is obviously not acceptable as a discretised fermion 
action. 

The occurrence of fermion doublers can also be understood as a consequence of
the symmetries of the na\"{\i}ve action. In the continuum, the flavour symmetry
of the massless action on the classical level is 
$U(1)_{\mathrm{L}}\times U(1)_{\mathrm{R}}\times SU(\Nf)_{\mathrm{L}}
\times SU(\Nf)_{\mathrm{R}}$, where $\Nf$ is the number of flavours. As a consequence, one
has the symmetries $U(1)_{\mathrm{V}}$ and $SU(\Nf)_{\mathrm{V}}$, 
which are valid even in the case
of non--vanishing masses, and the axial symmetries 
$U(1)_{\mathrm{A}}$ and $SU(\Nf)_{\mathrm{A}}$.
While $SU(\Nf)_{\mathrm{V}}$ and $U(1)_{\mathrm{V}}$ 
remain exact symmetries of the quantised 
theory giving rise to isospin and baryon number conservation, 
the $SU(\Nf)_{\mathrm{A}}$
is spontaneously broken, leading to $(\Nf^2-1)$ Goldstone bosons, which
explains the relative lightness of the pions. The $U(1)_{\mathrm{A}}$ is broken 
in the quantised
theory by the axial anomaly. The crucial point here is that the na\"{\i}ve
fermion action~(\ref{eq:naivefermact}), instead of 
$U(1)_{\mathrm{L}}\times U(1)_{\mathrm{R}}$,
has a much larger symmetry group $U(4)_{\mathrm{L}}\times U(4)_{\mathrm{R}}$, 
where the spurious
symmetry transformations exchange the corners of the Brillouin cell.

Several ways have been tried to circumvent the problem of fermion doubling.
One possibility is to define only one spin component per lattice site, thus
reducing the number of fermions to four, which may then be interpreted as
different flavours. These \emph{staggered fermions} introduced by Kogut
and Susskind~\cite{Kogut:1975ag,Susskind:1977jm} have a major drawback in the
fact that gauge interactions break the flavour symmetry at finite lattice spacing.
As a consequence, the 16 degrees of freedom on the lattice become a mixture of
spin and flavour, making the interpretation of operators in terms of spin and
flavour non trivial.

Another possibility is to add a dimension five operator to the fermion action.
This term will vanish in the continuum limit and can be chosen such that the
15 spurious flavours become infinitely heavy. A convenient choice was proposed
by Wilson~\cite{Wilson:1975id}. Instead of taking the na\"{\i}ve Dirac 
operator~(\ref{eq:naivefermact}), the action of \emph{Wilson fermions} 
is constructed
using the Dirac--Wilson operator
\beq
D = \frac{1}{2}\left\{\gamma_{\mu}(\nabla^{\ast}_{\mu}+\nabla_{\mu})
-a\nabla^{\ast}_{\mu}\nabla_{\mu}\right\}.
\label{eq:DiracWilson}
\eeq
Using this action, the spurious fermion states get a mass proportional to
$1/a$ and thus decouple in the continuum limit.

While the problem of fermion doubling is solved by the Wilson action, it 
introduces a new difficulty. The chiral symmetry of the massless theory is
explicitly broken by the additional term in~(\ref{eq:DiracWilson}). This is
easily seen by stating that
\beq
 \gamma_5 D+D\gamma_5 \neq 0.
\eeq
This is, of course, a major drawback, because it disables us from
treating theories in which chiral eigenstates play a crucial role on the lattice.
For these purposes, one would like to have a theory free of doublers and preserving
chiral symmetry. The question whether such a theory does exist is answered by
the \emph{Nielsen--Ninomiya theorem}~\cite{Nielsen:1981rz}. If one uses an
action which
\begin{itemize}
\item
is translation invariant,
\item
has continuous lattice momenta in the range $[0,2\pi]$ for $L\rightarrow\infty$,
\item
has only local interactions,
\item
gives the correct continuum propagator in the continuum limit,
\item
and preserves chiral symmetry at finite lattice spacing $a$,
\end{itemize}
then the theory will have doublers. This means 
that one has the choice either to get rid of the doublers or to preserve 
chiral symmetry, but one can not have both at the same time.

A recent development in this field is the resurrection~\cite{Hasenfratz:1998jp}
of the Ginsparg--Wilson relation~\cite{Ginsparg:1982bj} for a chirally invariant
formulation. If the Dirac operator satisfies the condition
\beq
\gamma_5 D+D\gamma_5 = aD\gamma_5 D,
\eeq
then it has a symmetry that becomes the chiral symmetry in the continuum limit.
L\"{u}scher has presented a detailed analysis how to avoid the implications of
the Nielsen--Ninomiya theorem using 
Ginsparg--Wilson fermions~\cite{Luscher:1998pq}.

For our calculations, which do not depend on the chiral symmetry of the theory,
it is however simpler to use the Dirac--Wilson operator. The only implications
of broken chiral symmetry will be a finite scale independent renormalisation
of the axial current and an additional renormalisation of the quark mass.

\section{Symanzik's improvement programme\label{sect:improvement}}
\sectionmark{Symanzik's improvement program}

The main purpose of lattice gauge theory is to calculate physical 
quantities on the lattice and then extrapolate to the continuum limit.
However, the closer one gets to the continuum limit, the more expensive
are Monte Carlo simulations due to critical slowing down. Therefore, it
is desirable to make the theory converge faster. This is possible due to the
fact that the lattice discretisation of the action is not unique. One may add
irrelevant terms that vanish in the continuum limit. This property may be
used to cancel $\rmO(a)$ effects. At small lattice spacing
$a$, one gets an effective action
\beq
S_{\mathrm{eff}} = S_0 +aS_1 +\rmO(a^2)
\eeq 
by Taylor expansion. The principle idea of 
Symanzik~\cite{Symanzik:1983dc,Symanzik:1983gh} was to subtract
the order $a$ term, thus making the theory converge at a rate proportional
to $a^2$.

The pure gauge action can be shown to reach its continuum limit at $\rmO(a^2)$
without any additional terms, but the quark action does need improvement.
For on--shell quantities,
the $\rmO(a)$ contribution can be cancelled by the \emph{Sheikholeslami--Wohlert
term}~\cite{Sheikholeslami:1985ij}
\beq
\delta S_{\mathrm{V}}[U,\bar{\psi},\psi] = a^5\sum_{x}
\bar\psi(x)\delta D_{\mathrm{V}}\psi(x),
\label{eq:sheikho}
\eeq
with 
\beq
\delta D_{\mathrm{V}} =
\csw\frac{i}{4}\sigma_{\mu\nu}\hat{F}_{\mu\nu}(x).
\eeq
The Dirac structure of the Sheikholeslami--Wohlert term is given by
\beq
\sigma_{\mu\nu} = \frac{i}{2}[\gamma_{\mu},\gamma_{\nu}],
\eeq
and $\hat{F}_{\mu\nu}(x)$ is the lattice field tensor given by
\beq
\hat{F}_{\mu\nu}(x) = 
  \frac{1}{8a^2}\left\{Q_{\mu\nu}(x) - Q_{\nu\mu}(x)\right\},
\eeq
with
\beqn
  Q_{\mu\nu}(x) &=& \Bigl\{ 
    U(x,\mu) U(x+a\muhat,\nu) 
    U(x+a\nuhat,\mu)^{-1} U(x,\nu)^{-1} \nonumber\\
&&
    +U(x,\nu) U(x+a\nuhat-a\muhat,\mu)^{-1} 
    U(x-a\muhat,\nu)^{-1} U(x-a\muhat,\mu) \nonumber\\
&&
    +U(x-a\muhat,\mu)^{-1} U(x-a\nuhat-a\muhat,\nu)^{-1} 
    U(x-a\nuhat-a\muhat,\mu) U(x-a\nuhat,\nu) \nonumber\\
&&
    +U(x-a\nuhat,\nu)^{-1} U(x-a\nuhat,\mu)
    U(x-a\nuhat+a\muhat,\nu) U(x,\mu)^{-1} \Bigr\}.
\eeqn
\begin{figure}
\begin{center}
\includegraphics{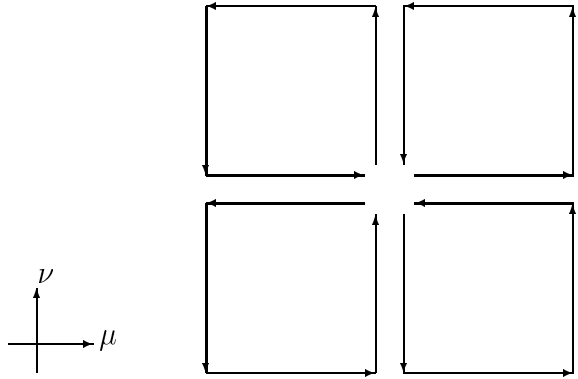}
\end{center}
\caption[The Sheikholeslami--Wohlert term]
{\sl The Sheikholeslami--Wohlert term \label{fig:clover}}
\end{figure}
The Sheikholeslami--Wohlert term may be visualised by the characteristic
shape in figure~\ref{fig:clover}. Therefore, it is often called the
\emph{clover term.}
The improvement coefficient $\csw$ may be computed in perturbation 
theory. Up to 1--loop order, the result originally obtained by 
Wohlert~\cite{Wohlert:1987rf} is
\beq
\csw = 1 + 0.26590(7)g_0^2 + \rmO(g_0^4),
\eeq
independent of $\Nf$. Values obtained later using the Schr\"{o}dinger
functional~\cite{Luscher:1996vw,Weisz:1996csw} are completely compatible 
with this result. For $\Nf=0$ and $\Nf=2$, $\csw$ has also been computed
non--perturbatively~\cite{Luscher:1997ug,Jansen:1998mx}. For
$0\leq g_0\leq 1$, the results can be represented in good approximation by
\beqn
  \csw(g_0)\big|_{\Nf=0} &=& 
    \frac{1-0.656 g_0^2-0.152 g_0^4-0.054 g_0^6}
         {1-0.922 g_0^2},  \\
  \csw(g_0)\big|_{\Nf=2} &=&
    \frac{1-0.454 g_0^2-0.175 g_0^4+0.012 g_0^6+0.045 g_0^8}
         {1-0.720 g_0^2}. 
\eeqn
For the case of $\Nf=-2$ (``bermions''), one gets good results by linear 
extrapolation of the results for $\Nf=2$ and $\Nf=0$~\cite{Gehrmann:2001yn}.

However, the improvement of the action is not sufficient to improve the
expectation value of every composite operator. The operator has to be
improved in addition. To achieve this improvement, one has to expand the
operator in powers of $a$ and then remove the $\rmO(a)$ contribution. To
this end, one has to write down a basis of operators with the correct 
dimensions and the symmetries required and subtract them with appropriate
improvement coefficients. Often, the number of operators can
be reduced by using the equations of motion. An important example is the
improvement of the axial current~\cite{Luscher:1996sc}, which will play
an important role in later chapters.
\vfill

\section{Renormalised parameters}
\sectionmark{Renormalised parameters}

It is well known that, in order to get finite results, quantum field
theories have to be renormalised, leading to a redefinition of the parameters
of the theory, like couplings and fermion masses. The renormalisation is done
by first introducing a regularisation, then doing the renormalisation cancelling
divergences, and finally removing the cutoff again. This situation does also
occur in lattice QCD, where the cutoff is given by the inverse lattice
spacing $a^{-1}$. In the following, the renormalisation of the coupling and 
the masses will be briefly summarised.

\subsection{The renormalised coupling}

Since, in the continuum, the coupling can only be calculated perturbatively, one
has to apply a perturbative renormalisation scheme to compute the renormalised
coupling $\bar{g}$. Such schemes are the MOM--scheme~\cite{Celmaster:1979dm},
the MS--scheme~\cite{'tHooft:1973mm}, and the 
$\msbar$--scheme~\cite{Bardeen:1978yd} of dimensional regularisation. 
The last two schemes differ by the 
subtraction of a constant in the $\msbar$--scheme in addition to the divergence, 
while in the MS--scheme, only the divergence itself is removed. 
This variety of renormalisation schemes is in
contrast to the case of QED, where one naturally chooses the scheme such that the
charges and masses of the leptons get their physical values. This is however not
possible in QCD, because free quarks cannot be observed.

On the lattice, one can define a renormalised coupling
non--perturbatively~\cite{Montvay:Book}. One possibility is to consider a pair of
a static quark and a static antiquark
separated by the distance $r$. The force $F(r)$ acting between
these quarks is the derivative of the static potential $V(r)$, which can be
computed from Wilson loops. Defining $\alpha=\bar{g}^2/4\pi$, a physical
coupling is given by
\beq
\aqq(q)=\frac{1}{\cf}r^2 F(r),
\label{eq:alphaqq}
\eeq
where $q=1/r$, and $\cf$ is the eigenvalue of the quadratic Casimir operator of the
gauge group in the fundamental representation. For $SU(N)$, it is
\beq
\cf = \frac{N^2 -1}{2N}.
\eeq
For QCD, one thus has $\cf=4/3$.

What all couplings computed in different renormalisation schemes have in 
common is the fact that they depend on a renormalisation scale, like
$q$ in~(\ref{eq:alphaqq}). The running
of the coupling is described by the Callan--Symanzik equation
\beq
q\frac{\partial\gbar}{\partial q} =\beta({\gbar}),
\label{eq:beta}
\eeq
where the $\beta$--function has an asymptotic expansion
\beq
\beta(\gbar)\stackrel{\gbar\rightarrow 0}{\sim} -\gbar^3
\left[b_0 + \gbar^2 b_1 +\rmO(\gbar^4)\right].
\eeq
In general, the expansion coefficients will be scheme dependent. The
first two, however, are the same in any two mass--independent renormalisation
schemes. In these schemes, the renormalisation conditions are imposed at
zero quark mass, thus avoiding an implicit dependence of the renormalised
coupling and fields on the quark mass.
Then $b_0$ and $b_1$ are given by
\beqn
b_0 &=& \frac{1}{(4\pi)^2}\left(11 -\frac{2}{3}\Nf\right),
\label{eq:b0}\\
b_1 &=& \frac{1}{(4\pi)^4}\left(102 -\frac{38}{3}\Nf\right).
\label{eq:b1}
\eeqn
For energies high enough to make perturbation theory possible and for sufficiently
small flavour number $\Nf$, one has asymptotic freedom, which means that the
coupling approaches zero in the limit of infinite energy. 
The asymptotic solution is
\beq
\gbar^2 \stackrel{q\rightarrow\infty}{\sim}
\frac{1}{b_0\ln(q^2/\Lambda^2)}
-\frac{b_1\ln[\ln(q^2/\Lambda^2)]}{b_0^3[\ln(q^2/\Lambda^2)]^2}
+\rmO\left(\frac{\{\ln[\ln(q^2/\Lambda^2)]\}^2}{[\ln(q^2/\Lambda^2)]^3}\right),
\eeq
where $\Lambda$ is a scheme dependent integration constant. In the high energy
regime, it may be used to relate different renormalisation schemes to each
other.

\subsection{The renormalised quark masses}

Like the coupling, also the quark masses have to be renormalised. In
continuum perturbation theory, their renormalisation is incorporated in the
MOM--, MS--, or $\msbar$--scheme. On the lattice, a good choice is the 
\emph{hadronic scheme}, in which the bare quark masses are eliminated in
favour of physical hadron masses. First, one chooses certain values for the
bare coupling $g_0$ and the bare masses $am_0^{\mathrm{f}}$, 
where the index f labels
the different quark flavours u,d,s,c,b. Neglecting isospin breaking, one
may assume the light quarks to be degenerate and define 
$m_0^\mathrm{l}\equiv m_0^{\mathrm{u}}=m_0^{\mathrm{d}}$. Next,
one calculates the masses of five different hadrons H containing quarks of
all flavours that are to be renormalised, for example H=p,$\pi$,K,D,B. The
hadron masses will, of course, depend on the bare parameters,
\beq
am_{\mathrm{H}} = am_{\mathrm{H}}(g_0,am_0^{\mathrm{l}},
am_0^{\mathrm{s}},am_0^{\mathrm{c}},am_0^{\mathrm{b}}).
\eeq
To renormalise the theory, one first sets the proton mass $m_\mathrm{p}$ to its
experimental value $m_{\mathrm{p}}^{\mathrm{exp}}$, determining the lattice spacing
by
\beq
a = \frac{am_\mathrm{p}}{m_{\mathrm{p}}^{\mathrm{exp}}},
\label{eq:lattspacing}
\eeq
Next, one has to choose the parameters $am_0^\mathrm{f}$ such that the hadrons get
the masses known from experiment. Equivalently, one can fix the bare quark masses
at a given value of $g_0$ from the condition
\beq
\frac{am_{\mathrm{H}}}{am_{\mathrm{p}}} =
\frac{am_{\mathrm{H}}^{\mathrm{exp}}}{am_{\mathrm{p}}^{\mathrm{exp}}}
\eeq
with H = $\pi$,K,D,B. The bare coupling then determines the lattice spacing
via~\ref{eq:lattspacing}.
Once having renormalised the masses this 
way and the coupling according to~\ref{eq:alphaqq}, the theory is completely 
defined in terms of physical observables.

Like the renormalised coupling, the renormalised masses are scale dependent. To
describe their running, one may use the $\tau$--function,
\beq
q\frac{\partial\mbar}{\partial q} =\tau(\gbar)\mbar.
\label{eq:tau}
\eeq
It has an asymptotic expansion
\beq
\tau(\gbar)\stackrel{\gbar\rightarrow 0}{\sim} -\gbar^2
\left[d_0 +\gbar^2 d_1 +\rmO(\gbar^4)\right],
\eeq
where all expansion coefficients are scheme dependent except the 1--loop
anomalous dimension $d_0$.
It is given by
\beq
d_0 = \frac{6\cf}{(4\pi)^2}.
\label{eq:d0}
\eeq
The 2--loop anomalous dimension $d_1$ is known in the $\msbar$ scheme. For
the gauge group $SU(N)$ it is given by~\cite{Sint:1998iq}
\beq
d_1^{\msbar} = \frac{\cf}{(4\pi)^4}
\left\{\frac{203}{6}N-\frac{3}{2}N^{-1}-\frac{10}{3}\Nf\right\}.
\label{eq:d1msbar}
\eeq
As an analogy to the $\Lambda$--parameter, one may introduce a 
\emph{renormalisation group invariant quark mass} $M$, defined by
\beq 
M = \lim_{q\rightarrow\infty}\mbar(2b_0\bar{g}^2)^{-d_0/2b_0^2}.
\eeq
It can be shown to be independent of the renormalisation scheme. It can be
used to obtain the running mass in any scheme by first computing it in one
scheme and then inserting the proper $\beta$-- and $\tau$--functions of the
other schemes into the renormalisation group equations~\cite{Capitani:1998mw}.

$\Lambda$ and $M$ are fundamental parameters of QCD, which are not fixed 
by the theory, but have to be determined from experimental results.

\subsection{Finite renormalisations\label{subsect:finren}}

Having renormalised the coupling and the masses, the running parameters at
the momentum scale $q$ can be obtained by integrating~(\ref{eq:beta})
and~(\ref{eq:tau}) with the boundary conditions
\beq
\bar{g}(\mu)=\gren,\qquad \mbar^f(\mu) = \mren^f,\qquad f=1,\ldots,\Nf.
\eeq

Any two mass--independent renormalisation schemes are related by a finite 
renormalisation of the parameters,
\beqn
\mu' &=& c\mu,\qquad c>0,\\
\gren' &=& \gren\sqrt{\chig(\gren)},\\
{\mren^{f}}' &=& \mren^f\,\chim(\gren).
\eeqn
The finite renormalisation constants $\chig$ and $\chim$ may be expanded
in perturbation theory,
\beq
\chi(\gren)=\stackrel{\gren\rightarrow 0}{\sim} 1+\sum_{k=1}^{\infty}
\chi^{(k)}\gren^{2k}.
\eeq
With this expansion, one finds that the 2--loop anomalous dimensions in
both schemes are related by
\beq
d'_1 = d_1 +2b_0\chim^{(1)}-d_0\chig^{(1)}.
\label{eq:andimswitch}
\eeq
The crucial point of this equation is that $\chim^{(1)}$ and $\chig^{(1)}$
are 1--loop coefficients. This means that, knowing the 2--loop anomalous dimension
in one scheme, one may compute it in a different scheme by a 1--loop calculation.
This will be used later in order to get the 2--loop anomalous dimension in
the Schr\"{o}dinger functional scheme from the 2--loop anomalous dimension in the
$\msbar$--scheme.
\cleardoublepage
\chapter{The Schr\"odinger functional \label{chap:schrodinger}}
\chaptermark{The Schr\"odinger functional}

\section{Motivation}
\sectionmark{Motivation}

In the last chapter, it was seen that one has two kinds of renormalised couplings:
The perturbatively defined couplings like $\amsbar$ extracted from
high energy experiments, and $\aqq$, defined non--perturbatively at the scale
of hadron physics. Now, an obvious task for lattice QCD is to connect these
couplings. This amounts to calculating the coupling non--perturbatively at
low energy scales and then evolving it to high energies, where one may
compare it with high energy experiments like the determination of jet cross 
sections. Thus one should be able to verify that the hadron spectrum and the
properties of jets really are described by the same theory.

To do this calculation, one has to fulfil several requirements at the
same time.
\begin{itemize}
\item
One has to calculate $\aqq(q)$ at energy scales $q\sim 10\,\mathrm{GeV}$ or higher,
in order to make the connection to other schemes with sufficiently small
perturbative errors.
\item
One has to keep the renormalisation scale $q$ sufficiently far from the 
cutoff $a^{-1}$ in order to keep the discretisation errors small. Otherwise,
a safe extrapolation to the continuum limit might not be possible.
\item
One has to keep the box size $L$ large compared to the confinement scale in order
to avoid large finite size effects.
\end{itemize}
These criteria can be summarised by
\beq
L \gg\frac{1}{0.4\,\mathrm{GeV}}\gg\frac{1}{q}\sim\frac{1}{10\,\mathrm{GeV}}
\gg a.
\eeq
This means that one has to perform a Monte Carlo simulation on a four dimensional
lattice with a size of $L/a\gg 25$. Even with modern computer technique, lattices
of this size are far from being accessible.

However, there is a solution to this problem. The difficulties can be avoided by
identifying the two physical scales~\cite{Luscher:1991wu}, 
\beq
q = \frac{1}{L}.
\eeq
This means one takes a finite size effect as the physical observable. 
\begin{figure}
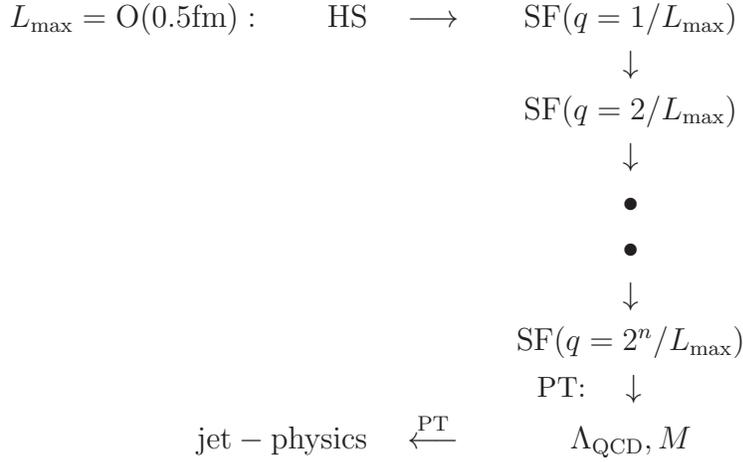

\begin{center}
\begin{eqnarray*}
 { L_{\rm max}}={\rm O}(0.5{\rm fm}): \qquad 
 {\rm HS} \quad \longrightarrow \quad
      &{\rm SF} (q=1/{ L_{\rm max}})& \quad 
               \nonumber \\
      &\downarrow&  \nonumber \\
      &{\rm SF} (q=2/{ L_{\rm max}})&  \nonumber \\ 
      &\downarrow&  \nonumber \\
      &\bullet&  \nonumber \\
      &\bullet&  \nonumber \\
      &\downarrow&  \nonumber \\
      &{\rm SF} (q=2^n/{ L_{\rm max}})& \nonumber \\
   &\mbox{ \small PT:} \quad  \downarrow \qquad \quad &  \nonumber \\
{\rm jet-physics} \quad \stackrel{\rm  PT}{\longleftarrow} 
     \quad   &\Lambda_{\rm QCD}, M & \nonumber
\end{eqnarray*}
\end{center}
\caption[Strategy for evolving the renormalised coupling to large
energy scales]
{\sl Strategy for evolving the renormalised coupling to large
energy scales \label{fig:strategy}}
\end{figure}
The strategy one may use now is presented in figure~\ref{fig:strategy}. First,
the hadronic scheme HS can be related to a finite volume scheme denoted by SF
at a low energy scale $q=1/L_{\mathrm{max}}$.
As a finite volume scheme, we will use the Schr\"{o}dinger functional, which
will be defined in the next section. Within this scheme, one may now compute
the evolution up to the desired scale $q=2^n /L_{\mathrm{max}}$ step by step.
With this recursive procedure, one can thus avoid using quantities defined
at energy scales very wide apart.
At an energy scale large enough for perturbation theory, 
one can then compute the $\Lambda$--parameter and
the renormalisation group invariant quark masses in order to make the connection
to renormalisation schemes defined perturbatively.

\section{Definition of the Schr\"{o}dinger functional}
\sectionmark{Definition of the Schr\"{o}dinger functional}

The purpose of this section is to define the Schr\"{o}dinger functional and
introduce the basic notation. This will be done in the continuum. In the
next section, these definitions will be carried over to the lattice.

\subsection{Formal definition}
We start with the definition of the Schr\"{o}dinger functional in pure gauge
theory and later add the fermions to this framework.

The Schr\"{o}dinger functional is basically defined by the Hamiltonian evolution
of the gauge fields. To this end, one has to specify the theory at $x_0=0$
(or any other fixed time) and write down the Hamiltonian of the theory.

Since the main purpose of the Schr\"{o}dinger functional method is to study
the scaling properties of QCD at a finite volume, the theory is set up in
an $L\times L\times L$ box where $L$ serves as the scale, and one assumes
periodic boundary conditions for the gauge fields. The gauge fields are
represented by vector potentials with values in the Lie algebra of $SU(N)$.
The calculations in this thesis will be done for the case of $SU(3)$, but the
definition is applicable for arbitrary $N$. In the temporal gauge, one is thus
left with Lie algebra valued vector components $A_k(\vx)$ which have to be
periodic.\footnote{Here and in the following, bold letters like $\vx$ denote
vectors in the three dimensional space.} 
Under gauge transformations $\Lambda(\vx)$, they transform like
\beq
A_k(\vx) \rightarrow A^{\Lambda}_k(\vx) = \Lambda (\vx)A_k(\vx)\Lambda (\vx)^{-1}
+\Lambda(\vx)\partial_k\Lambda(\vx)^{-1}.
\eeq
In order to preserve the periodicity of the gauge fields, only periodic gauge
functions $\Lambda(\vx)$ may be allowed, which can be regarded as continuous
Lie group valued functions on a 3--dimensional torus. These functions are 
topologically non--trivial and fall into disconnected classes that are distinguished
by an integer winding number. If one wrote down the Schr\"{o}dinger
functional in the functional integral representation, where one has to integrate
over all gauge configurations, one would have to sum over all topological classes.
However, since this difficulty does not occur on the lattice, it will not be
studied in detail here.

Now, the quantum mechanical states may be defined as wave functionals $\psi[A]$
acting on the gauge fields. Then the quantity
\beq
\langle\psi | \chi\rangle = \int D[A]\psi[A]^{\ast}\chi[A]
\eeq
defines a scalar product on these states. The measure $D[A]$ is meant to be
\beq
D[A] = \prod_{\vx,k,a}dA^a_k(\vx),
\eeq
where $A^a_k(\vx)$ are the components of $A_k(\vx)$ in a basis of the Lie
algebra $su(n)$. The basis used in this thesis is explained in 
appendix~\ref{app:group}. However, not all states defined
this way are physical. Physical states have to be gauge invariant, which means
they have to satisfy
\beq
\psi[A^{\Lambda}] = \psi[A].
\eeq
Obviously, these physical states form a subspace of the space of states. Any state
can be projected onto this subspace by a projector $\mathbb{P}$, given by
\beq
\mathbb{P}\psi[A] = \int D[\Lambda]\psi[A^{\Lambda}],
\eeq
where the measure $D[\Lambda]$ is defined by
\beq
D[\Lambda] = \prod_{\vx}d\Lambda (\vx).
\eeq
The canonically conjugate field of the gauge field is the colour electric field
\beq
F^a_{0k}(\vx) = -i\frac{\delta}{\delta A^a_k(\vx)}.
\eeq
It is part of the colour field tensor, with the magnetic components given
by
\beq
F^a_{kl}(\vx) = \partial_k A^a_l(\vx) -\partial_l A^a_k(\vx)
+ f^{abc}A^b_k(\vx)A^c_l(\vx),
\eeq
where $f^{abc}$ are the structure constants of $SU(N)$. Using this tensor,
one can write down the Hamilton operator $\mathbb{H}$, 
\beq
\mathbb{H} = \int_0^L d^3\vx\left\{\frac{g_0^2}{2}F^a_{0k}(\vx)F^a_{0k}(\vx)
+\frac{1}{4g_0^2}F^a_{kl}(\vx)F^a_{kl}(\vx)\right\}.
\eeq
Now, for any smooth gauge field $C_k$, a state $|C\rangle$ may be defined such
that
\beq
\langle C|\psi\rangle = \psi[C]
\eeq
holds for all wave functionals $\psi[A]$. This state is not necessarily physical,
but of course it can be made so by projecting it onto the physical subspace
using the projector $\mathbb{P}$.
The Schr\"{o}dinger functional $\mathcal{Z}[C',C]$
is then defined by
\beq
\mathcal{Z}[C',C] = \langle C'|e^{-\mathbb{H}\,T}\mathbb{P}|C\rangle.
\eeq
The projector $\mathbb{P}$ makes sure that the Schr\"{o}dinger functional is
invariant under gauge transformations of $C$ and $C'$, since only gauge invariant
intermediate states contribute.

\begin{figure}
  \noindent
  \begin{center}
    \includegraphics[width=.4\linewidth]{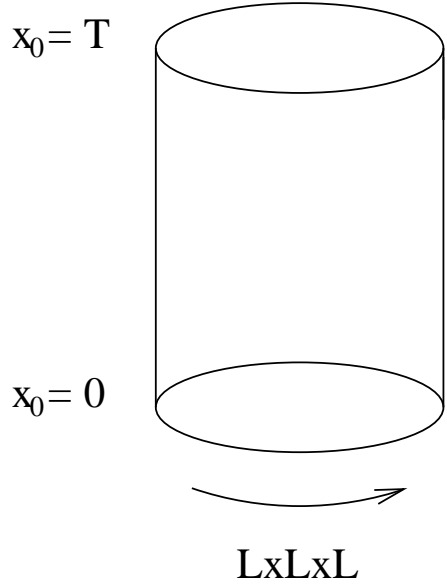}
  \end{center}
  \caption[The space time box of the Schr\"{o}dinger functional]
          {\label{fig:schroedinger}
           \sl The space time box of the Schr\"{o}dinger functional}
\end{figure}

An alternative way to write down the Schr\"{o}dinger functional is to set up
the theory on a 4--dimensional box $L\times L\times L\times T$ and express the
time evolution operator $e^{-\mathbb{H}\,T}$ by a functional integral, using 
the action
\beq
S[A] = -\frac{1}{2g_0^2}\int_0^L d^4x\,\tr\{F_{\mu\nu}F_{\mu\nu}\}.
\eeq
The fields
$C$ and $C'$ then define the boundary conditions on the gauge field at $x_0=0$
and $x_0=T$. This situation may be depicted by figure~\ref{fig:schroedinger}. 
On the lattice, this representation will be chosen.

\subsection{Fermions in the Schr\"{o}dinger functional}
Though mostly used on a lattice, also the fermion part of the Schr\"{o}dinger
functional may formally be defined in the continuum~\cite{Sint:1995rb,Sint:1996ch}.
The easiest way to do this is in the functional integral representation with
the action
\beqn
S_{\mathrm{F}}[A,\bar{\psi},\psi] &=&
\int_0^T dx_0\int_0^L d^3\vx\,\bar{\psi}(x)\left(
\gamma_{\mu}D_{\mu}+m_0\right)\psi(x)
\nonumber\\
& & 
-\int_0^L d^3\vx\,\left[\bar{\psi}(x)P_-\psi(x)\right]_{x_0=0}\nonumber\\
& &
-\int_0^L d^3\vx\,\left[\bar{\psi}(x)P_+\psi(x)\right]_{x_0=T},
\eeqn
where $D_{\mu}=\partial_{\mu}+A_{\mu}$ is the covariant derivative of the
fermion fields. The projectors $P_+$ and $P_-$ are defined by
\beq
P_{\pm} = \frac{1}{2}\left(1\pm\gamma_0\right).
\eeq
In time direction, Dirichlet boundary conditions are imposed for the quark
fields
\beq
P_{+}\psi(x)|_{x_0=0} = \rho(\vx), \;\;\;\;\;
P_{-}\psi(x)|_{x_0=T} = \rho'(\vx),
\eeq
and for the antiquark fields
\beq 
\bar{\psi}(x)P_{-}|_{x_0=0} =\bar{\rho}(\vx), \;\;\;\;\;
\bar{\psi}(x)P_{+}|_{x_0=T} = \bar{\rho}'(\vx).
\eeq
In space direction, the quark fields are chosen to be periodic up to a phase
factor,
\beq
\psi(x+L\hat{k}) = e^{i\theta}\psi(x),\quad 
\bar{\psi}(x+L\hat{k}) = e^{-i\theta}\bar{\psi}(x).
\eeq
For the time being, $\theta$ will be left arbitrary. Later, several values
for $\theta$ will be considered.

\section{Lattice formulation}
\sectionmark{Lattice formulation}

\subsection{The Schr\"{o}dinger functional action on the lattice}
Since the Schr\"{o}dinger functional is characterised by its boundary 
conditions, these boundary conditions have to be carried over to the
lattice. This means that the link variables have to take fixed boundary
values in the $x_0=0$ and $x_0=T$ planes, 
\beq
U(x,k)|_{x_0=0} = W(\vx,k),\;\;\;\;\; U(x,k)|_{x_0=T} = W'(\vx,k).
\eeq
In order to make contact to the continuum definition of the Schr\"{o}dinger
functional, one has to relate the lattice boundary fields $W$ and $W'$ to
the continuum boundary fields $C$ and $C'$. This relationship is naturally
established by recalling that the link variables are parallel transporters 
for colour vector fields. So it seems natural to define
\beq
W(\vx,k) = \mathcal{P}\exp\left\{a\int_0^1 dt\,C_k(\vx+a\khat -ta\khat)
\right\},
\eeq
and analogously for $W'$ using the continuum field $C'$ instead of $C$.
Here, $\mathcal{P}$ denotes the path ordering along the link.

The gauge field action in the Schr\"{o}dinger functional is given by the
Wilson action 
\beq
S_{\rmg}[U] = \frac{1}{g_0^2}\sum_p w(p)\tr\left\{1-U(p)\right\},
\label{eq:gaugeactionschrod}
\eeq
only modified by the weight factors $w(p)$. In the unimproved theory, one has
$w(p) =1/2$ for the spatial plaquettes at $x_0=0$ and $x_0=T$ and $w(p)=1$
in all other cases. In the improved theory, these weight factors will be
modified at the boundaries. Here again, the sum in~(\ref{eq:gaugeactionschrod})
is to be taken over all \emph{oriented} plaquettes.

For the quarks, we take the usual Dirac--Wilson action,
\beq
S_{\rmf}[U,\bar{\psi},\psi] = a^4\sum_x\bar{\psi}(x)(D+m_0)\psi(x),
\eeq
with the Dirac--Wilson operator~(\ref{eq:DiracWilson})
\beq
D = \frac{1}{2}\left\{\gamma_{\mu}(\nabla^{\ast}_{\mu}+\nabla_{\mu})
-a\nabla^{\ast}_{\mu}\nabla_{\mu}\right\}.
\eeq
The Schr\"{o}dinger functional on the lattice is then given by
\beq
\mathcal{Z}[C',\bar{\rho}',\rho',C,\bar{\rho},\rho] 
= \int D[\psi]D[\bar{\psi}]D[U]e^{-S[U,\bar{\psi},\psi]},
\eeq
where the action $S$ is the sum of the gauge field action and the fermionic
action,
$S = S_{\rmg} + S_{\rmf}$.
The expectation value of any product $\mathcal{O}$ of fields is now given
by
\beq
\langle\mathcal{O}\rangle = \left\{\frac{1}{\mathcal{Z}}
\int D[U]D[\psi]D[\bar{\psi}]\mathcal{O}e^{-S[U,\bar{\psi},\psi]}
\right\}_{\bar{\rho}'=\rho'=\bar{\rho}=\rho=0}.
\label{eq:schrodfint}
\eeq
Apart from the quark and gauge fields, the operator $\mathcal{O}$ may also 
contain the boundary fields
\beqn
\zeta(\vx)=\frac{\delta}{\delta\bar{\rho}(\vx)}&,&\quad
 \zeta'(\vx)=\frac{\delta}{\delta\bar{\rho}'(\vx)},\label{eq:zetadef}\\
\bar{\zeta}(\vx)=-\frac{\delta}{\delta\rho(\vx)}&,&\quad
 \bar{\zeta}'(\vx)=-\frac{\delta}{\delta\rho'(\vx)}.\label{eq:zetabardef}
\eeqn
These derivatives have to be understood in the sense that they act on the 
Boltzmann factor in~(\ref{eq:schrodfint}). They will be used to define 
correlation functions which play an important role in later chapters. 

\subsection{The background field}

Unless they are zero, the boundary fields $C$ and $C'$ obviously impose a
background field which is given by a solution of the field equations with
the correct boundary values. It will be explained later that a non--zero 
background field is inevitably needed to define the renormalised coupling
in the Schr\"{o}dinger functional. Since only small lattices are accessible
for numerical simulations, one cannot make the lattice spacing arbitrarily
small. Thus one has to deal with considerable cutoff effects, which requires
a strategy to keep the discretisation errors as small as possible. In particular,
one has to look for a background field leading to minimal cutoff effects.
  
It can be shown that spatially constant Abelian boundary
fields give good results~\cite{Luscher:1992an}. So one may choose
\beq
C_k = \frac{i}{L}\left(
\begin{array}{ccc}
\phi_{k1} & 0 & 0 \\
0 & \phi_{k2} & 0 \\
0 & 0 & \phi_{k3} \\
\end{array}
\right),
\eeq
and analogously for $C'$. In the following, $C_k$ and $C'_k$ will be chosen
to be independent of $k$, i.e.~$C_1=C_2=C_3$ and $C'_1=C'_2=C'_3$.
An obvious solution of the field equations with these
boundary conditions is
\beq
B_0 = 0,\quad B_k = \left[x_0 C'_k +(T-x_0)C_k\right]/T.\label{eq:backcont}
\eeq
This solution is a constant colour electric field, which can easily be
seen from the corresponding field tensor
\beq
G_{0k} = i\mathcal{E},\quad G_{kl} = 0,
\eeq
with
\beq
\mathcal{E} = -i[C'_k - C_k]/T.
\eeq
On the lattice, the background field must of course be given in terms of the
link variables. A good candidate for the lattice background field is
\beq
V(x,\mu) = \exp\{aB_{\mu}(x)\},
\eeq
which, after inserting~(\ref{eq:backcont}), gives a field that only in the spatial
directions is different from $1$,
\beq
V(x,0) = 1,\quad V(x,k) = V(x_0),
\eeq
with
\beq
V(x_0) = \exp\left\{i[\mathcal{E}x_0 -iC]\right\}.
\eeq
This field trivially fulfils the boundary conditions and it can easily be shown
to be a solution of the field equations. 
Yet, this field is still only a candidate for the background field. It has to
be shown that $V$ really is a configuration of least action and that the same
action is only given by fields that are gauge equivalent to $V$. According
to~\cite{Luscher:1992an}, this is the case for the gauge group $SU(N)$
if the $\phi_{k\alpha}$ and the
$\phi'_{k\alpha}$ lie in the so called fundamental domain
\beq
\phi_{k1}<\phi_{k2}<\ldots <\phi_{kN},\quad |\phi_{k\alpha}-\phi_{k\beta}|<2\pi,
\quad \sum_{\alpha=1}^N\phi_{k\alpha}=0,
\eeq
and the relation 
\beq
TL/a^2 > (N-1)\pi^2\mbox{max}\{1,N/16\} 
\eeq
holds. This relation is only needed in the proof of the theorem for technical 
reasons, it is of no physical significance. For $N\leq 3$ and the lattices
of interest, it is unimportant.

A convenient choice of boundary fields obeying the conditions mentioned above
is
\beq
C_k = \frac{i}{L}\left(
\begin{array}{ccc}
\eta-\frac{\pi}{3} & 0 & 0 \\
0 & \eta(-\frac{1}{2}+\nu) & 0 \\
0 & 0 & -\eta(\frac{1}{2}+\nu)+\frac{\pi}{3} \\
\end{array}
\right)\label{eq:gaugebound0}
\eeq
for the boundary field at $x_0=0$ and
\beq
C'_k = \frac{i}{L}\left(
\begin{array}{ccc}
-\eta-\pi & 0 & 0 \\
0 & \eta(\frac{1}{2}+\nu)+\frac{\pi}{3} & 0 \\
0 & 0 & \eta(\frac{1}{2}-\nu)+\frac{2\pi}{3} \\
\end{array}
\right)\label{eq:gaugeboundT}
\eeq
for the field at $x_0=T$. Here, we will choose $\nu=0$. Later, also $\eta$
will be set to zero, but first it is needed as a variable in order to define
the renormalised coupling in section~\ref{sect:coupling}.

\section[$\rmO(a)$ improvement of the Schr\"{o}dinger functional]
{\boldmath $\rmO(a)$ \unboldmath improvement of the 
Schr\"{o}dinger functional}
\sectionmark{\boldmath $\rmO(a)$ \unboldmath improvement of the 
Schr\"{o}dinger functional}

In the Schr\"{o}dinger functional, one has two kinds of counter-terms for
$\rmO(a)$ improvement. The first one is the volume term, improving the
fields in the interior of the lattice. The second kind are the 
boundary counter-terms
arising from the Schr\"{o}dinger functional boundary conditions in the
time direction. The improved
action thus becomes a sum
\beq
S_{\mathrm{impr}}[U,\bar{\psi},\psi] = S[U,\bar{\psi},\psi]
+\delta S_{\mathrm{V}}[U,\bar{\psi},\psi] +\delta S_{\mathrm{G,b}}[U]
+ \delta S_{\mathrm{F,b}}[U,\bar{\psi},\psi],
\eeq
where the volume term 
$\delta S_{\mathrm{V}}$ is the Sheikholeslami--Wohlert term familiar from
section~\ref{sect:improvement}.

On the boundaries at $x_0=0$ and $x_0=T$, composite fields of dimension 4
have to be subtracted. There are contributions of order $a$
both from the gauge fields and the quark fields~\cite{Luscher:1996sc}. 
The gauge field contribution is
cancelled by the counter-term
\beqn
\delta S_{\mathrm{G,b}}[U] &=& \frac{1}{2g_0^2}(\cs -1)\sum_{\pls}
\tr\{1-U(\pls)\} \nonumber\\
& & +\frac{1}{g_0^2}(\ct -1)\sum_{\plt}\tr\{1-U(\plt)\},
\eeqn
where the sum runs over all oriented plaquettes at the boundaries, both
time-like ($\plt$) and space-like ($\pls$). This boundary improvement term
simply amounts to changing the weights $w(p)$ in (\ref{eq:gaugeactionschrod})
to $w(\plt)=\ct$ and $w(\pls)=\cs/2$.
The boundary improvement coefficients have
perturbative expansions
\beqn
\ct &=& 1+\ct^{(1)}g_0^2 + \ct^{(2)}g_0^4 + \rmO(g_0^6),\\
\cs &=& 1+\cs^{(1)}g_0^2 + \cs^{(2)}g_0^4 + \rmO(g_0^6).
\eeqn
The 1--loop coefficient for the time-like plaquettes
can be decomposed into a part independent of the
flavour number and a term linear in $\Nf$,
\beq
\ct^{(1)} = \ct^{(1,0)} + \ct^{(1,1)}\Nf,
\eeq
where the coefficients are given by~\cite{Luscher:1994gh,Sint:1996ch}
\beqn
\ct^{(1,0)} &=& -0.08900(5), \\
\ct^{(1,1)} &=& 0.0191410(1).
\eeqn
The 2--loop coefficient has a decomposition
\beq
\ct^{(2)} = \ct^{(2,0)} + \ct^{(2,1)}\Nf + \ct^{(2,2)}\Nf^2,
\eeq
with~\cite{Bode:1998uj,Bode:1999sm}
\beqn
\ct^{(2,0)} &=& -0.0294(3),\\
\ct^{(2,1)} &=& 0.002(1),\\
\ct^{(2,2)} &=& 0.0000(1).
\eeqn
With the special choice of spatially constant Abelian boundary fields, the
improvement term for the space-like plaquettes vanishes and $\cs$ may be
disregarded.

For the boundary counter-terms depending on the quark fields, all possible
composite fields of dimension 4 and how most of them are eliminated are presented
in~\cite{Luscher:1996sc}. It can be shown that one is left with four terms which,
ensuring time reversal invariance, are given by
\beqn
\delta S_{\mathrm{F,b}}[U,\bar{\psi},\psi] &=&
a^4\sum_{\vx}\Bigl\{(\cst -1)[\hat{\mathcal{O}}_{\mathrm{s}}(\vx)
+\hat{\mathcal{O}}'_{\mathrm{s}}(\vx)] \nonumber\\
& &
(\ctt -1)[\hat{\mathcal{O}}_{\mathrm{t}}(\vx)
+\hat{\mathcal{O}}'_{\mathrm{t}}(\vx)]
\Bigr\},
\eeqn
with the fields
\beqn
\hat{\mathcal{O}}_{\mathrm{s}}(\vx) &=& \frac{1}{2}\bar{\rho}(\vx)\gamma_k
(\nabla_k^{\ast}+\nabla_k)\rho(\vx), \\
\hat{\mathcal{O}}'_{\mathrm{s}}(\vx) &=& \frac{1}{2}\bar{\rho}'(\vx)\gamma_k
(\nabla_k^{\ast}+\nabla_k)\rho'(\vx), \\
\hat{\mathcal{O}}_{\mathrm{t}}(\vx) 
&=& \left\{\bar{\psi}(y)P_+\nabla_0^{\ast}\psi(y)
+\bar{\psi}(y)\overleftarrow{\nabla}_0^{\ast}P_-\psi(y)\right\}_{y=(a,\vx)},\\
\hat{\mathcal{O}}'_{\mathrm{t}}(\vx) &=& \left\{\bar{\psi}(y)P_-\nabla_0\psi(y)
+\bar{\psi}(y)\overleftarrow{\nabla}_0P_+\psi(y)\right\}_{y=(T-a,\vx)}.
\eeqn
Setting the boundary quark fields to zero, one is left only with 
$\hat{\mathcal{O}}_{\mathrm{t}}(\vx)$ and 
$\hat{\mathcal{O}}'_{\mathrm{t}}(\vx)$. The volume term
and the quark boundary term may then be rewritten as a change in the
Dirac--Wilson operator,
\beq
\delta D = \delta D_{\mathrm{V}} +\delta D_{\mathrm{b}},
\eeq
with the volume term
\beq
\delta D_{\mathrm{V}}\psi(x) = \csw\frac{i}{4}a\sigma_{\mu\nu}
\hat{F}_{\mu\nu}(x)\psi(x),
\eeq
and the boundary term
\beqn
\delta D_{\mathrm{b}}\psi(x) &=& (\ctt -1)\frac{1}{a}\biggl\{\delta_{x_0,a}
\Bigl[\psi(x) -U(x-a\hat{0},0)^{-1}P_+ \psi(x-a\hat{0})\Bigr] \nonumber\\
& & +\delta_{x_0,T-a}
\Bigl[\psi(x) -U(x,0)P_- \psi(x+a\hat{0})\Bigr]\biggr\}.
\eeqn
The improvement coefficient $\ctt$ can be computed in perturbation 
theory~\cite{Luscher:1996vw,Sint:1997jx} and is, up to 1--loop order, given
by
\beq
\ctt = 1 -0.01795(2)g_0^2 +\rmO(g_0^4).
\eeq

Having improved the action, the expectation value of any improved field polynomial
$\mathcal{O}$ converges to its continuum limit at a rate proportional to
$a^2$.
First integrating over the quark fields, this expectation value may be decomposed
into a quark field average and a gauge field average,
\beq
\langle\mathcal{O}\rangle 
= \left\langle[\mathcal{O}]_{\mathrm{F}}\right\rangle_{\mathrm{G}},
\eeq
where $[\ldots]_{\mathrm{F}}$ is the quark field average at a given gauge
field configuration $U$ and $\langle\ldots\rangle_{\mathrm{G}}$ denotes
the gauge field average with a probability density proportional to
\beq
\det(D+\delta D+m_0)\exp\left\{-S_{\mathrm{G}}[U] -\delta S_{\mathrm{G,b}}\right\}.
\label{eq:gaugefieldaverage}
\eeq
The quark field average $[\mathcal{O}]_{\mathrm{F}}$ can be defined by
a generating functional,
\beqn
\mathcal{Z}_{\mathrm{F}}[\bar{\rho}',\rho';\bar{\rho},\rho;\bar{\eta},\eta;U]
&=& \int D[\psi]D[\bar{\psi}]\exp\biggl\{-S_{\mathrm{F,impr}}[U,\bar{\psi},\psi]
\nonumber\\
& & +a^4\sum_{x}\left[\bar{\psi}(x)\eta(x)+\bar{\eta}(x)\psi(x)\right]\biggr\},
\eeqn
where $S_{\mathrm{F,impr}}$ is the improved fermion action and $\eta(x)$ and
$\bar{\eta}(x)$, $0<x_0<T$, are source fields for the quark and antiquark
fields in the interior of the space--time box. Substituting
\beq
\psi(x)\rightarrow\frac{\delta}{\delta\bar{\eta}(x)}, \qquad
\bar{\psi}(x)\rightarrow -\frac{\delta}{\delta\eta(x)},
\eeq
one gets the quark field average by
\beq
[\mathcal{O}]_{\mathrm{F}} = \left\{\frac{1}{\mathcal{Z}_{\mathrm{F}}}
\mathcal{O}\mathcal{Z}_{\mathrm{F}}\right\}_{\bar{\rho}'=\ldots =\eta=0}.
\eeq
This decomposition into quark and gauge field averages will be used in later
calculations.

\section{The renormalised coupling in the Schr\"{o}\-din\-ger functional
\label{sect:coupling}}
\sectionmark{The renormalised coupling in the Schr\"{o}\-dinger functional}

\subsection{Definition of the coupling}
Having introduced the Schr\"{o}dinger functional, we can now turn to defining
the renormalised coupling in this framework. In order to make practical use 
of this method, certain criteria have to be met by
the coupling.

Since a central point in the Schr\"{o}dinger functional method is the wish
to compute a renormalised coupling non--perturbatively, the coupling should
be easy to measure in Monte Carlo simulations. Furthermore, the results
obtained in these simulations
should be easy to extrapolate to the continuum limit without too large
uncertainties. For this reason, discretisation errors should be small. And,
last but not least, the coupling should have an easy perturbative expansion,
such that the beta function can be computed to sufficient order and a matching
can be done to perturbatively defined renormalisation schemes in the high energy
regime.

Considering all these criteria leads to a definition based on the effective
action
\beq
\Gamma[V] = -\ln\mathcal{Z}[C',C].
\eeq
Since the background field $V$ depends on the parameter $\eta$ in the boundary
gauge fields~(\ref{eq:gaugebound0}) and~(\ref{eq:gaugeboundT}), the effective
action will be a function of $\eta$, too. So one may take the derivative
\beq
\Gamma'[V] = \frac{\partial}{\partial\eta}\Gamma[V],
\eeq
and expand it in powers of the bare coupling $g_0$,
\beq
\Gamma'[V] = g_0^{-2}\Gamma'_0[V] +\Gamma'_1[V] +g_0^2\Gamma'_2[V]+\rmO(g_0^4).
\eeq
$\Gamma'$ can be shown to be finite when it is expressed in terms of a 
renormalised coupling like $\bar{g}_{\msbar}$. Properly normalised, it
may itself serve as a renormalised coupling. Thus, the Schr\"{o}dinger 
functional coupling is defined as
\beq
\bar{g}^2(L) = \left.\frac{\Gamma'_0[V]}{\Gamma'[V]}\right|_{\eta=0}.
\eeq
The normalisation factor $\Gamma'_0$ makes sure that the coupling defined this
way coincides with the bare coupling at tree level of perturbation theory.
This coupling satisfies all requirements mentioned above. As the expectation
value
\beq
\Gamma'[V] = \left\langle\frac{\partial S}{\partial\eta}\right\rangle,
\eeq
it is easily accessible to Monte Carlo simulations. The discretisation errors
turn out to be tolerable. And it is relatively easy to expand in perturbation
theory. The problem of the perturbative expansion of the Schr\"{o}dinger functional
will be addressed in chapter~\ref{chapt:schroedpert}.

\subsection{The step scaling function and its lattice artefacts}

By definition, the Schr\"{o}dinger functional coupling defined in this chapter
runs with the box size $L$. 
In order to study its scaling behaviour, one needs a recipe for
evolving the coupling from one length scale to another, which means that some
kind of discrete $\beta$--function is required. An appropriate function in the
Schr\"{o}dinger functional is the
step scaling function. First of all, one has to start from a given coupling
$u=\bar{g}^2(L)$. Then the length scale is changed by a factor 2. (In principle,
one may choose a different scale factor, but 2 is commonly used for obvious 
reasons. If one took a larger factor, one would have to perform simulations on
larger lattices.) Then, keeping
the bare coupling fixed, the coupling at the length scale $2L$ has a value
$\bar{g}^2(2L)=u'$. The step scaling function $\sigma$ is then defined as
\beq
\sigma(u) = u'.\label{eq:stepcont}
\eeq
Obviously, $\sigma$ is an integral over the $\beta$--function. In fact, it
has a perturbative expansion
\beq
\sigma(u) = u +2b_0\ln(2)\, u^2 +\ldots
\eeq
However, this is only true in the continuum limit. On a lattice with finite
lattice spacing $a$, the step scaling function will depend on the lattice
resolution $a/L$. So, instead of~(\ref{eq:stepcont}), one has to define
\beq
\Sigma(u,a/L) = u'.
\eeq
The continuum step
scaling function $\sigma$ is the continuum limit of $\Sigma$, 
i.e.~$\sigma(u)=\lim_{a/L\rightarrow 0}\Sigma(u,a/L)$. 
In practice, the step scaling function
is obtained by simulations on pairs of lattices
to determine $\Sigma(u,a/L)$. One then gets
$\sigma(u)$ by ex\-tra\-po\-la\-ting these results to the continuum limit. 
The deviation of the step scaling function from
its continuum limit,
\beq
\delta(u,a/L) = \frac{\Sigma(u,a/L)-\sigma(u)}{\sigma(u)},
\label{eq:defdelta}
\eeq
may be expanded in perturbation theory to give an idea of the size of the
cutoff effects. Details of the perturbative expansion will be presented in
chapter~\ref{chapt:schroedpert}.
\cleardoublepage
\chapter{Perturbation theory in the Schr\"{o}dinger functional
\label{chapt:schroedpert}}
\chaptermark{Perturbation theory in the Schr\"{o}dinger functional}

\section{The gauge fixed action}
\sectionmark{The gauge fixed action}

\subsection{Preliminaries}

It is a well known fact that, in order to set up perturbation theory, one
has to fix the gauge. Otherwise, zero modes would occur, since the minimum
of the action
is degenerate in the directions of gauge transformations.

The Schr\"{o}dinger functional on the lattice is invariant under all gauge
transformations that leave the boundary fields unchanged. This condition is
only satisfied by gauge functions $\Omega(x)$ that are constant and diagonal
on the boundaries $x_0=0$ and $x_0=T$. These gauge functions form a group
$\hat{\mathcal{G}}$. However, not the whole group $\hat\mathcal{G}$ needs to be
fixed. Those gauge functions $\Omega(x)$ which are constant and diagonal not
only on the boundaries but on the whole lattice act trivially on the background
field. They form a subgroup of
$\hat{\mathcal{G}}$ isomorphic to the Cartan subgroup $C_N$ 
of $SU(N)$, which may be
factored out and then survives as a global symmetry of the theory. The gauge
fixing has thus to be done on the group 
\beq
\mathcal{G} = \hat{\mathcal{G}}/C_N.
\eeq
$\mathcal{G}$ may be identified with the subgroup of transformations 
$\Omega\in\hat{\mathcal{G}}$ that are equal to a fixed constant diagonal
matrix at $x_0=T$. Here we choose $\Omega(x)|_{x_0=T}=1$.

The Lie algebra $\mathcal{L}_{\mathcal{G}}$ of $\mathcal{G}$
consists of fields $\omega$
with $\omega(x)\in su(N)$. The gauge functions $\Omega$ may be parametrised
as
\beq
\Omega(x)=\exp\{-g_0\omega(x)\}=1-g_0\omega(x)+\rmO(g_0^2).
\eeq
Obviously, $\omega$ must obey the boundary conditions
\beq
\omega(0,\vx)=\kappa,\qquad \omega(T,\vx)=0,
\eeq
where $\kappa$ is constant and diagonal.

Analogously, one may expand the gauge field $U$ around the background field.
Let $\mathcal{H}$ be the set of gauge fields satisfying the Schr\"{o}dinger
functional boundary conditions. In a neighbourhood of the background field,
any gauge field $U$ may be parametrised by
\beqn
U(x,\mu) &=& \exp\{g_0 aq_{\mu}(x)\}V(x,\mu) \nonumber\\
&=& \left\{1 + g_0 aq_{\mu}(x) +\frac{1}{2}g_0^2 a^2 q_{\mu}(x)^2
+\rmO(g_0^3)\right\}V(x,\mu),
\eeqn
with $q_{\mu}(x)\in su(N)$. In perturbation theory, the vector fields $q$
become the gluons. They form a linear space $\mathcal{L}_{\mathcal{H}}$ and
obey the boundary conditions
\beq
q_k(0,\vx)=0,\qquad q_k(T,\vx)=0.
\eeq
The time components $q_0(x)$ are unconstrained and defined for $0\leq x_0<T$.
We will however get boundary conditions for $q_0$ later by extending the
lattice.

For later use, it is useful to define a scalar product on 
$\mathcal{L}_{\mathcal{G}}$ and $\mathcal{L}_{\mathcal{H}}$. For two vectors
$q$ and $r$ in $\mathcal{L}_{\mathcal{H}}$, an obvious choice is
\beq
(q,r) = -2a^4\sum_{x,\mu}\tr\{q_{\mu}(x)r_{\mu}(x)\}.
\eeq
The scalar product on $\mathcal{L}_{\mathcal{G}}$ is defined similarly.

For the definition of a suitable gauge fixing function, we will need the covariant
forward and backward derivatives on the lattice,
\beqn
D_{\mu}f(x) &=& \frac{1}{a}\left[V(x,\mu)f(x+a\hat{\mu})V(x,\mu)^{-1}
-f(x)\right],\\
D^{\ast}_{\mu}f(x) &=& \frac{1}{a}\left[f(x) -V(x-a\hat{\mu},\mu)^{-1}
f(x-a\hat{\mu})V(x-a\hat{\mu},\mu)\right].
\eeqn
Using the forward covariant derivative, one may define a linear operator
$d:\slg\rightarrow\slh$ by
\beq
(d\omega)_{\mu}(x)=D_{\mu}\omega(x).
\eeq
Another operator that will be used later is $d^{\ast}$. It is defined as
minus the adjoint of $d$,
\beq
(d^{\ast}q,\omega) = -(q,d\omega)
\eeq
for all $q\in\slh$ and $\omega\in\slg$. Starting from this definition, $d^{\ast}$
is explicitly given by
\beq
(d^{\ast}q)_{\alpha\beta}(x)
=\left\{
\begin{array}{l}
\sum_{\mu}(D^{\ast}_{\mu}q_{\mu})_{\alpha\beta}(x)\quad\mbox{if}\quad 0<x_0<T,\\
(a^2/L^3)\sum_{\vy}q_0(0,\vy)_{\alpha\beta}\quad\mbox{if}\quad 
\alpha = \beta \quad \mbox{and} \quad
x_0=0,\\
0\quad\mbox{otherwise}.\\
\end{array}
\right.
\eeq
This rather complicated formulation may be simplified to
\beq
d^{\ast}q(x) = D^{\ast}_{\mu}q_{\mu}(x)
\eeq
for all $x_0$ by extending the field $q$ to $x_0=-1$ and $x_0=T$ and imposing
appropriate boundary conditions on the time component $q_0$. The new components
are physically irrelevant and only used for technical reasons. Due to the special
choice of background field used here, a complication does occur in this procedure.
If one used, for example, the self dual background field considered in the
continuum formulation in~\cite{Luscher:1992an}, one would get Neumann boundary
conditions for all $q_{0}$. This is also the case with the Abelian background
field used here, except for one case. If $q_0$ has got a part which is diagonal 
and spatially constant, this part
satisfies Dirichlet boundary conditions at $x_0=-1$. For this reason, 
the boundary 
conditions are easier to define in momentum space, where spatially constant
fields $q$ simply become gluons of zero momentum. So we switch to momentum
space by taking the Fourier transform
\beq
q_0(\vp,x_0) = \sum_{\vx}e^{-i\vp\vx}q_0(x),
\eeq
which may be decomposed in a basis of the Lie algebra of $SU(N)$,
\beq
q_0(\vp,x_0) = \tilde{q}_0^a(\vp,x_0) I^a.
\eeq
The special choice of the basis $I^a$ for the case of $SU(3)$ used in later
calculations
can be found in appendix~\ref{app:group}.

Now the boundary conditions are given by
\beqn
\tilde{q}_0^a(\vp,-1) &=& 0\quad \mbox{if} \quad I^a\in C_N\quad\mbox{and}
\quad \vp=\vo,\nonumber\\
\partial_0^{\ast}\tilde{q}_0^a(\vp,0) &=& 0 \quad\mbox{else},\nonumber\\
\partial_0^{\ast}\tilde{q}_0^a(\vp,T) &=& 0.
\label{eq:gluonbound}
\eeqn

\subsection{The gauge fixing procedure}

In the following, we will use the gauge fixing procedure outlined 
in~\cite{Luscher:1988sd}. To this end, we will need a gauge fixing function
$F(U)$, which is conveniently chosen to be a linear mapping from $\mathcal{H}$
to $\mathcal{L}_{\mathcal{G}}$. 
This function is defined in a neighbourhood
$\mathcal{N}$ of the background field $V$ and
has to fulfil several conditions. First of all, it has to vanish on the 
background field,
\beq
F(V) = 0.
\eeq
The first order variation of $F$ under a gauge transformation generated by
$\omega$ is a linear operator $L(U):\slg\rightarrow\slg$,
\beq
L(U)\omega = \delta_{\omega}F(U).
\eeq
For a gauge fixing function, the determinant of $L(U)$ is required not to
vanish. 

A suitable function that fulfils all conditions and is relatively easy to
handle is
\beq
F(U) = d^{\ast}q.
\eeq
Let now $f(U)$ be a function that is non--zero only in the neighbourhood 
$\mathcal{N}$.
It can be shown~\cite{Luscher:1988sd} that
\beq
\int D[U] f(U) = k\int_{\mathcal{N}}D[U]f(U)\delta(F(U))\det(L(U)).
\label{eq:gaugefixingint}
\eeq
This relation stays valid if one replaces the delta function by
$\delta(F(U)-Z)$, where $Z$ is an element of $\slg$ with $(Z,Z)<\epsilon$,
where $\epsilon$ is chosen to be appropriately small. Since the left hand 
side of~(\ref{eq:gaugefixingint}) is independent of $Z$, one can integrate
over $Z$ with a Gaussian measure, resulting in the gauge fixing term
\beq
S_{\mathrm{gf}}[V,q] = \frac{\lambda_0}{2}(d^{\ast}q,d^{\ast}q)
\eeq 
in the action.

In the usual way, the determinant in~(\ref{eq:gaugefixingint}) may be rewritten
in terms of Lie algebra valued Grassmann variables,
the Faddeev--Popov ghost fields $c$ and $\bar{c}$ 
with the action
\beq
S_{\rmfp}[V,q,\bar{c},c] = -(\bar{c},d^{\ast}\delta_{c}q).
\label{eq:ghostaction}
\eeq
Here, $\delta_{c}q$ denotes the first order variation of $q$ under the gauge
transformation generated by $c$. Expanding it to order $g_0^2$
yields
\beqn
\delta_c q_{\mu} &=& D_{\mu}c+g_0\mbox{Ad}q_{\mu}c\nonumber\\
& & +\left[\frac{1}{2}g_0 a\mbox{Ad}q_{\mu}
+\frac{1}{12}(g_0 a\mbox{Ad}q_{\mu})^2 +\ldots\right]D_{\mu}c,
\eeqn
with no sum over $\mu$ implied. 

Also he ghost fields $c$ and $\bar{c}$ have to satisfy certain boundary conditions.
Like $q$, also the ghost field $c$ may be Fourier transformed and decomposed
in the basis $I^a$ of the Lie algebra of $SU(N)$, resulting in the momentum
dependent components $\tilde{c}^a(\vp,x_0)$.
By an analysis similar to the gluon case, one then gets the boundary
conditions~\cite{Narayanan:1995ex}
\beqn
\partial_0^{\ast}\tilde{c}^a(\vp,0) &=& 0
\quad \mbox{if} \quad I^a\in C_N\quad\mbox{and}\quad \vp=\vo,\nonumber\\
\tilde{c}^a(\vp,0) &=& 0 \quad\mbox{else},\nonumber\\
\tilde{c}^a(\vp,T) &=& 0.
\label{eq:ghostbound}
\eeqn  
The same boundary conditions are valid for the ghost field $\bar{c}$.

\subsection{The total action}

Having fixed the gauge, we are now able to write down the action that will
be used for perturbation theory in the Schr\"{o}dinger functional.

The gluonic part of the action may be summarised as
\beq
S_{\rmg}[V,q] = \frac{1}{g_0^2}\sum_p w(p)\tr\left\{1-U(p)\right\} 
+ S_{\mathrm{gf}}[V,q],
\label{eq:gluonaction}
\eeq
where the weights $w(p)$ are chosen such that the action is $\rmO(a)$ improved,
i.e. $w(p_{\mathrm{t}})=\ct$ and $w(p_{\mathrm{s}})=\cs/2$ for the 
time-like and space-like plaquettes
at the boundaries.\\
The ghosts contribute to the action with $S_{\rmfp}$ given 
in~(\ref{eq:ghostaction}).
The fermionic part of the total action is given by the Dirac--Wilson action
including the volume and boundary improvement terms,
\beq
S_{\rmf}[V,q,\bar{\psi},\psi] = a^4\sum_{x}\bar{\psi}(x)(D+\delta D +m_0)\psi(x).
\eeq
Apart from these contributions, there is an additional part of the action arising
from the change of integration variables. Before, the gauge field integral was
formulated using the measure $D[U]$. Now, we want to express the Schr\"{o}dinger
functional as an integral over the gluon fields $q$. For this purpose, it is
useful to use
\beq
D[q]=\prod_{x,\mu,a}dq_{\mu}^a(x)
\eeq
as a new measure. It may be obtained from the old measure by
\beq
D[U]=D[q]e^{-S_{\mathrm{m}}[q]},
\eeq
where, in the case of $SU(3)$, the measure part of the action is given by
\beq
S_{\mathrm{m}}[q]=\frac{g_0^2}{8}\sum_{x,\mu,a}q_{\mu}^{a}(x)q_{\mu}^{\bar{a}}(x)
+\rmO(g_0^4).
\eeq
In the case of $SU(2)$, one would have to replace the factor $\frac{1}{8}$
by $\frac{1}{12}$. So the additional contribution to the action due to the change
of integration variables turns out to be of order $g_0^2$. For the later
calculations in this thesis, the total action will only be needed up to order
$g_0$, so $S_{\mathrm{m}}$ may be ignored here. 
It has, however, to be taken into account
in the computation of the running coupling at 2--loop order.

The total action is the sum of all contributions,
\beq
S_{\rmt}[V,q,\bar{c},c,\bar{\psi},\psi] = S_{\rmg}[V,q] + 
S_{\rmfp}[V,q,\bar{c},c] + S_{\rmf}[V,q,\bar{\psi},\psi] +S_{\mathrm{m}}[q].
\eeq

The Schr\"{o}dinger functional now becomes
\beq
\mathcal{Z}=\int D[q]D[c]D[\bar{c}]D[\psi]D[\bar{\psi}]e^{-S_{\rmt}}.
\eeq

Let now $\mathcal{O}$ be a product of link variables. (We assume that 
$\mathcal{O}$ does not
depend on the quark fields, which is the case after integrating over them.)
The expectation value is then to be computed using the probability 
density~(\ref{eq:gaugefieldaverage}). Writing the quark determinant as an
integral over Grassmann variables, one gets
\beq
\langle\mathcal{O}\rangle_{\mathrm{G}} =\frac{1}{\mathcal{Z}}
\int D[q]D[c]D[\bar{c}]D[\psi]D[\bar{\psi}]\mathcal{O}e^{-S_{\rmt}}.
\eeq
As a product of link variables, $\mathcal{O}$ may be written as a series in $g_0$,
\beq
\mathcal{O} =\mathcal{O}^{(0)}+g_0\mathcal{O}^{(1)}+g_0^2\mathcal{O}^{(2)}
+\rmO(g_0^3),
\eeq
where $\mathcal{O}^{(n)}$ contains products of $n$ gluon fields $q_{\mu}$.
Since $\mathcal{O}^{(0)}$ is merely a constant, the functional integral becomes
trivial,
\beq
\left\langle\mathcal{O}^{(0)}\right\rangle_{\mathrm{G}} = \mathcal{O}^{(0)}.
\eeq
One thus gets the expansion
\beq
\left\langle\mathcal{O}\right\rangle_{\mathrm{G}} =
\mathcal{O}^{(0)}
+g_0\left\langle\mathcal{O}^{(1)}\right\rangle_{\mathrm{G}}
+g_0^2\left\langle\mathcal{O}^{(2)}\right\rangle_{\mathrm{G}} +\rmO(g_0^3).
\eeq
However, this is not yet the whole expansion, because the total action and thus
the expectation value of $\mathcal{O}^{(n)}$ still depends on the bare coupling.
In order to get the correct series, one has to expand the action too,
\beq
S_{\rmt}=\frac{1}{g_0^2}S_{\rmt}^{(-2)}
+S_{\rmt}^{(0)}+g_0 S_{\rmt}^{(1)}+g_0^2 S_{\rmt}^{(2)} +\rmO(g_0^3).
\eeq
The term $S_{\rmt}^{(-2)}$ comes from the gluon action and contains only the
background field,
\beqn
S_{\rmt}^{(-2)} &=& \sum_p \tr\left\{1-V(p)\right\}\nonumber\\
&=& 12\frac{TL^3}{a^4}\left[\sin^2(\gamma) +2\sin^2(\gamma/2)\right],
\eeqn
with
\beq
\gamma = \frac{1}{LT}\left(\eta + \frac{\pi}{3}\right).
\eeq
Since $S_{\rmt}^{(-2)}$ does not depend on the variables integrated over in
the functional integral, it cancels in all expectation values and may be
neglected here.
With this expansion, the exponential factor in the functional integral becomes
\beq
e^{-S_{\rmt}} = \left[1-g_0 S_{\rmt}^{(1)}+g_0^2\Bigl(
\frac{1}{2}S_{\rmt}^{(1)^2}-S_{\rmt}^{(2)}\Bigr)\right]e^{-S_{\rmt}^{(0)}}.
\label{eq:expactionexpansion}
\eeq
Let now $\left\langle\ldots\right\rangle_{0}$ denote the gauge field
average taken at zero bare coupling, which means that this average is computed
using $S_{\rmt}^{(0)}$ instead of the whole action. 
Inserting~(\ref{eq:expactionexpansion}) into the functional integrals 
and integrating out the quark fields yields
\beq
\left\langle\mathcal{O}\right\rangle_{\rmg} = 
\mathcal{O}^{(0)}
+ g_0^2\Biggl[\left\langle\mathcal{O}^{(2)}\right\rangle_{0}
-\left\langle\mathcal{O}^{(1)}\Bigl[S_{\mathrm{tot}}^{(1)}\Bigr]_{\mathrm{F}}
\right\rangle_{0}
\Biggr] +\rmO(g_0^4).
\eeq
Note that no terms proportional to $g_0$ or $g_0^3$ occur, because this would
involve an integral over an odd number of gluon fields, giving zero. Another
observation worth mentioning is that $S_{\mathrm{tot}}^{(1)}$ only is non--zero
due the presence of the background field. Later, this term will lead to some
Feynman diagrams that do not occur in the case of a vanishing background field.

\section{Perturbative expansion of the coupling}
\sectionmark{Perturbative expansion of the coupling}

\subsection{The coupling at 1-- and 2--loop order}

The Schr\"{o}dinger functional coupling $\bar{g}^2$ can be expanded in powers of
the bare coupling $g_0$,
\beq
\bar{g}^2 = g_0^2 +p_1(L/a) g_0^4 +p_2(L/a) g_0^6 +\rmO(g_0^8).
\eeq
At tree level, both couplings coincide due to the normalisation of
$\bar{g}^2$. The 1--loop coefficient $p_1$ may be decomposed in an $\Nf$ dependent
and an $\Nf$ independent part,
\beq
p_1 = p_{10}+p_{11}\Nf.
\eeq
Both $p_{10}$ and $p_{11}$ are easy to calculate by expanding the total
action. From the quadratic parts of the gluon, ghost, and quark actions
one gets the inverse of the propagators. The coefficients $p_{10}$ and
$p_{11}$ are obtained by differentiating the determinants of the inverse
propagators with respect to the parameter $\eta$. The coefficient $p_{10}$
has been computed in~\cite{Luscher:1994gh} (and in~\cite{Luscher:1992an} for the
case of $SU(2)$), the calculation of $p_{11}$ can
be found in~\cite{Sint:1996ch}.

At 2--loop level, one has a decomposition up to order 
$\Nf^2$~\cite{Bode:1999sm,Bode:1999dn},
\beq
p_2 = p_{20}+p_{21}\Nf+p_{22}\Nf^2.
\eeq
Here, the computation is more complicated and requires the calculation of
Feynman diagrams. The coefficient $p_{20}$ for the quenched case has been
computed in~\cite{Bode:1998hd}, and in~\cite{Narayanan:1995ex} for the
case of $SU(2)$. The remaining coefficients $p_{21}$ and $p_{22}$ can be
found in~\cite{Bode:1999sm}.

\subsection{The step scaling function and its lattice artefacts at 
1-- and 2--loop order}
Having expanded the coupling, one easily gets the expansion of the step scaling
function, since it is given by the difference between the coupling at length
scale $2L$ and $L$ at the required order of perturbation theory. This means one
has to calculate the difference
\beq
\Delta p_{ij}(L/a)=p_{ij}(2L/a)-p_{ij}(L/a).
\eeq
Also the deviation $\delta$ of the step scaling function from its continuum 
limit~(\ref{eq:defdelta}) can be expanded in perturbation theory,
\beq
\delta(u,a/L) = 
   \left[\delta_{10} + \delta_{11}\Nf\right]u
 + \left[\delta_{20}+\delta_{21}\Nf+\delta_{22}\Nf^2\right]u^2 
 + \rmO(u^3).
\eeq
The continuum limit of the step scaling function is determined by the beta
function. In order to get the coefficients $\delta_{ij}$, one has to decompose
the coefficients of the beta function,~(\ref{eq:b0}) and~(\ref{eq:b1}), into
the $\Nf$ independent parts $b_{00}$ and $b_{10}$ and the $\Nf$ dependent parts
$b_{01}$ and $b_{11}$. The coefficients $\delta_{ij}$ are then given by
\beqn
\delta_{10} &=& \Delta p_{10}-2b_{00}\ln(2),
\label{eq:delta10}\\
\delta_{11} &=& \Delta p_{11}-2b_{01}\ln(2),
\label{eq:delta11}\\
\delta_{20} &=& \Delta p_{20}-2b_{10}\ln(2)-2\Delta p_{10}(p_{10}+b_{00}\ln(2)),
\label{eq:delta20}\\
\delta_{21} &=& \Delta p_{21}-2b_{11}\ln(2)-2\Delta p_{11}(p_{10}+b_{00}\ln(2))
\nonumber\\
& &-2\Delta p_{10}(p_{11}+b_{01}\ln(2)),
\label{eq:delta21}\\
\delta_{22} &=& \Delta p_{22}-2\Delta p_{11}(p_{11}+b_{01}\ln(2)).
\label{eq:delta22}
\eeqn
The 1--loop coefficients $\delta_{1j}$ were first listed 
in~\cite{sommerunpublished}, while the 2--loop coefficients $\delta_{2j}$ 
have been estimated in~\cite{Bode:1999sm}. However, at vanishing renormalised
quark mass, the
two--loop coefficient $\delta_2$ contains the critical quark mass at 1--loop order,
which only in the continuum limit reaches the value used in~\cite{Bode:1999sm}.
In order to do a precise calculation, one has to specify the zero mass
condition with the cutoff in place. So the results of~\cite{Bode:1999sm} could
only give a first idea of the size of the cutoff effects. For more accurate
results, the critical quark mass has to be expanded up to 1--loop order at finite
$a/L$ and
inserted in the formulae for $\delta_{2j}$ in~\cite{Bode:1999sm}. This will
happen in chapter~\ref{chap:results}.

According to the estimation in~\cite{Bode:1999sm}, the cutoff effects seem
to be small. In order to compute the step scaling function non--perturbatively,
one has to simulate a sequence of lattice pairs with decreasing lattice 
spacing and fixed coupling $u$ and extrapolate the Monte Carlo data to the
continuum limit. In this procedure, the perturbative expansion of 
$\delta(u,a/L)$ may be used to remove the cutoff effects up to 2--loop
order from the non--perturbative values of the step scaling function 
$\Sigma(u,a/L)$~\cite{Gehrmann:2001yn}.

\cleardoublepage
\chapter{The current quark mass \label{chapt:curr_mass}}
\chaptermark{The current quark mass}

\section{The PCAC relation}
\sectionmark{The PCAC relation}
Several ways to define a renormalised quark mass have been sketched in
chapter~\ref{chapt:latt}. For the Schr\"odinger functional, a suitable definition
has been given in~\cite{Luscher:1996sc}, using the partial conservation of
the isovector axial current. The axial current and density are defined by
\beqn
A_{\mu}^a(x) &=& \bar{\psi}(x)\gamma_{\mu}\gamma_5\frac{1}{2}\tau^{a}\psi(x),\\
P^a(x) &=& \bar{\psi}(x)\gamma_5\frac{1}{2}\tau^a\psi(x).
\eeqn
As already mentioned in chapter~\ref{chapt:latt}, a deficit of Wilson fermions is
that the chiral symmetry of the theory is explicitly broken and only restored
in the continuum limit. Thus only in the continuum limit, the isovector axial
current $A^a_{\mu}(x)$ will satisfy the PCAC relation
\beq
\partial_{\mu}A^a_{\mu}(x) = 2mP^a(x),
\eeq
while at finite lattice spacing, this relation will be violated by terms of
order $a$.

In order to make the axial current converge faster towards its continuum limit,
one may apply Symanzik's improvement programme by adding an appropriate 
improvement term
\beq
(A_\mathrm{I})_{\mu}^a(x) = A_{\mu}^a(x) + a\delta A_{\mu}^a(x) \label{eq:Aimproved}.
\eeq
The improvement term $\delta A^a_{\mu}$ turns out to be~\cite{Luscher:1996sc}
\beq
\delta A_{\mu}^a(x) = \ca \frac{1}{2}(\partial_{\mu}+\partial_{\mu}^{\ast})P^a(x).
\label{eq:deltaA}
\eeq
This $\rmO(a)$ correction is proportional to the improvement coefficient $\ca$,
which is
\beq
\ca(g_0) = -0.00756(1) g_0^2 +\rmO(g_0^4)
\eeq
to 1--loop order of perturbation theory~\cite{Luscher:1996vw}. The axial density
$P^a$ can be shown to converge to its continuum limit with a rate proportional
to $a^2$, hence it does not need to be improved.

In the case of a non--vanishing physical quark mass, 
the quantities defined this way are still not fully improved.
In order to improve $A^a_{\mu}$ and $P^a$ completely, one still has
to subtract a mass dependent counter-term. However, this amounts to a
mass dependent multiplicative renormalisation, so it seems more natural to
include this factor in the definition of the renormalised quantities. 

In general, the quark mass will get an additive renormalisation, which means that,
in the plane of bare parameters, there will be a critical line
\beq
m_0=\mcrit(g_0),
\eeq
where the renormalised quark mass vanishes. For convenience, one may define
the subtracted mass
\beq
\mq = m_0 - \mcrit.
\eeq
Then this mass will only have to be renormalised multiplicatively. \\
Including the already mentioned mass dependent factors needed for improvement,
the renormalised axial current and density may now be written as
\beqn
(A_\mathrm{R})_{\mu}^a &=& Z_{\mathrm{A}} (1+b_{\mathrm{A}} a\mq) 
\{A_{\mu}^a + a\ca \frac{1}{2}(\partial_{\mu}+\partial_{\mu}^{\ast})P^a\},\\
(P_\mathrm{R})^a &=& Z_{\mathrm{P}}(1+b_{\mathrm{P}} a\mq) P^a.
\eeqn
Here, $b_{\mathrm{A}}$ and $b_{\mathrm{P}}$ cancel mass dependent cutoff effects. At tree level one 
has~\cite{Heatlie:1991kg,Luscher:1996vw}
\beq
b_{\mathrm{A}}^{(0)} = b_{\mathrm{P}}^{(0)} = 1.
\eeq
Since all calculations in this thesis are done at vanishing renormalised mass,
which is equivalent to setting $\mq = 0$, these coefficients are not needed here.
In the following, they are therefore ignored.

The renormalisation constant $Z_{\mathrm{A}}$ would be equal to one 
in the continuum. This
is a consequence of the $SU(\Nf)_{\mathrm{L}}\times SU(\Nf)_{\mathrm{R}}$ 
symmetry, leading to
chiral ward identities, which may be used to normalise the 
currents~\cite{Bochicchio:1985xa,Maiani:1986yj}.
In the regularised theory, this symmetry
is violated by terms of order $a$. This means that the PCAC relation and the
chiral ward identities are only
valid up to $\rmO(a)$ corrections, resulting in a renormalisation of the
axial current. It stays,
however, finite and scale independent. Up to 1--loop order of perturbation
theory, one gets~\cite{Gabrielli:1991us,Luscher:1997jn}
\beq
Z_{\mathrm{A}}(g_0) = 1+ Z_{\mathrm{A}}^{(1)}g_0^2 +\rmO(g_0^4),
\eeq
with
\beq
Z_{\mathrm{A}}^{(1)} = -0.087344(1)\times \cf.
\label{eq:ZA1}
\eeq
In contrast to $Z_{\mathrm{A}}$, the renormalisation constant $Z_{\mathrm{P}}$
is scale dependent and thus responsible for the running of the renormalised
mass. This issue will be addressed in chapter~\ref{chapt:renmass}.

A renormalised quark mass $\mren$ may now be defined as the proportionality
constant in
\beq
\left\langle\frac{1}{2}(\partial^{\ast}_{\mu}+\partial_{\mu})(A_{\mathrm{R}})^a_{\mu}(x)
\mathcal{O}\right\rangle = 2\mren \left\langle (P_{\mathrm{R}})^a(x)\mathcal{O}\right\rangle
+ \rmO(a^2)
\label{eq:pcacop}
\eeq
for any product $\mathcal{O}$ of renormalised improved fields located at a
non--zero distance from each other and from $x$. The lattice artefacts of
order $a^2$ depend on the choice of $\mathcal{O}$.

\section{The current mass and its lattice artefacts}
\sectionmark{The current mass and its lattice artefacts}
In order to compute the renormalised mass, one needs to choose an operator
$\mathcal{O}$. Using this operator, one then has to construct bare 
correlation functions 
containing the axial current and density and then renormalise both these
quantities and the operator $\mathcal{O}$.

\begin{figure}
  \noindent
  \begin{center}
    \includegraphics[width=0.3\linewidth]{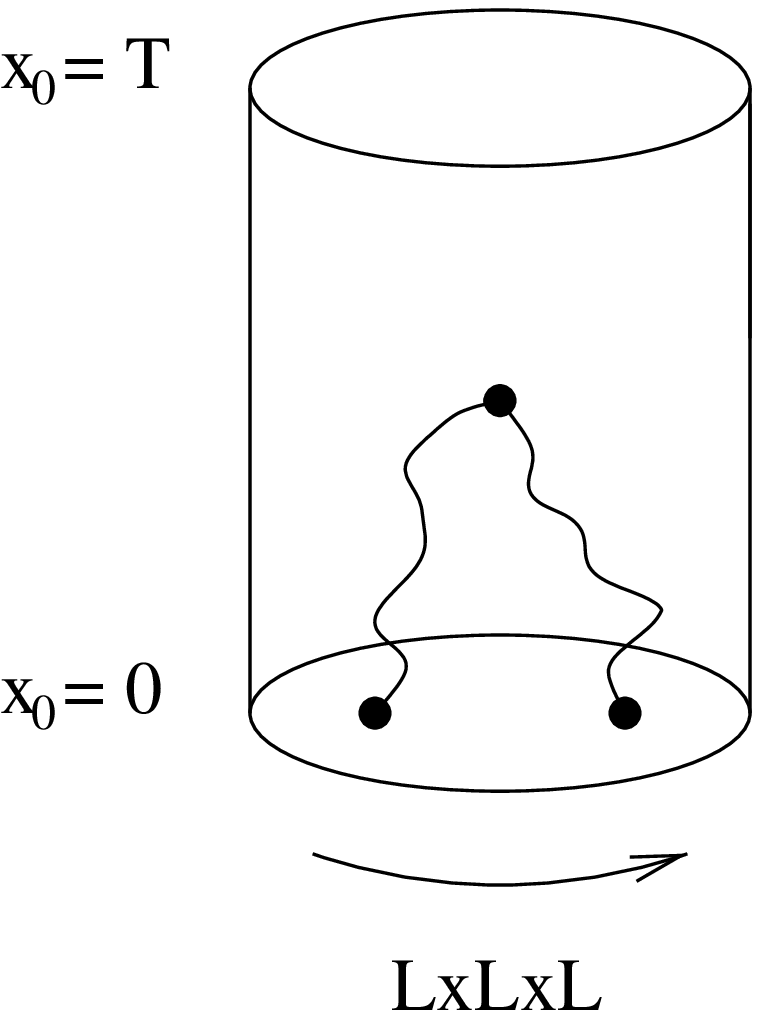}
  \end{center}
  \caption[The correlation functions $f_{\mathrm{A}}$ and $f_{\mathrm{P}}$]
          {\label{fig:cyl_fa}
           \sl The correlation functions $f_{\mathrm{A}}$ and $f_{\mathrm{P}}$}
\end{figure}

One possible choice is to introduce the bare correlation functions
\beqn
\fa(x_0) &=& -a^6\sum_{\mathbf{y},\mathbf{z}}\frac{1}{3}
\left\langle A_0^a(x)\bar{\zeta}(\mathbf{y})\gamma_5\frac{1}{2}
\tau^a\zeta(\mathbf{z})\right\rangle,\label{eq:fadef}\\
\fp(x_0) &=& -a^6\sum_{\mathbf{y},\mathbf{z}}\frac{1}{3}
\left\langle P^a(x)\bar{\zeta}(\mathbf{y})\gamma_5\frac{1}{2}
\tau^a\zeta(\mathbf{z})\right\rangle,\label{eq:fpdef}
\eeqn
where $\zeta(\vx)$ is the functional derivative with respect to the boundary
quark fields at $x_0=0$ defined in~(\ref{eq:zetadef}) 
and~(\ref{eq:zetabardef}). These correlation functions are proportional to
the probability amplitude that a quark antiquark pair created at $x_0=0$
propagates into the interior of the lattice and annihilates at the point
$x$. This situation may be depicted by figure~\ref{fig:cyl_fa}. 

Now one can define
the $x_0$ dependent current quark mass
\beq
m(x_0) = \frac{\frac{1}{2}(\partial_0^{\ast} + \partial_0)\fa(x_0)
 + \ca a\partial_0^{\ast}\partial_0\fp(x_0)}{2\fp(x_0)},
\label{eq:mdef}
\eeq
which, taken at a certain $x_0$, may serve as an unrenormalised quark mass.
A convenient choice is to take $m$ in the centre of the lattice, i.e.
\beq
  m_1 = \left\{ \begin{array}{ll}
      m\left(\frac T 2\right) 
      & \mbox{for even $T/a$,} \\
      \frac 1 2 \left( m\left(\frac{T-a}{2}\right)
                      +m\left(\frac{T+a}{2}\right)\right)
      & \mbox{for odd $T/a$.}
      \end{array} \right.
\label{eq:m1def}
\eeq
This current quark mass will, of course, depend on the lattice size $L/a$.
It is then straightforward to show that the renormalised quark mass defined
as the proportionality factor in the PCAC relation is given by
\beq
\mren = \frac{Z_{\mathrm{A}} (1+b_{\mathrm{A}} a\mq)}
{Z_{\mathrm{P}} (1+b_{\mathrm{A}} a\mq)}m_1 +\rmO(a^2).
\label{eq:renmass}
\eeq
One may of course take different operators $\mathcal{O}$ which will give 
quark masses differing from $m_1$ by terms that are expected to be
of order $a^2$. These $\rmO(a^2)$--terms may serve as a direct check of the size
of the lattice effects.

A possible choice is to take the boundary quark fields at $x_0=T$ instead of
$x_0=0$, i.e. to use the correlation functions
\beqn
\fa'(T - x_0) &=& -a^6\sum_{\mathbf{y},\mathbf{z}}\frac{1}{3}
\left\langle A_0^a(x)\bar{\zeta}'(\mathbf{y})\gamma_5\frac{1}{2}
\tau^a\zeta'(\mathbf{z})\right\rangle,\\
\fp'(T - x_0) &=& -a^6\sum_{\mathbf{y},\mathbf{z}}\frac{1}{3}
\left\langle P^a(x)\bar{\zeta}'(\mathbf{y})\gamma_5\frac{1}{2}
\tau^a\zeta'(\mathbf{z})\right\rangle
\eeqn
instead of $\fa$ and $\fp$. Let $m'$ be defined in the same way as $m_1$,
but with $\fa'$ and $\fp'$ instead of $\fa$ and $\fp$. Now, a new unrenormalised 
mass may be defined by
\beq
m_2 = \frac{1}{2}(m_1 + m').
\eeq
Then the lattice artefact 
\beq
d(L/a) = m_2(L/a) - m_1(L/a)
\label{eq:ddef}
\eeq
should be of order $a^2$.

A different possibility to estimate the size of the lattice effects is to
take the difference
\beq
e(L/a) = m_1(2L/a) - m_1(L/a).
\label{eq:edef}
\eeq
One aim of this work is to compute these lattice artefacts up to 1--loop
order of perturbation theory. To this end, the 1--loop expansion of the
correlation functions $\fa$ and $\fp$ is needed.

\section{1--loop expansion of $\fa$ and $\fp$}
\sectionmark{1--loop expansion of $\fa$ and $\fp$}
For convenience, the calculation in this section is done in lattice units,
i.e. \mbox{$a=1$}.

The principle ideas of the expansion of $\fa$ and $\fp$ have been outlined
in~\cite{Luscher:1996vw}, a more detailed discussion has been given 
in~\cite{Weisz:1996csw} (and in~\cite{Weisz:1995ca} for the case of a vanishing
background field). In this section, the details of the calculation shall be
presented following the lines of~\cite{Weisz:1996csw}.

\subsection{Preliminaries}
In order to expand $\fa$ and $\fp$, they have to be brought into a form more
suitable for perturbation theory. Applying Wick's theorem on~(\ref{eq:fadef})
and~(\ref{eq:fpdef}), one gets
\beqn
\fa(x_0) &=& \sum_{\vy,\vz}\frac{1}{2}\biggl\langle\Tr\Bigl\{
[\zeta(\vz)\bar{\psi}(x)]_{\mathrm F}\gamma_0\gamma_5 
[\psi(x)\bar{\zeta}(\vy)]_{\mathrm F}
\gamma_5\Bigr\}\biggr\rangle_{\rmg}, \\
\fp(x_0) &=& \sum_{\vy,\vz}\frac{1}{2}\biggl\langle\Tr\Bigl\{
[\zeta(\vz)\bar{\psi}(x)]_{\mathrm F}\gamma_5 [\psi(x)\bar{\zeta}(\vy)]_{\mathrm F}
\gamma_5\Bigr\}\biggr\rangle_{\rmg},
\eeqn
where the trace is to be taken over Dirac and colour indices only.\\
Let $\psi_{\mathrm{cl}}$ be a solution of the Dirac equation,
\beq
(D + \delta D + m_0)\psi_{\mathrm{cl}}(x) = 0.
\eeq
Then it can be shown~\cite{Luscher:1996vw} that
\beq
[\psi(x)\bar{\zeta}(\vy)]_{\mathrm F} = 
\frac{\delta\psi_{\mathrm{cl}}(x)}{\delta\rho(\vy)}.
\eeq
So using the matrix
\beq
H(x) = \sum_{\vy}\frac{\delta\psi_{\mathrm{cl}}(x)}{\delta\rho(\vy)},
\eeq
$\fa$ and $\fp$ may be written as
\beqn
\label{eq:fAwithH}
\fa(x_0) &=& -\frac{1}{2}\left\langle\Tr\{H(x)^{\dagger}\gamma_0 H(x)\}
\right\rangle_{\rmg},\\
\fp(x_0) &=& \frac{1}{2}\left\langle\Tr\{H(x)^{\dagger}H(x)\}
\right\rangle_{\rmg} .
\label{eq:fPwithH}
\eeqn
The matrix $H$ has an expansion
\beq
H(x) = H^{(0)}(x) +g_0 H^{(1)}(x) +g_0^2 H^{(2)}(x) + \rmO(g_0^3),
\eeq
resulting in the 1--loop expansion of $f_{\mathrm{A}}$,
\beqn
f_{\mathrm{A}}(x_0) &=& -\frac{1}{2}\Tr\Bigl\{H^{(0)}(x)^{\dagger}\gamma_0 H^{(0)}(x)
\Bigr\}
-g_0^2\frac{1}{2}\biggl[\left\langle 
\Tr\Bigl\{H^{(0)}(x)^{\dagger}\gamma_0 H^{(2)}(x)\Bigr\}\right\rangle_{0}
\nonumber\\
& &
+\left\langle\Tr\Bigl\{H^{(1)}(x)^{\dagger}\gamma_0 H^{(1)}(x)\Bigr\}
\right\rangle_{0}
+\left\langle\Tr\Bigl\{H^{(2)}(x)^{\dagger}\gamma_0 H^{(0)}(x)\Bigr\}
\right\rangle_{0}
\nonumber\\
& &
-\left\langle\Tr\Bigl\{H^{(0)}(x)^{\dagger}\gamma_0 H^{(1)}(x)\Bigr\}
\Bigl[S^{(1)}_{\mathrm{tot}}\Bigr]_{\mathrm{F}}
\right\rangle_{0}\nonumber\\
& &
-\left\langle\Tr\Bigl\{H^{(1)}(x)^{\dagger}\gamma_0 H^{(0)}(x)\Bigr\}
\Bigl[S^{(1)}_{\mathrm{tot}}\Bigr]_{\mathrm{F}}
\right\rangle_{0}\biggr],
\label{eq:Hexpansion}
\eeqn
and analogously for $\fp$ with $-1$ instead of $\gamma_0$. So, in order to
compute the 1--loop coefficients of $\fa$ and $\fp$, one has to
expand $S_{\mathrm{tot}}$ up to order $g_0$ and 
\beq
H(x) = \ctt\sum_{\vy}S(x,y)U(y-\hat{0},0)^{-1}P_+|_{y_0=1}
\label{eq:Hexplicitly}
\eeq
up to order $g_0^2$. Here, $S(x,y)$ denotes the quark propagator, which
is the inverse of the Dirac--Wilson operator
$(D+\delta D +m_0)$. So one has to expand the propagator $S(x,y)$
and the link variables. Since the overall factor $\ctt$ cancels in the
definition of the current mass, it may be ignored here. There will however
be a contribution proportional to $\ctt^{(1)}$ due to the boundary counter-term
in the quark action, leading to a corresponding term in the $g_0^2$ coefficient
in the expansion of the propagator.

Following the lines of~\cite{Weisz:1996csw}, the calculation is done in
momentum space. First, the gluon fields have to be decomposed in a basis
of the Lie--algebra of $SU(3)$
\beq
q(x) = \sum_a \tilde{q}^a(x)I^a,
\eeq
with the basis $I^a$ chosen such that the star operation defined in
appendix~\ref{app:group} and the covariant
derivative act diagonally:
\beqn
\cosh G_{0k}\star I^a &=& C_a I^a,\label{eq:coshg0k}\\
\sinh G_{0k}\star I^a &=& S_a I^a,\label{eq:sinhg0k}\\
(D_k f)(x) &=& \sum_a \Bigl[ \Omega_a(x_0)f^a(x+\hat{k})-f^a(x)\Bigr]I^a,\\
(D^{\ast}_k f)(x) &=& \sum_a \Bigl[
  f^a(x) - \Omega_a(x_0)^{\ast}f^a(x-\hat{k})\Bigr]I^a.
\eeqn
The $I^a$ are explicitly given in appendix~\ref{app:group}.
The coefficients $\Omega_a(x_0)$ can be parametrised as
\beq
\Omega_a(x_0) = e^{i\phi_a(x_0)},
\eeq
with the $\phi_a(x_0)$ given in appendix~\ref{app:group}.
Then one may introduce the Fourier transformed fields
\beqn
\tilde{q}^a_0(x) &=& \frac{1}{L^3}\sum_{\vp}e^{i\vp\vx}\tilde{q}^a_0(\vp,x_0),\\
\tilde{q}^a_k(x) &=& \frac{1}{L^3}\sum_{\vp}e^{i\vp\vx}
e^{(p_k + \phi_a(x_0))/2}\tilde{q}^a_k(\vp,x_0).
\eeqn
Furthermore, the $\phi_a$ 
may be used to define the quantities
\beqn
s^a_k(\vp,x_0) &=& 2\sin\left[\frac{1}{2}\Bigl(p_k + \phi_a(x_0)\Bigr)\right],\\
c^a_k(\vp,x_0) &=& 2\cos\left[\frac{1}{2}\Bigl(p_k + \phi_a(x_0)\Bigr)\right],
\eeqn
and
\beq
R_a = (C_a - S_a)e^{i\partial_0\phi_a(x_0)/2},
\eeq
which will be used to make the expressions for the propagators and vertices
more compact. Here, $C_a$ and $S_a$ are the components of $\cosh G_{0k}$
and $\sinh G_{0k}$ according to~(\ref{eq:coshg0k}) and~(\ref{eq:sinhg0k}).
They and the $R_a$ are explicitly given in appendix~\ref{app:group}.

The same is done for the ghost fields
\beqn
c(x) &=& \frac{1}{L^3}\sum_{\vp}e^{i\vp\vx}c(\vp,x_0)
= \frac{1}{L^3}\sum_{\vp}e^{i\vp\vx}\sum_a\tilde{c}^a(\vp,x_0)I^a,\\
\bar{c}(x) &=& \frac{1}{L^3}\sum_{\vp}e^{i\vp\vx}\bar{c}(\vp,x_0)
= \frac{1}{L^3}\sum_{\vp}e^{i\vp\vx}\sum_a\tilde{\bar{c}}^a(\vp,x_0)I^a,
\eeqn
and also the quark fields are Fourier transformed
\beqn
\psi(x) &=& \frac{1}{L^3}\sum_{\vp}e^{i\vp\vx}\psi(\vp,x_0),\\
\bar{\psi}(x) &=& \frac{1}{L^3}\sum_{\vp}e^{i\vp\vx}\bar{\psi}(\vp,x_0).
\eeqn

\subsection{The propagators}
In momentum space, the quadratic part of the pure gluonic action takes
the form
\beq
S_{\mathrm{G}}^{(0)}
=\frac{1}{2}\frac{1}{L^3}\sum_{\vp}\sum_{x_0,y_0=0}^{T-1}\sum_{a}
\tilde{q}^{\bar{a}}_{\mu}(-\vp,x_0)K^a_{\mu\nu}(\vp;x_0,y_0)
\tilde{q}^a_{\nu}(\vp,y_0),
\eeq
where
\beqn
K^a_{kl}(\vp;x_0,y_0) &=& \delta_{x_0,y_0}\biggl[
\delta_{kl}\vs^a(\vp,x_0)^2 - s^a_k(\vp,x_0)s^a_l(\vp,x_0)(1-\lambda_0)\biggr]
\nonumber\\
& & + \delta_{kl}\biggl[ 2C_a\delta_{x_0,y_0} - R_a(\delta_{x_0+1,y_0}
+ \delta_{x_0-1,y_0})\biggr],
\label{eq:gluonquad1}
\eeqn
\beqn
K^a_{k0}(\vp;x_0,y_0) &=& iR_a\biggl[\delta_{x_0,y_0}s^a_k(\vp,x_0+1)
-\delta_{x_0-1,y_0}s^a_k(\vp,y_0)\biggr]\nonumber\\
& & -i\lambda_0 s^a_k(\vp,x_0)\biggl[\delta_{x_0,y_0}-\delta_{x_0-1,y_0}\biggr],
\label{eq:gluonquad2}\\
K^a_{0k}(\vp;x_0,y_0) &=& -K^a_{k0}(\vp;y_0,x_0),
\label{eq:gluonquad3}\\
K^a_{00}(\vp;x_0,y_0) &=& R_a\delta_{x_0,y_0}\vs^a(\vp,x_0)\cdot\vs^a(\vp,x_0+1)
\nonumber\\
& & +\lambda_0\biggl[ 2\delta_{x_0,y_0} -\delta_{x_0+1,y_0} -\delta_{x_0-1,y_0}
\biggr] \nonumber\\
& & -\lambda_0\delta_{x_0,y_0}\biggl[\delta_{x_0,0}(1-\chi_a\delta_{\vp,\vo})
+\delta_{x_0,T-1}\biggr],\label{eq:gluonquad4}
\eeqn
where $\chi_a=1$ for $a=3,8$ and $\chi_a=0$ otherwise.
Then the free gluon propagator is given by
\beq
\langle\tilde{q}^a_{\mu}(\vp,x_0)\tilde{q}^b_{\nu}(\vp',y_0)\rangle_{0}
= \delta_{b\bar{a}}L^3\delta_{\vp +\vp'}D^a_{\mu\nu}(\vp;x_0,y_0),
\eeq
where $D^a$ is the inverse of $K^a$.

The ghost action takes the form
\beq
S_{\mathrm{gh}} = \sum_{n=0}^{\infty}\frac{1}{n!}g_0^n 
S_{\mathrm{gh}}^{(n)},
\eeq
where the quadratic part is given by
\beq
S_{\mathrm{gh}}^{(0)} 
= \frac{1}{L^3}\sum_{\vp}\sum_{s_0,t_0}\sum_{a}\tilde{\bar{c}}^{\bar{a}}(-\vp,s_0)
F^a(\vp;s_0,t_0)\tilde{c}^a(\vp,t_0).
\eeq
For $a\neq 3,8$ one has
\beq
F^a(\vp;s_0,t_0) = \delta_{s_0,t_0}\left[2 + \vs^a(\vp,s_0)^2\right]
-\delta_{s_0+1,t_0} - \delta_{s_0-1,t_0}.
\label{eq:ghostquad}
\eeq
The computation aimed at here only involves closed loops of charged ghost fields
coupled to a neutral gluon. Hence the quadratic ghost action for \mbox{$a=3,8$} is
not needed here. Now we get the free ghost propagator
\beq
\langle \tilde{c}^a(\vp,s_0)\tilde{\bar{c}}^b(\vp',t_0)\rangle_{0} 
= \delta_{b\bar{a}}
L^3 \delta_{\vp +\vp'}D^a(\vp;s_0,t_0),
\eeq
where $D^a$ is the inverse of $F^a$.

The free part of the quark action is given by
\beq
S_{\mathrm{F}}^{(0)} = \frac{1}{L^3}\sum_{\vp}\sum_{x_0,y_0} 
\bar{\psi}(-\vp,x_0)\tilde{D}(\vp;x_0,y_0)\psi(\vp,y_0),
\label{eq:freewilsonaction}
\eeq
where $\tilde{D}$ is the improved Dirac--Wilson operator $(D+\delta D+m_0)$
at lowest order of $g_0$, namely
\beq
\tilde{D}(\vp;x_0,y_0) = 
-P_-\delta_{x_0+1,y_0} + B(\vp^+,x_0)\delta_{x_0,y_0}
- P_+\delta_{x_0-1,y_0},
\label{eq:dwithb}
\eeq
with $p_k^+ = p_k + \theta/L$ and
\beqn
B(\vp,x_0) &=& 4+m_0 - \sum_k\left[ \frac{1}{2}(1+\gamma_k)e^{-ip_k}
V(x_0)^{\dagger} + \frac{1}{2}(1-\gamma_k)e^{ip_k}V(x_0)\right]\nonumber\\
& & + iH\gamma_0\sum_k\gamma_k.
\eeqn
Here $H$ is a diagonal matrix in colour space with the elements
\beq
H_{\alpha\alpha} = -\frac{1}{2}\csw^{(0)}\sin\mathcal{E}_{\alpha}.
\eeq
The quark propagator $S$ is now defined as 
the inverse of the operator $\tilde{D}$.

\subsection{The vertices}

The expansion of the pure gluonic action gives the triple gluon vertex.
At 1--loop order we have
\beqn
S_{\mathrm{G}}^{(1)} &=& \frac{1}{3!}g_0\frac{1}{L^6}\sum_{\vq_1,\vq_2,\vq_3}
\delta_{\mathrm{P}}(\vq_1 + \vq_2 + \vq_3)\nonumber\\
& & \cdot\sum_{\mu_1,\mu_2,\mu_3}\sum_{t_1,t_2,t_3}\sum_{a_1,a_2,a_3}
V^{a_1 a_2 a_3}_{\mu_1\mu_2\mu_3}(\vq_1,\vq_2,\vq_3;t_1,t_2,t_3)\nonumber\\
& &\cdot\prod_j
\tilde{q}^{a_j}_{\mu_j}(-\vq_j,t_j),
\eeqn
where $\delta_{\mathrm{P}}$ denotes the periodic delta function, i.e. the delta 
function modulo $2\pi$. In this expression, a term proportional to
$\ct^{(1)}$ is missing, which will be treated later.
In order to make the somewhat complicated expressions for the vertex $V$ 
slightly more compact, we introduce the following notations for traces and
permutations of the basis of the Lie algebra and the background field:
\beq
c_{abc} = -2i\;\tr I^a [I^b,I^c],
\eeq
\beq
e_{abc} = -2i\;\tr \Bigl( e^{i\mathcal{E}}I^aI^bI^c 
- e^{-i\mathcal{E}}I^cI^bI^a \Bigr).
\eeq
Furthermore, we use the shorthand notations
\beq
E_k = e^{\frac{i}{2}(p_k + q_k + r_k)}
\eeq
and $\phi'_c = \partial_0 \phi_c(x_0)$. Since all non--zero $\phi_c$ are
linear in $x_0$, $\phi'_c$ does not depend on $x_0$.

Using these notations, one gets for the vertex
\beqn
& &V^{abc}_{klm}(\vp,\vq,\vr;s_0,t_0,u_0) =  c_{abc}\delta_{s_0,t_0}
\delta_{s_0,u_0}\Bigl\{\delta_{kl}E_k E_m \nonumber\\
& & \cdot \sin\frac{1}{2}\bigl[ p_m -\phi_a(s_0) -q_m +\phi_b(s_0)\bigr]
c^c_k(-\vr,s_0) + 2 \;\mbox{permutations}\Bigr\} \nonumber\\
& & -\frac{i}{4}\exp\frac{i}{2}\bigl[\phi_a(s_0)+\phi_b(t_0)+\phi_c(u_0)\bigr]
\delta_{klm}E_k 
\Bigl\{\delta_{s_0,t_0}\bigl[ (e_{abc}+e_{bac})\delta_{s_0+1,u_0}\nonumber\\
& & -(e_{cab}+e_{cba})\delta_{s_0-1,u_0}\bigr] +2\;\mbox{permutations}\Bigr\}. 
\eeqn
Here, ``2 permutations'' is meant such that one has to take cyclic permutations
of $k,l,m$ and $a,b,c$, respectively. The vertex is thus made totally symmetric
under interchange of labels. 
Then the other parts of the vertex become
\beqn
& &V^{abc}_{000}(\vp,\vq,\vr;s_0,t_0,u_0) = -\frac{i}{4}\delta_{s_0,t_0}
\delta_{s_0,u_0}\sum_k\Bigl\{ (e_{abc}+e_{bac})e^{i[r_k-\phi_c(s_0)]}
\nonumber\\
& & -(e_{cab}+e_{cba})e^{-i[r_k-\phi_c(s_0)]} +2\;\mbox{permutations}\Bigr\},
\eeqn
\beqn
& &V^{abc}_{kl0}(\vp,\vq,\vr;s_0,t_0,u_0) = -\frac{i}{2}\delta_{kl}E_k\nonumber\\
& &\cdot\biggl(\frac{1}{2}\delta_{s_0,t_0}\delta_{s_0,u_0}\Bigl\{
(e_{abc}+e_{bac})e^{\frac{i}{2}[r_k-\phi_c(u_0)]}
-(e_{cab}+e_{cba})e^{-\frac{i}{2}[r_k-\phi_c(u_0)]}\Bigr\}\nonumber\\
& &-\frac{i}{2}\delta_{s_0,t_0}\delta_{s_0-1,u_0}(e_{cab}+e_{cba})
e^{-\frac{i}{2}\phi'_c}s^c_k(-\vr,u_0) \nonumber\\
& & +\Bigl\{ e^{\frac{i}{2}\phi'_b}\Bigl[
e_{cab}e^{-\frac{i}{2}[r_k-\phi_c(u_0)]} -e_{acb}e^{\frac{i}{2}
[r_k-\phi_c(u_0)]}\Bigr]\delta_{s_0,t_0-1}\delta_{s_0,u_0}\nonumber\\
& & +1\;\mbox{permutation}\Bigr\}\biggr),
\eeqn
\beqn
& &V^{abc}_{00k}(\vp,\vq,\vr;s_0,t_0,u_0) = -\frac{i}{2}\delta_{s_0,t_0}
\nonumber\\
& & \cdot\biggl(
-\frac{1}{2}\delta_{s_0,u_0}\Bigl\{ 
(e_{cab}+e_{cba})e^{-\frac{i}{2}[r_k-\phi_c(u_0)]}
+(e_{abc}+e_{bac})e^{\frac{i}{2}[r_k-\phi_c(u_0)]}\Bigr\}\nonumber\\
& & +\frac{1}{2}\delta_{s_0,u_0-1}e^{\frac{i}{2}\phi'_c}
(e_{abc}+e_{bac})c^c_k(-\vr,s_0) \nonumber\\
& & +\Bigl\{ e^{\frac{i}{2}[-r_k-2p_k+2\phi_a(s_0)+\phi_c(u_0)]}
\bigl[e_{acb}\delta_{s_0,u_0} -e_{abc}\delta_{s_0,u_0-1}\bigr]\nonumber\\
& & + 1\;\mbox{permutation}\Bigr\}\biggr).
\eeqn

The gluon--ghost vertex is given by the 1--loop expansion of the ghost action.
Here one gets
\beqn
S_{\mathrm{gh}}^{(1)} &=& \frac{1}{L^6}\sum_{\vp,\vp',\vq}
\delta_{\mathrm{P}}(\vp+\vp'+\vq)\sum_{s_0,t_0,u_0}\sum_{\mu}\nonumber\\
& & \sum_{a,b,c} \tilde{\bar{c}}^a(-\vp',s_0)
F^{abc}_{\mu}(\vp',\vp,\vq;s_0,t_0,u_0)\tilde{c}^b(-\vp,t_0)
\tilde{q}^c_{\mu}(-\vq,u_0).
\eeqn
As already mentioned, only the vertex of a neutral gluon (i.e.~$c=3,8$)
and charged ghosts is needed here. In this case, the vertex becomes
\beqn
F_0^{abc}(\vp',\vp,\vq;s_0,t_0,u_0) &=&
-\frac{i}{2}c_{abc}\Bigl\{ \delta_{s_0,u_0}
[\delta_{s_0+1,t_0}+\delta_{s_0,t_0}] \nonumber\\
& & -\delta_{s_0-1,u_0}[\delta_{s_0,t_0}+\delta_{s_0-1,t_0}]\Bigr\},
\eeqn

\beqn
F_k^{abc}(\vp',\vp,\vq;s_0,t_0,u_0) &=&
c_{abc}\delta_{s_0,t_0}\delta_{s_0,u_0}c^b_k(-\vp,s_0)\nonumber\\
& & \cdot\sin\frac{1}{2}[-p_k -q_k +\phi_b(s_0) +\phi_c(s_0)].
\eeqn

The quark action has to be expanded up to order $g_0^2$ to get the 
quark--quark--gluon and the 2 quark -- 2 gluon vertices. Both vertices have
two terms, one coming from the Wilson part and one coming from the 
Sheikholeslami--Wohlert part of the action.

The Wilson part of the action has the expansion
\beq
S_{\mathrm{F,Wilson}} = S_{\mathrm{F,Wilson}}^{(0)} 
+\sum_{n=1}^{\infty}\frac{1}{n!}g_0^n 
S_{\mathrm{F,Wilson}}^{(n)},
\eeq
where $S_{\mathrm{F,Wilson}}^{(0)}$ has been given 
in~(\ref{eq:freewilsonaction}) and the
n-th order expansion coefficient may be written as
\beqn
S_{\mathrm{F,Wilson}}^{(n)} &=&
\left(\frac{1}{L^3}\right)^{n+1}\sum_{\vp,\vp',\vq_1,\ldots,\vq_n}
\delta_{\mathrm{P}}\left(\vp + \vp' +\sum_{j=1}^n\vq_j\right)
\sum_{\mu}\sum_{x_0,y_0,z_0}\sum_{a_1,\ldots,a_n}\nonumber\\
& & \bar{\psi}(-\vp',x_0)
V^{a_1\ldots a_n}_{\mu}(\vs;x_0,y_0,z_0)\psi(-\vp,y_0)
\prod_{j=1}^n \tilde{q}^{a_j}_{\mu}(-\vq_j,z_0),
\eeqn
with $\vs = \frac{1}{2}(\vp' - \vp)$. Using the notation
\beq
I^{a_1\ldots a_n} = \frac{1}{n!}\sum_{\mbox{\scriptsize perms}\;\sigma}
I^{\sigma(a_1)}\ldots I^{\sigma(a_n)},
\eeq
the time components of the vertex $V$ become
\beqn
V_0^{a_1\ldots a_n}(\vs;x_0,y_0,z_0) &=&
-I^{a_1\ldots a_n}\Bigl\{ P_- \delta_{x_0+1,y_0}\delta_{x_0,z_0}
\nonumber\\
& & +(-1)^n P_+ \delta_{x_0-1,y_0}\delta_{y_0,z_0}\Bigr\}.
\eeqn
In contrast to the time components, the space components
\beqn
V_k^{a_1\ldots a_n}(\vs;x_0,y_0,z_0) &=&
-I^{a_1\ldots a_n}\frac{1}{2}\delta_{x_0,y_0}\delta_{x_0,z_0}\Bigl\{
W_k^{a_1\ldots a_n}(\vs,x_0)(1-\gamma_k)
\nonumber\\
& & +(-1)^n W_k^{a_1\ldots a_n}(\vs,x_0)^{-1}(1+\gamma_k)\Bigr\}
\eeqn
depend on the background field via
\beq
W_k^{a_1\ldots a_n}(\vs,x_0) = V(x_0)\exp\Bigl(i[s_k^+ 
+ \frac{1}{2}\sum_j \phi_{a_j}(x_0)]\Bigr).
\eeq
While the Sheikholeslami--Wohlert term~(\ref{eq:sheikho}) does not 
contribute to the free 
action~(\ref{eq:freewilsonaction}), it has to be taken into account for higher
orders,
\beq
\delta S_{\mathrm{V}} = \csw(g_0)\sum_{n=1}^{\infty}\frac{1}{n!}g_0^n 
\delta S_{\mathrm{V}}^{(n)},
\eeq
where the n-th order expansion coefficient is given by
\beqn
\delta S_{\mathrm{V}}^{(n)} &=& \left(\frac{1}{L^3}\right)^{n+1}
\sum_{\vp,\vp',\vq_1,\ldots,\vq_n}\delta_{\mathrm{P}}
\left(\vp + \vp' +\sum_{j=1}^{n}\vq_j\right)
\sum_{x_0,z_{10},\ldots,z_{n0}}\sum_{\mu_1,\ldots,\mu_n}\sum_{a_1,\ldots,a_n}
\nonumber\\
& & \bar{\psi}(-\vp',x_0)S^{a_1\ldots a_n}_{\mu_1\ldots\mu_n}
(\vq_1,\ldots,\vq_n;x_0,z_{10},\ldots,z_{n0})\psi(-\vp,x_0)\nonumber\\
& & \cdot\prod_{j=1}^n\tilde{q}^{a_j}_{\mu_j}(-\vq_j,z_{j0}).
\eeqn
For $n=1$, the vertex $S$ may easily be written down as
\beqn
S_0^a(\vq;x_0,z_0) &=& \frac{1}{16}\sum_j\sigma_{0j}c_j^a(-\vq,x_0)\nonumber\\
& & \cdot\Bigl( (\delta_{x_0,z_0} + \delta_{x_0-1,z_0})\{I^a,\cos\mathcal{E}\}
s_j^a(-\vq,x_0)\nonumber\\
& & -(\delta_{x_0,z_0} - \delta_{x_0-1,z_0})[I^a,\sin\mathcal{E}]
c_j^a(-\vq,x_0)\Bigr),
\eeqn

\beqn
S_k^a(\vq;x_0,z_0) &=& \frac{i}{16}\sigma_{0k}\Bigl(
(\delta_{x_0+1,z_0} - \delta_{x_0-1,z_0})\{I^a,\cos\mathcal{E}\}c_k^a(-\vq,z_0)
\nonumber\\
& & -(\delta_{x_0+1,z_0} + 2\delta_{x_0,z_0} + \delta_{x_0-1,z_0})
[I^a,\sin\mathcal{E}]s_k^a(-\vq,z_0)\Bigr)\nonumber\\
& & -\frac{1}{8}I^a\delta_{x_0,z_0}c_k^a(-\vq,z_0)\sum_j \sigma_{jk}
s_j^a(-\vq,z_0)c_j^a(-\vq,z_0).
\eeqn
For $n=2$, the vertex is much more complicated. Fortunately, it is not needed
completely for the calculation of $\fa$ and $\fp$ at one loop order. The
only thing which really is needed is $\langle\mathcal{F}_{\mu\nu}(x)\rangle$
at order $g_0^2$, which is given by
\beq
\langle\mathcal{F}_{jk}(x)\rangle = \rmO (g_0^3),
\eeq
and
\beqn
& & \langle\mathcal{F}_{0k}(x)\rangle =
g_0^2\frac{1}{4}\frac{1}{L^3}\sum_{\vq}\sum_c I^cI^{\bar{c}}\Biggl\{ \nonumber\\
& & 2i\cos [\mathcal{E} - \frac{1}{2}(q_k + \phi_c(x_0 +1))]
\biggl(D^c_{00}(\vq;x_0,x_0)s^c_k(\vq,x_0 +1) \nonumber\\
& & + 2iD^c_{k0}(\vq;x_0 +1,x_0)\biggl) \nonumber\\
& & -2i\cos [\mathcal{E} +\frac{1}{2}(q_k +\phi_c(x_0 -1))]
\biggl(D^c_{00}(\vq;x_0 -1,x_0 -1)s^c_k(\vq,x_0 -1)\nonumber\\
& & -2iD^c_{k0}(\vq;x_0 -1,x_0 -1)\biggr) \nonumber\\
& &+ i\sin\mathcal{E}\biggl(D^c_{kk}(\vq;x_0 +1,x_0 +1) + 2D^c_{kk}(\vq;x_0,x_0)
\nonumber\\
& & + D^c_{kk}(\vq;x_0 -1,x_0 -1)\biggr) \nonumber\\
& &-2i\sin [\mathcal{E} -\frac{1}{2}\phi'_c]
\biggl(D^c_{kk}(\vq;x_0 +1,x_0) + D^c_{kk}(\vq;x_0 -1,x_0) \nonumber\\
& & +is^c_k(\vq,x_0 +1)D^c_{k0}(\vq;x_0,x_0) 
-is^c_k(\vq,x_0 -1)D^c_{k0}(\vq;x_0,x_0 -1)\biggr)\Biggr\} \nonumber\\
& & + \rmO (g_0^3).
\eeqn
 
The complete vertices may now be computed by combining the Wilson and
Sheikholeslami--Wohlert parts. The quark--quark--gluon vertex is then
given by
\beq
V^a_{\mu}(\vp',\vp,\vq;s_0,t_0,u_0) = V^a_{\mu}(\vs;s_0,t_0,u_0)
+ \csw^{(0)}S^a_{\mu}(\vq;s_0,u_0)\delta_{s_0,t_0},
\eeq
and the 2 quark -- 2 gluon vertex by
\beqn
V^{ab}_{\mu\nu}(\vp',\vp,\vq,\vq';s_0,t_0,u_0,u_0') &=&
V^{ab}_{\mu}(\vs;s_0,t_0,u_0)\delta_{\mu\nu}\delta_{u_0,u_0'} \nonumber\\
& & + \csw^{(0)}S^{ab}_{\mu\nu}(\vq,\vq';s_0,u_0,u_0')\delta_{s_0,t_0}.
\eeqn

\subsection{The diagrams}

Having calculated the propagators and vertices, one may now expand $\fa$ and
$\fp$. The calculation can be made slightly more general by defining the
function
\beqn
f(\Gamma;\vp,x_0) &=& 
\sum_{\mathbf{y},\mathbf{z}}e^{i\vp(\vy - \vz)}\frac{1}{2}
\Bigl\langle \mbox{Tr}\{P_+\Gamma P_- U(z-\hat{0},0)S(z,x) \nonumber\\
& & \;\;\;\;\;\left. \cdot\Gamma S(x,y)U(y-\hat{0},0)^{-1}
\}\Bigr\rangle_{\mathrm{G}}\right|_{y_0=z_0=1}.
\label{eq:fgeneral}
\eeqn
According to (\ref{eq:fAwithH}),(\ref{eq:fPwithH}), and~(\ref{eq:Hexplicitly}),
one has
\beqn
\fa(x_0) &=& \ctt^2 f(\gamma_0\gamma_5;\vo,x_0), \label{eq:fawithf}\\
\fp(x_0) &=& \ctt^2 f(\gamma_5;\vo,x_0). \label{eq:fpwithf}
\eeqn

\begin{figure}
  \noindent
  \begin{center}
  \begin{minipage}[b]{.3\linewidth}
     \centering\includegraphics[width=.8\linewidth]{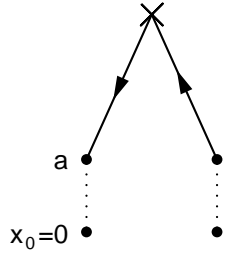}
  \end{minipage}
  \end{center}
  \caption[Diagram for $\fa$ and $\fp$ at tree level]
          {\sl Diagram for $\fa$ and $\fp$ at tree level. The dotted lines denote
           the link variables between $x_0=0$ and $x_0=a$, and the cross represents
           the insertion of the axial current or density.\label{fig:faptree}}
\end{figure}

Calculating the function $f$ at tree level only amounts to taking the tree
level values of the quark propagators and Fourier
transforming everything, leading to
\beq
f(\Gamma;\vp,x_0)^{(0)} = \frac{1}{2}\Tr\Bigl\{\pgp S(\vp;1,x_0)\Gamma 
  S(\vp;x_0,1)\Bigr\}.\label{eq:ftree}
\eeq
Since the quark propagator is diagonal in colour space, $f(\Gamma;\vp,x_0)^{(0)}$ 
may be decomposed into colour components
\beq
f(\Gamma;\vp,x_0)^{(0)} = \sum_{\alpha=1}^N f(\Gamma;\vp,x_0)^{(0)}_{\alpha},
\eeq
with the colour components given by
\beq
f(\Gamma;\vp,x_0)^{(0)}_{\alpha} = \frac{1}{2}\tr\Bigl\{\pgp S(\vp;1,x_0)_{\alpha}
 \Gamma S(\vp;x_0,1)_{\alpha}\Bigr\},
\eeq
where tr denotes the trace over the Dirac indices only.
Thus one gets a tree level coefficient which can be depicted by the diagram
in figure~\ref{fig:faptree}.

For the 1--loop expansion, the propagators and link variables 
in~(\ref{eq:fgeneral}) have to be expanded according to
\beq
S(x,y) = S^{(0)}(x,y) +g_0 S^{(1)}(x,y) +g_0^2 S^{(2)}(x,y) + \rmO(g_0^3),
\eeq
and
\beqn
U(x,\mu) &=& \exp\{g_0q_{\mu}(x)\}V(x,\mu)\nonumber\\ 
&=&
\left\{1 + g_0q_{\mu}(x) + \frac{1}{2}g_0^2q_{\mu}(x)^2 
+ \rmO(g_0^3)\right\}V(x,\mu).
\eeqn
In order to compute $f$ at 1--loop order, one has to insert these expansions,
gather all contributions of order $g_0^2$ and contract the gluon fields to
gluon propagators. Apart from these contributions, there are improvement
terms proportional to the 1--loop improvement coefficients. In order to 
compute $f$ at $m_1=0$, one has to do the calculation at the critical quark
mass $\mcrit$, which has an expansion
\beq
\mcrit = \mcrit^{(0)} + g_0^2\mcrit^{(1)} + \rmO(g_0^4).
\eeq
The 1--loop coefficient of $f$ then gets a contribution proportional to
$\mcrit^{(1)}$. 
Thus we get the sum 
\beqn
f(\Gamma;\vp,x_0)^{(1)} &=& \sum_n f(\Gamma;\vp,x_0)^{(1)}_n
+\csw^{(1)}f(\Gamma;\vp,x_0)^{(1)}_{\mathrm{V}}\nonumber\\
& &+\ctt^{(1)}f(\Gamma;\vp,x_0)^{(1)}_{\mathrm{Fb}}
+\ct^{(1)}f(\Gamma;\vp,x_0)^{(1)}_{\mathrm{Gb}}\nonumber\\
& &+\mcrit^{(1)}\frac{\partial}{\partial m_0}f(\Gamma;\vp,x_0)^{(0)}.
\eeqn
Here, all expansion coefficients have to be calculated at $m_0=\mcrit^{(0)}$,
which will be obtained by a tree level calculation explained in
section~\ref{sec:critmass}. For $f_{\mathrm{A}}$, one gets the additional
contribution $f_{\delta A}^{(1)}$ proportional to $\ca^{(1)}$.

\begin{figure}
  \noindent
  \begin{center}
  \begin{minipage}[b]{.3\linewidth}
     \centering\includegraphics[width=.8\linewidth]{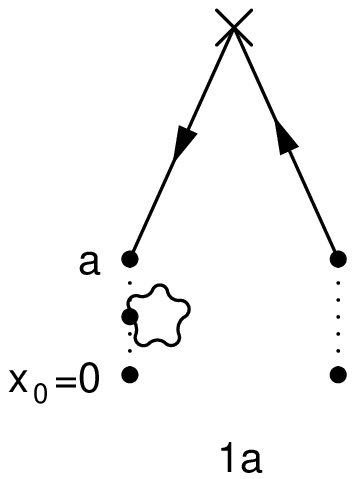}
  \end{minipage}
  \begin{minipage}[b]{.3\linewidth}
     \centering\includegraphics[width=.8\linewidth]{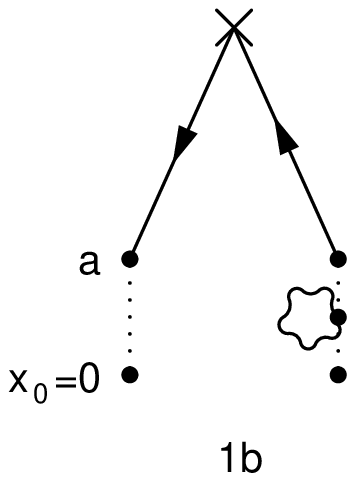}
  \end{minipage}
  \begin{minipage}[b]{.3\linewidth}
     \centering\includegraphics[width=.8\linewidth]{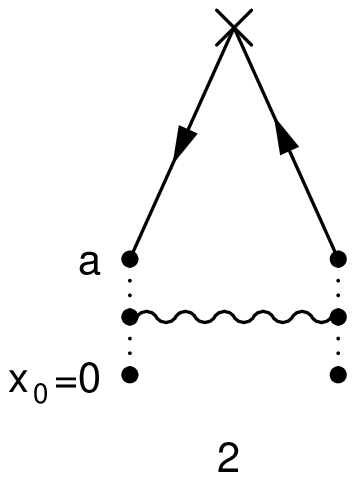}
  \end{minipage}\\
  \begin{minipage}[b]{.3\linewidth}
     \centering\includegraphics[width=.8\linewidth]{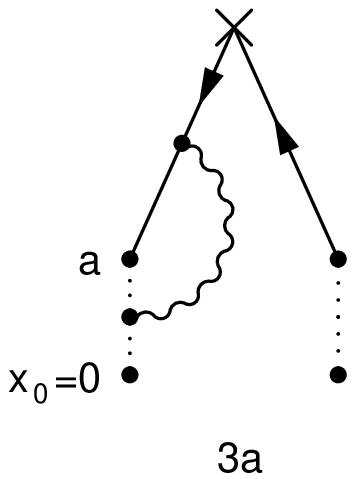}
  \end{minipage}
  \begin{minipage}[b]{.3\linewidth}
     \centering\includegraphics[width=.8\linewidth]{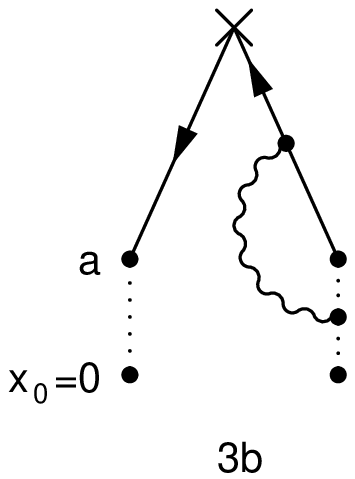}
  \end{minipage}
  \begin{minipage}[b]{.3\linewidth}
     \centering\includegraphics[width=.8\linewidth]{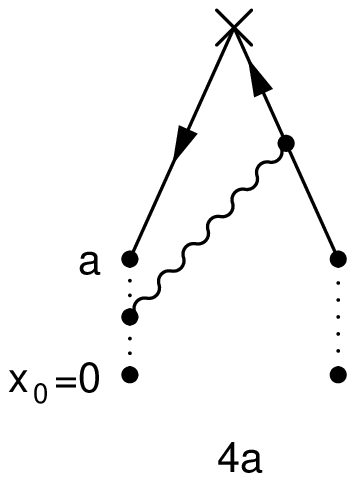}
  \end{minipage}\\
  \begin{minipage}[b]{.3\linewidth}
     \centering\includegraphics[width=.8\linewidth]{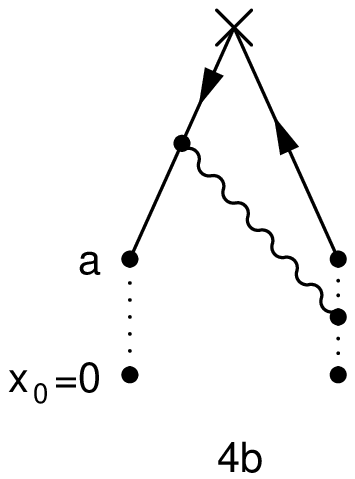}
  \end{minipage}
  \begin{minipage}[b]{.3\linewidth}
     \centering\includegraphics[width=.8\linewidth]{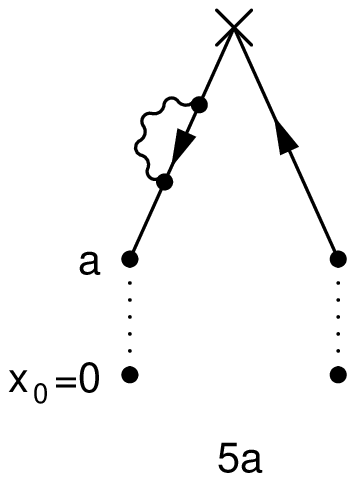}
  \end{minipage}
  \begin{minipage}[b]{.3\linewidth}
     \centering\includegraphics[width=.8\linewidth]{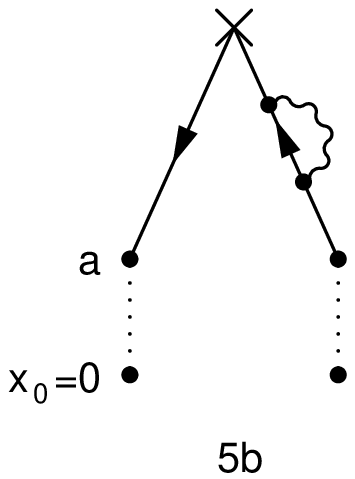}
  \end{minipage}\\
  \begin{minipage}[b]{.3\linewidth}
     \centering\includegraphics[width=.8\linewidth]{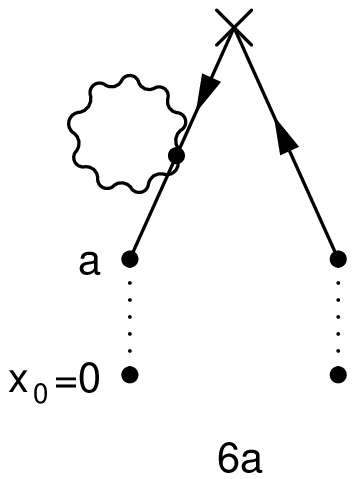}
  \end{minipage}
  \begin{minipage}[b]{.3\linewidth}
     \centering\includegraphics[width=.8\linewidth]{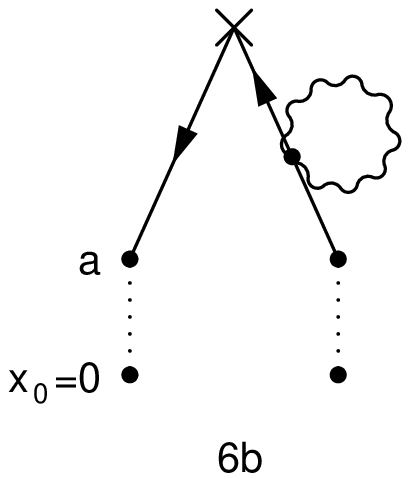}
  \end{minipage}
  \begin{minipage}[b]{.3\linewidth}
     \centering\includegraphics[width=.8\linewidth]{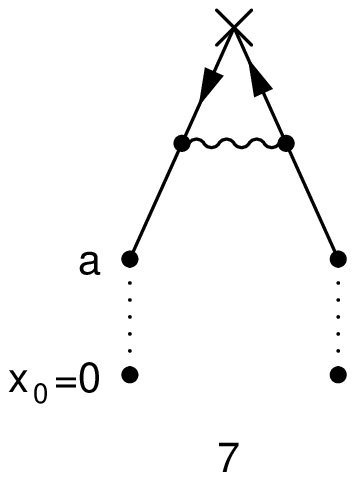}
  \end{minipage}
  \end{center}
  \caption[Diagrams contributing to $\fa(x_0)$ and $\fp(x_0)$ at
           1--loop order of perturbation theory both at vanishing and
           non--vanishing background field]
          {\label{fig:faponel}
           \sl Diagrams contributing to $\fa(x_0)$ and $\fp(x_0)$ at
           1--loop order of perturbation theory both at vanishing and
           non--vanishing background field.}
\end{figure}

The contributions $f(\Gamma;\vp,x_0)^{(1)}_n$ may be depicted by the diagrams in
figure~\ref{fig:faponel}.
The second order contributions of the link variables give diagrams 1a and
1b,
\beq
f(\Gamma;\vp,x_0)^{(1)}_{1a} = -\frac{1}{2}\frac{1}{L^3}\sum_{\alpha}
  f(\Gamma;\vp,x_0)^{(0)}_{\alpha}\sum_{\vq}\sum_{a}D^{a}_{00}(\vq;0,0)
  \mathcal{C}^a_{\alpha},
\eeq  
where
\beq
\mathcal{C}^a_{\alpha} = -\left(I^a I^{\bar{a}}\right)_{\alpha\alpha}.
\eeq
Diagram 1b is simply given by
\beq
f(\Gamma;\vp,x_0)^{(1)}_{1b} = f(\Gamma;\vp,x_0)^{(1)}_{1a}.
\eeq
In the following, we define $\vr = \vp + \vq$ for convenience.
Then the contraction of the first order terms of both link variables yields
diagram 2,
\beq
f(\Gamma;\vp,x_0)^{(1)}_{2} = \frac{1}{L^3}\sum_{\alpha}\sum_{\vq}
f(\Gamma;\vr,x_0)^{(0)}_{\alpha}\sum_a D^a_{00}(\vq;0,0)\mathcal{C}^a_{\alpha}.
\label{eq:fapdiagram2}
\eeq
Obviously, there are four different ways to combine the first order terms of
the link variables with the first order terms of the propagators. Contracting
the first order term of the link variables with the first order terms of the
propagators on the same side gives diagrams 3a and 3b,
\beqn
f(\Gamma;\vp,x_0)^{(1)}_{3a} &=& \frac{1}{2}\frac{1}{L^3}\sum_{\vq}
\sum_{\mu}\sum_{s_0,t_0,u_0}\sum_a D^a_{0\mu}(\vq;0,u_0)\Tr\Bigl\{
I^a\pgp S(\vp;1,x_0) \nonumber\\
& & \cdot \Gamma S(\vp;x_0,s_0) V^{\bar{a}}_{\mu}(\vp,-\vr,\vq;
s_0,t_0,u_0)S(\vr;t_0,1)\Bigl\},
\eeqn
\beqn
f(\Gamma;\vp,x_0)^{(1)}_{3b} &=& -\frac{1}{2}\frac{1}{L^3}\sum_{\vq}
\sum_{\mu}\sum_{s_0,t_0,u_0}\sum_a D^a_{\mu 0}(\vq;u_0,0)\Tr\Bigl\{
I^{\bar{a}}\pgp S(\vr;1,s_0) \nonumber\\
& & \cdot V^{a}_{\mu}(\vr,-\vp,-\vq;s_0,t_0,u_0)S(\vp;t_0,x_0)
\Gamma S(\vp;x_0,1)\Bigl\},
\eeqn
while combining them with the first order terms of the propagators on the
opposite side results in diagrams 4a and 4b,
\beqn
f(\Gamma;\vp,x_0)^{(1)}_{4a} &=& \frac{1}{2}\frac{1}{L^3}\sum_{\vq}\sum_{\mu}
\sum_{s_0,t_0,u_0}\sum_{a}D^a_{0\mu}(\vq;0,u_0)\Tr\Bigl\{I^a\pgp S(\vp;1,s_0)
\nonumber\\
& & \cdot V^{\bar{a}}_{\mu}(\vp,-\vr,\vq;s_0,t_0,u_0)
S(\vr;t_0,x_0)\Gamma S(\vr;x_0,1)\Bigr\},
\label{eq:fapdiagram4a}
\eeqn
\beqn
f(\Gamma;\vp,x_0)^{(1)}_{4b} &=& -\frac{1}{2}\frac{1}{L^3}\sum_{\vq}
\sum_{\mu}\sum_{s_0,t_0,u_0}\sum_a D^a_{\mu 0}(\vq;u_0,0)\Tr\Bigl\{
I^{\bar{a}}\pgp S(\vr;1,x_0) \nonumber\\
& & \cdot \Gamma S(\vr;x_0,s_0) V^a_{\mu}(\vr,-\vp,-\vq;
s_0,t_0,u_0)S(\vp;t_0,1)\Bigl\}.
\label{eq:fapdiagram4b}
\eeqn
The second order terms of the quark propagators consist of two parts: One
containing two quark--quark--gluon vertices with a quark line in between,
and one containing a 2 quark--2 gluon vertex. The part with the two 
vertices leads to diagrams 5a and 5b,
\beqn
f(\Gamma;\vp,x_0)^{(1)}_{5a} &=& \frac{1}{2}\frac{1}{L^3}\sum_{\vq}
\sum_{\mu,\nu}\sum_{s_0,t_0,u_0}\sum_{s_0',t_0',u_0'}\sum_{a}
D^a_{\nu\mu}(\vq;u_0',u_0)\Tr\Bigl\{\pgp \nonumber\\
& & \cdot S(\vp;1,x_0)\Gamma S(\vp;x_0,s_0)
V^{\bar{a}}_{\mu}(\vp,-\vr,\vq;s_0,t_0,u_0)
S(\vr;t_0,s_0') \nonumber\\
& &\cdot V^a_{\nu}(\vr,-\vp,-\vq;s_0',t_0',u_0')S(\vp;t_0',1)\Bigr\},
\eeqn
\beqn
f(\Gamma;\vp,x_0)^{(1)}_{5b} &=& \frac{1}{2}\frac{1}{L^3}\sum_{\vq}
\sum_{\mu,\nu}\sum_{s_0,t_0,u_0}\sum_{s_0',t_0',u_0'}\sum_{a}
D^a_{\nu\mu}(\vq;u_0',u_0)\Tr\Bigl\{\pgp\nonumber\\
& &\cdot S(\vp;1,s_0) V^{\bar{a}}_{\mu}(\vp,-\vr,\vq;s_0,t_0,u_0)
S(\vr;t_0,s_0') \nonumber\\
& & \cdot V^a_{\nu}(\vr,-\vp,-\vq;s_0',t_0',u_0')
S(\vp;t_0',x_0)\Gamma S(\vp;x_0,1)
\Bigr\},
\eeqn
while the part containing the 2 quark--2 gluon vertex gives diagrams 6a and 6b,
\beqn
f(\Gamma;\vp,x_0)^{(1)}_{6a} &=& -\frac{1}{4}\frac{1}{L^3}\sum_{\vq}
\sum_{\mu,\nu}\sum_{s_0,t_0}\sum_{u_0,u_0'}\sum_{a}
D^a_{\nu\mu}(\vq;u_0',u_0)\Tr\Bigl\{\pgp \nonumber\\
& & \cdot S(\vp;1,x_0)\Gamma S(\vp;x_0,s_0)
V^{\bar{a}a}_{\mu\nu}(\vp,-\vp,\vq,-\vq;s_0,t_0,u_0,u_0') \nonumber\\
& & \cdot S(\vp;t_0,1)\Bigr\},
\eeqn
\beqn
f(\Gamma;\vp,x_0)^{(1)}_{6b} &=& -\frac{1}{4}\frac{1}{L^3}\sum_{\vq}
\sum_{\mu,\nu}\sum_{s_0,t_0}\sum_{u_0,u_0'}\sum_{a}
D^a_{\nu\mu}(\vq;u_0',u_0)\Tr\Bigl\{\pgp\nonumber\\
& &\cdot S(\vp;1,s_0) V^{\bar{a}a}_{\mu\nu}(\vp,-\vp,\vq,-\vq;s_0,t_0,u_0,u_0')
S(\vp;t_0,x_0) \nonumber\\
& & \cdot \Gamma S(\vp;x_0,1)
\Bigr\}.
\eeqn
In the improved theory, these diagrams do not give the total contribution 
of the second order terms of the quark propagators. There are two more parts,
one proportional to $\ctt^{(1)}$ and one proportional to $\csw^{(1)}$. These
contributions will be dealt with separately.
Finally, the combination of the first order terms of both propagators results
in diagram~7,
\beqn
f(\Gamma;\vp,x_0)^{(1)}_{7} &=& \frac{1}{2}\frac{1}{L^3}\sum_{\vq}
\sum_{\mu,\nu}\sum_{s_0,t_0,u_0}\sum_{s_0',t_0',u_0'}\sum_{a}
D^a_{\nu\mu}(\vq;u_0',u_0)\Tr\Bigl\{\pgp\nonumber\\
& &\cdot S(\vp;1,s_0) V^{\bar{a}}_{\mu}(\vp,-\vr,\vq;s_0,t_0,u_0)
S(\vr;t_0,x_0) \nonumber\\
& & \cdot \Gamma S(\vr;x_0,s_0') V^a_{\nu}(\vr,-\vp,-\vq;s_0',t_0',u_0')
S(\vp;t_0',1)
\Bigr\}.
\label{eq:fapdiagram7}
\eeqn

\begin{figure}
  \noindent
  \begin{center}
  \begin{minipage}[b]{.3\linewidth}
     \centering\includegraphics[width=.8\linewidth]{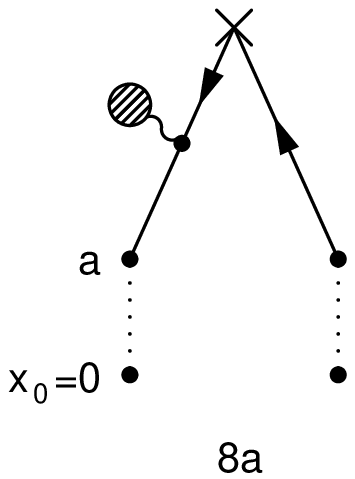}
  \end{minipage}
  \begin{minipage}[b]{.3\linewidth}
     \centering\includegraphics[width=.8\linewidth]{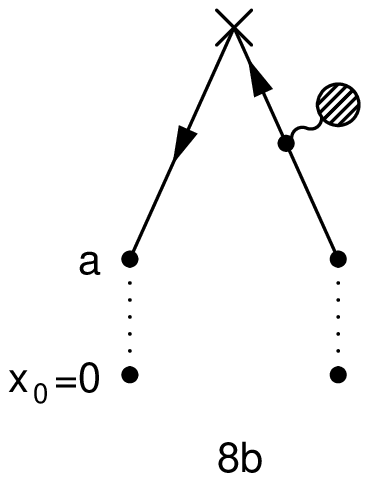}
  \end{minipage}\\
  \begin{minipage}[b]{.3\linewidth}
     \centering\includegraphics[width=\linewidth]{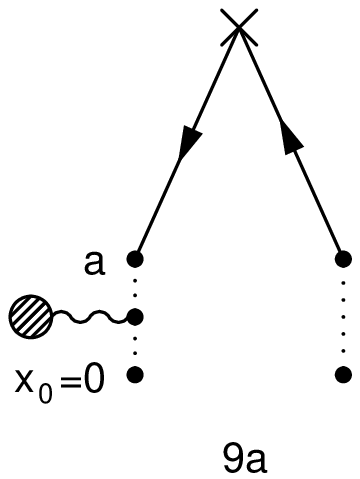}
  \end{minipage}\hspace{4mm}
  \begin{minipage}[b]{.3\linewidth}
     \centering\includegraphics[,width=\linewidth]{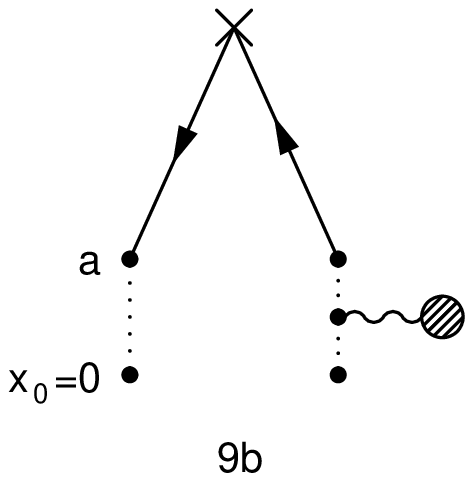}
  \end{minipage}\\
  \vspace{4mm}
  \begin{minipage}[b]{.6\linewidth}
     \centering\includegraphics[width=\linewidth]{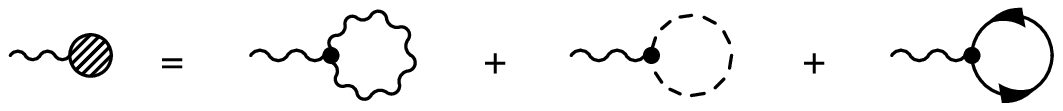}
  \end{minipage}
  \end{center}
  \caption[Tadpole diagrams contributing to $\fa(x_0)$ and $\fp(x_0)$
           at 1--loop order of perturbation theory]
          {\label{fig:faptad} 
           \sl Tadpole diagrams contributing to $\fa(x_0)$ and $\fp(x_0)$
           at 1--loop order of perturbation theory. These terms are only 
           present with a non--vanishing background field.}
\end{figure}

So far, the calculation has been completely analogous to the case of a
vanishing background field. However, with the non--zero background field
considered here, one has to take the terms in~(\ref{eq:Hexpansion}) containing
$S^{(1)}_{\mathrm{tot}}$ into account. This means one has to contract the
first order terms of the link variables and the propagators with the first
order terms of the total action, leading to the diagrams in figure~\ref{fig:faptad}.
Starting with the first order term of a 
propagator, we get
\beqn
f(\Gamma;\vp,x_0)^{(1)}_{8a} &=& -\frac{1}{2}
\sum_{\mu,\nu}\sum_{s_0,t_0}\sum_{u_0,u_0'}\sum_{a}
D^a_{\nu\mu}(\vo;u_0',u_0)\Tr\Bigl\{\pgp \nonumber\\
& & \cdot S(\vp;1,x_0)\Gamma S(\vp;x_0,s_0)
V^a_{\nu}(\vp,-\vp,\vo;s_0,t_0,u_0') \nonumber\\
& & \cdot S(\vp;t_0,1)\Bigr\}T^{\bar{a}}_{\mu}(u_0),
\eeqn
where $T^a_{\mu}$ denotes the sum of the closed gluon, ghost, and quark loops,
\beq
T^a_{\mu}(u_0) = T^a_{\mu,\mbox{\scriptsize gluon}}(u_0) 
+ T^a_{\mu,\mbox{\scriptsize ghost}}(u_0)
+ \Nf T^a_{\mu,\mbox{\scriptsize quark}}(u_0).
\eeq
Note that the time component $T^a_{0}$ vanishes due to CP invariance.
The gluon loop is given by
\beq
T^a_{\mu,\mbox{\scriptsize gluon}}(u_0) = -\frac{1}{2}\frac{1}{L^3}\sum_{\vq}
\sum_{\nu,\rho}\sum_{s_0,t_0}\sum_{c} 
V^{\bar{c}ca}_{\nu\rho\mu}(\vq,-\vq,\vo;s_0,t_0,u_0)
D^c_{\rho\nu}(\vq;t_0,s_0),
\eeq
the ghost loop by
\beq
T^a_{\mu,\mbox{\scriptsize ghost}}(u_0) 
= \frac{1}{L^3}\sum_{\vq}\sum_{s_0,t_0}\sum_c
F^{\bar{c}ca}_{\mu}(\vq,-\vq,\vo;s_0,t_0,u_0)D^c(\vq;t_0,s_0),
\eeq
and the quark loop by
\beq
T^a_{\mu,\mbox{\scriptsize quark}}(u_0) = \frac{1}{L^3}\sum_{\vq}\sum_{s_0,t_0}
\Tr\Bigl\{V^a_{\mu}(\vq,-\vq,\vo;s_0,t_0,u_0)S(\vq;t_0,s_0)\Bigr\}.
\eeq 
Analogously, one gets diagram 8b,
\beqn
f(\Gamma;\vp,x_0)^{(1)}_{8b} &=& -\frac{1}{2}
\sum_{\mu,\nu}\sum_{s_0,t_0}\sum_{u_0,u_0'}\sum_{a}
D^a_{\nu\mu}(\vo;u_0',u_0)\Tr\Bigl\{\pgp\nonumber\\
& &\cdot S(\vp;1,s_0) V^a_{\nu}(\vp,-\vp,\vo;s_0,t_0,u_0')
S(\vp;t_0,x_0) \nonumber\\
& & \cdot \Gamma S(\vp;x_0,1)
\Bigr\}T^{\bar{a}}_{\mu}(u_0).
\eeqn
The combination of $S^{(1)}_{\mathrm{tot}}$ and the first order terms of the link
variables yields diagrams 9a and 9b, which are given by
\beqn
f(\Gamma;\vp,x_0)^{(1)}_{9a} &=& -\frac{1}{2}\sum_{\mu}\sum_{u_0}\sum_a
D^a_{0\mu}(\vo;0,u_0) \Tr\Bigl\{ I^a \pgp S(\vp;1,x_0) \nonumber\\
& & \cdot\Gamma S(\vp;x_0,1)\Bigr\}T^{\bar{a}}_{\mu}(u_0)
\eeqn
and
\beq
f(\Gamma;\vp,x_0)^{(1)}_{9b} = - f(\Gamma;\vp,x_0)^{(1)}_{9a}.
\eeq
Since they are of opposite sign, they cancel in the sum and may be ignored. This
leaves diagrams 8a and 8b as the only diagrams depending on the number of
flavours. Note that these terms are not present in the case of a vanishing
background field. This means that with a non--vanishing background field, the
critical quark mass is dependent on $\Nf$ at 1--loop order, while with a
vanishing background field it is not.

\subsection{The improvement terms}
It was already mentioned that diagrams 6a and 6b get additional contributions
from the volume and boundary counter-terms in the quark action.\\
The volume term gives corrections proportional to $\csw^{(0)}$ ,which have
already been taken into account in the vertices, and corrections proportional
to $\csw^{(1)}$ which still have to be computed. These corrections to the
propagator at order $g_0^2$ are imposed by the Sheikholeslami--Wohlert part
of the Dirac operator, which at this order is
\beq
\delta D_{\mathrm{V}}^{(2)} = -\frac{i}{2}\csw^{(1)}\gamma_0\sum_k\gamma_k 
\sin\mathcal{E}.
\eeq
In diagram 6a, this leads to the correction
\beqn
f(\Gamma;\vp,x_0)^{(1)}_{6a,\mathrm{V}} &=& \frac{i}{4}\csw^{(1)}
\sum_{\alpha}\sum_{s_0=1}^{T-1}
\tr \Bigl\{\pgp S(\vp;1,x_0)_{\alpha}\Gamma S(\vp;x_0,s_0)_{\alpha}\nonumber\\
& & \cdot
\gamma_0\sum_k\gamma_k S(\vp;x_0,1)_{\alpha}\Bigr\}\sin\mathcal{E}_{\alpha}.
\eeqn
Using the symmetry of the quark propagator
\beq
S(\vp;x_0,y_0) = \gamma_5 S(\vp;y_0,x_0)^{\dagger}\gamma_5,
\eeq
one can show that the contribution to diagram 6b is the same. Thus,
the total contribution proportional to $\csw^{(1)}$ becomes
\beq
f(\Gamma;\vp,x_0)^{(1)}_{\mathrm{V}} = 2f(\Gamma;\vp,x_0)^{(1)}_{6a,\mathrm{V}}.
\label{eq:fimprvol}
\eeq
Another way to get this improvement term is by Taylor expansion,
\beq
f(\Gamma;\vp,x_0)^{(1)}_{\mathrm{V}}
= \left.\frac{\partial}{\partial \csw} f(\Gamma;\vp,x_0)^{(0)}
\right|_{\csw =1}.
\eeq
So, the volume term could, in principle, also be obtained from the
tree level result by numerical differentiation. While this procedure
turned out not to be suitable for the calculation, since it gives the
result with less accuracy, it was successfully used to check the result
obtained according to~(\ref{eq:fimprvol}).

Apart from the volume term, diagrams 6a and 6b get a contribution from the boundary
counter-term of the quark action, which acts on the propagator as
\beq
\delta D_{\mathrm b}^{(2)}S(x,y) = \ctt^{(1)}\left\{\delta_{x_0,1}S(x,y)
+\delta_{x_0,T-1}S(x,y)\right\}.
\eeq
In diagram 6a, this leads to the contribution
\beqn
f(\Gamma;\vp,x_0)^{(1)}_{6a,\mathrm b} &=& -\frac{1}{2}\ctt^{(1)}\Tr
\Bigl\{\pgp S(\vp;1,x_0)\Gamma\Bigl[S(\vp;x_0,1)S(\vp;1,1) \nonumber\\
& &
+S(\vp;x_0,T-1)S(\vp;T-1,1)\Bigr]\Bigr\}.
\eeqn
Here again, the corresponding contribution to diagram 6b can be shown to be
the same. One thus gets
\beq
f(\Gamma;\vp,x_0)^{(1)}_{\mathrm{Fb}} = 2f(\Gamma;\vp,x_0)^{(1)}_{6a,\mathrm b}.
\label{eq:fimprbound}
\eeq
Strictly speaking, this does not give the whole term of $\fa$ and $\fp$ proportional
to $\ctt$, because there is still the overall factor of $\ctt^2$ 
in~(\ref{eq:fawithf})
and~(\ref{eq:fpwithf}) to be taken into account. As stated before, this
factor cancels in the current quark mass and may be neglected in this 
calculation. It may, however, be used as a consistency check, since,
as an improvement term, 
$(f^{(1)}_{\mathrm{A/P,Fb}}+2f^{(0)}_{\mathrm{A/P}})$
has to approach zero in the continuum limit, which $f^{(1)}_{\mathrm{A/P,Fb}}$
and $f^{(0)}_{\mathrm{A/P}}$ do not do separately.

\begin{figure}
  \noindent
  \begin{center}
  \begin{minipage}[b]{.3\linewidth}
     \centering\includegraphics[width=.8\linewidth]{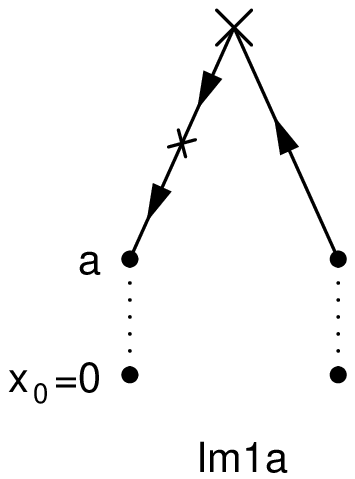}
  \end{minipage}
  \begin{minipage}[b]{.3\linewidth}
     \centering\includegraphics[width=.8\linewidth]{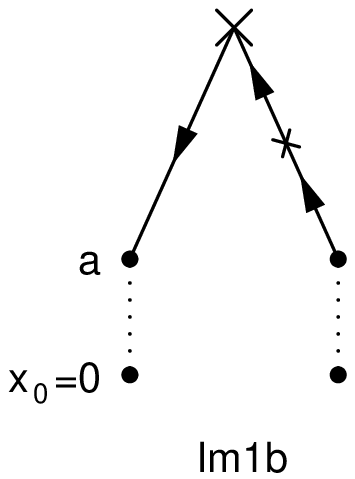}
  \end{minipage}
  \end{center}
  \caption[Diagrams for $f(\Gamma;\vp,x_0)^{(1)}_{\mathrm{V}}$,
           $f(\Gamma;\vp,x_0)^{(1)}_{\mathrm{Fb}}$, and
           $\frac{\partial}{\partial m_0}f(\Gamma;\vp,x_0)^{(0)}$]
          {\sl Diagrams for $f(\Gamma;\vp,x_0)^{(1)}_{\mathrm{V}}$,
           $f(\Gamma;\vp,x_0)^{(1)}_{\mathrm{Fb}}$, and
           $\frac{\partial}{\partial m_0}f(\Gamma;\vp,x_0)^{(0)}$
           \label{fig:fapim}}
\end{figure}

Both improvement terms calculated so far come from the insertion of
$\delta D^{(2)}$ into the quark line and may thus be depicted by the
diagrams in figure~\ref{fig:fapim}.

Another term that may be depicted by the same diagrams is the 
derivative of $f$ at tree level
with respect to the bare mass.
To this end, one has to take the derivative
\beq
\frac{\partial}{\partial m_0} S(\vp;x_0,y_0)
= -\sum_{s_0}S(\vp;x_0,s_0)S(\vp;s_0,y_0)
\eeq
of the quark propagators in~(\ref{eq:ftree}). The two contributions coming from
differentiating the two propagators can be shown to be the same, leading to
\beq
\frac{\partial}{\partial m_0}f(\Gamma;\vp,x_0)^{(0)} = -\sum_{s_0}\Tr
\left\{\pgp S(\vp;1,x_0)\Gamma S(\vp;x_0,s_0)S(\vp;s_0,1)\right\}.
\eeq

\begin{figure}
  \noindent
  \begin{center}
  \begin{minipage}[b]{.3\linewidth}
     \centering\includegraphics[width=\linewidth]{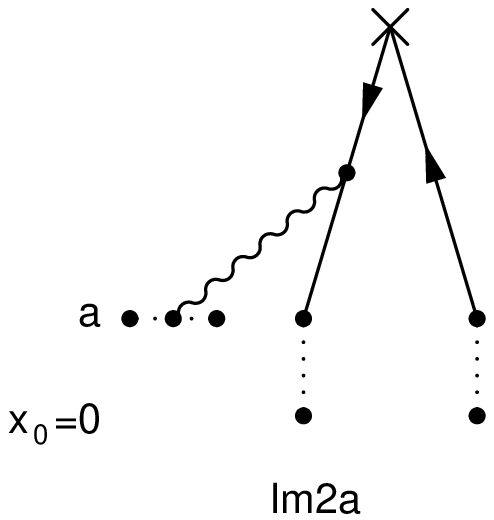}
  \end{minipage}
  \begin{minipage}[b]{.3\linewidth}
     \centering\includegraphics[width=\linewidth]{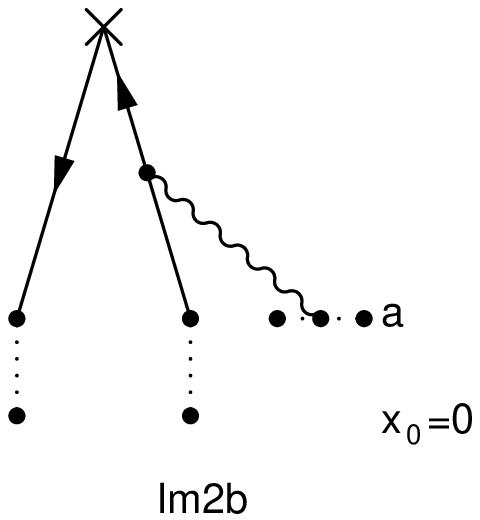}
  \end{minipage}\\
  \begin{minipage}[b]{.3\linewidth}
     \centering\includegraphics[width=\linewidth]{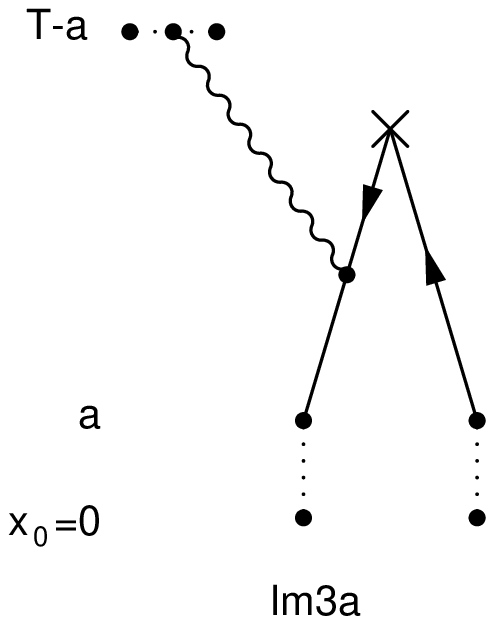}
  \end{minipage}
  \begin{minipage}[b]{.3\linewidth}
     \centering\includegraphics[width=\linewidth]{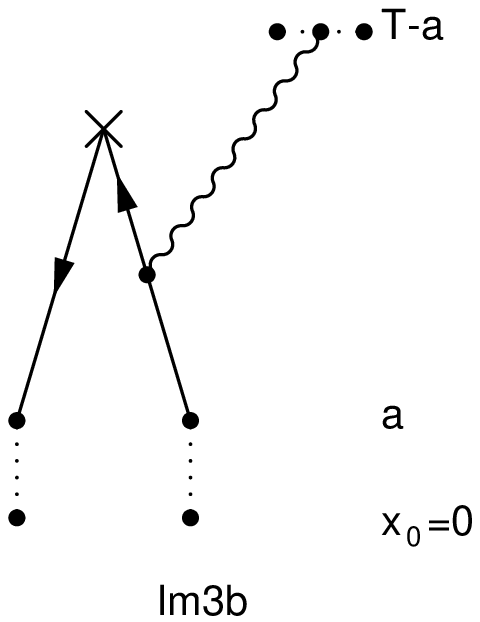}
  \end{minipage}
  \end{center}
  \caption[Diagrams contributing to the improvement term proportional to
           $\ct^{(1)}$ for $f_{\mathrm{A}}$ and $f_{\mathrm{P}}$
           at 1--loop order]
          {\sl Diagrams contributing to the improvement term proportional to
           $\ct^{(1)}$ for $f_{\mathrm{A}}$ and $f_{\mathrm{P}}$
           at 1--loop order
           \label{fig:fapimct}}
\end{figure}

With a non--vanishing background field, there is also an improvement term
proportional to $\ct^{(1)}$, giving an additional contribution to 
diagram~8a and~8b. This term arises 
from a contribution proportional to $\ct^{(1)}$
in the total action, contributing to $\fa$ and $\fp$ via~(\ref{eq:Hexpansion}).
Apart from the gluon tadpole, the gluon part~(\ref{eq:gluonaction}) of the 
total action yields the boundary counter-term 
\beq
S_{\mathrm{tot,b}}^{(1)} = \frac{2}{\sqrt{3}} \ct^{(1)}\sum_k
\left[\tilde{q}^{8}_k(\vo,1) -\tilde{q}^8_k(\vo,T-1)\right]
\left[\sin(2\gamma)+\sin\gamma\right],
\label{eq:stot1}
\eeq
where the parameter $\gamma$ is given by
\beq
\gamma = \frac{1}{LT}\left(\eta +\frac{\pi}{3}\right).
\eeq
Contracting this part of the total action with the first order term of the
quark line gives the contribution
\beqn
f(\Gamma;\vp,x_0)^{(1)}_{\mathrm{Im2a/Im3a}} &=&
\frac{\ct^{(1)}}{\sqrt{3}}\sum_{s_0,t_0,u_0}\sum_{\mu}\sum_k
\Tr\Bigl\{\pgp S(\vp;1,x_0)\Gamma S(\vp;x_0,s_0) \nonumber\\
& & \cdot V^8_{\mu}(\vp,-\vp,\vo;s_0,t_0,u_0)S(\vp;t_0,1)
\Bigr\}\biggl[D^8_{\mu k}(\vo;u_0,1) \nonumber\\
& & \qquad - D^8_{\mu k}(\vo;u_0,T-1)\biggr]
\left[\sin(2\gamma)+\sin\gamma\right],
\eeqn
which may be depicted by diagrams~Im2a and~Im3a in figure~\ref{fig:fapimct}.
\begin{figure}
  \noindent
  \begin{center}
  \begin{minipage}[b]{.3\linewidth}
     \centering\includegraphics[width=\linewidth]{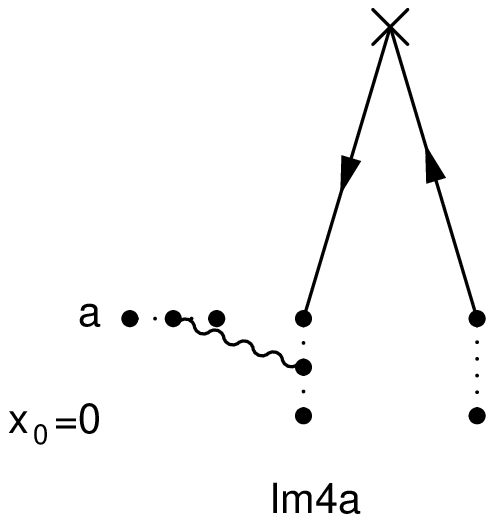}
  \end{minipage}
  \begin{minipage}[b]{.3\linewidth}
     \centering\includegraphics[width=\linewidth]{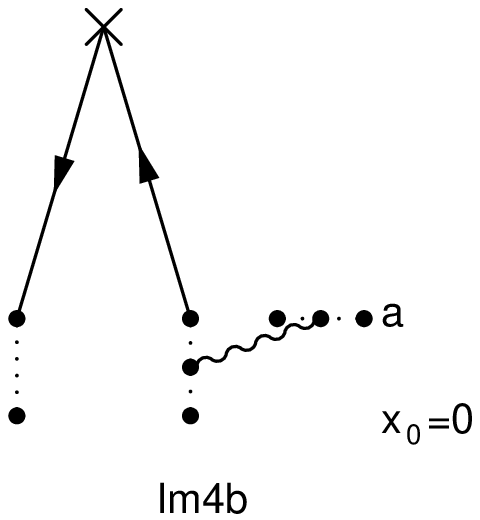}
  \end{minipage}\\
  \begin{minipage}[b]{.3\linewidth}
     \centering\includegraphics[width=\linewidth]{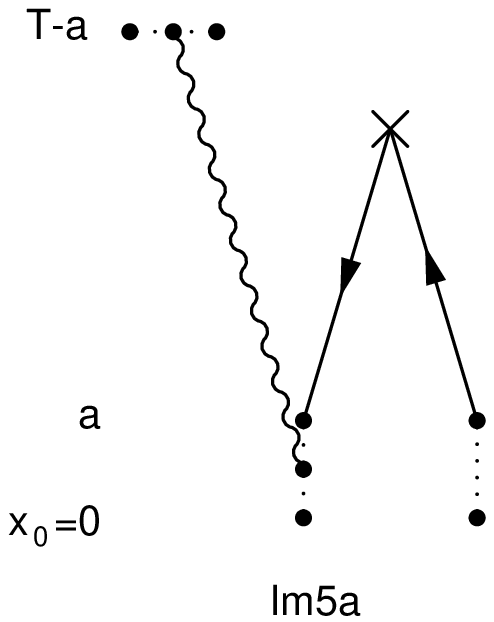}
  \end{minipage}
  \begin{minipage}[b]{.3\linewidth}
     \centering\includegraphics[width=\linewidth]{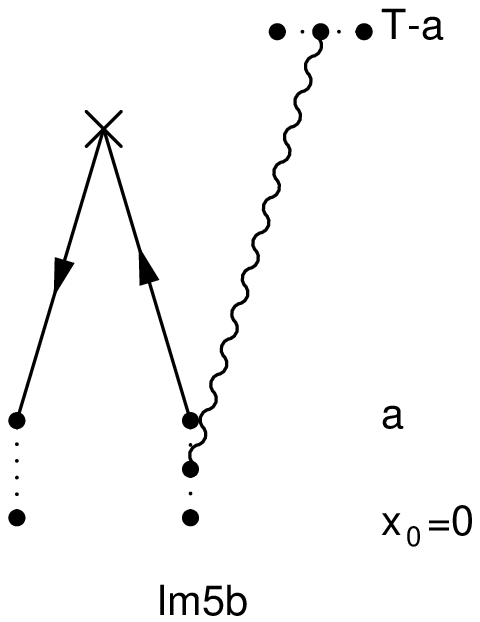}
  \end{minipage}
  \end{center}
  \caption[Further diagrams contributing to the 
           improvement term proportional to
           $\ct^{(1)}$ for $f_{\mathrm{A}}$ and $f_{\mathrm{P}}$
           at 1--loop order]
          {\sl Further diagrams contributing to the 
           improvement term proportional to
           $\ct^{(1)}$ for $f_{\mathrm{A}}$ and $f_{\mathrm{P}}$
           at 1--loop order
           \label{fig:fapimct2}}
\end{figure}
The corresponding correction depicted by diagrams~Im2b and~Im3b 
can be shown to be the same, such
that the total contribution proportional to $\ct^{(1)}$ becomes
\beq
f(\Gamma;\vp,x_0)^{(1)}_{\mathrm{Gb}}
= 2 f(\Gamma;\vp,x_0)^{(1)}_{\mathrm{Im2a/Im3a}}.
\label{eq:fimprglbound}
\eeq
Strictly speaking, there are also contributions proportional to
$\ct^{(1)}$ from the contractions of the time--like link variables at the
boundary with the gluons in~(\ref{eq:stot1}), depicted by diagrams~Im4a,
Im4b, Im5a, and Im5b in figure~\ref{fig:fapimct2}. 
But, like the diagrams~9a and~9b, these contributions are of
opposite sign and thus cancel in the sum.

The $\ct^{(1)}$--term was neglected in the computation of $\csw^{(1)}$ 
in~\cite{Weisz:1996csw}. This is permissible, because in the
current mass, the boundary
counter-terms only contribute at order $a^2$. This is due to the fact 
that~(\ref{eq:pcacop}) remains valid even if the boundary counter-terms
$\delta S_{\mathrm{G,b}}$ and $\delta S_{\mathrm{F,b}}$ are dropped. 
Further explanations concerning this matter can be found in~\cite{Luscher:1996sc}.
Furthermore, this term was missing in the perturbative results for the
discretisation errors in~\cite{Bode:2001jv}. To which extent this term 
changes the results published there will be discussed in 
chapter~\ref{chap:results}.

A further improvement term only present in $\fa$ 
arises from the $\rmO(a)$ correction of the axial
current~(\ref{eq:Aimproved}). Insertion of~(\ref{eq:deltaA}) into the correlation
function yields
\beq
f_{\delta A}^{(1)}(x_0) = \frac{1}{2}\ca^{(1)}(\partial_0^{\ast}+\partial_0)
\fp^{(0)}(x_0).
\eeq

\section{Calculation of the critical quark mass and the lattice artefacts
         \label{sec:critmass}}
\sectionmark{Calculation of the critical quark mass and the lattice artefacts}

With the one loop expansion of $\fa$ and $\fp$, one is now able to compute
the critical quark mass at 1--loop order. To this end, the tree level and 
1--loop coefficients of $m_1$ have to be computed at a given $m_0$.
From~(\ref{eq:mdef}) one gets the $x_0$ dependent mass $m(x_0)$ at tree level by
\beq
m^{(0)}(x_0) = \frac{(\partial_0^{\ast}+\partial_0)\fa^{(0)}(x_0)}{4\fp^{(0)}(x_0)},
\eeq
while at 1--loop level it is given by
\beqn
m^{(1)}(x_0) &=& \frac{(\partial_0^{\ast}+\partial_0)\fa^{(1)}(x_0)}{4\fp^{(0)}(x_0)}
-\fp^{(1)}(x_0)\frac{(\partial_0^{\ast}+\partial_0)\fa^{(0)}(x_0)}{4\fp^{(0)2}(x_0)}
\nonumber\\
& &
+c_{\mathrm{A}}^{(1)}\frac{\partial_0^{\ast}\partial_0 \fp^{(0)}(x_0)}
{2\fp^{(0)}(x_0)},
\eeqn
with $\fa^{(1)}$ and $\fp^{(1)}$ including the volume and boundary improvement 
terms (\ref{eq:fimprvol}), (\ref{eq:fimprbound}), and~(\ref{eq:fimprglbound}).
The current quark mass $m_1$ is then computed at tree and 1--loop level by
taking $m^{(0)}$ and $m^{(1)}$ in the centre of the lattice according 
to~(\ref{eq:m1def}). Of course, 
the expansion coefficients of $\fa$ and $\fp$ depend on
the bare quark mass at which they are computed. Thus one gets an expansion
\beq
m_1 = m_1^{(0)}(m_0) + m_1^{(1)}(m_0)g_0^2 + \rmO(g_0^4),
\eeq
where the expansion coefficients 
depend on the bare quark mass $m_0$.
To set up perturbation theory for the critical mass, 
we consider $m_0$ also to be a series,
\beq
m_0 = m_0^{(0)} + m_0^{(1)}g_0^2 + \rmO(g_0^4),
\eeq
and expand $m_1$ further as
\beq
m_1 = m_1^{(0)}\left(m_0^{(0)}\right) + \Biggl[m_1^{(1)}\left(m_0^{(0)}\right)
 + m_0^{(1)}\frac{\partial}{\partial m_0}m_1^{(0)}\left(m_0^{(0)}\right)
\Biggr]g_0^2 + \rmO(g_0^4).
\eeq
Therefore, the computation of $\mcrit^{(1)}$ has to be done in two steps. First we
compute $\mcrit^{(0)}$ by requiring
\beq
m_1^{(0)}\left(\mcrit^{(0)}\right) = 0,
\eeq
then we can determine $\mcrit^{(1)}$ from
\beq
m_1^{(1)}\left(\mcrit^{(0)}\right) 
+ \mcrit^{(1)}\frac{\partial}{\partial m_0}m_1^{(0)}\left(\mcrit^{(0)}\right) = 0.
\eeq
The first step is easily done numerically using a discretised version of the
Newton--Raphson method. The second step mainly amounts to expanding $\fa$ and
$\fp$ up to 1--loop order as described in the previous subsection. Results for
$\mcrit^{(0)}$ and $\mcrit^{(1)}$ can be found in chapter~\ref{chap:results}.

The lattice artefact $e$ may now be computed at tree and 1--loop level. In lattice
units, $e$ is given by
\beq
e(L) = m_1(2L) - m_1(L),
\eeq
where $m_1$ is to be taken at the critical bare mass at lattice size $L$, which
means that the bare mass is chosen such that
$m_1(L) = 0$. So, in order to get $e(L)$, 
one first has to compute the critical mass at lattice
size $L$ and then, using this bare mass, compute $m_1$ on a lattice twice as
large. The results for $e$ are shown in chapter~\ref{chap:results}.

The lattice artefact 
\beq
d(L) = m_2(L) - m_1(L)
\eeq
remains to be computed. To this end, one has to expand the functions $\fa'$ and
$\fp'$ up to 1--loop order. However, this turns out to be an easy task. It is
straightforward to show that $\fa'$ and $\fp'$ are the same as $\fa$ and $\fp$ 
with inverted background field. This means, in order to compute $m_2(L)$, one only
has to compute $m_1$ and then repeat the computation after exchanging the
boundary values $C$ and $C'$. (One has of course to take care of the $\phi_a(x_0)$
in the propagators and vertices, which depend crucially on the background field.)
The results for $d$ can be found in chapter~\ref{chap:results}.

\section{Numerical computation of $\fa^{(1)}$ and $\fp^{(1)}$}
\sectionmark{Numerical computation of $\fa^{(1)}$ and $\fp^{(1)}$}

Since most of the computation merely consists of matrix multiplications,
the structure of the program is rather simple. However, due to the 
complicated formulae for the propagators and vertices, there is a big risk
of errors. For this reason, a careful check of the results is necessary. To this
end, the gauge parameter $\lambda_0$ was left arbitrary, and it was checked
that the results do not depend on $\lambda_0$. While this only gives a
check on the total result, there is a possibility to check at least those diagrams
separately that come in two types, labelled a and b here. 
Due to the symmetries of the propagators
and vertices, these two types have to give the same results. The results using
the tree level improvement coefficients as well as the gluon boundary 
term have been compared to Monte Carlo results obtained by Juri Rolf.
Finally, two sets of programs have been written independently, 
one by Peter Weisz and one by the
author of this thesis, and the results have been compared.  

While the general structure of the program is relatively simple, the computation
of the propagators is technically somewhat more involved.
In contrast to the case of a vanishing background field, the propagators for
the gluon, ghost, and fermion fields cannot be computed analytically. 
For this reason, 
a numerical calculation is required. Of course it would be possible to invert
the operators $K^a$, $F^a$, and $D$ by a simple inversion routine. However, on
large lattices, this would mean inverting very large matrices, which would 
take a large amount of computer time. A more convenient way to compute the
propagators is a procedure using Wronskian forms described 
in~\cite{Narayanan:1995ex} for the gluon and ghost propagators and 
in~\cite{Bode:1999sm} for the fermions. This method, which is briefly outlined
in appendix~\ref{app:prop}, was used in the computation of $\fa$ and $\fp$
at 1--loop order. However, even with this method, computer
time is far from being negligible. On a 200 MHz Pentium PC, the computation
at $L=16$ took 16 hours of CPU time. Even on a \mbox{900 MHz PC}, lattices with
$L>32$ could not be computed in reasonable time. Since one
has to sum over three momentum components and two vertex times in the
diagrams, the time needed
scales asymptotically with $L^5$. This means that larger lattices up to
$L=64$ seem to be beyond reach.

\cleardoublepage
\chapter{The renormalised quark mass\label{chapt:renmass}}
\chaptermark{The renormalised quark mass}

\section{The renormalisation constant $\! Z_{\mathrm{P}}$ and its 
step scaling function}
\sectionmark{The renormalisation constant $Z_{\mathrm{P}}$ and its step scaling function}

\subsection{The renormalisation constant $Z_{\mathrm{P}}$}

\begin{figure}
  \noindent
  \begin{center}
    \includegraphics[width=0.3\linewidth]{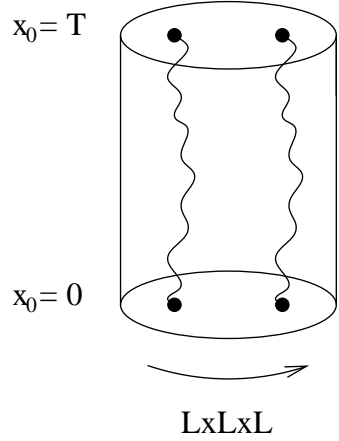}
  \end{center}
  \caption[The correlation function $f_1$]
          {\label{fig:cyl_f1}
           \sl The correlation function $f_1$}
\end{figure}

In chapter~\ref{chapt:curr_mass}, the renormalised quark mass defined by
the PCAC relation was found to be related to the current quark mass 
via~(\ref{eq:renmass}).
The actual renormalisation is done by renormalising the axial current and
density, leading to a running mass, since the renormalisation constants
depend on the renormalisation scale. This is, however, not true for 
$Z_{\mathrm{A}}$,
which is only present due to the violation of the chiral symmetry
and can be shown to be finite and scale independent.
Thus the renormalisation constant that actually makes the running 
is $Z_{\mathrm{P}}$.
In order to compute $Z_{\mathrm{P}}$, one may use the correlation function
$\fp$. However, in $\fp$, one does not only have to renormalise the pseudo
scalar density, but also the boundary fields $\zeta$ and $\bar{\zeta}$, such
that the renormalised correlation function becomes
\beq
(\fp)_{\mathrm{R}}(x_0)
= Z_{\mathrm{P}}Z_{\zeta}^{2}\fp(x_0).
\eeq
In order to cancel the unwelcome renormalisation factor $Z_{\zeta}^{2}$,
one may use the correlation function $f_1$, defined by
\beq
f_1 = -\frac{a^{12}}{L^6}\sum_{\vu,\vv,\vy,\vz}
\frac{1}{3}\left\langle\bar{\zeta}'(\vu)\gamma_5\frac{1}{2}\tau^a\zeta'(\vv)
\bar{\zeta}(\vy)\gamma_5\frac{1}{2}\tau^a\zeta(\vz)\right\rangle,
\eeq
which is renormalised by $Z_{\zeta}$ only,
\beq
(f_1)_{\mathrm{R}} = Z_{\zeta}^{4} f_1.
\eeq
This correlation function is proportional to the probability amplitude that
a quark antiquark pair created at $x_0=0$ propagates to $x_0=T$, a situation
which may be depicted by figure~\ref{fig:cyl_f1}. Like $\fa$ and $\fp$, 
$f_1$ may be expanded in powers of the bare coupling,
\beq
f_1 = f_1^{(0)} +g_0^2 f_1^{(1)} +\rmO(g_0^4).
\eeq

Now, $Z_{\mathrm{P}}^2$ can be computed as
\beq
Z_{\mathrm{P}}^2 = \frac{\left[\fp^{(0)}(T/2)\right]^2}{f_1^{(0)}}
\frac{f_1}{\left[\fp(T/2)\right]^2},
\label{eq:ZPdef}
\eeq
where the normalisation is chosen such that $Z_{\mathrm{P}}$ is 1 at 
tree level. 

The renormalisation constant $Z_{\mathrm{P}}$ may be expanded in powers of the bare
coupling,
\beq
Z_{\mathrm{P}}(g_0,L/a) = 1 + \sum_{k=1}^{\infty}Z_{\mathrm{P}}^{(k)}(L/a)g_0^{2k}.
\eeq
The 1--loop coefficient $Z_{\mathrm{P}}^{(1)}$ contains a logarithmic divergence. It may
hence be decomposed into a logarithmic term, a constant term, and terms that
vanish in the continuum limit,
\beq
Z_{\mathrm{P}}^{(1)} = \cf z_{\mathrm{p}}(\theta,\rho)-d_0\ln(L/a)+\rmO(a/L).
\label{eq:ZPoneldiv}
\eeq
Here, $z_{\mathrm{p}}$ denotes the term independent of $L/a$. It only depends on the phase
$\theta$ and on the ratio of the lattice extensions in time and space directions,
$\rho=T/L$.

\subsection{The step scaling function of the running mass}

In analogy to the coupling, the running of the mass may be described by a step
scaling function, giving the change of the running mass when changing the scale
$L$ by a certain scale factor. In the following, this scale factor is chosen
to be $2$, which is a convention widely used in numerical computations.
Since the running of the renormalised mass is completely described by the
running of $Z_{\mathrm{P}}$, a suitable definition of the step scaling function 
at vanishing renormalised quark mass is~\cite{Jansen:1996ck}
\beq
\Sigma_{\mathrm{P}}(\gren^2,a/L) = \left.\frac{Z_{\mathrm{P}}(g_0,2L/a)}{Z_{\mathrm{P}}(g_0,L/a)}\right|_{g_0=g_0(\gren,L/a)},
\eeq
with the continuum limit
\beq
\lim_{a/L\rightarrow 0}\Sigma_{\mathrm{P}}(\gren^2,a/L) = 
\sigma_{\mathrm{P}}(\gren^2).
\eeq
As the renormalised coupling, one here has to take the Schr\"{o}dinger 
functional coupling at length scale $L$, i.e.~$\gren=\bar{g}(L)$.
Like the step scaling function $\sigma$ is used as a discretised version of
the $\beta$--function, the step scaling function $\sigma_{\mathrm{P}}$ serves
as a discretised $\tau$--function.

The step scaling function may be expanded in perturbation theory,
\beq
\Sigma_{\mathrm{P}}(\gren^2,a/L) = 1 + k(L/a)\gren^2 + \rmO(\gren^4).
\eeq
Since the tree level value of $Z_{\mathrm{P}}$ is 1 at all scales, the 1--loop coefficient
$k(L/a)$ is simply given by
\beq
k(L/a) = Z_{\mathrm{P}}^{(1)}(2L/a) - Z_{\mathrm{P}}^{(1)}(L/a),
\eeq
where $Z_{\mathrm{P}}^{(1)}(L/a)$ and $Z_{\mathrm{P}}^{(1)}(2L/a)$ have to be taken at the same bare
quark mass. In the case of a vanishing physical quark mass, this is the critical
mass at length scale $L$. For the continuum limit, one gets 
from~(\ref{eq:ZPoneldiv})
\beq
k(\infty) = -d_0\ln(2).
\eeq
Like in the case of the renormalised coupling, one has lattice artefacts making
the step scaling function deviate from its continuum limit. These lattice 
artefacts can be estimated at 1--loop order by
\beq
\delta_k(L/a) = \frac{k(L/a)}{k(\infty)}-1.
\label{eq:massdelta}
\eeq
In~\cite{Sint:1998iq}, these lattice artefacts have been computed for several
choices of $\theta$ and $\rho$ in the case of a vanishing background field.
To do this calculation with a non vanishing background field is one of the
main aims of this thesis. The size of the lattice artefacts is also important
for the question whether it makes sense to compute the renormalised quark mass
non--perturbatively with a non--vanishing background field. So far, this
calculation has only be done at zero background field~\cite{Capitani:1998mq}.

\subsection{The 2--loop anomalous dimension}

It was already shown in subsection~\ref{subsect:finren} that one may obtain
the 2--loop anomalous dimension $d_1$ in the Schr\"{o}dinger functional scheme
from a 1--loop computation by using~(\ref{eq:andimswitch}) to convert the
known anomalous dimension in the $\msbar$--scheme to the Schr\"{o}dinger
functional scheme. In order to do this, one has to convert the coupling and
the masses. For the coupling, one has
\beq
\gsf^2(L) = \gbar^2_{\msbar}(q)\chig(\gbar_{\msbar}(q),qL),
\eeq
where the 1--loop coefficient of $\chig$ is given by
\beq
\chig^{(1)}=2b_0 \ln(qL) -\frac{1}{4\pi}(c_{1,0}+c_{1,1}\Nf).
\label{eq:chig}
\eeq
The coefficient $c_{1,0}$ has been computed in~\cite{Luscher:1994gh} and
$c_{1,1}$ in~\cite{Sint:1996ch}. For $\theta=\pi/5$ and $\rho=1$, one
obtains
\beq
c_{1,0} = 1.25563(4),\qquad c_{1,1} = 0.039863(2).
\eeq
In order to get the conversion of the renormalised mass from the $\msbar$--scheme
to the Schr\"{o}dinger functional scheme, one has to use the relation between
the renormalised mass and the subtracted bare mass. In the $\msbar$--scheme, it
has been computed in~\cite{Gabrielli:1991us} and since then been verified
numerous times. It turns out to be
\beq
\mbar_{\msbar}(q)=\mq\left\{1+g_0^2\Bigl[-d_0\ln(aq)+0.122282(1)\cf\Bigr]
+\mathrm{O}(g_0^4)\right\}.
\eeq
By calculating the renormalised mass in the Schr\"{o}dinger functional at 
a non--vanishing mass, one gets~\cite{Sint:1997jx}
\beq
\mbar_{\mathrm{SF}}(L) = \mq\left\{1 +g_0^2\Bigl[Z_{\mathrm{A}}^{(1)}-Z_{\mathrm{P}}^{(1)}(L/a)
+0.067886(1)\cf\Bigr]+\mathrm{O}(g_0^4)\right\}.
\eeq
Inserting the results for $Z_{\mathrm{A}}^{(1)}$ from~(\ref{eq:ZA1}) and $Z_{\mathrm{P}}^{(1)}$ 
from~(\ref{eq:ZPoneldiv}) then yields
\beq
\mbar_{\mathrm{SF}}(L) = \mq\left\{1 +g_0^2\Bigl[d_0\ln(L/a)
-\bigl(z_{\mathrm{p}}+0.019458(1)\bigr)\cf\Bigr]+\mathrm{O}(g_0^4)\right\}.
\eeq
Using these results for the masses in the $\msbar$ scheme and in the 
Schr\"{o}dinger functional scheme, one finds the relation between both
schemes,
\beq
\mbar_{\mathrm{SF}}(L) = \mbar_{\msbar}(q)\left\{1
 + g_0^2\Bigl[d_0\ln(qL) -\bigl(z_{\mathrm{p}} + 0.141740(2)\bigr)\cf\Bigr] 
+\rmO(g_0^4)\right\}.
\eeq
From this relation, one may easily read off the 1--loop coefficient 
$\chim^{(1)}$,
\beq
\chim^{(1)}=d_0\ln(qL)-\bigl(z_{\mathrm{p}}+0.141740(2)\bigr)\cf.
\label{eq:chim}
\eeq
Now, one can compute the 2--loop anomalous dimension in the Schr\"{o}dinger
functional scheme. From~(\ref{eq:andimswitch}), one has
\beq
d_1^{\mathrm{SF}} = d_1^{\msbar} -d_0\chig^{(1)} +2b_0\chim^{(1)},
\label{eq:d1sf}
\eeq
with $b_0$, $d_0$, and $d_1^{\msbar}$ from~(\ref{eq:b0}),
(\ref{eq:d0}), and~(\ref{eq:d1msbar}), and $\chig^{(1)}$ and $\chim^{(1)}$
from~(\ref{eq:chig}) and~(\ref{eq:chim}).

\section{The renormalisation constant $Z_{\mathrm{P}}$ at 1--loop order}
\sectionmark{The renormalisation constant $Z_{\mathrm{P}}$ at 1--loop order}

\subsection{Preliminaries}
The expansion of $f_1$ at 1--loop order has been outlined in~\cite{Sint:1997jx}
for the case of a vanishing background field. In the following, the calculation
will be done for the non--zero background field used to define the coupling.

\begin{figure}
  \noindent
  \begin{center}
  \begin{minipage}[b]{.3\linewidth}
     \centering\includegraphics[width=.8\linewidth]{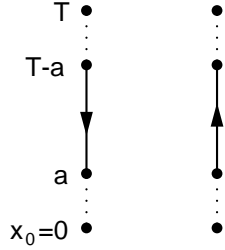}
  \end{minipage}
  \end{center}
  \caption[Diagram for $f_1$ at tree level]
          {\sl Diagram for $f_1$ at tree level\label{fig:f1tree}}
\end{figure}

In order to make contact to the calculation of $\fa$ and $\fp$ and thus simplify
the computation, it is useful to rewrite $f_1$ in a slightly different form.
To this end, we define
\beq
K = \ctt\frac{a^3}{L^3}\sum_{\vx}\left.\biggl\{
P_+ U(x,0)^{-1}H(x)\biggr\}\right|_{x_0=T-a}.
\label{eq:Kdef}
\eeq
Using this notation, $f_1$ may be written as
\beq
f_1 = \frac{1}{2}\left\langle\Tr\{K^{\dagger}K\}\right\rangle_{\rmg}.
\label{eq:f1withK}
\eeq
The expansion of $f_1$ thus amounts to expanding $K$. This makes part of the
calculation exceptionally easy, because all contributions in which the link
variables only appear at tree level may be reduced to $\fa$ and $\fp$, so
these diagrams may be computed using the results of chapter~\ref{chapt:curr_mass}.
This is easily seen at tree level, where one has
\beq
K^{(0)} = \frac{a^3}{L^3}\sum_{\vx}\left.\biggl\{
P_+ H^{(0)}(x)\biggr\}\right|_{x_0=T-a}.
\eeq
The clou to reducing $f_1$ to $\fa$ and $\fp$ lies in the fact that $H^{(0)}(x)$
does not depend on the spatial position $\vx$. This means that the sum over
$\vx$ is trivial, giving
\beq
K^{(0)}= P_+ \left. H^{(0)}(x)\right|_{x_0=T-a},
\label{eq:Ktree}
\eeq
or, inserted in~(\ref{eq:f1withK})
\beq
f_1^{(0)}= \frac{1}{2}\left.\Tr\Bigl\{H^{(0)\dagger}(x)
P_+ H^{(0)}(x)\Bigr\}\right|_{x_0=T-a}.
\eeq
Comparing this expression to~(\ref{eq:fAwithH}) and~(\ref{eq:fPwithH}) then
yields
\beq
f_1^{(0)}=\frac{1}{2}\left[\fp^{(0)}(T-a)-\fa^{(0)}(T-a)\right],
\label{eq:f1tree}
\eeq 
which may be decomposed into colour components,
\beq
f_{1\alpha}^{(0)}=\frac{1}{2}\left[f_{\mathrm{P}\alpha}^{(0)}(T-a)
-f_{\mathrm{A}\alpha}^{(0)}(T-a)\right].
\eeq 
The same will apply for several diagrams at 1--loop order. At tree level, $f_1$
may be visualised by the diagram in figure~\ref{fig:f1tree}.

\subsection{The correlation function $f_1$ at 1--loop order}
For simplicity, we return to lattice units and take
$a=1$ for the rest of this chapter.

At 1--loop order, $f_1$ can, just like $\fa$ and $\fp$, be written as
a sum over diagrams, improvement terms, and the mass derivative,
\beqn
f_1^{(1)} &=&\sum_{n} f^{(1)}_{1,n} +\csw^{(1)}f^{(1)}_{1,\mathrm{V}}
\nonumber\\
& & +\ctt^{(1)}\left(f^{(1)}_{1,\mathrm{Fb}}+4 f_1^{(0)}\right)
+\ct^{(1)}f^{(1)}_{1,\mathrm{Gb}}\nonumber\\
& & +\mcrit^{(1)}\frac{\partial}{\partial m_0}f_1^{(0)}.
\eeqn
The term $4 f_1^{(0)}$ is there because $f_1$ gets an overall factor $\ctt^4$
from $K$ and $H$. However, this term may be neglected, since it is cancelled
by the overall factor $\ctt^2$ of $\fp$.

For the 1--loop diagrams, one has to expand the link variables and $H$ 
in~(\ref{eq:Kdef}) up to order $g_0^2$. It was already stated that the diagrams
containing the tree level values of the link variables are obviously related
to the corresponding diagrams for $\fa$ and $\fp$. These diagrams either 
contain a combination of $H^{(0)}$ and $H^{(2)}$ or of $H^{(1)}$ and $H^{(1)}$.
The expression becomes particularly simple for the first case since $H^{(0)}$
is spatially constant and does not depend on the gluon fields. The gluonic
expectation value is thus only to be taken over $H^{(2)}$. This expectation
value $\langle H^{(2)}(x)\rangle_{0}$ is again spatially constant, 
which means that
the same argument may be used as in the tree level case. One thus gets 
for $n=1a,1b,3a,3b,5a,5b,6a,6b$
\beq
f_{1,n}^{(1)} = \frac{1}{2}\Bigl[ f_{\mathrm{P},n}^{(1)}(T-1) - f_{\mathrm{A},n}^{(1)}(T-1)\Bigr].
\eeq
\begin{figure}
  \noindent
  \begin{center}
  \begin{minipage}[b]{.3\linewidth}
     \centering\includegraphics[width=.8\linewidth]{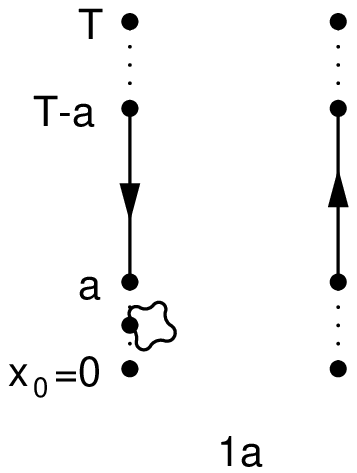}
  \end{minipage}
  \begin{minipage}[b]{.3\linewidth}
     \centering\includegraphics[width=.8\linewidth]{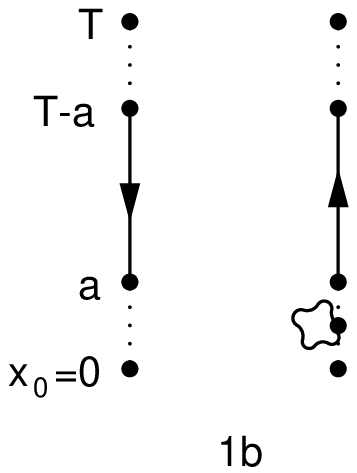}
  \end{minipage}
  \begin{minipage}[b]{.3\linewidth}
     \centering\includegraphics[width=.8\linewidth]{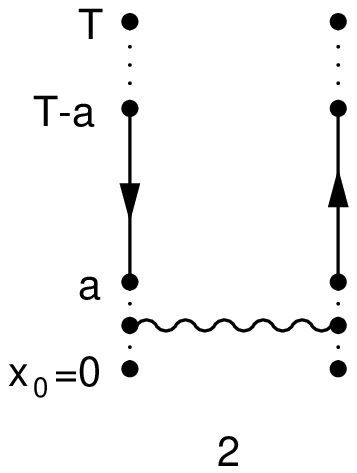}
  \end{minipage}\\
  \begin{minipage}[b]{.3\linewidth}
     \centering\includegraphics[width=.8\linewidth]{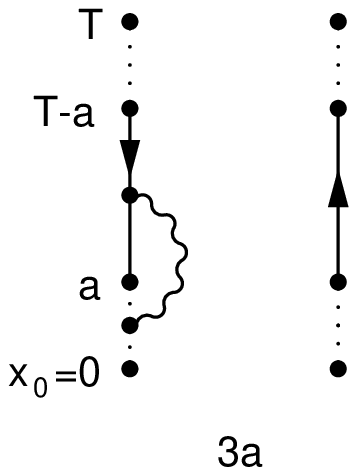}
  \end{minipage}
  \begin{minipage}[b]{.3\linewidth}
     \centering\includegraphics[width=.8\linewidth]{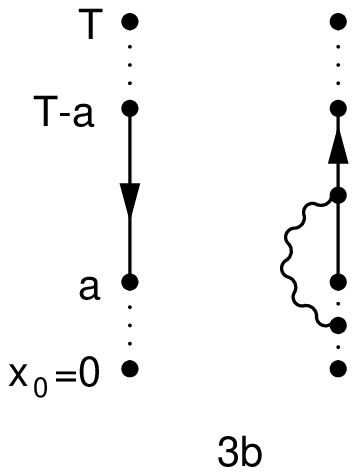}
  \end{minipage}
  \begin{minipage}[b]{.3\linewidth}
     \centering\includegraphics[width=.8\linewidth]{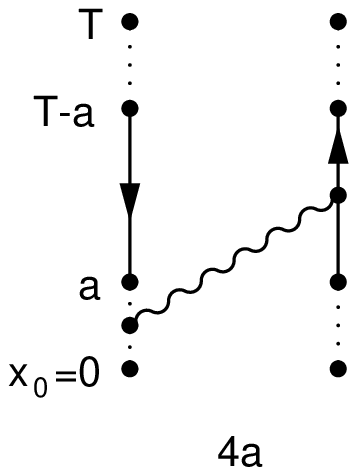}
  \end{minipage}\\
  \begin{minipage}[b]{.3\linewidth}
     \centering\includegraphics[width=.8\linewidth]{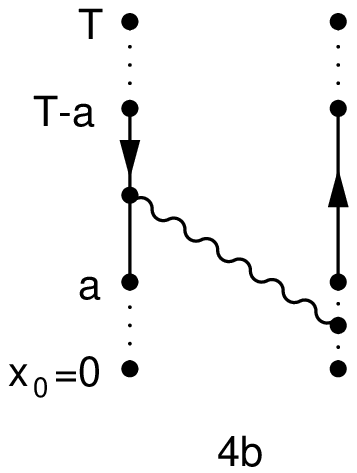}
  \end{minipage}
  \begin{minipage}[b]{.3\linewidth}
     \centering\includegraphics[width=.8\linewidth]{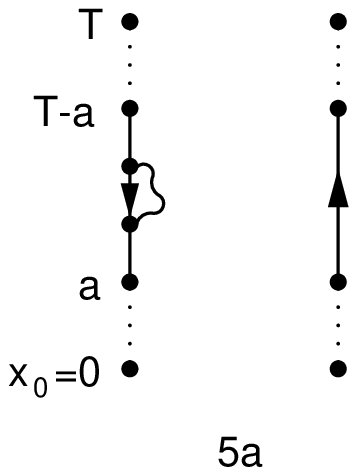}
  \end{minipage}
  \begin{minipage}[b]{.3\linewidth}
     \centering\includegraphics[width=.8\linewidth]{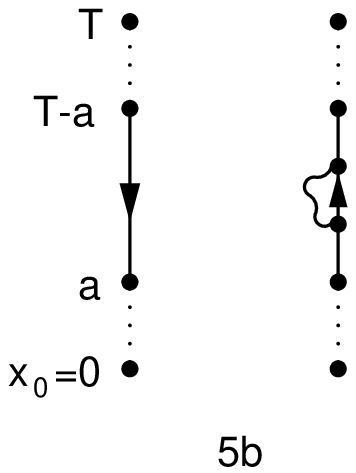}
  \end{minipage}\\
  \begin{minipage}[b]{.3\linewidth}
     \centering\includegraphics[width=.8\linewidth]{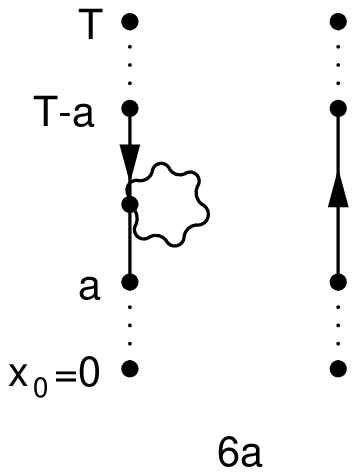}
  \end{minipage}
  \begin{minipage}[b]{.3\linewidth}
     \centering\includegraphics[width=.8\linewidth]{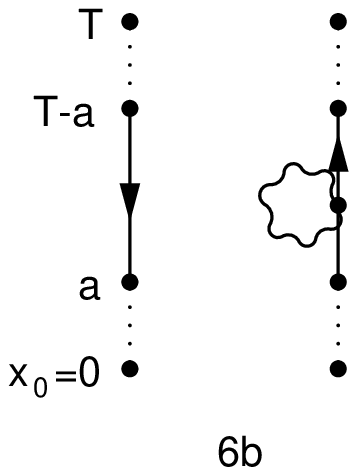}
  \end{minipage}
  \begin{minipage}[b]{.3\linewidth}
     \centering\includegraphics[width=.8\linewidth]{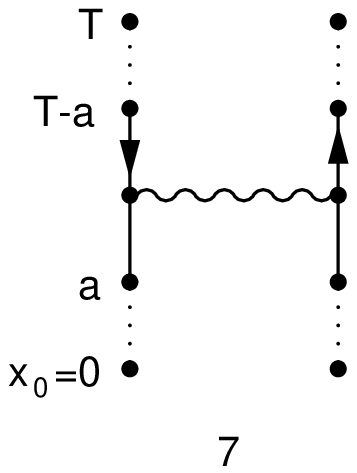}
  \end{minipage}
  \end{center}
  \caption[Diagrams contributing to $f_1$ at 1--loop order of perturbation theory]
          {\sl Diagrams contributing to $f_1$ at
           1--loop order of perturbation theory.\label{fig:f1onel1}}
\end{figure}

For the combination of $H^{(1)}$ with $H^{(1)}$, the situation is slightly more
complicated. Here, the expectation value 
$\langle H^{(1)}(x)^{\dagger}H^{(1)}(x)\rangle_0$ does depend on $\vx$ and the
sum has a non trivial effect. When transforming the expectation value to
the momentum representation, the sum together with the exponential factor
acts as a delta function, leaving only 
the contribution with zero loop momentum. One thus gets for 
$n=2,4a,4b,7$ 
\beq
f_{1,n}^{(1)} = \frac{1}{2}\Bigl[ f_{\mathrm{P},n}^{(1)}(T-1) - f_{\mathrm{A},n}^{(1)}(T-1)
\Bigr]_{\vq=\vo},
\eeq
where $[\ldots ]_{\vq=\vo}$ means that, instead of the sum over $\vq$ 
in~(\ref{eq:fapdiagram2}), (\ref{eq:fapdiagram4a}), (\ref{eq:fapdiagram4b}),
and~(\ref{eq:fapdiagram7}), one has to take the term with $\vq=\vo$ only. 
The 1--loop terms covered so far are represented by the diagrams in
figure~\ref{fig:f1onel1}.

For the tadpole diagrams, one finds that 
$\langle H^{(1)}(x)[S_{\mathrm{tot}}^{(1)}]_{\mathrm{F}}\rangle_{0}$
is independent of $\vx$. One thus gets
\beq
f_{1,n}^{(1)} = \frac{1}{2}\Bigl[ f_{\mathrm{P},n}^{(1)}(T-1) - f_{\mathrm{A},n}^{(1)}(T-1)\Bigr]
\eeq
for $n=8a,8b,9a,9b$.

\begin{figure}
  \noindent
  \begin{center}
  \begin{minipage}[b]{.3\linewidth}
     \centering\includegraphics[width=\linewidth]{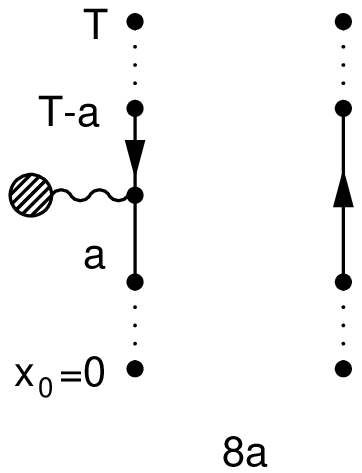}
  \end{minipage}
  \begin{minipage}[b]{.3\linewidth}
     \centering\includegraphics[width=\linewidth]{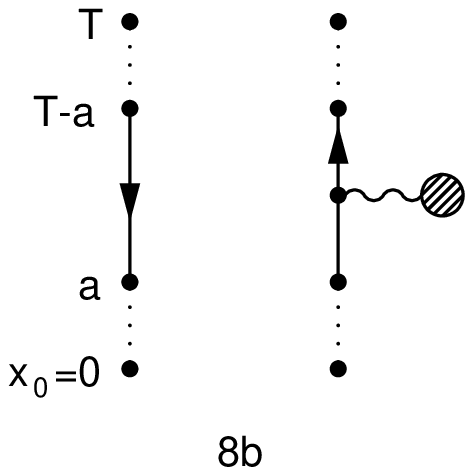}
  \end{minipage}
  \begin{minipage}[b]{.3\linewidth}
     \centering\includegraphics[width=\linewidth]{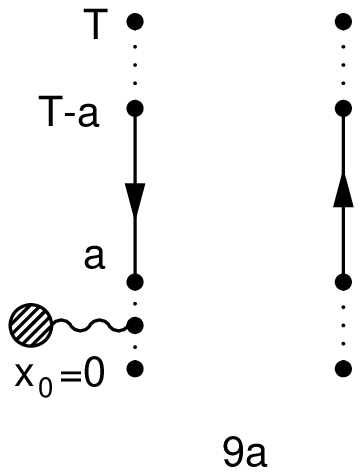}
  \end{minipage}\\
  \begin{minipage}[b]{.3\linewidth}
     \centering\includegraphics[width=\linewidth]{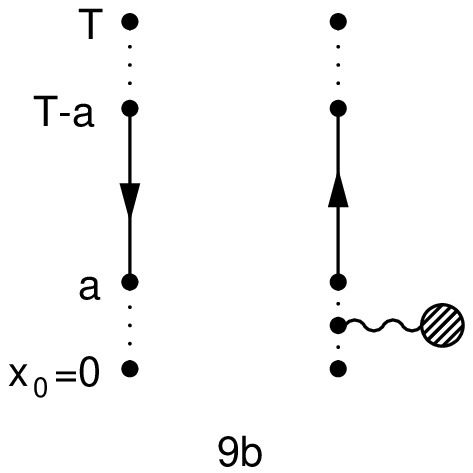}
  \end{minipage}
  \begin{minipage}[b]{.3\linewidth}
     \centering\includegraphics[width=.8\linewidth]{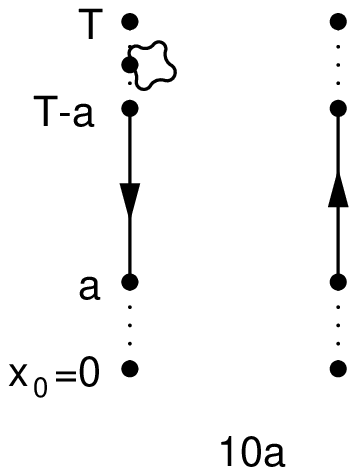}
  \end{minipage}
  \begin{minipage}[b]{.3\linewidth}
     \centering\includegraphics[width=.8\linewidth]{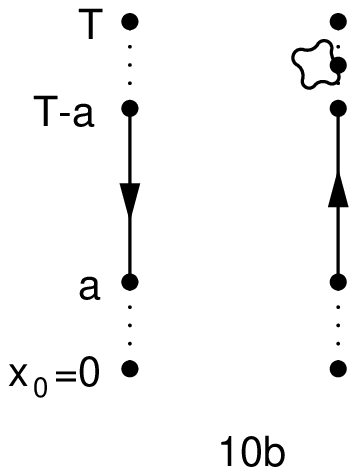}
  \end{minipage}\\
  \begin{minipage}[b]{.3\linewidth}
     \centering\includegraphics[width=.8\linewidth]{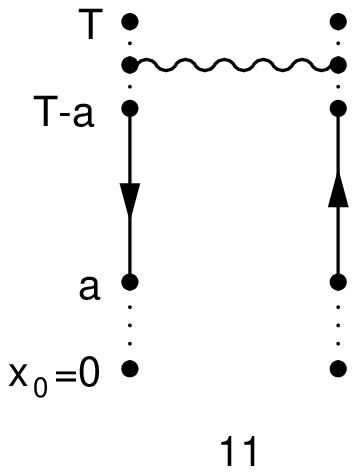}
  \end{minipage}
  \begin{minipage}[b]{.3\linewidth}
     \centering\includegraphics[width=.8\linewidth]{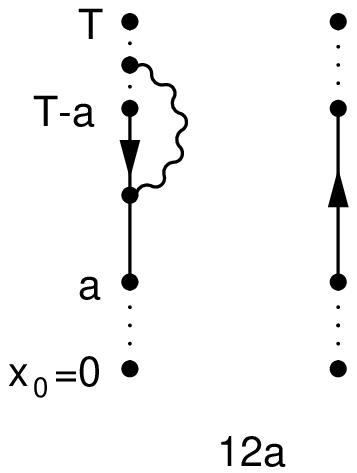}
  \end{minipage}
  \begin{minipage}[b]{.3\linewidth}
     \centering\includegraphics[width=.8\linewidth]{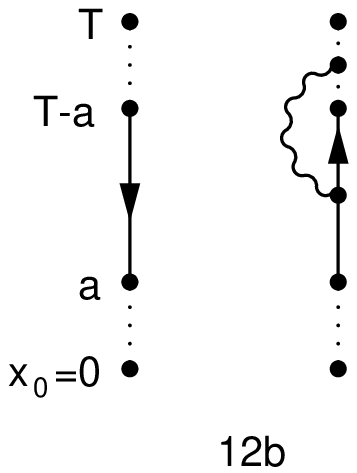}
  \end{minipage}\\
  \begin{minipage}[b]{.3\linewidth}
     \centering\includegraphics[width=.8\linewidth]{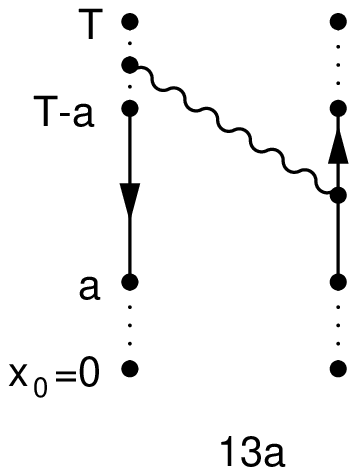}
  \end{minipage}
  \begin{minipage}[b]{.3\linewidth}
     \centering\includegraphics[width=.8\linewidth]{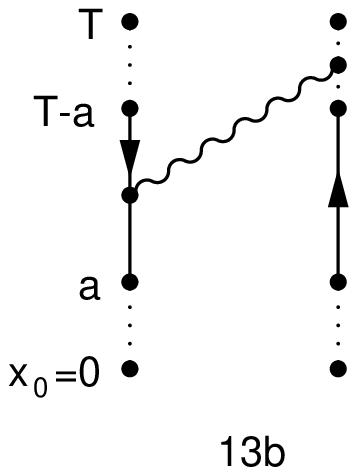}
  \end{minipage}
  \end{center}
  \caption[Further diagrams contributing to $f_1$
           at 1--loop order of perturbation theory]
          {\sl Further diagrams contributing to $f_1$
           at 1--loop order of perturbation theory.\label{fig:f1onel2}}
\end{figure}

Now one is left with the diagrams containing the link variables at the upper 
boundary of the box, i.e.~the link variables in~(\ref{eq:Kdef}),
at order $g_0$ and $g_0^2$. The $g_0^2$ terms can be treated in total analogy to
diagrams 1a and 1b, resulting in diagrams 10a and 10b of figure~\ref{fig:f1onel2}.
They are given by
\beq
f_{1,10a}^{(1)} = -\frac{1}{2}\frac{1}{L^3}\sum_{\alpha}f_{1\alpha}^{(0)}
\sum_{\vq}\sum_a D^a_{00}(\vq;T-1,T-1)\mathcal{C}^a_{\alpha},
\eeq
and 
\beq
f_{1,10b}^{(1)} = f_{1,10a}^{(1)}.
\eeq
Contracting the terms of both link variables proportional to $g_0$ yields 
diagram~11,
\beq
f_{1,11}^{(1)} = \frac{1}{L^3}\sum_{\alpha}f_{1\alpha}^{(0)}\sum_a
D^a_{00}(\vo;T-1,T-1)\mathcal{C}^a_{\alpha}.
\eeq

\begin{figure}
  \noindent
  \begin{center}
  \begin{minipage}[b]{.3\linewidth}
     \centering\includegraphics[width=.8\linewidth]{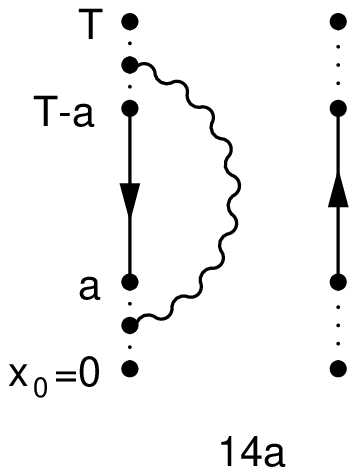}
  \end{minipage}
  \begin{minipage}[b]{.3\linewidth}
     \centering\includegraphics[width=.8\linewidth]{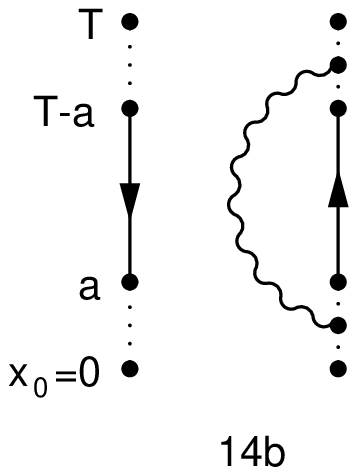}
  \end{minipage}
  \begin{minipage}[b]{.3\linewidth}
     \centering\includegraphics[width=.8\linewidth]{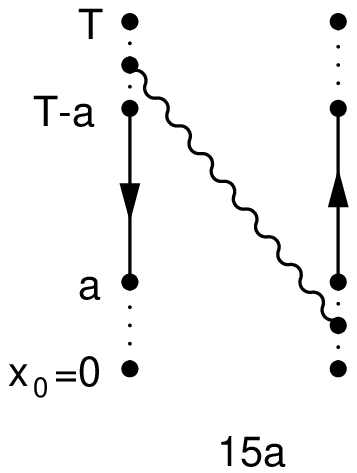}
  \end{minipage}\\
  \begin{minipage}[b]{.3\linewidth}
     \centering\includegraphics[width=.8\linewidth]{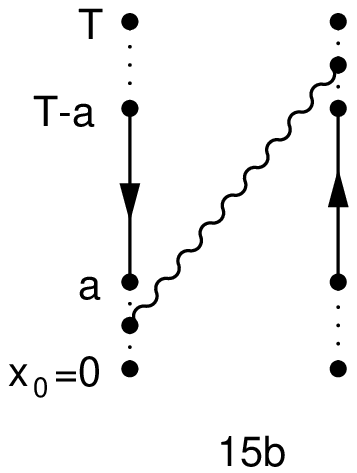}
  \end{minipage}
  \begin{minipage}[b]{.3\linewidth}
     \centering\includegraphics[width=\linewidth]{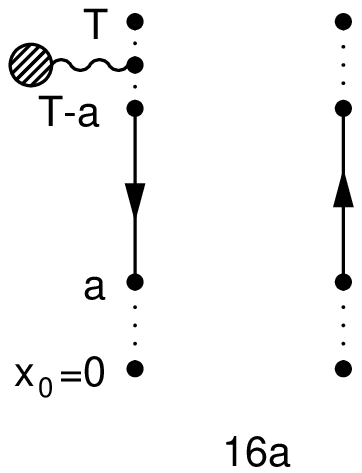}
  \end{minipage}
  \begin{minipage}[b]{.3\linewidth}
     \centering\includegraphics[width=\linewidth]{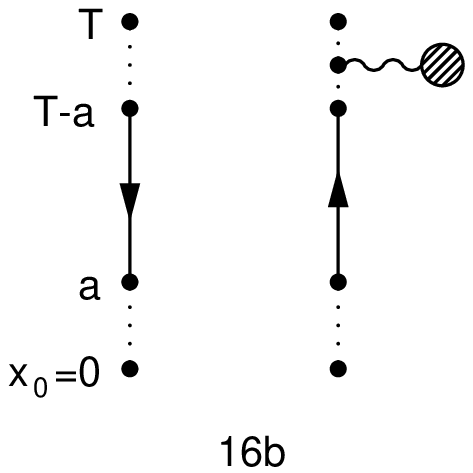}
  \end{minipage}
  \end{center}
  \caption[More diagrams contributing to $f_1$
           at 1--loop order of perturbation theory]
          {\sl More diagrams contributing to $f_1$
           at 1--loop order of perturbation theory.\label{fig:f1onel3}}
\end{figure}

Next, one has to contract the link variables at the upper boundary
with $H^{(1)}$ or, to be more precise,
with the first order terms of the propagators and link variables contained in
$H^{(1)}$.
Combining them with the quark lines on the same side gives diagrams 12a and 12b,
\beqn
f_{1,12a}^{(1)} &=& \frac{1}{2}\frac{1}{L^3}\sum_{\vq}\sum_{\mu}\sum_{s_0,t_0,u_0}
\sum_a D^a_{\mu 0}(\vq;u_0,T-1)
\Tr\Bigl\{\ppf S(\vo;1,T-1)\pfp \nonumber\\
& & \cdot I^{\bar{a}} S(\vq;T-1,s_0)
V^a_{\mu}(\vq,\vo,-\vq;s_0,t_0,u_0) S(\vo;t_0,1)\Bigr\},
\eeqn
\beqn
f_{1,12b}^{(1)} &=& -\frac{1}{2}\frac{1}{L^3}\sum_{\vq}\sum_{\mu}\sum_{s_0,t_0,u_0}
\sum_a D^a_{0\mu}(\vq;T-1,u_0)\Tr\Bigl\{\ppf S(\vo;1,s_0) \nonumber\\
& & \cdot V^{\bar{a}}_{\mu}(\vo,-\vq,\vq;s_0,t_0,u_0) S(\vq;t_0,T-1) I^a
\pfp S(\vo;T-1,1)\Bigr\},
\eeqn
while the combination with the quarks on the opposite side result in diagrams
13a and 13b,
\beqn
f_{1,13a}^{(1)} &=& \frac{1}{2}\frac{1}{L^3}\sum_{\mu}\sum_{s_0,t_0,u_0}
\sum_a D^a_{0\mu}(\vo;T-1,u_0)\Tr\Bigl\{\ppf S(\vo;1,s_0) \nonumber\\
& & \cdot V^{\bar{a}}_{\mu}(\vo,\vo,\vo;s_0,t_0,u_0) S(\vo;t_0,T-1) I^a
\pfp S(\vo;T-1,1)\Bigr\},
\eeqn
\beqn
f_{1,13b}^{(1)} &=& -\frac{1}{2}\frac{1}{L^3}\sum_{\mu}\sum_{s_0,t_0,u_0}
\sum_a D^a_{0\mu}(\vo;T-1,u_0)
\Tr\Bigl\{\ppf S(\vo;1,T-1)\pfp \nonumber\\
& & \cdot I^a S(\vo;T-1,s_0)
V^{\bar{a}}_{\mu}(\vo,\vo,\vo;s_0,t_0,u_0) S(\vo;t_0,1)\Bigr\}.
\eeqn
The remaining combinations are those of the link variables at the upper boundary
with the link variables at the lower boundary. They are depicted by the diagrams
in figure~\ref{fig:f1onel3} and can be computed via
\beqn
f_{1,14a}^{(1)} &=& \frac{1}{2}\frac{1}{L^3}\sum_{\vq}\sum_a
D^a_{00}(\vq;0,T-1)\Tr\Bigl\{\ppf S(\vo;1,T-1) \nonumber\\
& & \cdot\pfp I^{\bar{a}} S(\vq;T-1,1) I^a\Bigr\},
\eeqn
\beqn
f_{1,14b}^{(1)} &=& \frac{1}{2}\frac{1}{L^3}\sum_{\vq}\sum_a
D^a_{00}(\vq;T-1,0)\Tr\Bigl\{\ppf S(\vq;1,T-1) \nonumber\\
& & \cdot\pfp I^a S(\vo;T-1,1) I^{\bar{a}}\Bigr\},
\eeqn

\beqn
f_{1,15a}^{(1)} &=& -\frac{1}{2}\frac{1}{L^3}\sum_a
D^a_{00}(\vo;T-1,0)\Tr\Bigl\{\ppf S(\vo;1,T-1) \nonumber\\
& & \cdot\pfp I^a S(\vo;T-1,1) I^{\bar{a}}\Bigr\},
\eeqn
and
\beq
f_{1,15b}^{(1)} = f_{1,15a}^{(1)}.
\eeq
The only remaining diagrams are now the contractions of the first order terms
of the link variables at the upper boundary with the first order term of the
total action. In complete analogy to diagrams 9a and 9b one gets
\beqn
f_{1,16a}^{(1)} &=& -\frac{1}{2}\sum_{\mu}\sum_{u_0}\sum_a
D^a_{00}(\vo;T-1,u_0)\Tr\Bigl\{ \ppf S(\vo;1,T-1) \nonumber\\
& & \cdot\pfp I^a S(\vo;T-1,1)\Bigr\} T^{\bar{a}}_{\mu}(u_0),
\eeqn
and 
\beq
f_{1,16b}^{(1)} = -f_{1,16a}^{(1)}.
\eeq
So these diagrams cancel in the sum and leave diagrams 8a and 8b as the only
$\Nf$ dependent contributions to $f_1$.

\subsection{The improvement terms}

\begin{figure}
  \noindent
  \begin{center}
  \begin{minipage}[b]{.3\linewidth}
     \centering\includegraphics[width=.8\linewidth]{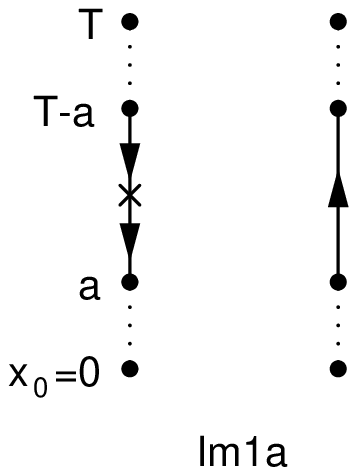}
  \end{minipage}
  \begin{minipage}[b]{.3\linewidth}
     \centering\includegraphics[width=.8\linewidth]{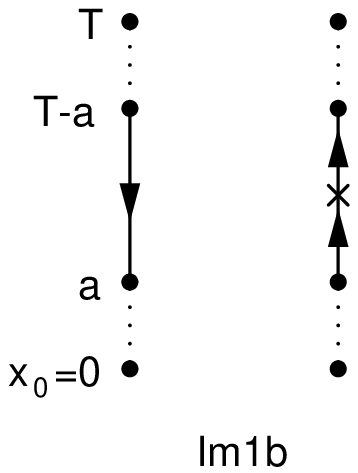}
  \end{minipage}
  \end{center}
  \caption[Diagrams for $f^{(1)}_{1,\mathrm{V}}$ and 
           $f^{(1)}_{1,\mathrm{Fb}}$]
          {\sl Diagrams for $f^{(1)}_{1,\mathrm{V}}$ and
           $f^{(1)}_{1,\mathrm{Fb}}$
           \label{fig:f1im}}
\end{figure}

In the computation of $f_1$, all improvement terms can be reduced 
to the corresponding terms
of $\fa$ and $\fp$. For the terms proportional to $\csw^{(1)}$ and $\ctt^{(1)}$,
which are additional contributions to diagrams~6a and~6b, one may use the same
argument applied there, namely the independence of $H^{(0)}(x)$  and
$\langle H^{(2)}(x)\rangle_{0}$ of the spatial position $\vx$. 
One thus gets
\beq
f_{1,\mathrm{V}}= \frac{1}{2}\left[f_{\mathrm{P,V}}(T-1)-f_{\mathrm{A,V}}(T-1)
\right]
\eeq
for the volume term and
\beq
f_{1,\mathrm{Fb}} = 
\frac{1}{2}\left[f_{\mathrm{P,Fb}}(T-1)-f_{\mathrm{A,Fb}}(T-1)\right]
\eeq
for the quark boundary term. These terms can be depicted by the diagrams
in figure~\ref{fig:f1im}. 

\begin{figure}
  \noindent
  \begin{center}
  \begin{minipage}[b]{.3\linewidth}
     \centering\includegraphics[width=\linewidth]{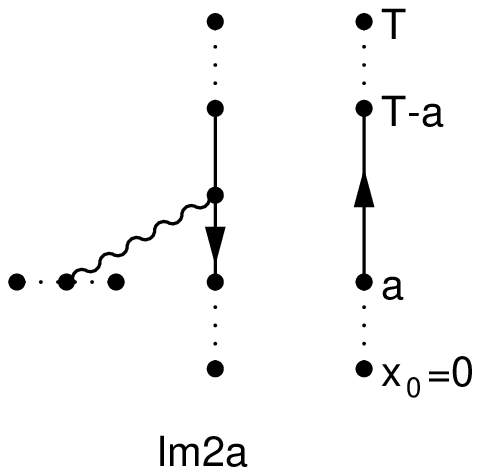}
  \end{minipage}
  \hspace{1cm}
  \begin{minipage}[b]{.3\linewidth}
     \centering\includegraphics[width=\linewidth]{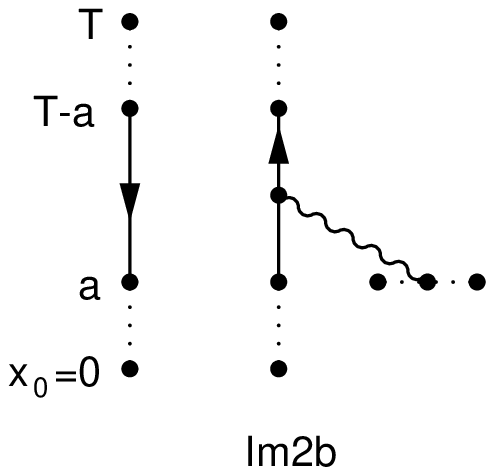}
  \end{minipage}\\
  \begin{minipage}[b]{.3\linewidth}
     \centering\includegraphics[width=\linewidth]{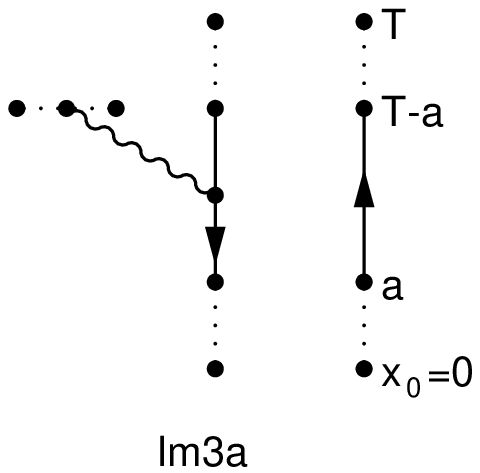}
  \end{minipage}
  \hspace{1cm}
  \begin{minipage}[b]{.3\linewidth}
     \centering\includegraphics[width=\linewidth]{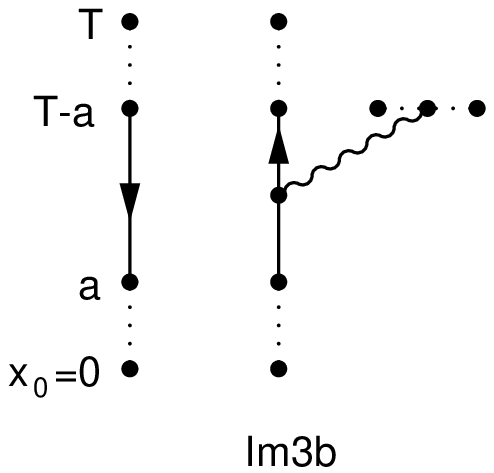}
  \end{minipage}
  \end{center}
  \caption[Diagrams contributing to $f_{1,\mathrm{Gb}}^{(1)}$]
          {Diagrams contributing to $f_{1,\mathrm{Gb}}^{(1)}$
           \label{fig:f1imct}}
\end{figure}

For the boundary counter-term proportional to $\ct^{(1)}$, one gets several
diagrams in total analogy to the calculation for $\fa$ and $\fp$.
For the diagrams shown in figure~\ref{fig:f1imct}, the same argument
applies as above, yielding
\beq
f_{1,\mathrm{Gb}} = 
\frac{1}{2}\left[f_{\mathrm{P,Gb}}(T-1)-f_{\mathrm{A,Gb}}(T-1)\right].
\eeq
\begin{figure}
  \noindent
  \begin{center}
  \begin{minipage}[b]{.3\linewidth}
     \centering\includegraphics[width=\linewidth]{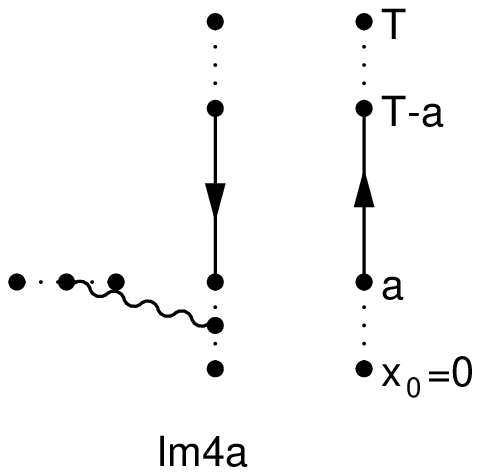}
  \end{minipage}
  \hspace{1cm}
  \begin{minipage}[b]{.3\linewidth}
     \centering\includegraphics[width=\linewidth]{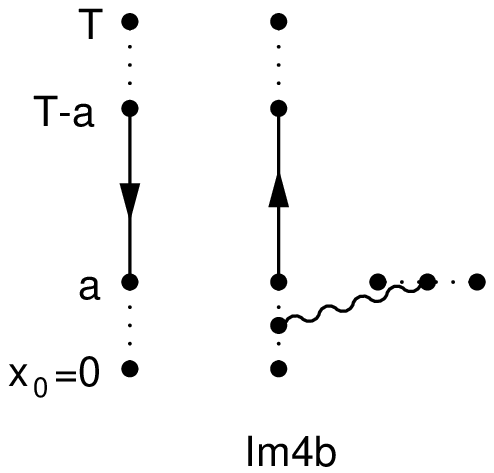}
  \end{minipage}\\
  \begin{minipage}[b]{.3\linewidth}
     \centering\includegraphics[width=\linewidth]{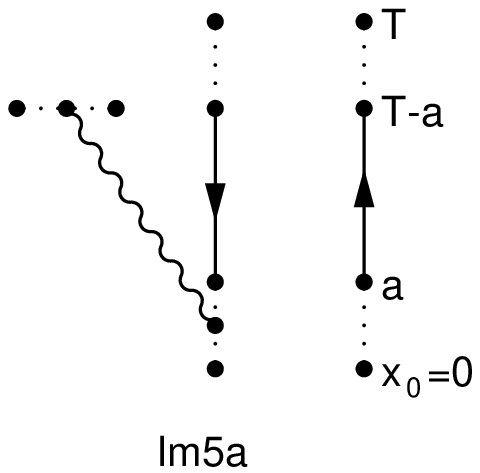}
  \end{minipage}
  \hspace{1cm}
  \begin{minipage}[b]{.3\linewidth}
     \centering\includegraphics[width=\linewidth]{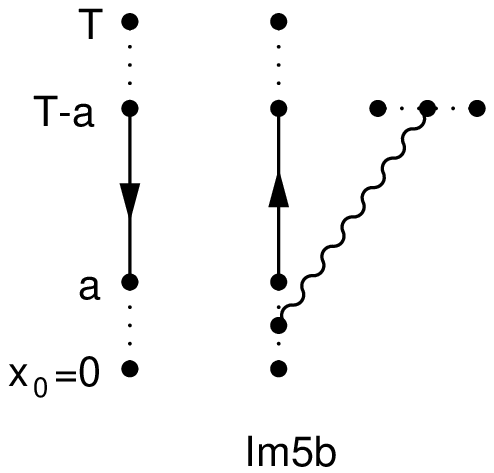}
  \end{minipage}\\
  \begin{minipage}[b]{.3\linewidth}
     \centering\includegraphics[width=\linewidth]{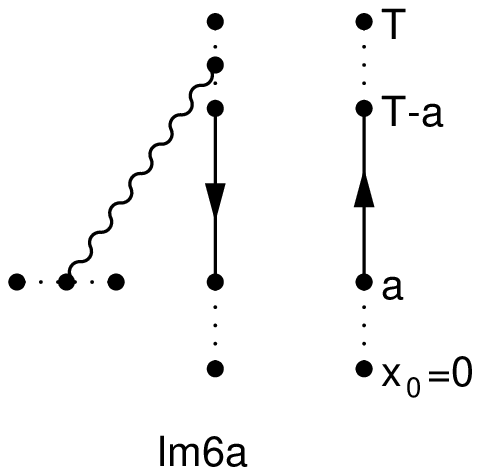}
  \end{minipage}
  \hspace{1cm}
  \begin{minipage}[b]{.3\linewidth}
     \centering\includegraphics[width=\linewidth]{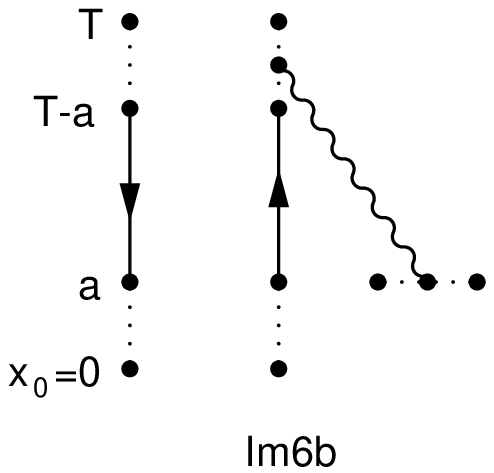}
  \end{minipage}\\
  \begin{minipage}[b]{.3\linewidth}
     \centering\includegraphics[width=\linewidth]{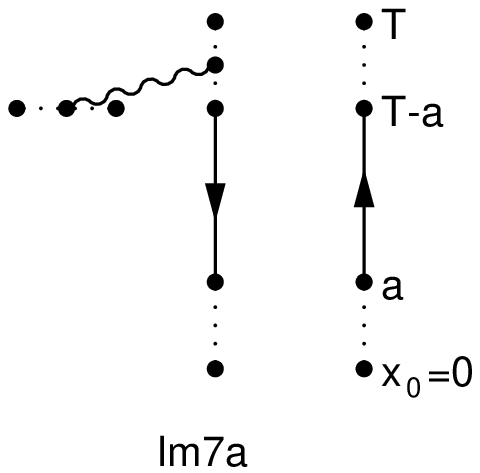}
  \end{minipage}
  \hspace{1cm}
  \begin{minipage}[b]{.3\linewidth}
     \centering\includegraphics[width=\linewidth]{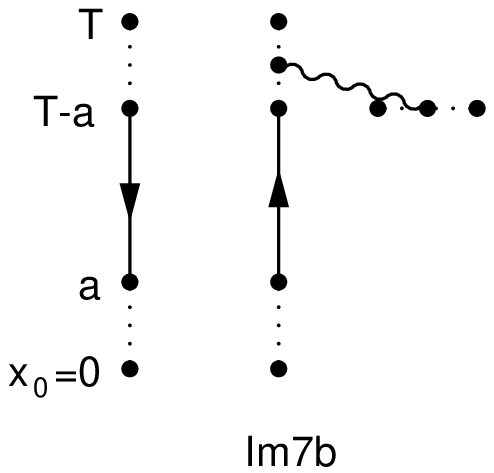}
  \end{minipage}
  \end{center}
  \caption[More diagrams for $f_{1,\mathrm{Gb}}^{(1)}$]
          {More diagrams for $f_{1,\mathrm{Gb}}^{(1)}$. These contributions
           cancel in the sum.\label{fig:f1imctvan}}
\end{figure}
Apart from these diagrams, one gets the contributions shown in 
figure~\ref{fig:f1imctvan}. However, like for $\fa$ and $\fp$, these 
contributions cancel pairwise in the sum and may thus be ignored.
So the improvement terms for $f_1$ can easily be computed 
using~(\ref{eq:fimprvol}), (\ref{eq:fimprbound}), and~(\ref{eq:fimprglbound}).

Also for $f_1$, we will need the derivative of the tree level coefficient
with respect to the bare mass. Differentiating~(\ref{eq:f1tree}) gives
\beq
\frac{\partial}{\partial m_0} f_1^{(0)}
= \frac{1}{2}\frac{\partial}{\partial m_0}
\left[\fp^{(0)}(T-1) - \fa^{(0)}(T-1)\right],
\eeq
where $\fp^{(0)}$ and $\fa^{(0)}$ can be taken from~(\ref{eq:ftree}).

\subsection{Computation of $Z_{\mathrm{P}}$ at 1--loop order}

The renormalisation constant $Z_{\mathrm{P}}$ at 1--loop order is obtained by expanding
$f_1$ and $\fp$ in~(\ref{eq:ZPdef}). One thus gets
\beq
Z_{\mathrm{P}}^{(1)} = \frac{f_1^{(1)}}{2f_1^{(0)}} -\frac{\fp^{(1)}(T/2)}{\fp^{(0)}(T/2)}.
\eeq
In order to get $Z_{\mathrm{P}}$ at vanishing renormalised quark mass, $f_1$ and $\fp$,
which depend on the bare quark mass, have to be computed at the critical
quark mass $\mcrit$. In perturbation theory, one has to use the expansion
\beq
\mcrit = \mcrit^{(0)} +\mcrit^{(1)}g_0^2 +\rmO(g_0^4),
\eeq
where the computation of $\mcrit^{(0)}$ and $\mcrit^{(1)}$ has been outlined
in chapter~\ref{chapt:curr_mass}. Taylor expansion then yields
\beq
f_1(\mcrit) = f_1(\mcrit^{(0)}) 
+ g_0^2 \mcrit^{(1)}\frac{\partial}{\partial m_0}f_1(\mcrit^{(0)})
+\rmO(g_0^4),
\eeq
which leads to the 1--loop coefficient
\beq
f_1^{(1)}(\mcrit) = f_1^{(1)}(\mcrit^{(0)})
+\mcrit^{(1)}\frac{\partial}{\partial m_0}f_1^{(0)}(\mcrit^{(0)}),
\eeq
and analogously for $\fp$.

Furthermore, $f_1^{(1)}$ and $\fp^{(1)}$ are meant to include all improvement
terms. Then the results obtained in the calculation should converge to
their continuum limits at a rate proportional to $a^2$.

By calculating all the diagrams, one is now able to compute the renormalisation
constant $Z_{\mathrm{P}}$ at different lattice sizes. In the case of $\theta=\pi/5$
and $\rho=1$, the coefficient $z_{\mathrm{p}}$ defined in~(\ref{eq:ZPoneldiv}) could be
obtained by extrapolation to the continuum limit, using the extrapolation 
procedure described in appendix~\ref{app:extra}. From this calculation, one
automatically gets the 2--loop anomalous dimension in the Schr\"{o}dinger
functional scheme by~(\ref{eq:d1sf}). This means that the 2--loop anomalous
dimension could be obtained by a 1--loop calculation, since the 2--loop 
anomalous dimension in the $\msbar$--scheme is already known.

Setting $m_1(L)=0$ and computing $Z_{\mathrm{P}}(L)$ and $Z_{\mathrm{P}}(2L)$ at this mass gives the
1--loop coefficient $k(L)$ of the step scaling function. One is thus able to
compute the size of the discretisation error~(\ref{eq:massdelta}). This 
computation was done in the case of $\theta=\pi/5$ and $\rho=1$. For some
smaller lattices, we did the calculation also at $\theta=0$ and $\theta=0.5$
and at $\rho=2$, in order to get a better comparison with the results obtained
in~\cite{Sint:1998iq}, which were computed at these values of $\theta$ and
$\rho$.

Since $f_1^{(1)}$ has not been computed with a non--vanishing background field
before, there are no results to compare the results obtained in this calculation
to. For this reason, careful checks of the results are necessary. For $f_1$,
all checks described for $\fa$ and $\fp$ are applicable. Moreover, it was checked
that the results both for the sum of the diagrams and for the improvement terms
are invariant when interchanging the boundary conditions in time direction,
as they should be.

All results can be found in chapter~\ref{chap:results}.

\cleardoublepage
\chapter{Results \label{chap:results}}
\chaptermark{Results}

The following results have been computed at vanishing renormalised mass,
which means setting $m_1=0$ or $m_0=m_{\mathrm{c}}$.
The results depend, of course, on the phase angle $\theta$ and the
ratio $\rho=T/L$. Unless stated otherwise, all results presented here
are for $\theta=\pi/5$ and $\rho=1$.

\section{The current quark mass}
\sectionmark{The current quark mass}

\subsection{The critical quark mass}

\begin{figure}
  \noindent
  \begin{center}
    \includegraphics[width=\linewidth]{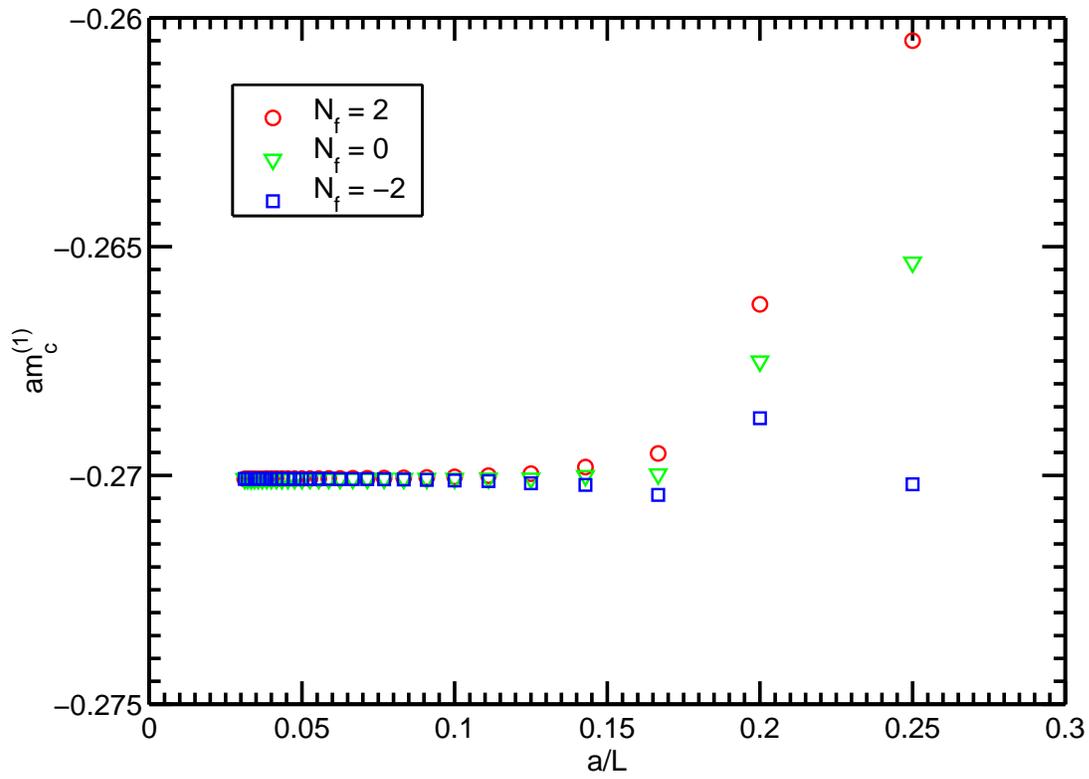}
  \end{center}
  \caption[The critical quark mass $\mcrit$ at 1--loop order]
          {\label{fig:critmass}
           \sl The critical quark mass $\mcrit$ 
               at 1--loop order}
\end{figure}

At tree level, the critical quark mass is independent of the flavour number,
while at 1--loop level, there is a linear dependence on $\Nf$ due to the
closed quark loops in diagrams 8a and 8b in figure~\ref{fig:faptad}. The 1--loop
coefficient may hence be written as
\beq
\mcrit^{(1)} = m_{\mathrm{c}0}^{(1)} + m_{\mathrm{c}1}^{(1)}\Nf.
\eeq
The results for the tree level coefficient $m_{\mathrm{c}}^{(0)}$ and the 1--loop
coefficients $m_{\mathrm{c}0}^{(1)}$ and $m_{\mathrm{c}1}^{(1)}$ can be found in 
table~\ref{tab:pertmass}. The critical mass at 1--loop level is also shown
in figure~\ref{fig:critmass} for different flavour numbers. The unusual case
$\Nf=-2$ refers to so called bermions, which were invented because Monte Carlo
simulations are considerably cheaper in this model than in full QCD. The 
original idea was to extrapolate from negative values to 
$\Nf=2$~\cite{deDivitiis:1995au,Anthony:1982fe}, which is, however, problematical.
Nowadays, bermions are more used as a toy model~\cite{Rolf:1999ih,Gehrmann:2001yn}.

A clear dependence of the 1--loop critical quark mass on the flavour number 
can only be seen on very small lattices. For larger $L/a$, the 1--loop 
coefficient can be seen to be practically constant at its continuum limit
value, which is~\cite{Wohlert:1987rf,Luscher:1996vw}
\beq
a\mcrit^{(1)} = -0.2700753495(2).
\eeq
\begin{table}[htbp]
  \begin{center}
    \begin{tabular}{|r|l|l|l|}
      \hline\hline
      $L/a$ 
      & \multicolumn{1}{c|}{$a\mcrit^{(0)}$} 
      & \multicolumn{1}{c|}{$am_{\mathrm{c}0}^{(1)}$} 
      & \multicolumn{1}{c|}{$am_{\mathrm{c}1}^{(1)}$} \\
      \hline
 4 &   -0.0015131 &   -0.2653473 &    0.0024231 \\
 5 &   -0.0016969 &   -0.2675073 &    0.0006225 \\
 6 &   -0.0006384 &   -0.2699739 &    0.0002269 \\
 7 &   -0.0005761 &   -0.2700126 &    0.0000978 \\
 8 &   -0.0003209 &   -0.2700683 &    0.0000522 \\
 9 &   -0.0002753 &   -0.2700656 &    0.0000300 \\
10 &   -0.0001835 &   -0.2700733 &    0.0000192 \\
11 &   -0.0001561 &   -0.2700708 &    0.0000129 \\
12 &   -0.0001145 &   -0.2700730 &    0.0000091 \\
13 &   -0.0000979 &   -0.2700720 &    0.0000066 \\
14 &   -0.0000761 &   -0.2700731 &    0.0000050 \\
15 &   -0.0000657 &   -0.2700729 &    0.0000039 \\
16 &   -0.0000531 &   -0.2700736 &    0.0000030 \\
17 &   -0.0000463 &   -0.2700736 &    0.0000024 \\
18 &   -0.0000385 &   -0.2700741 &    0.0000020 \\
19 &   -0.0000339 &   -0.2700742 &    0.0000016 \\
20 &   -0.0000288 &   -0.2700745 &    0.0000013 \\
21 &   -0.0000256 &   -0.2700746 &    0.0000011 \\
22 &   -0.0000221 &   -0.2700749 &    0.0000009 \\
23 &   -0.0000198 &   -0.2700750 &    0.0000008 \\
24 &   -0.0000173 &   -0.2700751 &    0.0000007 \\
25 &   -0.0000156 &   -0.2700752 &    0.0000006 \\
26 &   -0.0000138 &   -0.2700753 &    0.0000005 \\
27 &   -0.0000125 &   -0.2700754 &    0.0000005 \\
28 &   -0.0000112 &   -0.2700755 &    0.0000004 \\
29 &   -0.0000102 &   -0.2700755 &    0.0000004 \\
30 &   -0.0000092 &   -0.2700756 &    0.0000003 \\
31 &   -0.0000084 &   -0.2700756 &    0.0000003 \\
32 &   -0.0000076 &   -0.2700756 &    0.0000003 \\
      \hline\hline
    \end{tabular}
    \caption[Perturbative results for the critical quark mass $\mcrit$ 
    up to 1--loop order]
            {\sl Perturbative results for the critical quark mass $\mcrit$ 
    up to 1--loop order}
    \label{tab:pertmass}
  \end{center}
\end{table}

\subsection{Lattice artefacts of the current mass up to 1--loop order}

\begin{figure}
  \noindent
  \begin{center}
    \includegraphics[width=\linewidth]{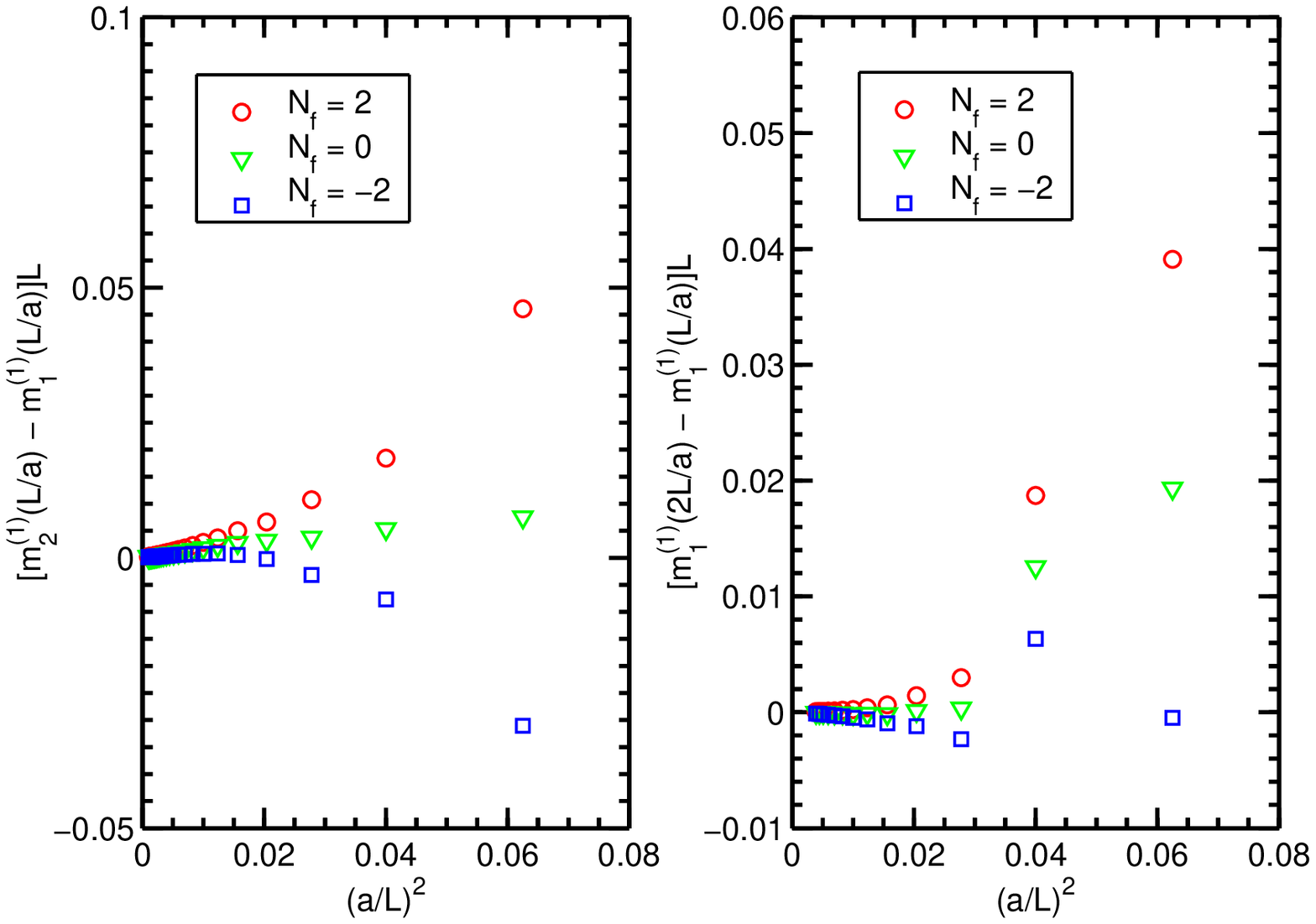}
  \end{center}
  \caption[The lattice artefacts $d=m_2(L/a)-m_1(L/a)$
               and $e=m_1(2L/a)-m_1(L/a)$ 
               at 1--loop order]
          {\label{fig:diserr}
           \sl The lattice artefacts $d=m_2(L/a)-m_1(L/a)$
               and $e=m_1(2L/a)-m_1(L/a)$ 
               at 1--loop order}
\end{figure}

Using the expansion of $\fa$ and $\fp$, it is possible to compute the
lattice artefacts $d$ and $e$ defined in~(\ref{eq:ddef}) and~(\ref{eq:edef}).
Here again, the 1--loop coefficients get a linear $\Nf$ dependence from
diagrams 8a and 8b in figure~\ref{fig:faptad} and may hence be written as
\beqn
d^{(1)} &=& d_0^{(1)} +d_1^{(1)}\Nf,\\
e^{(1)} &=& e_0^{(1)} +e_1^{(1)}\Nf.
\eeqn
The results at tree and 1--loop level can be found in table~\ref{tab:diserr},
the 1--loop coefficients are also shown in figure~\ref{fig:diserr} for
different values of $\Nf$. They are obviously small and converge quickly
to zero. The expected behaviour in the improved theory is a convergence at
rate proportional to $(a/L)^2$. Figure~\ref{fig:diserr}, where the 1--loop
coefficients are plotted with respect to $(a/L)^2$, does however show that,
for $\Nf\neq 0$, they are dominated by terms of higher order.
In~\cite{Bode:2001jv}, the perturbative results up to 1--loop order
are compared to results obtained by Monte Carlo simulations. While the
values for $e$ are approximated by perturbation theory remarkably well, the
perturbative values for $d$ are significantly larger than the simulation
results, especially at large values of the renormalised 
coupling~($\bar{g}^2(L)=3.334$).

\begin{table}[htbp]
  \begin{center}
    \begin{tabular}{|r|r|r|r||r|r|r|}
      \hline\hline
      $L/a$ 
      & \multicolumn{1}{c|}{$Ld^{(0)}$} 
      & \multicolumn{1}{c|}{$ Ld_0^{(1)} $} 
      & \multicolumn{1}{c||}{$ Ld_1^{(1)}$} 
      & \multicolumn{1}{c|}{$Le^{(0)}$} 
      & \multicolumn{1}{c|}{$ Le_0^{(1)} $} 
      & \multicolumn{1}{c|}{$ Le_1^{(1)}$}\\
      \hline
 4 &    0.1123 &    0.00752 &    0.01929 &   -0.004981 &    0.01931 &    0.00990 \\
 5 &    0.0522 &    0.00535 &    0.00653 &   -0.007783 &    0.01252 &    0.00309 \\
 6 &    0.0258 &    0.00378 &    0.00348 &   -0.003204 &    0.00032 &    0.00133 \\
 7 &    0.0140 &    0.00320 &    0.00172 &   -0.003550 &    0.00011 &    0.00066 \\
 8 &    0.0078 &    0.00277 &    0.00112 &   -0.002165 &   -0.00015 &    0.00040 \\
 9 &    0.0044 &    0.00224 &    0.00071 &   -0.002149 &   -0.00012 &    0.00025 \\
10 &    0.0025 &    0.00180 &    0.00053 &   -0.001558 &   -0.00013 &    0.00018 \\
11 &    0.0014 &    0.00149 &    0.00038 &   -0.001483 &   -0.00009 &    0.00013 \\
12 &    0.0007 &    0.00122 &    0.00030 &   -0.001172 &   -0.00008 &    0.00010 \\
13 &    0.0002 &    0.00104 &    0.00023 &   -0.001098 &   -0.00006 &    0.00008 \\
14 &   -0.0001 &    0.00087 &    0.00019 &   -0.000913 &   -0.00005 &    0.00006 \\
15 &   -0.0002 &    0.00076 &    0.00016 &   -0.000850 &   -0.00004 &    0.00005 \\
16 &   -0.0003 &    0.00065 &    0.00014 &   -0.000730 &   -0.00003 &    0.00004 \\
17 &   -0.0004 &    0.00057 &    0.00011 & & & \\
18 &   -0.0005 &    0.00050 &    0.00010 & & & \\
19 &   -0.0005 &    0.00045 &    0.00008 & & & \\
20 &   -0.0005 &    0.00040 &    0.00007 & & & \\
21 &   -0.0005 &    0.00036 &    0.00007 & & & \\
22 &   -0.0005 &    0.00032 &    0.00006 & & & \\
23 &   -0.0005 &    0.00029 &    0.00005 & & & \\
24 &   -0.0005 &    0.00026 &    0.00005 & & & \\
25 &   -0.0005 &    0.00024 &    0.00004 & & & \\
26 &   -0.0004 &    0.00022 &    0.00004 & & & \\
27 &   -0.0004 &    0.00020 &    0.00003 & & & \\
28 &   -0.0004 &    0.00019 &    0.00003 & & & \\
29 &   -0.0004 &    0.00017 &    0.00003 & & & \\
30 &   -0.0004 &    0.00016 &    0.00003 & & & \\
31 &   -0.0004 &    0.00015 &    0.00002 & & & \\
32 &   -0.0003 &    0.00014 &    0.00002 & & & \\
      \hline\hline
    \end{tabular}
    \caption[The lattice artefacts $d=m_2(L)-m_1(L)$ and $e=m_1(2L)-m_1(L)$  
    up to 1--loop order]
            {\sl The lattice artefacts $d=m_2(L)-m_1(L)$ and $e=m_1(2L)-m_1(L)$  
    up to 1--loop order}
    \label{tab:diserr}
  \end{center}
\end{table}

\begin{figure}
  \noindent
  \begin{center}
    \includegraphics[width=\linewidth]{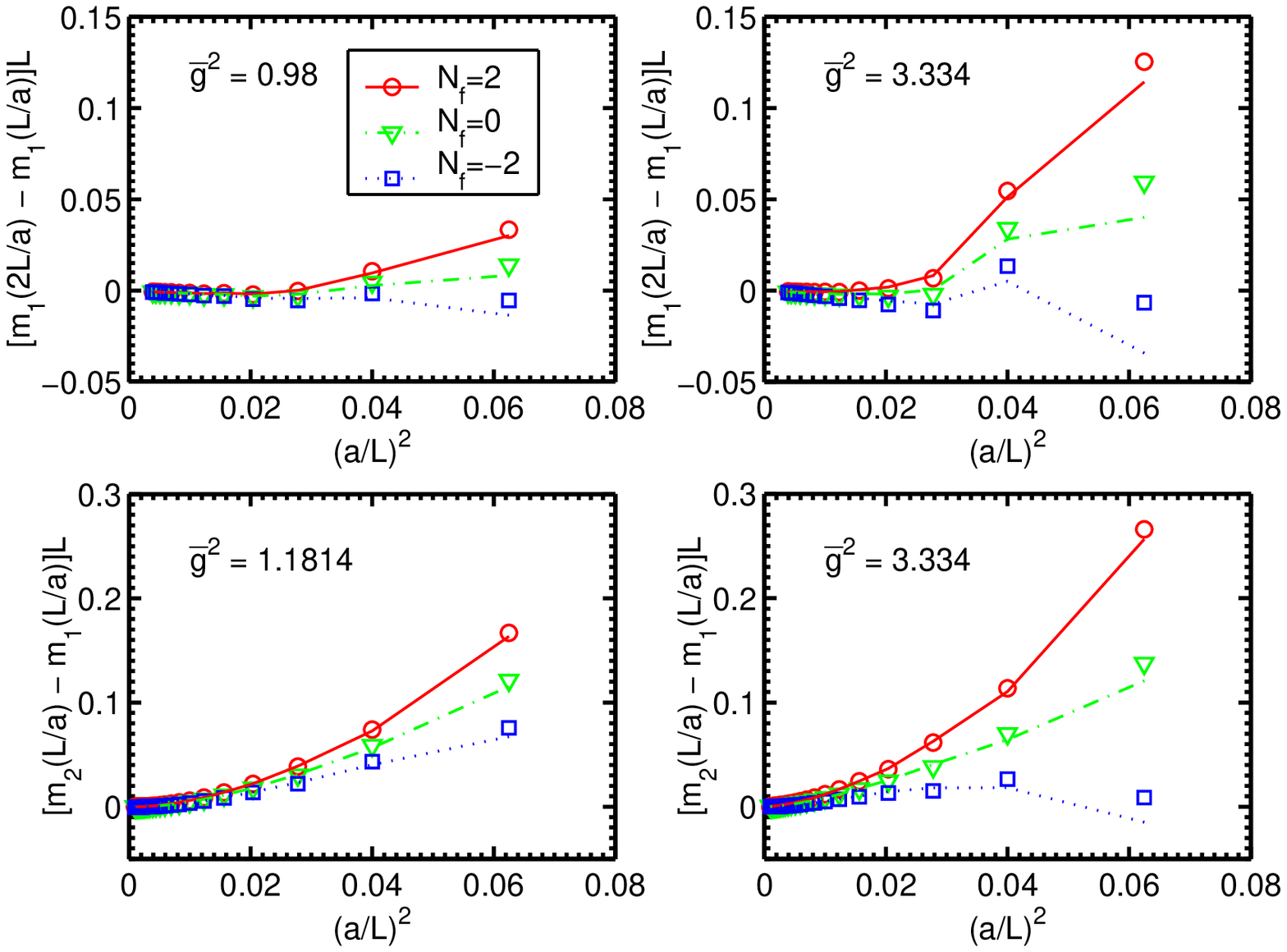}
  \end{center}
  \caption[The lattice artefacts $e=m_1(2L/a)-m_1(L/a)$
               and $d=m_2(L/a)-m_1(L/a)$ 
               up to 1--loop order at different values of the
               renormalised coupling $\bar{g}^2$]
          {\label{fig:decompare}
           \sl The lattice artefacts $e=m_1(2L/a)-m_1(L/a)$
               and $d=m_2(L/a)-m_1(L/a)$ up to 1--loop order
               at different values of the renormalised coupling
               $\bar{g}^2$. The lines show the values without the
               $\ct^{(1)}$--term, while in the symbols, it is 
               included.}
\end{figure}

An important question in this context is the dependence of the results on
the improvement term proportional to $\ct^{(1)}$, which was missing in
previous publications. As already stated in chapter~\ref{chapt:curr_mass},
this term is absent in zero background field calculations and irrelevant
for the extrapolation both of $\csw^{(1)}$ and $m_{\mathrm{c}}^{(1)}$. It 
has, however, to be taken into account in the results for $d(L/a)$ and
$e(L/a)$ published in~\cite{Bode:2001jv}. There, the perturbative results
including tree and 1--loop level are plotted for different values of the
renormalised coupling $\bar{g}^2$. The results for the same values of
$\bar{g}^2$ are shown in figure~\ref{fig:decompare}, where the symbols
denote the results with full improvement, while the lines show the results
without the $\ct^{(1)}$--term, i.e. the values presented in~\cite{Bode:2001jv}.
The picture shows that the additional term only gives considerable changes
in the case $\bar{g}^2=3.334$ for $\Nf=0$ and $\Nf=-2$. Since, for this coupling,
only the results for $\Nf=2$ are plotted in~\cite{Bode:2001jv}, the corrections
to the plots presented there are almost invisible. 

\section{Lattice artefacts of the step scaling function of the coupling
\label{sect:couparte}}
\sectionmark{Lattice artefacts of the step scaling function of the coupling}

\begin{figure}
  \noindent
  \begin{center}
    \includegraphics[width=\linewidth]{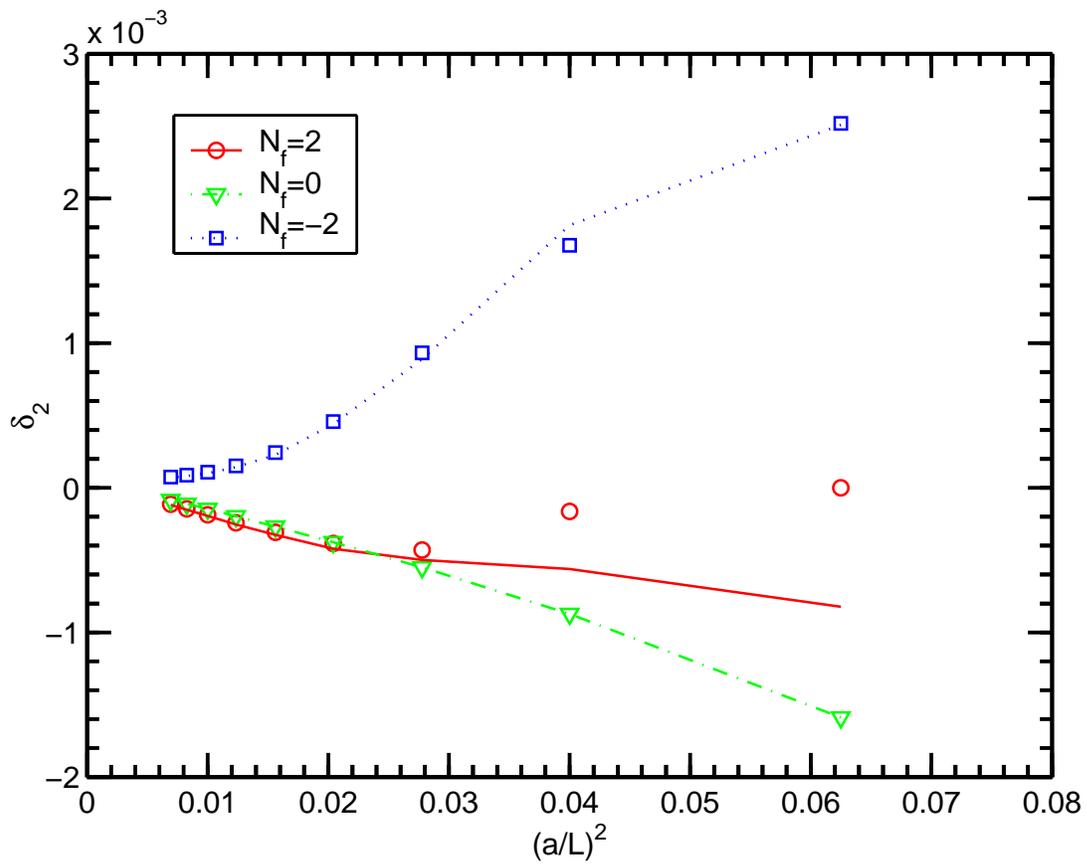}
  \end{center}
  \caption[The 2--loop discretisation error $\delta_2$]
          {\label{fig:delta2}
           \sl The 2--loop discretisation error $\delta_2$. The lines
            show the values for $\delta_2$ if the continuum limit of
            the critical mass is used.}
\end{figure}

One of the reasons to compute the critical quark mass up to 1--loop order
was to get the correct discretisation errors for the coupling up to 
2--loop order. To this end,
one has to use~(\ref{eq:delta10})--(\ref{eq:delta22}) with the current
mass properly set to zero, $m_1(L)=0$. Since the critical mass converges
quickly to its continuum limit, the discretisation errors $\delta_{ij}$,
shown in table~\ref{tab:deltas}
are expected to differ from the estimation in~\cite{Bode:1999sm}, which was
obtained using the continuum limit value of the critical mass, only for
small lattices. Since $\delta_{20}$ does not depend on the critical quark
mass, the 2--loop coefficient will remain unaltered for $\Nf=0$ when using the
correct critical mass instead of its continuum limit. From the
size of the critical mass (figure~\ref{fig:critmass}), one expects a small
deviation for $\Nf=-2$ at $L/a=6$, while for $\Nf=2$, larger deviations are
expected at $L/a=4$ and $L/a=6$. Figure~\ref{fig:delta2} shows that this 
indeed is the case. The lines are the estimates from~\cite{Bode:1999sm},
while the symbols denote the values obtained with the proper $\mcrit(L/a)$.
In any case, the 2--loop coefficient $\delta_2$ is small and does vanish
at a rate proportional to $(a/L)^2$, as one should expect in
the improved theory. Especially for $\Nf=2$, it does get even smaller by using
the correct critical mass.

\begin{table}[htbp]
  \begin{center}
    \begin{tabular}{|r|r|r|r|r|r|}
      \hline\hline
      $L/a$ 
      & \multicolumn{1}{c|}{$\delta_{10}$} 
      & \multicolumn{1}{c|}{$\delta_{11}$} 
      & \multicolumn{1}{c|}{$\delta_{20}$} 
      & \multicolumn{1}{c|}{$\delta_{21}$} 
      & \multicolumn{1}{c|}{$\delta_{22}$} \\
      \hline
 4 &   -0.01033 &    0.00002 &   -0.001588 &   -0.000630 &    0.000712 \\
 5 &   -0.00625 &   -0.00014 &   -0.000872 &   -0.000460 &    0.000407 \\
 6 &   -0.00394 &   -0.00014 &   -0.000550 &   -0.000341 &    0.000200 \\
 7 &   -0.00268 &   -0.00014 &   -0.000376 &   -0.000211 &    0.000103 \\
 8 &   -0.00194 &   -0.00011 &   -0.000268 &   -0.000138 &    0.000059 \\
 9 &   -0.00148 &   -0.00009 &   -0.000196 &   -0.000098 &    0.000038 \\
10 &   -0.00117 &   -0.00007 &   -0.000147 &   -0.000073 &    0.000027 \\
11 &   -0.00095 &   -0.00006 &   -0.000111 &   -0.000058 &    0.000020 \\
12 &   -0.00079 &   -0.00005 &   -0.000085 &   -0.000047 &    0.000016 \\
      \hline\hline
    \end{tabular}
    \caption[The discretisation errors $\delta_{ij}$ 
    up to 2--loop order]
            {\sl The discretisation errors $\delta_{ij}$ 
    up to 2--loop order
    \label{tab:deltas}}
  \end{center}
\end{table}

\section{The renormalised quark mass}
\sectionmark{The renormalised quark mass}

\subsection{The renormalisation constant $Z_{\mathrm{P}}$ up to 1--loop order}

\begin{table}[htbp]
  \begin{center}
    \begin{tabular}{|r|r|r|}
      \hline\hline
      $L/a$ 
      & \multicolumn{1}{c|}{$Y_{\mathrm{P}0}$} 
      & \multicolumn{1}{c|}{$Y_{\mathrm{P}1}$} \\ 
      \hline
 4 &   -0.13990 &    0.025124 \\
 6 &   -0.14008 &    0.009000 \\
 8 &   -0.13757 &    0.005284 \\
10 &   -0.13590 &    0.004070 \\
12 &   -0.13476 &    0.003539 \\
14 &   -0.13395 &    0.003254 \\
16 &   -0.13336 &    0.003081 \\
18 &   -0.13291 &    0.002966 \\
20 &   -0.13256 &    0.002886 \\
22 &   -0.13228 &    0.002828 \\
24 &   -0.13205 &    0.002785 \\
26 &   -0.13187 &    0.002751 \\
28 &   -0.13172 &    0.002724 \\
30 &   -0.13159 &    0.002703 \\
32 &   -0.13148 &    0.002686 \\
      \hline\hline
    \end{tabular}
   \caption[The $\Nf$ independent and $\Nf$ dependent part of $Y_{\mathrm{P}}(L/a)$]
           {\sl The $\Nf$ independent and $\Nf$ dependent part of 
             $Y_{\mathrm{P}}(L/a)$
  \label{tab:zp}}
  \end{center}
\end{table}

\begin{figure}
  \noindent
  \begin{center}
    \includegraphics[width=\linewidth]{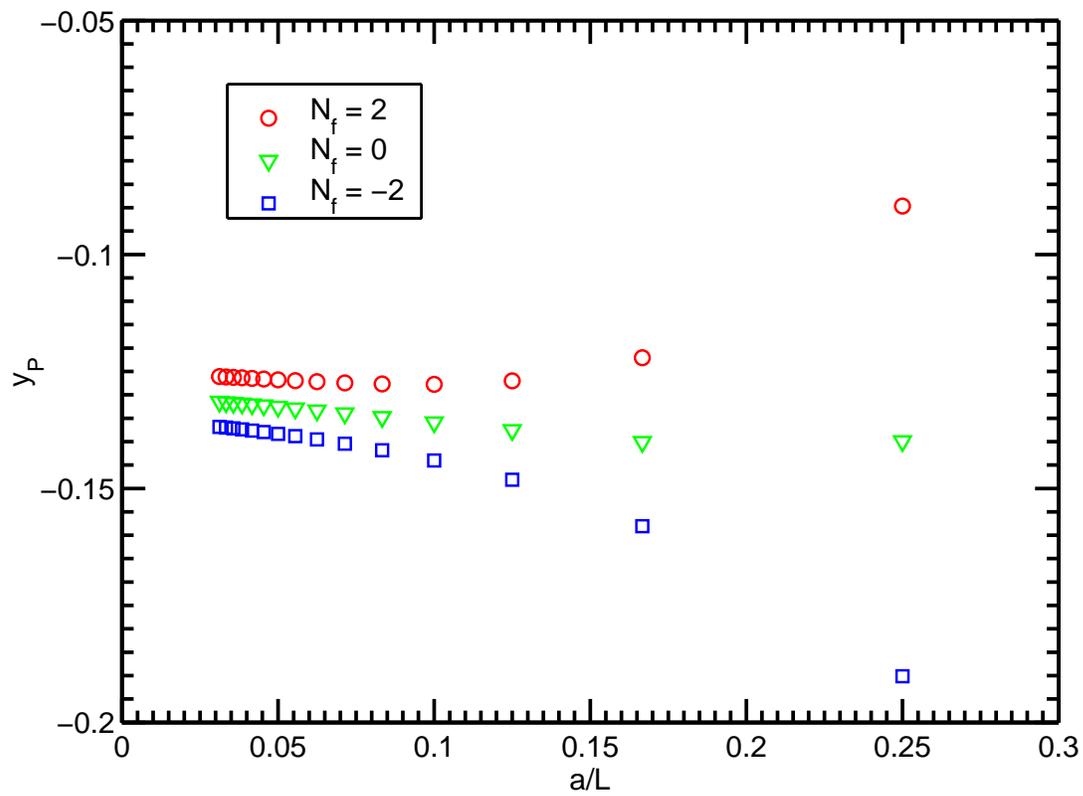}
  \end{center}
  \caption[The finite part $Y_{\mathrm{p}}$ of the renormalisation constant $Z_{\mathrm{P}}$ 
               at 1--loop order]
          {\label{fig:zp}
           \sl The finite part $Y_{\mathrm{P}}$ of the renormalisation 
             constant $Z_{\mathrm{P}}$ at 1--loop order}
\end{figure}

In order to obtain the finite part $z_{\mathrm{p}}$ of $Z_{\mathrm{P}}$
defined in~(\ref{eq:ZPoneldiv}), one has to extrapolate the quantity
\beq
Y_{\mathrm{P}}(L/a) = \frac{1}{\cf}\left(Z_{\mathrm{P}}^{(1)}(L/a)
+d_0\ln(L/a)\right)
\eeq
to the continuum limit.
Due to the $\Nf$ dependent tadpole graphs, one has a linear $\Nf$ dependence
of $Y_{\mathrm{P}}$,
\beq
Y_{\mathrm{P}} = Y_{\mathrm{P}0} +\Nf Y_{\mathrm{P}1},
\eeq
and thus the same dependence for $z_{\mathrm{p}}$,
\beq
z_{\mathrm{p}} = z_{\mathrm{p}0} +\Nf z_{\mathrm{p}1}.
\eeq
The results for different lattice sizes are shown in table~\ref{tab:zp}.
The behaviour of $Y_{\mathrm{P}}$ is shown in figure~\ref{fig:zp} for 
different values of $\Nf$.

For the extrapolation, the procedure described in appendix~\ref{app:extra} may be
used. 
Following~\cite{Luscher:1986wf}, one expects an asymptotic expansion
\beq
Y_{\mathrm{P}}(L/a) = a_0 + \sum_{k=1}^{\infty}
\left[a_k + b_k\ln(a/L)\right](a/L)^k
\eeq
to hold, where the terms linear in $(a/L)$ are cancelled by improvement.
Taking the functions $1$, $(a/L)^2$, and $(a/L)^2\ln(a/L)$ into the fit
and using $(a/L)^3$ and $(a/L)^3\ln(a/L)$ for the estimation of the systematic
error yields the results
\beq
z_{\mathrm{p}0}=-0.13044(2),\qquad  z_{\mathrm{p}1}=0.002565(10).
\label{eq:zpextra}
\eeq
Since $Y_{\mathrm{P}}$ depends on the improvement coefficients $\csw^{(1)}$,
$\ctt^{(1)}$, $\ct^{(1)}$, and via the critical mass also on $\ca^{(1)}$,
the errors of these coefficients have to be taken into account. However,
varying the improvement coefficients within their error ranges shows that
the change in $z_{\mathrm{p}0}$ and $z_{\mathrm{p}1}$ is smaller than the
errors quoted in~(\ref{eq:zpextra}). The error due to the uncertainty in the
improvement coefficients is thus negligible.

Taking $(a/L)^3$ and $(a/L)^3\ln(a/L)$ into the fit and using the functions 
$(a/L)^4$
and $(a/L)^4\ln(a/L)$ for the error estimation does not change the picture
significantly. While one gets smaller systematic errors in the extrapolation,
the errors of the improvement coefficients become more important. Taking
everything into account, the errors quoted in~(\ref{eq:zpextra}) seem to 
be reasonably conservative. Including $(a/L)$ and $(a/L)\ln(a/L)$ in the fit
shows that the coefficients of these functions are compatible with zero, as
one should expect in the $\rmO(a)$ improved theory.

Inserting $z_{\mathrm{p}}$ into~(\ref{eq:chim}) and using~(\ref{eq:d1sf}),
one now gets the 2--loop anomalous dimension $d_1^{\mathrm{SF}}$. Due to
the linear dependence of $z_{\mathrm{p}}$ on $\Nf$, the anomalous dimension
will get an $\Nf$
dependence up to order $\Nf^2$, in contrast to order $\Nf$ in the case
of a vanishing background field. Now, $d_1^{\mathrm{SF}}$ becomes
\beq
d_1^{\mathrm{SF}}/d_0 = 0.16508(8) - 0.00724(4)\Nf + 0.000570(2)\Nf^2.
\eeq
This result for $\theta = \pi/5$ and $\rho=1$ is small, but larger than
the anomalous dimension with a vanishing background field computed
in~\cite{Sint:1998iq}, which is
$d_1^{\mathrm{SF}}/d_0=0.1251+0.0046\Nf$ for $\theta=0$ and
$d_1^{\mathrm{SF}}/d_0=0.0271+0.0105\Nf$ for $\theta=0.5$.

\subsection{Lattice artefacts of $\Sigma_{\mathrm{P}}$ up to 1--loop order}

By computing the renormalisation constant $Z_{\mathrm{P}}$ both at
$L/a$ and $2L/a$ with $m_1(L/a)=0$, one obtains the one loop coefficient
$k(L/a)$ of the step scaling function
$\Sigma_{\mathrm{P}}$,
\beq
k(L/a) = Z_{\mathrm{P}}^{(1)}(2L/a) - Z_{\mathrm{P}}^{(1)}(L/a),
\eeq
and its deviation $\delta_k$ from the continuum limit,
\beq
\delta_k(L/a) = \frac{k(L/a)}{k(\infty)}-1,
\eeq
with $k(\infty) = -d_0\ln(2)$.
In contrast to the zero background field case, the tadpole graphs give a
dependence on $\Nf$,
\beq
\delta_k = \delta_{k0} +\delta_{k1}\Nf,
\eeq
\begin{table}[tbp]
  \begin{center}
    \begin{tabular}{|r|r|r|}
      \hline\hline
      $L/a$ 
      & \multicolumn{1}{c|}{$\delta_{k0}$} 
      & \multicolumn{1}{c|}{$\delta_{k1}$} \\ 
      \hline
 4 &   -0.3258 &    0.6248 \\
 6 &   -0.1998 &    0.1916 \\
 8 &   -0.1548 &    0.0790 \\
10 &   -0.1229 &    0.0428 \\
12 &   -0.0998 &    0.0274 \\
14 &   -0.0826 &    0.0193 \\
16 &   -0.0696 &    0.0144 \\
      \hline\hline
    \end{tabular}
   \caption[The $\Nf$ independent and $\Nf$ dependent part of $\delta_k(L/a)$
            for $\theta=\pi/5$ and $\rho=1$]
           {\sl The $\Nf$ independent and $\Nf$ dependent part of $\delta_k(L/a)$
             for $\theta=\pi/5$ and $\rho=1$ \label{tab:deltak}}
  \end{center}
\end{table}
\begin{figure}
  \noindent
  \begin{center}
    \includegraphics[width=\linewidth]{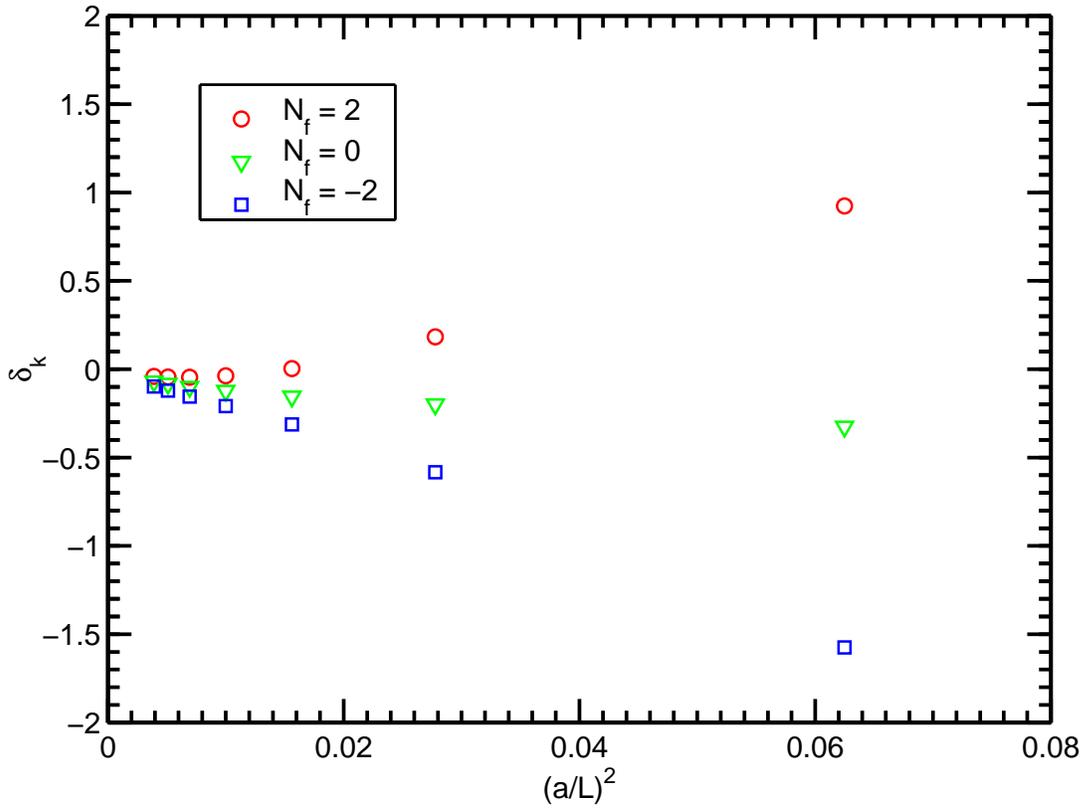}
  \end{center}
  \caption[The 1--loop order discretisation error $\delta_k$ of the step
           scaling function $\Sigma_{\mathrm{P}}$ for $\theta=\pi/5$
           and $\rho=1$]
          {\label{fig:deltak}
           \sl The 1--loop order discretisation error $\delta_k$ of the step
           scaling function $\Sigma_{\mathrm{P}}$ for $\theta=\pi/5$
           and $\rho=1$}
\end{figure}
where the coefficients $\delta_{k0}$ and $\delta_{k1}$ (not to be confused
with the $\delta_{ij}$ from section~\ref{sect:couparte}) are expected to
vanish at a rate proportional to $a^2$.
The results for $\delta_{k0}$ and $\delta_{k1}$ can be found in
table~\ref{tab:deltak}. In figure~\ref{fig:deltak}, $\delta_k$ is 
shown for several flavour numbers. Compared to the results obtained with
a vanishing background field in~\cite{Sint:1998iq}, these discretisation
errors are very large, especially for $L/a=4$ and $L/a=6$. However, for
$\Nf=2$ and $L/a\geq 8$, the error seems to be reasonably small.

In order to get a direct impression of the change of the discretisation
errors due to the presence of the background field, one should compare
the values with and without background field at the same values of $\theta$
and $\rho$.
\begin{table}[htbp]
  \begin{center}
    \begin{tabular}{|r|r|r|l||r|r|l|}
      \hline\hline
      $L/a$ 
      & \multicolumn{1}{c|}{$\left.\delta_{k0}\right|_{\theta=0}$}
      & \multicolumn{1}{c|}{$\left.\delta_{k1}\right|_{\theta=0}$}
      & \multicolumn{1}{c||}{$\left.\delta_k\right|_{\theta=0,V=1}$}
      & \multicolumn{1}{c|}{$\left.\delta_{k0}\right|_{\theta=0.5}$}
      & \multicolumn{1}{c|}{$\left.\delta_{k1}\right|_{\theta=0.5}$}
      & \multicolumn{1}{c||}{$\left.\delta_k\right|_{\theta=0.5,V=1}$}\\
      \hline
      \multicolumn{7}{|c|}{$\rho=1$}\\
      \hline
 4 & $\,\!$ 0.1425 & 0.2688 & $\,\!$ 0.2040 & -0.1697 & 0.5765 & $\,\!$ 0.2136 \\
 6 & $\,\!$ 0.0538 & 0.1046 &   -0.0121 &   -0.1167 &    0.1891 & $\,\!$ 0.0208 \\
 8 & $\,\!$ 0.0164 & 0.0442 &   -0.0253 &   -0.0983 &    0.0791 &   -0.0026 \\
10 & $\,\!$ 0.0033 &    0.0231 &   -0.0215 &   -0.0815 &    0.0425 &   -0.0062 \\
12 &   -0.0019 &    0.0144 &   -0.0171 &   -0.0680 &    0.0270 &   -0.0064 \\
14 &   -0.0040 &    0.0100 &   -0.0137 &   -0.0573 &    0.0190 &   -0.0058 \\
16 &   -0.0049 &    0.0074 &   -0.0111 &   -0.0489 &    0.0141 &   -0.0052 \\
      \hline\hline
      \multicolumn{7}{|c|}{$\rho=2$}\\
      \hline
 4 &    0.0963 &    0.5651 &   -0.3084 &   -0.5116 &    3.1960 &   -0.2456 \\
 6 &   -0.0020 &    0.1768 &   -0.2292 &   -0.3020 &    1.1311 &   -0.1499 \\
 8 &   -0.0144 &    0.0880 &   -0.1449 &   -0.2007 &    0.4744 &   -0.0893 \\
10 &   -0.0155 &    0.0546 &   -0.0974 &   -0.1472 &    0.2502 &   -0.0584 \\
12 &   -0.0142 &    0.0377 &   -0.0696 &   -0.1137 &    0.1567 &   -0.0412 \\
      \hline\hline
    \end{tabular}
    \caption[The $\Nf$ independent and $\Nf$ dependent part of $\delta_k(L/a)$
            for various values of $\theta$ and $\rho$ compared to the zero
            background field values]
            {The $\Nf$ independent and $\Nf$ dependent part of $\delta_k(L/a)$
            for various values of $\theta$ and $\rho$ compared to the zero
            background field ($V=1$) values from~\cite{Sint:1998iq}
            \label{tab:compare}}
  \end{center}
\end{table}
\begin{figure}
  \noindent
  \begin{center}
    \includegraphics[width=\linewidth]{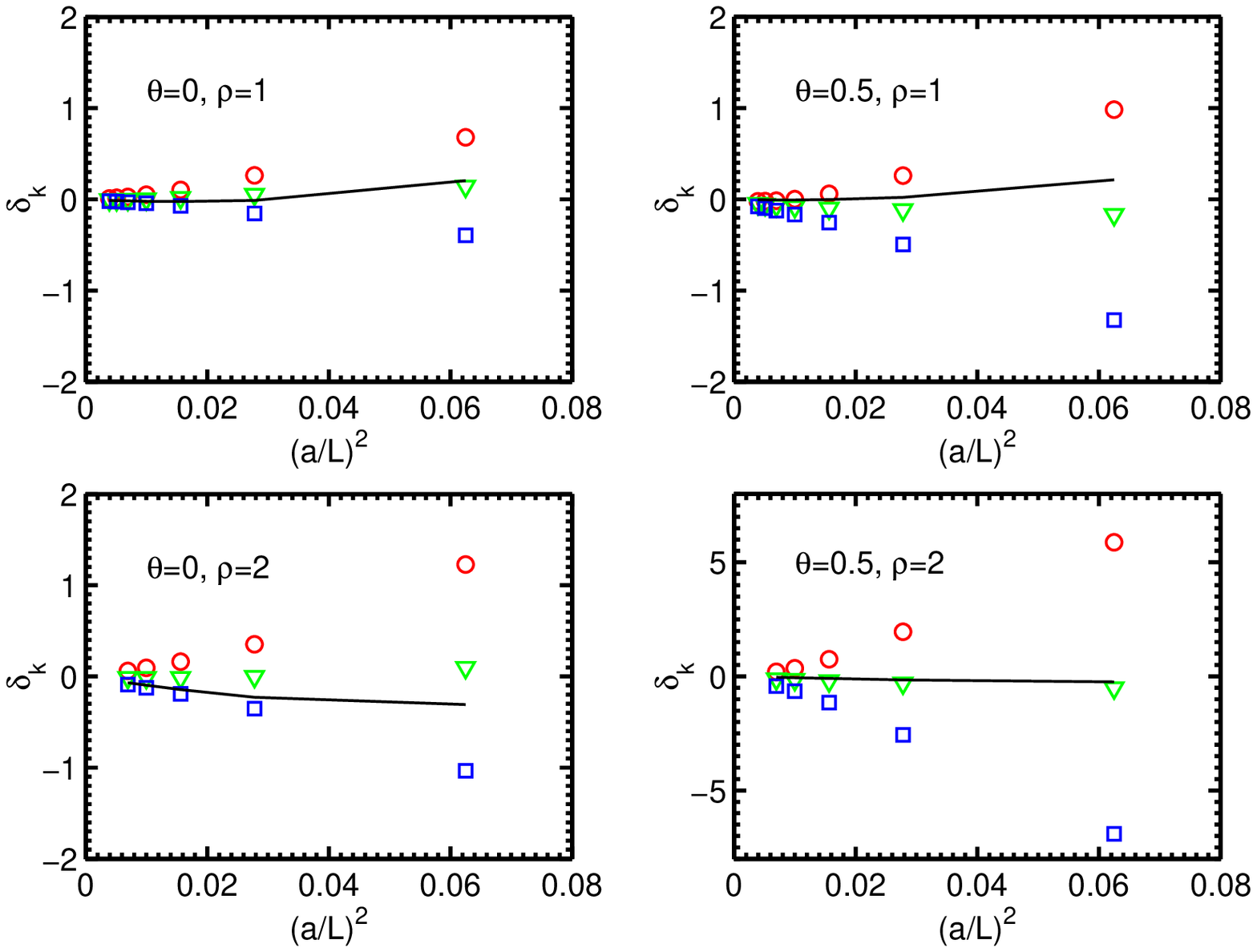}
  \end{center}
  \caption[The 1--loop order discretisation error $\delta_k$ of the step
           scaling function $\Sigma_{\mathrm{P}}$ for various values
           of $\theta$ and $\rho$]
          {\label{fig:compare}
           \sl The 1--loop order discretisation error $\delta_k$ of the step
           scaling function $\Sigma_{\mathrm{P}}$ for various values of
           $\theta$ and $\rho$ at $\Nf=2$ (\textcolor{red}{$\circ $}), $\Nf=0$
           (\textcolor{green}{\small\mbox{$\triangledown $}\normalsize}), 
           and $\Nf=-2$ 
           (\textcolor{blue}{\scriptsize\mbox{$\square $}\normalsize}). 
           The solid line (---) shows the values for a vanishing
           background field from~\cite{Sint:1998iq}.}
\end{figure}
This is done in table~\ref{tab:compare} and in figure~\ref{fig:compare}, where
the black lines represent the $\Nf$ independent results with a vanishing
background field from~\cite{Sint:1998iq}, while the symbols denote the results
obtained with the non--vanishing background field for several flavour numbers.
For $\Nf=0$, the discretisation errors with a non--vanishing background field 
are smaller than
those with zero background field at $\theta=0$, while at $\theta=0.5$, they
are larger. For $\Nf\neq 0$, the discretisation errors become drastically larger
due to the large $\Nf$--term, which is not present in the case of a vanishing
background field. This effect is particularly large at $\theta=0.5$ and
$\rho=2$.

\section{Numerical checks of the results}
\sectionmark{Numerical checks of the results}

\begin{figure}[htb]
  \noindent
  \begin{center}
    \includegraphics[width=\linewidth]{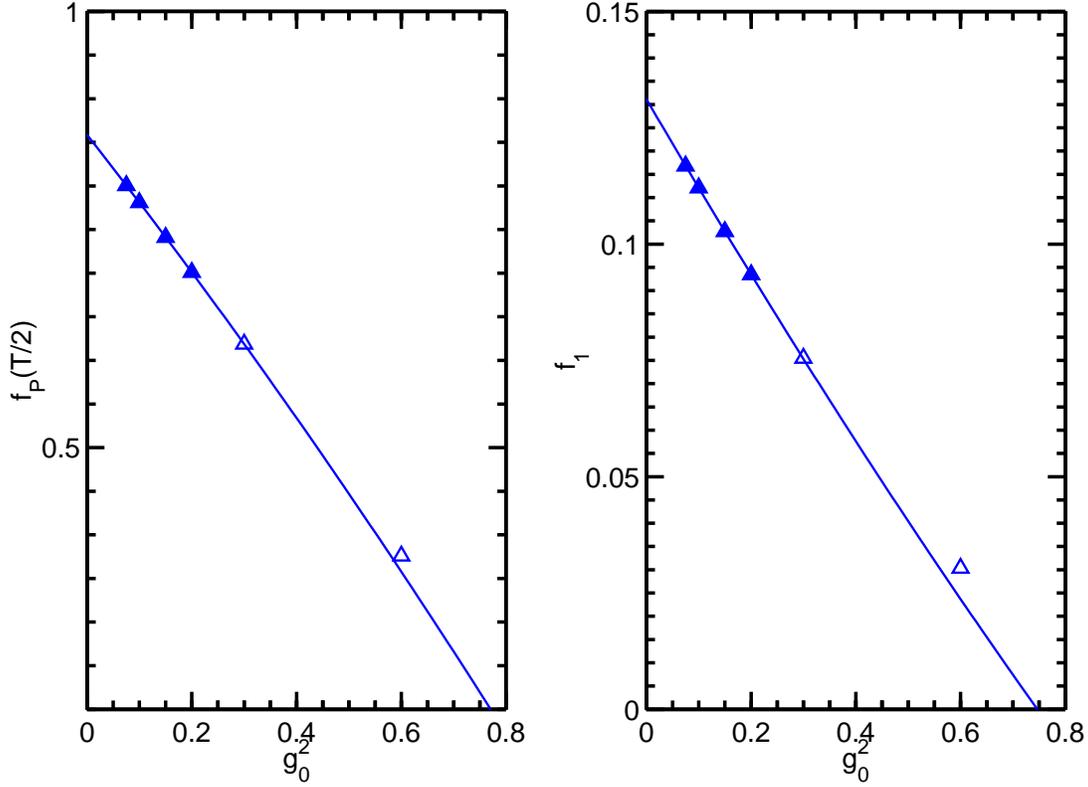}
  \end{center}
  \caption[Monte Carlo results for $f_{\mathrm{P}}(T/2)$ and $f_1$ at $L/a=6$]
          {\label{fig:check1}
           \sl  Monte Carlo results for $f_{\mathrm{P}}(T/2)$ 
           and $f_1$ at $L/a=6$. The error bars are smaller than the
           symbols. The lines
           are polynomial fits up to order $g_0^4$ including the filled
           symbols, while the open symbols were disregarded.}
\end{figure}

In order to check the programs, they  were used to compute the tree
level and 1--loop coefficients at $m_0=0$ and $L/a=6$, where the results can 
be compared to non perturbative results obtained in Monte Carlo simulations
by Juri Rolf. The simulations were done at small bare couplings ranging
from 0.075 to 0.6 using the tree level values for the improvement coefficients.
The results are shown in figure~\ref{fig:check1}.
The tree level values of $f_{\mathrm{P}}(T/2)$ and $f_1$ 
are obtained as the constant
parts in polynomial fits up to order $g_0^4$. (The results do, of course, depend
on the fit ansatz. Using a polynomial of the form $a_0+a_1 g_0^2 +a_2 g_0^4$ turned
out to give results with reasonable errors.) Fitting the polynomials to the
filled symbols in figure~\ref{fig:check1} yields
\beq
f_{\mathrm{P}}^{(0)}(T/2) = 0.8586(11) , \qquad f_1^{(0)} = 0.1312(3) ,
\eeq
compared to the perturbative results
\beq
f_{\mathrm{P}}^{(0)}(T/2) = 0.8587 , \qquad f_1^{(0)} = 0.1310.
\eeq
Obviously, these results agree perfectly well.

For the one loop coefficients, one could in principle use the coefficient
of $g_0^2$ obtained in the fit. However, assuming the tree level values
of $f_{\mathrm{P}}$ and $f_1$ to be correct, 
one achieves more precise results doing the
fit on $(f_{\mathrm{P}}(T/2)-f_{\mathrm{P}}^{(0)}(T/2))/g_0^2$
and $(f_1-f_1^{(0)})/g_0^2$, where the 1--loop coefficients get the constant
parts in the fit.
\begin{figure}[htb]
  \noindent
  \begin{center}
    \includegraphics[width=\linewidth]{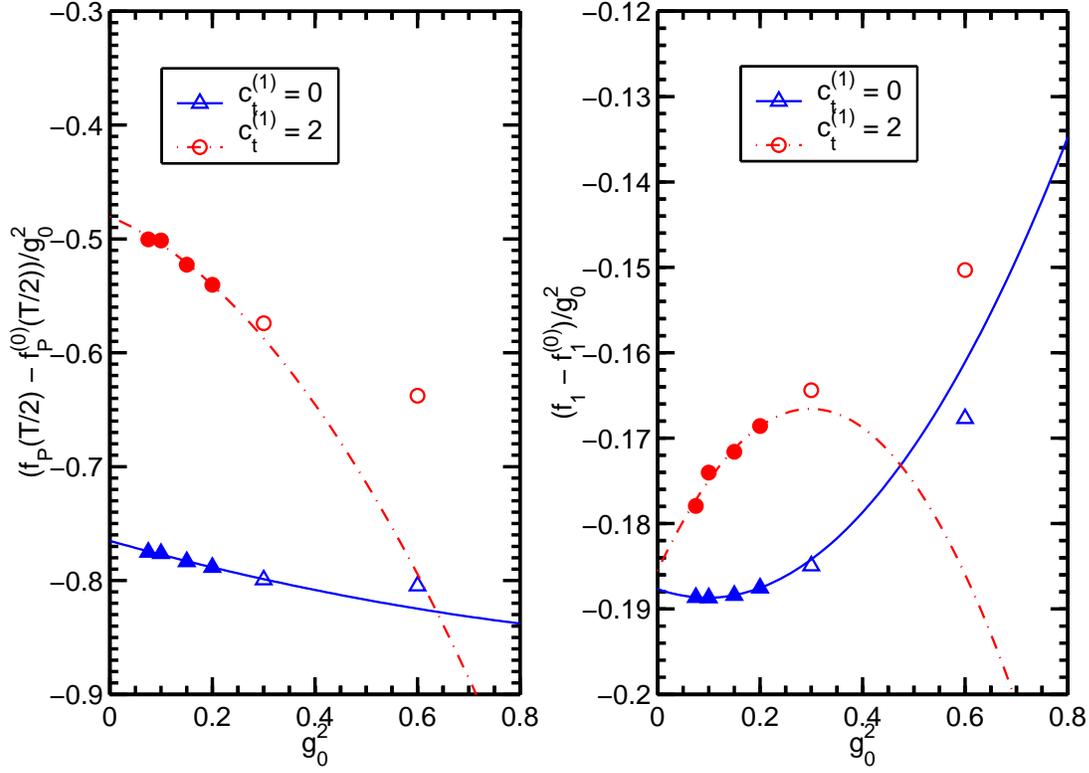}
  \end{center}
  \caption[Monte Carlo results for 
           $(f_{\mathrm{P}}(T/2)-f_{\mathrm{P}}^{(0)}(T/2))/g_0^2$
           and $(f_1-f_1^{(0)})/g_0^2$ at $L/a=6$]
          {\label{fig:check2}
           \sl Monte Carlo results for 
           $(f_{\mathrm{P}}(T/2)-f_{\mathrm{P}}^{(0)}(T/2))/g_0^2$
           and $(f_1-f_1^{(0)})/g_0^2$ at $L/a=6$. The error bars are
           smaller than the symbols. The lines
           are polynomial fits up to order $g_0^4$ including the filled
           symbols.}
\end{figure}
These functions are shown in figure~\ref{fig:check2} both for tree level
improvement and for $\ct^{(1)}=2$, which was used to check the
$\ct^{(1)}$-term. In this case, the results for bigger $g_0^2$ seem to be
dominated by terms of higher order than $g_0^4$.

For $\ct^{(1)}=0$, the extrapolation yields 
\beq
f_{\mathrm{P}}^{(1)}(T/2) = -0.765(11)  , \qquad f_1^{(1)} = -0.188(3),
\eeq
compared to
\beq
f_{\mathrm{P}}^{(1)}(T/2) = -0.760 , \qquad f_1^{(1)} = -0.189,
\eeq
obtained by perturbation theory, while for $\ct^{(1)}=2$ one gets
\beq
f_{\mathrm{P}}^{(1)}(T/2) =  -0.480(10)   , \qquad f_1^{(1)} = -0.186(3) ,
\eeq
compared to the perturbative results
\beq
f_{\mathrm{P}}^{(1)}(T/2) = -0.466 , \qquad f_1^{(1)} = -0.183.
\eeq
These results seem to agree well enough to state that the perturbative
results could be confirmed by the simulations.
\cleardoublepage
\chapter{Summary \label{chapt:summary}}
\chaptermark{Summary}

In this thesis, a part of the programme of the ALPHA collaboration to
renormalise QCD has been presented.

The basic ideas of the Schr\"{o}dinger functional as a method to compute 
renormalised quantities have been outlined and it has been explained how
this scheme avoids computations on very large lattices by using the length
of the space time box as the renormalisation scale. Furthermore, 
Symanzik's improvement programme has been introduced as a method 
to reduce discretisation errors to $\rmO(a^2)$.

The main purpose of the Schr\"{o}dinger functional method, the renormalisation
of parameters, has been studied in detail. The definition of the renormalised
coupling has been outlined, and it was shown how a renormalised quark mass
can be introduced via the PCAC relation.

In order to do several 1--loop calculations, perturbation theory in the
Schr\"{o}\-ding\-er functional has been discussed. Perturbation theory has then
been used to do a variety of calculations with a non--vanishing background field,
which so far had only been done in the zero background field case. 
In particular, the following calculations have been done:

\begin{itemize}
\item
The correlation functions $f_{\mathrm{A}}$, $f_{\mathrm{P}}$, and $f_1$
have been expanded up to 1--loop order.
\item
In these expansions, it was found that, with a non--vanishing background
field, an improvement term proportional to $\ct^{(1)}$ arises, which
was missing in previous publications.
\item
The critical quark mass has been computed up to 1--loop order and was
used to calculate the discretisation errors of the coupling's step scaling
function up to 2--loop order.
\item
Several discretisation errors of the current quark mass have been
calculated.
\item
The renormalisation constant $Z_{\mathrm{P}}$ has been computed at
1--loop order, giving the 2--loop anomalous dimension.
\item
The discretisation error of the step scaling function of the quark mass has
been computed at 1--loop order.
\end{itemize}

These computations turned out to be time consuming, so that lattices 
larger than $L/a=32$ could not be reached. Up to
$L/a=32$, calculations could however be done in reasonable time.
In these calculations, the following results have been obtained:

\begin{itemize}
\item
The critical quark mass was shown to converge quickly towards its continuum
limit. Considerable deviations from this limit were only observed for very
small lattices.
\item
The discretisation errors of the coupling's step scaling function 
turned out to be small and close
to those computed with the continuum limit of the critical quark mass. Where
the results using the correct critical mass deviate from the previous results,
they were shown to make the discretisation errors smaller.
\item
The discretisation errors of the current quark mass turned out to be small.
\item
The effect of the previously missing $\ct^{(1)}$--term on results published
so far was shown to be negligible.
\item
The 2--loop anomalous dimension was shown to be small but larger than in the
case of a vanishing background field.
\item
The discretisation error of the step scaling function of the renormalised quark
mass turned out to be much larger than with a vanishing
background field for $\Nf\neq 0$, while for $\Nf=0$, it is smaller or larger, 
depending on $\theta$. For the case of $\theta=\pi/5$, $\rho=1$, and $\Nf=2$,
it turned however out to be reasonably small for $L/a\geq 8$.
\item
The results obtained perturbatively could be shown to agree well with
non--perturbative results from Monte Carlo simulations in the small coupling
limit on small lattices ($L/a = 6$).
\end{itemize}

With these calculations, the perturbative treatment of the quark mass
renormalisation up to 1--loop order at vanishing renormalised mass is 
completed. The programs are, however, written such that they can be used
for calculations at different quark masses without too much effort.

What has not yet been done are non--perturbative calculations of the renormalised
mass with a non--vanishing background field. However, due to the large 
discretisation errors found in this thesis for small lattices, safe extrapolations
to the continuum limit could turn out to be difficult. 
\cleardoublepage
\begin{appendix}
\chapter{Notations and conventions \label{app:group}}
\chaptermark{Notations and conventions}

\section{The Dirac matrices}
\sectionmark{The Dirac matrices}

The Dirac matrices in Euclidean space may be obtained from those in
Minkowski space by
\beqn
\gamma_{0}^{\rm{Euclidean}} &=& \gamma_{0}^{\rm{Minkowski}},\\
\gamma_{1,2,3}^{\rm{Euclidean}} &=& -i\gamma_{1,2,3}^{\rm{Minkowski}},
\eeqn
giving
\beqn
\gamma_0 &=&
\left(
\begin{array}{cc}
1 & 0\\
0 & -1\\
\end{array}
\right),\\
\gamma_{1,2,3} &=&
\left(
\begin{array}{cc}
0 & -i\sigma_{1,2,3}\\
i\sigma_{1,2,3} & 0\\
\end{array}
\right),
\eeqn
with the Pauli matrices
\beq
\sigma_1 =
\left(
\begin{array}{cc}
0 & 1\\
1 & 0\\
\end{array}
\right),
\quad
\sigma_2 =
\left(
\begin{array}{cc}
0 & -i\\
i & 0\\
\end{array}
\right),
\quad
\sigma_3 =
\left(
\begin{array}{cc}
1 & 0\\
0 & -1\\
\end{array}
\right).
\eeq
In this representation, $\gamma_5=\gamma_1\gamma_2\gamma_3\gamma_4$ is
given by
\beq
\gamma_5 = \left(
\begin{array}{cc}
0 & -1\\
-1 & 0\\
\end{array}
\right).
\eeq
The projectors $P_+=\frac{1}{2}(1+\gamma_0)$ and
$P_-=\frac{1}{2}(1-\gamma_0)$ become
\beq
P_+ = \left(
\begin{array}{cc}
1 & 0\\
0 & 0\\
\end{array}
\right),
\qquad
P_- = \left(
\begin{array}{cc}
0 & 0\\
0 & 1\\
\end{array}
\right).
\eeq

\section[The basis of the Lie algebra $su(3)$]
{The basis of the Lie algebra $\mathbf{su(3)}$}
\sectionmark{The basis of the Lie algebra $su(3)$}

One possible choice for a basis of $su(3)$ are the well known Gell-Mann
matrices~\cite{Gell-Mann:1962xb}. For the calculations in this thesis,
it is however more convenient to use the same basis as in~\cite{Weisz:1996csw}.
To this end, we introduce colour matrices $\tilde{\lambda}_a$, which coincide
with the Gell-Mann matrices $\lambda_a$ in all cases except for the two
diagonal matrices, for which we choose
\beq
\tilde{\lambda}_3 = -\frac{1}{2}\lambda_3 +\frac{\sqrt{3}}{2}\lambda_8,
\eeq
\beq
\tilde{\lambda}_8 = \frac{\sqrt{3}}{2}\lambda_3 +\frac{1}{2}\lambda_8.
\eeq
Then, the matrices $\tilde{\lambda}_a$ are 
\beqn
\tilde{\lambda}_1 = \left(
\begin{array}{ccc}
0 & 1 & 0 \\
1 & 0 & 0 \\
0 & 0 & 0 \\
\end{array}
\right), &\qquad &
\tilde{\lambda}_2 = \left(
\begin{array}{ccc}
0 & -i & 0 \\
i & 0 & 0 \\
0 & 0 & 0 \\
\end{array}
\right), \nonumber\\
\tilde{\lambda}_4 = \left(
\begin{array}{ccc}
0 & 0 & 1 \\
0 & 0 & 0 \\
1 & 0 & 0 \\
\end{array}
\right), &\qquad & 
\tilde{\lambda}_5 = \left(
\begin{array}{ccc}
0 & 0 & -i \\
0 & 0 & 0 \\
i & 0 & 0 \\
\end{array}
\right), \nonumber\\
\tilde{\lambda}_6 = \left(
\begin{array}{ccc}
0 & 0 & 0 \\
0 & 0 & 1 \\
0 & 1 & 0 \\
\end{array}
\right), &\qquad & 
\tilde{\lambda}_7 = \left(
\begin{array}{ccc}
0 & 0 & 0 \\
0 & 0 & -i \\
0 & i & 0 \\
\end{array}
\right), \nonumber\\
\tilde{\lambda}_3 = \left(
\begin{array}{ccc}
0 & 0 & 0 \\
0 & 1 & 0 \\
0 & 0 & -1 \\
\end{array}
\right), &\qquad& 
\tilde{\lambda}_8 = \frac{1}{\sqrt{3}}\left(
\begin{array}{ccc}
2 & 0 & 0 \\
0 & -1 & 0 \\
0 & 0 & -1 \\
\end{array}
\right).
\eeqn
After normalisation,
\beq
T_a = \frac{1}{2i}\tilde{\lambda}_a,
\eeq
these matrices may be used to define a new basis $I^a$, which is given by
\beqn
I^1 = T_+ = \frac{1}{\sqrt{2}}(T_1+iT_2), &\qquad &
I^2 = T_- = \frac{1}{\sqrt{2}}(T_1-iT_2), \nonumber\\
I^4 = U_+ = \frac{1}{\sqrt{2}}(T_4+iT_5), &\qquad &
I^5 = U_- = \frac{1}{\sqrt{2}}(T_4-iT_5), \nonumber\\
I^6 = V_+ = \frac{1}{\sqrt{2}}(T_6+iT_7), &\qquad &
I^7 = V_- = \frac{1}{\sqrt{2}}(T_6-iT_7), 
\eeqn
for the non--diagonal matrices and
\beq
I^3 = T_3, \qquad  I^8 =T_8
\eeq
for the diagonal ones. 
For this basis, one has
\beq
I^{a\dagger} = -I^{\bar{a}},
\eeq
where $\bar{1}=2$, $\bar{4}=5$, $\bar{6}=7$, and vice versa. For the diagonal
matrices, one has $\bar{3}=3$ and $\bar{8}=8$. The normalisation is chosen
such that
\beq
\Tr \left(I^a I^b\right) = -\frac{1}{2}\delta_{b\bar{a}}.
\eeq

\section{The background field}
\sectionmark{The background field}

In the chosen basis, the constant colour electric background field
$\mathcal{E}$ is proportional to $I^8$. It can be written as
\beq
\mathcal{E} = -\gamma\left(
\begin{array}{ccc}
2 & 0 & 0\\
0 & -1 & 0\\
0 & 0 & -1\\
\end{array}
\right) = -\gamma\sqrt{3}\tilde{\lambda}_8,
\eeq
with
\beq
\gamma = \frac{1}{LT}\left(\eta + \frac{\pi}{3}\right).
\eeq
A special feature of the chosen basis is that the star operation defined by
\beq
M\star G =\frac{1}{2}\left[ (MG +GM^{\dagger})
-\frac{1}{N}\tr(MG +GM^{\dagger})\right]
\eeq
acts diagonally,
\beqn
\cosh G_{0k}\star I^a &=& C_a I^a,\\
\sinh G_{0k}\star I^a &=& S_a I^a.
\eeqn
The values of $C_a$ and $S_a$ can be found in table~\ref{tab:casa}.
\begin{table}[tbp]
  \begin{center}
    \begin{tabular}{ccc}
      \hline\hline
$a$ & $C_a$ & $S_a$ \\
\hline
1 & $\frac{1}{2}(\cos 2\gamma +\cos\gamma)$ 
& $-i\frac{1}{2}(\sin 2\gamma +\sin\gamma)$ \\
3 & $\cos\gamma$ & 0 \\
4 & $\frac{1}{2}(\cos 2\gamma +\cos\gamma)$ 
& $-i\frac{1}{2}(\sin 2\gamma +\sin\gamma)$ \\
6 & $\cos\gamma$ & 0 \\
8 & $\frac{1}{3}(2\cos 2\gamma +\cos\gamma)$ & 0\\
\hline
    \end{tabular}
    \caption[$C_a$ and $S_a$ for the gauge group $SU(3)$]
            {\sl $C_a$ and $S_a$ for the gauge group $SU(3)$. The other 
              coefficients are $C_2=C_1$, $C_5=C_4$, $C_7=C_6$,
              $S_2=-S_1$, $S_5=-S_4$, and $S_7=-S_6$.
    \label{tab:casa}}          
  \end{center}
\end{table}

The covariant derivative may be decomposed in the basis $I^a$, giving
\beqn
(D_k f)(x) &=& \sum_a \Bigl[ \Omega_a(x_0)f^a(x+\hat{k})-f^a(x)\Bigr]I^a,\\
(D^{\ast}_k f)(x) &=& \sum_a \Bigl[
  f^a(x) - \Omega_a(x_0)^{\ast}f^a(x-\hat{k})\Bigr]I^a,
\eeqn
where $\Omega_a(x_0)$ is a phase factor and can thus be written as
\beq
\Omega_a(x_0) = e^{i\phi_a(x_0)}.
\eeq
The values of $\phi_a$ are shown in table~\ref{tab:phiara}. 
Another quantity useful for the calculation of the gluon propagator is
\beq
R_a = (C_a - S_a)e^{i\partial_0\phi_a(x_0)/2}.
\eeq
Also the values of $R_a$ can be found in table~\ref{tab:phiara}.

\begin{table}[tbp]
  \begin{center}
    \begin{tabular}{ccc}
      \hline\hline
$a$ & $\phi_a(x_0)$ & $R_a$ \\
\hline
1 & $-3\gamma x_0 +\frac{1}{L}(\eta[\frac{3}{2}-\nu] -\frac{\pi}{3})$ 
& $\cos\frac{\gamma}{2}$ \\
3 & 0 & $\cos\gamma$ \\
4 & $-3\gamma x_0 +\frac{1}{L}(\eta[\frac{3}{2}+\nu] -\frac{2\pi}{3})$ 
& $\cos\frac{\gamma}{2}$ \\
6 & $\frac{1}{L}(2\eta\nu -\frac{\pi}{3})$ & $\cos\gamma$ \\
8 & 0 & $\frac{1}{3}(2\cos 2\gamma +\cos\gamma)$\\
\hline
    \end{tabular}
    \caption[$\phi_a(x_0)$ and $R_a$ for the gauge group $SU(3)$]
            {\sl $\phi_a(x_0)$ and $R_a$ for the gauge group $SU(3)$. The other 
              coefficients are $\phi_2=-\phi_1$, $\phi_5=-\phi_4$, $\phi_7=-\phi_6$,
              $R_2=R_1$, $R_5=R_4$, and $R_7=R_6$.
    \label{tab:phiara}}          
  \end{center}
\end{table}
\cleardoublepage
\chapter{The numerical construction of the propagators \label{app:prop}}
\chaptermark{The numerical construction of the propagators}

\section{The gluon and ghost propagators}
\sectionmark{The gluon and ghost propagators}

In order to obtain the gluon and ghost propagators, one has to invert
the operators $K^a$ and $F^a$ from~(\ref{eq:gluonquad1}), (\ref{eq:gluonquad2}),
(\ref{eq:gluonquad3}), (\ref{eq:gluonquad4}), and~(\ref{eq:ghostquad}). A very
convenient way to do this is described in~\cite{Narayanan:1995ex} and will be
briefly outlined.

For the gluon propagator, one has to solve the difference equation 
\beq
\sum_{y_0}\sum_{\nu}K^a_{\mu\nu}(\vp;x_0,y_0)D^a_{\nu\sigma}(\vp;y_0,z_0)
= \delta_{x_0,z_0}\delta_{\mu\sigma}.
\label{eq:kinv}
\eeq
This is particularly simple for $a=3,8$ with $\vp=\vo$. 
In this case, one simply has
\beq
D^a_{\mu\nu}(\vo;x_0,y_0) =\left\{
\begin{array}{ll}
\frac{1}{\lambda_0}\bigl(1+\min(x_0,y_0)\bigr) & \mbox{if $\mu=\nu=0$},\\
\frac{1}{R_a}\bigl(\min(x_0,y_0)-\frac{x_0 y_0}{T}\bigr)
& \mbox{if $\mu=\nu=k$}, \\
0 & \mbox{if $ \mu\neq\nu$}.
\end{array}\right.
\eeq
In all other cases, the difference equation has to be solved numerically. To
simplify the notation, we denote the propagator at fixed values for $\vp$ and
$a$ by
\beq
D^a_{\mu\nu}(\vp;x_0,y_0)=Q_{\mu\nu}(x_0,y_0).
\eeq
The operator $K^a$ may be decomposed according to
\beq
K^a_{\mu\nu}(\vp;x_0,y_0) =
\acal_{\mu\nu}(x_0)\delta_{x_0+1,y_0}
+ \bcal_{\mu\nu}(x_0)\delta_{x_0,y_0}
+ \acal_{\nu\mu}(x_0-1)\delta_{x_0-1,y_0}.
\eeq
Using this notation, the difference equation~(\ref{eq:kinv}) becomes
\beqn
\acal_{\mu\sigma}(x_0)Q_{\sigma\nu}(x_0+1,y_0)
+ \bcal_{\mu\sigma}(x_0)Q_{\sigma\nu}(x_0,y_0) & & \nonumber\\
+ \acal_{\sigma\mu}(x_0-1)Q_{\sigma\mu}(x_0-1,y_0) 
&=& \delta_{\mu\nu}\delta_{x_0,y_0}.
\label{eq:dif}
\eeqn
In order to solve this equation, we now construct two solutions $\psi^f(x_0)$
and $\psi^b(x_0)$ of the homogeneous equation by two step recursion 
forward and backward in time. The
forward solution $\psi^f$ starts from
\beqn
\psi^f_{0\nu}(-1) = \psi^f_{0\nu}(0) &=& \delta_{0\nu},\\
\psi^f_{k\nu}(0) &=& 0,\\
\psi^f_{k\nu}(1) &=& \delta_{k\nu},
\eeqn
while the backward solution $\psi^b$ starts from
\beqn
\psi^b_{0\nu}(T) = \psi^b_{0\nu}(T-1) &=& \delta_{0\nu},\\
\psi^b_{k\nu}(T) &=& 0,\\
\psi^b_{k\nu}(T-1) &=& \delta_{k\nu}.
\eeqn
(\ref{eq:dif}) expresses the fact that Q is the right--inverse of $K^a$ 
expressed in terms of $\acal$ and $\bcal$. Then, $Q$ is also the left--inverse,
leading to the difference equation
\beqn
Q_{\mu\sigma}(x_0,y_0+1)\acal_{\nu\sigma}(y_0)
+ Q_{\mu\sigma}(x_0,y_0)\bcal_{\sigma\nu}(y_0) & & \nonumber\\
+ Q_{\mu\sigma}(x_0,y_0-1)\acal_{\sigma\nu}(y_0-1) 
&=& \delta_{\mu\nu}\delta_{x_0,y_0}.
\label{eq:difleft}
\eeqn
The homogeneous solutions of~(\ref{eq:difleft}) are also given 
by $\psi^f$ and $\psi^b$. Assuming the validity of~(\ref{eq:dif}) 
and~(\ref{eq:difleft}) for $x_0\neq y_0$ and demanding the symmetry
\beq
Q_{\mu\nu}(x_0,y_0) = Q_{\nu\mu}(y_0,x_0)
\eeq
leads to
\beq
Q_{\mu\nu}(x_0,y_0) = \left\{
\begin{array}{ll}
\psi^f_{\mu\sigma}(x_0)\wcal^{-1}_{\lambda\sigma}\psi^b_{\nu\lambda}(y_0)
& \mbox{for $x_0\leq y_0$} \\
\psi^b_{\mu\sigma}(x_0)\wcal^{-1}_{\sigma\lambda}\psi^f_{\nu\lambda}(y_0)
& \mbox{for $x_0\geq y_0$}
\end{array}
\right.
\label{eq:propwithq}
\eeq
We now demand that the matrix $\wcal$ makes this definition consistent and 
also for $x_0=y_0$ 
solves~(\ref{eq:dif}) and~(\ref{eq:difleft}). These 
requirements can be shown to imply
\beq
\wcal_{\mu\nu} = 
\psi^f_{\sigma\mu}(x_0)\acal_{\sigma\lambda}(x_0)\psi^b_{\lambda\nu}(x_0+1)
-\psi^f_{\sigma\mu}(x_0+1)\acal_{\lambda\sigma}(x_0)\psi^b_{\lambda\nu}(x_0).
\eeq
This is a Wronskian form which is easy to compute. Using~(\ref{eq:dif}) and
the fact that 
$\psi^f_{\sigma\mu}(x_0)\acal_{\sigma\lambda}(x_0)\psi^b_{\lambda\nu}(x_0+1)$
is symmetric in $\mu$ and $\nu$, $\wcal$ can easily be shown to be independent of
$x_0$. A particularly simple choice is $x_0=0$, which gives after some algebra
\beqn
\wcal_{0\nu} &=& \acal_{0\sigma}(0)\psi^b_{\sigma\nu}(1)
+\bigl(\bcal_{00}(0)+\acal_{00}(-1)\bigr)\psi^b_{0\nu}(0),\\
\wcal_{k\nu} &=& -\acal_{jk}(0)\psi^b_{j\nu}(0).
\eeqn
Now one only has to insert $\wcal$ into~(\ref{eq:propwithq}) to get the
propagator.

The computation of the ghost propagator proceeds along the same lines. Due
to the absence of the spin matrix structure, it is, however, much simpler.

\section{The quark propagator}
\sectionmark{The quark propagator}

The quark propagator is computed by carrying over the methods used for
the gluon and ghost fields. The propagator $S$ is the inverse of the 
Dirac--Wilson operator $\tilde{D}$ from~(\ref{eq:dwithb}), 
which means one has to solve the equation
\beq
\sum_{y_0}\tilde{D}(\vp;x_0,y_0)S(\vp;y_0,z_0)=\delta_{x_0,z_0},
\label{eq:diracinv}
\eeq
with the boundary conditions
\beqn
\left. P_+ S(\vp;x_0,y_0)\right|_{x_0=0}
&=& \left. P_- S(\vp;x_0,y_0)\right|_{x_0=T} = 0, \\
\left. S(\vp;x_0,y_0)P_- \right|_{y_0=0}
&=& \left. S(\vp;x_0,y_0)P_+ \right|_{y_0=T} = 0,
\eeqn
which are obtained by setting the boundary quark fields to zero.

In analogy to the gluon case, one first constructs solutions of the homogeneous
equation by forward and backward recursion. This means one solves
\beq
\sum_{y_0}\tilde{D}(\vp;x_0,y_0)\psi^f(\vp,y_0)=0
\eeq
and
\beq
\sum_{y_0}\tilde{D}(\vp;x_0,y_0)\psi^b(\vp,y_0)=0,
\eeq 
starting from
\beq
\left. P_+\psi^f(\vp,x_0)\right|_{x_0=0}=0
\eeq
and
\beq
\left. P_-\psi^b(\vp,x_0)\right|_{x_0=T}=0.
\eeq
Here, only one starting value per solution is required, since one only has to
solve a first order difference equation. This is easily seen by defining
\beq
F^{f/b}(x_0) = P_- \psi^{f/b}(\vp,x_0) +P_+ \psi^{f/b}(\vp,x_0-1).
\eeq
Writing down the Dirac--Wilson operator in the form of~(\ref{eq:dwithb}),
\beq
\tilde{D}(\vp;x_0,y_0) = -P_-\delta_{x_0+1,y_0} + B(\vp^+,x_0)\delta_{x_0,y_0}
- P_+\delta_{x_0-1,y_0},
\eeq
then gives
\beq
\left[ B(\vp^+,x_0)P_+ -P_-\right] F^{f/b}(x_0+1)
+\left[ B(\vp^+,x_0)P_- -P_+\right] F^{f/b}(x_0)=0,
\eeq
which is solvable by one--step recursion.

Having computed the solutions $\psi^f$ and $\psi^b$, one can construct the
propagator for $x_0\neq y_0$,
\beq
S(\vp;x_0,y_0) = \left\{
\begin{array}{ll}
\psi^f(\vp,x_0)N^f(\vp,y_0)\gamma_5 & \mbox{for $x_0<y_0$},\\
\psi^b(\vp,x_0)N^b(\vp,y_0)\gamma_5 & \mbox{for $x_0>y_0$}.
\end{array}\right.
\eeq
In order to determine $N^f$ and $N^b$, one has to impose the required symmetry
of the propagator,
\beq
\gamma_5 S(\vp;x_0,y_0)\gamma_5 = S(\vp;y_0,x_0)^{\dagger},
\label{eq:quarkpropsym}
\eeq
which yields
\beqn
N^f(\vp,y_0) &=& \vcal(\vp,y_0) \psi^b(\vp,y_0)^{\dagger},\\
N^b(\vp,y_0) &=& \vcal(\vp,y_0)^{\dagger}\psi^f(\vp,y_0)^{\dagger}.
\eeqn
Knowing that~(\ref{eq:diracinv}) is valid especially for $z_0=x_0+1$
and $z_0=x_0-1$ and using~(\ref{eq:quarkpropsym}), one now concludes that,
for the case $x_0=y_0$, one has
\beqn
S(\vp;x_0,x_0) &=& 
P_- \psi^f(\vp,x_0)\vcal(\vp,x_0)\psi^b(\vp,x_0)^{\dagger}\gamma_5\nonumber\\
& &+ P_+\psi^b(\vp,x_0)\vcal(\vp,x_0)^{\dagger}\psi^f(\vp,x_0)^{\dagger}\gamma_5.
\eeqn
Now, one is left with the task to determine $\vcal(\vp,x_0)$. A lengthy 
calculation shows that one has
\beqn
\left[\vcal(\vp,x_0)^{\dagger}\right]^{-1} &=&
[P_-\psi^f(\vp,x_0)]^{\dagger}\gamma_5 P_+\psi^b(\vp,x_0-1) \nonumber\\
& & -[P_+\psi^f(\vp,x_0-1)]^{\dagger}\gamma_5 P_-\psi^b(\vp,x_0).
\label{eq:vexpr}
\eeqn
Furthermore, this expression can be shown to be independent of $x_0$,
\beq
\vcal(\vp,x_0) = \vcal(\vp).
\eeq
(\ref{eq:vexpr}) may hence be evaluated at an arbitrarily chosen $x_0$.
Due to the boundary conditions, the expression becomes particularly simple
for $x_0=1$, where one has
\beq
\left[\vcal(\vp)^{\dagger}\right]^{-1} =
[P_-\psi^f(\vp,1)]^{\dagger}\gamma_5 P_+\psi^b(\vp,0).
\eeq
Having computed $\vcal$, one can now construct the whole propagator,
\beq
S(\vp;x_0,y_0) = \left\{
\begin{array}{ll}
\psi^f(\vp,x_0)\vcal(\vp)\psi^b(\vp,y_0)^{\dagger}\gamma_5
& \mbox{for $x_0<y_0$},\\
\psi^b(\vp,x_0)\vcal(\vp)^{\dagger}\psi^f(\vp,y_0)^{\dagger}\gamma_5
& \mbox{for $x_0>y_0$},\\
P_- \psi^f(\vp,x_0)\vcal(\vp)\psi^b(\vp,x_0)^{\dagger}\gamma_5 & \\
\;\; + P_+\psi^b(\vp,x_0)\vcal(\vp)^{\dagger}\psi^f(\vp,x_0)^{\dagger}\gamma_5
& \mbox{for $x_0=y_0$}.
\end{array}\right.
\eeq

\cleardoublepage
\chapter{The extrapolation procedure \label{app:extra}}
\chaptermark{The extrapolation procedure}

For the extrapolation of the 1--loop data to the continuum limit, the
method described in~\cite{Bode:1999sm} was used, which is a generalisation
of the blocking technique used in~\cite{Luscher:1986wf}.

In order to simplify the notation, the following description is given in lattice
units, i.~e.~$a=1$. The results which are to be extrapolated are 
$n$ numbers $F(L)$
for a given range $L_1 <L_2\ldots <L_n$. The aim is to extract the leading 
coefficient for $L\rightarrow\infty$.

One source of errors are roundoff effects. Optimistically, one may assume that
the values $F(L)$ are correct up to machine precision,
\beq
\delta_F(L) = \epsilon |F(L)|,
\label{eq:roundoff}
\eeq
with $\epsilon\sim 10^{-14}$ for double precision arithmetic, which was used
for all computations in this thesis. Since the calculation of the Feynman 
diagrams involves large sums of terms of different sign, this is clearly an
underestimation. In order to get a more realistic idea of the size of the
roundoff errors, one could compare double precision to single (or quadruple)
precision results and estimate the $L$--dependence of $\epsilon$. This was
done, for example, with the 2--loop results for the coupling in~\cite{Bode:1999sm},
where one found a growth of the roundoff error proportional to $L^3$ for most
contributions. However, experience shows that the error of the extrapolation
usually is completely dominated by the systematic error, so that this procedure
for estimating the roundoff error did not seem to be necessary for our 
calculations. 

For the extrapolation, we assume $F(L)$ to have an asymptotic expansion in
functions $f_k(L)$ with $k=1,2,\ldots,n_f$,
\beq
F(L) = \sum_{k=1}^{n_f}\alpha_k f_k(L) +R(L),
\label{eq:Fassum}
\eeq
where the rest $R(L)$ behaves like
\beq
\left|\frac{R(L)}{f_{n_f}(L)}\right|\rightarrow 0
\quad\mbox{as}\quad L\rightarrow\infty.
\eeq
For improved 1--loop quantities, one would choose the functions $f_k(L)$ to
be $1, \ln L/L^2, 1/L^2, \ln L/L^3, 1/L^3 \ldots$. 
Writing the $n$ data values $F(L)$
as an $n$--dimensional column vector, (\ref{eq:Fassum}) becomes 
\beq
F = f\alpha +R,
\eeq
where $f$ is an $n\times n_f$ matrix and $\alpha$ is the $n_f$ dimensional vector
one wants to determine. 

In order to get $\alpha$, one has to minimise the quadratic form
\beq
\chi^2 = (F-f\alpha)^{\top}W^2(F-f\alpha).
\label{eq:chisq}
\eeq
Here, $W^2$ is an $n\times n$ matrix of positive weights. It can be used to put
an emphasis on small or large $L$. While the values at small $L$ are less affected
by roundoff errors, the asymptotic expansion is expected to hold to a better
degree at large $L$, resulting in a smaller systematic error. In this thesis,
all extrapolations where done with $W=1$, and the behaviour of the systematic
errors was examined by taking an $L$ range from $L_{\mathrm{min}}$ to 
$L_{\mathrm{max}}$ and varying $L_{\mathrm{min}}$ at fixed $L_{\mathrm{max}}$.  
However, for reasons of generality, the procedure will be outlined here for
arbitrary $W^2$. 

Minimisation of~(\ref{eq:chisq}) yields
\beq
f^{\top}W^2 f\alpha = f^{\top}W^2 F.
\eeq
The columns of $Wf$, which are assumed to be linearly independent, span an 
$n_f$--dimensional subspace. Denoting the projector onto this subspace by
$P$, one gets
\beq
W f \alpha = P W F.
\eeq
This equation has to be solved for $\alpha$. As a suitable way to do this, the
singular value decomposition~\cite{Press:Book} for $Wf$ turned out to give 
stable results. To this end, one uses the factorisation
\beq
W f = U S V^{\top},
\eeq
where $U$ is an $n\times n_f$ matrix with orthonormal columns and
\beq
U^{\top}U=1,\qquad UU^{\top} = P.
\eeq
$S$ and $V$ are both $n_f\times n_f$ matrices, $S$ being diagonal and $V$
orthonormal. Thus one gets the solution for $\alpha$,
\beq
\alpha = VS^{-1}U^{\top}WF.
\eeq
Now, one has to determine the error of $\alpha_k$. The roundoff error
$\delta_{\alpha_k}$ is obtained from~(\ref{eq:roundoff}) by simple error
propagation,
\beq
\delta_{\alpha_{k}}^2 = \sum_{L}(VS^{-1}U^{\top})^2_{kL}\delta^2_{F(L)}.
\eeq

The determination of the systematic error is a more delicate problem. 
In~\cite{Bode:1999sm}, it was found that a convenient method is the 
following.

One assumes that the remainder $R$ can be modelled by a linear combination
of $n_r$ functions $f_{n_f +1},\ldots, f_{n_f+n_r}$. For the 1--loop calculations
in this thesis, one may choose the functions $1,\ln L/L,1/L,\ldots,\ln L/L^m,
1/L^m$ with some positive integer $m$ for the fit and take the $n_r=2$ functions
$\ln L/L^{m+1}$ and $1/L^{m+1}$ to model the remainder. Now, one does $n_r$
separate fits including the $n_f$ functions used before and \emph{one} of
the $n_r$ extra functions. In these fits, one gets coefficients 
$A_1,\ldots A_{n_r}$ for the extra functions. In order to estimate the error,
one repeats the original fit, this time not using the data $F(L)$ but the
function $A_1 f_{n_f +1}(L)$. In this fit, each of the $n_f$ functions $f_k$
will get a coefficient $\beta_k$. The same is done with the other $n_r-1$
extra functions. For each function $f_k$, one thus gets $n_r$ coefficients
$\beta_k$. The largest of these $\beta_k$ is then taken as the systematic
error $d_{\alpha_k}$ of the coefficient $\alpha_k$.

\cleardoublepage
\chapter{Tables of expansion coefficients \label{app:tables}}
\chaptermark{Tables of expansion coefficients}

The 1--loop coefficient $m_1^{(1)}$, which is set to zero in order to 
compute the critical bare mass, can be decomposed into
\beq
am_1^{(1)} = u_1+\Nf u_2+\csw^{(1)}u_3+\ctt^{(1)}u_4+\ca^{(1)} u_5+a\mcrit^{(1)}u_6
+\ct^{(1)}u_7.
\eeq
The coefficients $u_i$ are shown in tables~\ref{tab:m1parts123}
and~\ref{tab:m1parts456} for $\theta=\pi/5$ and $\rho=1$ and in
tables~\ref{tab:m1parts123var} and~\ref{tab:m1parts456var} for the other
values of $\theta$ and $\rho$.

In the same way, $m_1^{(1)}(2L)$ and $m'$ (defined using the upper boundary
quark fields) may be decomposed according to
\beq
am_1^{(1)}(2L/a) = v_1+\Nf v_2+\csw^{(1)}v_3+\ctt^{(1)}v_4+\ca^{(1)} v_5
+a\mcrit^{(1)}(L/a)v_6
+\ct^{(1)}v_7,
\eeq
with the coefficients $v_i$ for $\theta=\pi/5$ and $\rho=1$
in tables~\ref{tab:m1parts2L123}
and~\ref{tab:m1parts2L456}, and
\beq
am'^{(1)} = w_1+\Nf w_2+\csw^{(1)}w_3+\ctt^{(1)}w_4+\ca^{(1)} w_5+a\mcrit^{(1)}w_6
+\ct^{(1)}w_7,
\eeq
with the coefficients $w_i$ for $\theta=\pi/5$ and $\rho=1$
in tables~\ref{tab:mprparts123}
and~\ref{tab:mprparts456}.

The renormalisation constant $Z_{\mathrm{P}}$ has the coefficients
\beq
Z_{\mathrm{P}}^{(1)} = z_1 +\Nf z_2 +\csw^{(1)}z_3 +\ctt^{(1)}z_4 +a\mcrit^{(1)}z_6 
+\ct^{(1)}z_7,
\eeq
where the coefficients $z_i$ can be found in tables~\ref{tab:zpparts123}
and~\ref{tab:zpparts467} for $\theta=\pi/5$ and $\rho=1$ and in 
tables~\ref{tab:zpparts123var} and~\ref{tab:zpparts467var} for different 
values of $\theta$ and $\rho$.
Finally, the renormalisation constant $Z_{\mathrm{P}}$ at $2L$ such that
$m_1(L)=0$ may be expanded according to
\beq
Z_{\mathrm{P}}^{(1)}(2L/a) = \tilde{z}_1 +\Nf\tilde{z}_2 
+\csw^{(1)}\tilde{z}_3 +\ctt^{(1)}\tilde{z}_4 
+a\mcrit^{(1)}(L/a)\tilde{z}_6 
+\ct^{(1)}\tilde{z}_7,
\eeq
with the coefficients $\tilde{z}_i$ in tables~\ref{tab:zptparts123}
and~\ref{tab:zptparts467} for $\theta=\pi/5$ and $\rho=1$ and in
tables~\ref{tab:zptparts123var} and~\ref{tab:zptparts467var} for the
other values of $\theta$ and $\rho$.

In the tables, up to ten digits of the coefficients are shown. Since they
have been computed using double precision arithmetic and experience shows
that up to $L/a=32$, one looses two to three digits due to roundoff errors,
all digits shown here should be significant. To get a better estimate for
the roundoff errors, one would have to compare the results for some small
lattices with single or quadruple precision results and extract the scaling
behaviour of the errors for increasing $L/a$. In this thesis, however, we
are only interested in discretisation errors, where the last digits are
irrelevant, and in extrapolations, which are dominated by the systematic
errors due to higher order terms in the expansion. Therefore, a precise
determination of the roundoff errors did not seem to be necessary.
\begin{table}[htbp]
  \begin{center}
    \begin{tabular}{|r|r|r|r|}
      \hline\hline
      $L/a$ 
      & \multicolumn{1}{c|}{$u_1$} 
      & \multicolumn{1}{c|}{$u_2$} 
      & \multicolumn{1}{c|}{$u_3$} \\
      \hline
 4 &    0.3602839939 &   -0.003225474813 &   -0.09605428870 \\
 5 &    0.3232002646 &   -0.000774641219 &   -0.05739586792 \\
 6 &    0.3037717960 &   -0.000214828175 &   -0.03886722333 \\
 7 &    0.2945227639 &   -0.000090081829 &   -0.02788612762 \\
 8 &    0.2886435020 &   -0.000047062385 &   -0.02116169433 \\
 9 &    0.2846484024 &   -0.000027131359 &   -0.01653724923 \\
10 &    0.2818166195 &   -0.000017284660 &   -0.01334401178 \\
11 &    0.2797403581 &   -0.000011720940 &   -0.01096120300 \\
12 &    0.2781681231 &   -0.000008284863 &   -0.00919331010 \\
13 &    0.2769535573 &   -0.000006129109 &   -0.00780391809 \\
14 &    0.2759917826 &   -0.000004608993 &   -0.00672232221 \\
15 &    0.2752204783 &   -0.000003596909 &   -0.00584115805 \\
16 &    0.2745896768 &   -0.000002820717 &   -0.00513109826 \\
17 &    0.2740694316 &   -0.000002283224 &   -0.00453708978 \\
18 &    0.2736334509 &   -0.000001846627 &   -0.00404578595 \\
19 &    0.2732659914 &   -0.000001535330 &   -0.00362632773 \\
20 &    0.2729521158 &   -0.000001271725 &   -0.00327224614 \\
21 &    0.2726829531 &   -0.000001079336 &   -0.00296502249 \\
22 &    0.2724494848 &   -0.000000911236 &   -0.00270139289 \\
23 &    0.2722464289 &   -0.000000786142 &   -0.00246962788 \\
24 &    0.2720680657 &   -0.000000674154 &   -0.00226804856 \\
25 &    0.2719111055 &   -0.000000589418 &   -0.00208889322 \\
26 &    0.2717717705 &   -0.000000512088 &   -0.00193130204 \\
27 &    0.2716479341 &   -0.000000452721 &   -0.00178994940 \\
28 &    0.2715370145 &   -0.000000397695 &   -0.00166441225 \\
29 &    0.2714375933 &   -0.000000354905 &   -0.00155092337 \\
30 &    0.2713478540 &   -0.000000314740 &   -0.00144929720 \\
31 &    0.2712668249 &   -0.000000283144 &   -0.00135679872 \\
32 &    0.2711931957 &   -0.000000253175 &   -0.00127337243 \\
      \hline\hline
    \end{tabular}
    \caption[Parts of the 1--loop coefficient $m_1^{(1)}$ for $\theta=\pi/5$
             and $\rho=1$]
            {\sl Parts of the 1--loop coefficient $m_1^{(1)}$ for $\theta=\pi/5$
              and $\rho=1$
    \label{tab:m1parts123}}
  \end{center}
\end{table}

\begin{table}[htbp]
  \begin{center}
    \begin{tabular}{|r|r|r|r|r|}
      \hline\hline
      $L/a$ 
      & \multicolumn{1}{c|}{$u_4$} 
      & \multicolumn{1}{c|}{$u_5$} 
      & \multicolumn{1}{c|}{$u_6$} 
      & \multicolumn{1}{c|}{$u_7$}\\
      \hline
 4 &    0.7316730490 &    0.6084363759 &    1.188647276 &    0.01803922007 \\
 5 &    0.3210591192 &    0.3878385759 &    1.117262157 &    0.00413349286 \\
 6 &    0.0042189812 &    0.2631884549 &    1.079778023 &   -0.00157603839 \\
 7 &    0.0024050333 &    0.1935221158 &    1.057964363 &   -0.00069723702 \\
 8 &    0.0012919050 &    0.1464623871 &    1.043886577 &   -0.00038756353 \\
 9 &    0.0008286417 &    0.1158988293 &    1.034483484 &   -0.00020611832 \\
10 &    0.0005124163 &    0.0932417601 &    1.027741909 &   -0.00012753286 \\
11 &    0.0003555287 &    0.0771715444 &    1.022853029 &   -0.00007616582 \\
12 &    0.0002404411 &    0.0645591187 &    1.019116087 &   -0.00005074611 \\
13 &    0.0001763427 &    0.0550783445 &    1.016254342 &   -0.00003280704 \\
14 &    0.0001268701 &    0.0473445009 &    1.013970372 &   -0.00002306893 \\
15 &    0.0000969510 &    0.0412863218 &    1.012152254 &   -0.00001579511 \\
16 &    0.0000729793 &    0.0362045275 &    1.010655337 &   -0.00001157037 \\
17 &    0.0000575581 &    0.0320993397 &    1.009428779 &   -0.00000827330 \\
18 &    0.0000448460 &    0.0285822227 &    1.008394882 &   -0.00000625754 \\
19 &    0.0000362650 &    0.0256723716 &    1.007528515 &   -0.00000462840 \\
20 &    0.0000290320 &    0.0231377112 &    1.006784665 &   -0.00000359173 \\
21 &    0.0000239575 &    0.0210003357 &    1.006150100 &   -0.00000272952 \\
22 &    0.0000196031 &    0.0191135425 &    1.005597146 &   -0.00000216309 \\
23 &    0.0000164500 &    0.0174975213 &    1.005118494 &   -0.00000168054 \\
24 &    0.0000137045 &    0.0160552118 &    1.004696293 &   -0.00000135523 \\
25 &    0.0000116632 &    0.0148037736 &    1.004326361 &   -0.00000107234 \\
26 &    0.0000098641 &    0.0136765405 &    1.003996718 &   -0.00000087756 \\
27 &    0.0000084961 &    0.0126877084 &    1.003704907 &   -0.00000070513 \\
28 &    0.0000072780 &    0.0117900311 &    1.003442618 &   -0.00000058430 \\
29 &    0.0000063338 &    0.0109951461 &    1.003208384 &   -0.00000047564 \\
30 &    0.0000054857 &    0.0102686695 &    1.002996274 &   -0.00000039839 \\
31 &    0.0000048172 &    0.0096201313 &    1.002805407 &   -0.00000032793 \\
32 &    0.0000042121 &    0.0090239315 &    1.002631448 &   -0.00000027722 \\
      \hline\hline
    \end{tabular}
    \caption[Parts of the 1--loop coefficient $m_1^{(1)}$ 
             for $\theta=\pi/5$ and $\rho=1$ (cont.)]
            {\sl Parts of the 1--loop coefficient $m_1^{(1)}$ 
             for $\theta=\pi/5$ and $\rho=1$ (cont.)
    \label{tab:m1parts456}}
  \end{center}
\end{table}

\begin{table}
  \begin{center}
    \begin{tabular}{|r|r|r|r|}
      \hline\hline
      $L/a$  
      & \multicolumn{1}{c|}{$u_1$}
      & \multicolumn{1}{c|}{$u_2$}
      & \multicolumn{1}{c|}{$u_3$}\\
      \hline   
      \multicolumn{4}{|c|}{$\theta=0$, $\rho=1$}\\
      \hline
 4 &    0.3117291700 &   -0.001413989874 &   -0.06325439280 \\
 6 &    0.2851814233 &   -0.000072348720 &   -0.02653692904 \\
 8 &    0.2786313238 &   -0.000015427643 &   -0.01462634537 \\
10 &    0.2755771628 &   -0.000005279182 &   -0.00927252482 \\
12 &    0.2739097717 &   -0.000002238838 &   -0.00640604311 \\
14 &    0.2729004050 &   -0.000001074282 &   -0.00469179577 \\
16 &    0.2722432007 &   -0.000000559106 &   -0.00358486772 \\
      \hline   
      \multicolumn{4}{|c|}{$\theta=0.5$, $\rho=1$}\\
      \hline
 4 &    0.3527716036 &   -0.003134326041 &   -0.08882286240 \\
 6 &    0.3009793922 &   -0.000203934067 &   -0.03600623562 \\
 8 &    0.2872085538 &   -0.000044256296 &   -0.01960752669 \\
10 &    0.2809505241 &   -0.000016156714 &   -0.01236285801 \\
12 &    0.2775900272 &   -0.000007687118 &   -0.00851637431 \\
14 &    0.2755788098 &   -0.000004247030 &   -0.00622673813 \\
16 &    0.2742799815 &   -0.000002584214 &   -0.00475246807 \\
      \hline   
      \multicolumn{4}{|c|}{$\theta=0$, $\rho=2$}\\
      \hline
 4 &    0.2890122981 &   -0.000466276200 &   -0.02966169499 \\
 6 &    0.2776917924 &   -0.000096885717 &   -0.01285486035 \\
 8 &    0.2742017998 &   -0.000032373848 &   -0.00717204102 \\
10 &    0.2726668499 &   -0.000014117091 &   -0.00457353269 \\
12 &    0.2718544505 &   -0.000007249144 &   -0.00317000300 \\
      \hline   
      \multicolumn{4}{|c|}{$\theta=0.5$, $\rho=2$}\\
      \hline
 4 &    0.3230287075 &    0.004646415684 &   -0.04091643990 \\
 6 &    0.2928573371 &    0.000682358234 &   -0.01682956807 \\
 8 &    0.2827493514 &    0.000177104745 &   -0.00920123428 \\
10 &    0.2781432425 &    0.000066678908 &   -0.00581015860 \\
12 &    0.2756603756 &    0.000030966376 &   -0.00400514977 \\
      \hline\hline
    \end{tabular}
    \caption[Parts of the 1--loop coefficient $m_1^{(1)}$ for various values
             of $\theta$ and $\rho$]
            {\sl Parts of the 1--loop coefficient $m_1^{(1)}$ for various values
             of $\theta$ and $\rho$
    \label{tab:m1parts123var}}
  \end{center}
\end{table}

\begin{table}
  \begin{center}
    \begin{tabular}{|r|r|r|r|r|}
      \hline\hline
      $L/a$  
      & \multicolumn{1}{c|}{$u_4$}
      & \multicolumn{1}{c|}{$u_5$}
      & \multicolumn{1}{c|}{$u_6$}
      & \multicolumn{1}{c|}{$u_7$}\\
      \hline   
      \multicolumn{5}{|c|}{$\theta=0$, $\rho=1$}\\
      \hline
 4 &    0.5887651776 &    0.2773460352 &    1.060808688 &    0.01414891282 \\
 6 &    0.0007741031 &    0.1200813871 &    1.027928002 &   -0.00039927726 \\
 8 &    0.0001985526 &    0.0670949370 &    1.016024884 &   -0.00007249879 \\
10 &    0.0000659113 &    0.0428310908 &    1.010376100 &   -0.00001642414 \\
12 &    0.0000255381 &    0.0297073108 &    1.007259112 &   -0.00000361335 \\
14 &    0.0000108268 &    0.0218109132 &    1.005360129 &   -0.00000027458 \\
16 &    0.0000047721 &    0.0166920181 &    1.004118703 &    0.00000058142 \\
      \hline   
      \multicolumn{5}{|c|}{$\theta=0.5$, $\rho=1$}\\
      \hline
 4 &    0.7136493082 &    0.5749737393 &    1.168810060 &    0.01714141379 \\
 6 &    0.0030661542 &    0.2495987793 &    1.072695582 &   -0.00138698280 \\
 8 &    0.0009267006 &    0.1391978462 &    1.040386927 &   -0.00033684366 \\
10 &    0.0003656376 &    0.0887236823 &    1.025677627 &   -0.00010975186 \\
12 &    0.0001711788 &    0.0614750484 &    1.017759437 &   -0.00004324006 \\
14 &    0.0000902366 &    0.0451034411 &    1.013012168 &   -0.00001945044 \\
16 &    0.0000518894 &    0.0345013971 &    1.009943004 &   -0.00000964499 \\
      \hline   
      \multicolumn{5}{|c|}{$\theta=0$, $\rho=2$}\\
      \hline
 4 &    0.0032392824 &    0.0936449462 &    1.036188030 &   -0.00040229454 \\
 6 &    0.0005702345 &    0.0394953561 &    1.014454343 &   -0.00004485954 \\
 8 &    0.0001630428 &    0.0217769672 &    1.007693781 &   -0.00000952668 \\
10 &    0.0000612800 &    0.0138044243 &    1.004761213 &   -0.00000285504 \\
12 &    0.0000273938 &    0.0095357430 &    1.003232426 &   -0.00000105727 \\
      \hline   
      \multicolumn{5}{|c|}{$\theta=0.5$, $\rho=2$}\\
      \hline
 4 &    0.0011902517 &    0.4959636094 &    1.137886597 &   -0.00272630910 \\
 6 &    0.0001742293 &    0.2178875276 &    1.060855811 &   -0.00034621022 \\
 8 &    0.0000433483 &    0.1220946720 &    1.034142311 &   -0.00008143737 \\
10 &    0.0000147043 &    0.0780067469 &    1.021817535 &   -0.00002664492 \\
12 &    0.0000060319 &    0.0541217346 &    1.015134789 &   -0.00001071394 \\
      \hline\hline
    \end{tabular}
    \caption[Parts of the 1--loop coefficient $m_1^{(1)}$ for various values
             of $\theta$ and $\rho$ (cont.)]
            {\sl Parts of the 1--loop coefficient $m_1^{(1)}$ for various values
             of $\theta$ and $\rho$ (cont.)
    \label{tab:m1parts456var}}
  \end{center}
\end{table}

\begin{table}[htbp]
  \begin{center}
    \begin{tabular}{|r|r|r|r|}
      \hline\hline
      $L/a$ 
      & \multicolumn{1}{c|}{$v_1$} 
      & \multicolumn{1}{c|}{$v_2$} 
      & \multicolumn{1}{c|}{$v_3$} \\
      \hline
 4 &    0.2888594021 &   -0.00005058908107 &   -0.02117043086 \\
 5 &    0.2820887015 &   -0.00001972144488 &   -0.01335256814 \\
 6 &    0.2782616600 &   -0.00000882525879 &   -0.00919576341 \\
 7 &    0.2760809582 &   -0.00000497218858 &   -0.00672431079 \\
 8 &    0.2746373747 &   -0.00000296561503 &   -0.00513202789 \\
 9 &    0.2736756316 &   -0.00000194609439 &   -0.00404651418 \\
10 &    0.2729796791 &   -0.00000132369711 &   -0.00327267396 \\
11 &    0.2724733713 &   -0.00000094810144 &   -0.00270172931 \\
12 &    0.2720853970 &   -0.00000069646118 &   -0.00226827213 \\
13 &    0.2717867718 &   -0.00000052844286 &   -0.00193148046 \\
14 &    0.2715486018 &   -0.00000040853487 &   -0.00166454014 \\
15 &    0.2713579394 &   -0.00000032292580 &   -0.00144940101 \\
16 &    0.2712013150 &   -0.00000025894756 &   -0.00127345074 \\
      \hline\hline
    \end{tabular}
    \caption[Parts of the 1--loop coefficient $m_1^{(1)}(2L)$ with $m_1(L) = 0$
             for $\theta=\pi/5$ and $\rho=1$]
            {\sl Parts of the 1--loop coefficient $m_1^{(1)}(2L)$ with $m_1(L) = 0$
             for $\theta=\pi/5$  and $\rho=1$
    \label{tab:m1parts2L123}}
  \end{center}
\end{table}

\begin{table}[htbp]
  \begin{center}
    \begin{tabular}{|r|r|r|r|r|}
      \hline\hline
      $L/a$ 
      & \multicolumn{1}{c|}{$v_4$} 
      & \multicolumn{1}{c|}{$v_5$} 
      & \multicolumn{1}{c|}{$v_6$} 
      & \multicolumn{1}{c|}{$v_7$}\\
      \hline
 4 &    0.001291461837 &    0.1463868229 &    1.045074777 &   -0.0003869872089 \\
 5 &    0.000511742035 &    0.0931663422 &    1.029247500 &   -0.0001269718170 \\
 6 &    0.000240252491 &    0.0645364689 &    1.019635452 &   -0.0000506118783 \\
 7 &    0.000126734771 &    0.0473260174 &    1.014466186 &   -0.0000229801784 \\
 8 &    0.000072924806 &    0.0361957674 &    1.010920783 &   -0.0000115365918 \\
 9 &    0.000044809374 &    0.0285753372 &    1.008629730 &   -0.0000062357257 \\
10 &    0.000029013497 &    0.0231336427 &    1.006938179 &   -0.0000035810441 \\
11 &    0.000019590549 &    0.0191103368 &    1.005730206 &   -0.0000021559966 \\
12 &    0.000013697243 &    0.0160530751 &    1.004792851 &   -0.0000013512045 \\
13 &    0.000009858993 &    0.0136748331 &    1.004080301 &   -0.0000008747879 \\
14 &    0.000007274769 &    0.0117888051 &    1.003507183 &   -0.0000005825718 \\
15 &    0.000005483328 &    0.0102676735 &    1.003052472 &   -0.0000003971558 \\
16 &    0.000004210539 &    0.0090231792 &    1.002676692 &   -0.0000002763990 \\
      \hline\hline
    \end{tabular}
    \caption[Parts of the 1--loop coefficient $m_1^{(1)}(2L)$ 
     with $m_1(L) = 0$ for $\theta=\pi/5$ and $\rho=1$ (cont.)]
            {\sl Parts of the 1--loop coefficient $m_1^{(1)}(2L)$ 
     with $m_1(L) = 0$ for $\theta=\pi/5$ and $\rho=1$ (cont.)
    \label{tab:m1parts2L456}}
  \end{center}
\end{table}

\begin{table}[htbp]
  \begin{center}
    \begin{tabular}{|r|r|r|r|}
      \hline\hline
      $L/a$ 
      & \multicolumn{1}{c|}{$w_1$} 
      & \multicolumn{1}{c|}{$w_2$} 
      & \multicolumn{1}{c|}{$w_3$} \\
      \hline
 4 &    0.2971376852 &    0.007159453631 &    0.07069378144 \\
 5 &    0.2930551099 &    0.002002507543 &    0.04435489541 \\
 6 &    0.2845876866 &    0.000904887361 &    0.03097531507 \\
 7 &    0.2820563435 &    0.000384152314 &    0.02227017749 \\
 8 &    0.2796089169 &    0.000228377566 &    0.01706219846 \\
 9 &    0.2779500101 &    0.000127950492 &    0.01329286954 \\
10 &    0.2764995647 &    0.000087862653 &    0.01077351137 \\
11 &    0.2755037279 &    0.000057206374 &    0.00882189295 \\
12 &    0.2746365666 &    0.000042530198 &    0.00741823947 \\
13 &    0.2740124775 &    0.000030217871 &    0.00628082481 \\
14 &    0.2734635897 &    0.000023601409 &    0.00541982387 \\
15 &    0.2730515248 &    0.000017755117 &    0.00469970451 \\
16 &    0.2726849153 &    0.000014352239 &    0.00413374337 \\
17 &    0.2724001308 &    0.000011245712 &    0.00364919780 \\
18 &    0.2721440627 &    0.000009327024 &    0.00325731535 \\
19 &    0.2719396253 &    0.000007535561 &    0.00291573391 \\
20 &    0.2717540932 &    0.000006376663 &    0.00263317208 \\
21 &    0.2716026412 &    0.000005276671 &    0.00238335648 \\
22 &    0.2714640928 &    0.000004537982 &    0.00217290423 \\
23 &    0.2713489030 &    0.000003828045 &    0.00198468504 \\
24 &    0.2712427988 &    0.000003336311 &    0.00182372283 \\
25 &    0.2711532200 &    0.000002859114 &    0.00167838569 \\
26 &    0.2710702125 &    0.000002519843 &    0.00155251931 \\
27 &    0.2709992153 &    0.000002188062 &    0.00143795523 \\
28 &    0.2709330829 &    0.000001946840 &    0.00133766988 \\
29 &    0.2708758835 &    0.000001709481 &    0.00124576235 \\
30 &    0.2708223594 &    0.000001533525 &    0.00116456364 \\
31 &    0.2707756142 &    0.000001359513 &    0.00108970680 \\
32 &    0.2707316956 &    0.000001228300 &    0.00102303872 \\
      \hline\hline
    \end{tabular}
    \caption[Parts of $m'$ at 1--loop order with $m_1(L) = 0$
             for $\theta=\pi/5$ and $\rho=1$]
            {\sl Parts of $m'$ at 1--loop order with $m_1(L) = 0$ 
             for $\theta=\pi/5$ and $\rho=1$ 
    \label{tab:mprparts123}}
  \end{center}
\end{table}

\begin{table}[htbp]
  \begin{center}
    \begin{tabular}{|r|r|r|r|r|}
      \hline\hline
      $L/a$ 
      & \multicolumn{1}{c|}{$w_4$} 
      & \multicolumn{1}{c|}{$w_5$} 
      & \multicolumn{1}{c|}{$w_6$} 
      & \multicolumn{1}{c|}{$w_7$}\\
      \hline
 4 &    0.6670524594 &    0.8104561493 &    1.111943322 &   -0.01095404633 \\
 5 &    0.3903966722 &    0.5388856057 &    1.091349905 &   -0.00357819396 \\
 6 &   -0.0022730281 &    0.3747386559 &    1.069419226 &    0.00063957047 \\
 7 &   -0.0010901031 &    0.2781810491 &    1.055379436 &    0.00017422876 \\
 8 &   -0.0006270025 &    0.2119249521 &    1.043698522 &   -0.00008757089 \\
 9 &   -0.0003743296 &    0.1681735583 &    1.035783207 &   -0.00008690205 \\
10 &   -0.0002493311 &    0.1356020041 &    1.029322006 &   -0.00009872911 \\
11 &   -0.0001683210 &    0.1123440306 &    1.024688617 &   -0.00007582581 \\
12 &   -0.0001217515 &    0.0940684807 &    1.020845628 &   -0.00006576171 \\
13 &   -0.0000885627 &    0.0802868921 &    1.017957986 &   -0.00005052598 \\
14 &   -0.0000674800 &    0.0690416443 &    1.015515375 &   -0.00004195232 \\
15 &   -0.0000515736 &    0.0602178847 &    1.013611180 &   -0.00003293300 \\
16 &   -0.0000407400 &    0.0528164319 &    1.011971652 &   -0.00002726629 \\
17 &   -0.0000322524 &    0.0468314608 &    1.010655815 &   -0.00002188842 \\
18 &   -0.0000261695 &    0.0417044433 &    1.009505714 &   -0.00001826479 \\
19 &   -0.0000212731 &    0.0374600536 &    1.008560908 &   -0.00001495738 \\
20 &   -0.0000176252 &    0.0337634280 &    1.007724628 &   -0.00001261428 \\
21 &   -0.0000146275 &    0.0306449645 &    1.007024441 &   -0.00001050889 \\
22 &   -0.0000123250 &    0.0278924569 &    1.006398071 &   -0.00000895924 \\
23 &   -0.0000104015 &    0.0255343162 &    1.005865307 &   -0.00000757449 \\
24 &   -0.0000088871 &    0.0234298978 &    1.005384399 &   -0.00000652389 \\
25 &   -0.0000076047 &    0.0216036230 &    1.004969907 &   -0.00000558578 \\
26 &   -0.0000065743 &    0.0199587646 &    1.004592861 &   -0.00000485624 \\
27 &   -0.0000056916 &    0.0185156623 &    1.004264203 &   -0.00000420373 \\
28 &   -0.0000049703 &    0.0172056983 &    1.003963238 &   -0.00000368576 \\
29 &   -0.0000043461 &    0.0160456199 &    1.003698336 &   -0.00000322116 \\
30 &   -0.0000038285 &    0.0149854514 &    1.003454342 &   -0.00000284588 \\
31 &   -0.0000033768 &    0.0140389494 &    1.003237764 &   -0.00000250813 \\
32 &   -0.0000029975 &    0.0131688820 &    1.003037259 &   -0.00000223120 \\
      \hline\hline
    \end{tabular}
    \caption[Parts of $m'$ at 1--loop order with $m_1(L) = 0$ for $\theta=\pi/5$ 
             and $\rho=1$ (cont.)]
            {\sl Parts of $m'$ at 1--loop order with $m_1(L) = 0$ 
             for $\theta=\pi/5$ and $\rho=1$ (cont.)
    \label{tab:mprparts456}}
  \end{center}
\end{table}

\begin{table}[htbp]
  \begin{center}
    \begin{tabular}{|r|r|r|r|}
      \hline\hline
      $L/a$ 
      & \multicolumn{1}{c|}{$z_1$} 
      & \multicolumn{1}{c|}{$z_2$} 
      & \multicolumn{1}{c|}{$z_3$} \\
      \hline
 4 &    0.09550891946 &    0.03423528831 &   -0.1341134935 \\
 6 &    0.16565857330 &    0.01473258948 &   -0.0932504829 \\
 8 &    0.22869553805 &    0.00949691345 &   -0.0711779334 \\
10 &    0.28916821812 &    0.00753246481 &   -0.0574507258 \\
12 &    0.34887762674 &    0.00654228932 &   -0.0481191321 \\
14 &    0.40849882024 &    0.00594073122 &   -0.0413752997 \\
16 &    0.46828819102 &    0.00553322566 &   -0.0362791400 \\
18 &    0.52833626903 &    0.00523794769 &   -0.0322950506 \\
20 &    0.58866515448 &    0.00501397159 &   -0.0290961208 \\
22 &    0.64926855814 &    0.00483827767 &   -0.0264717741 \\
24 &    0.71012902785 &    0.00469683169 &   -0.0242803679 \\
26 &    0.77122555828 &    0.00458056303 &   -0.0224231782 \\
28 &    0.83253695525 &    0.00448334105 &   -0.0208293183 \\
30 &    0.89404326241 &    0.00440087155 &   -0.0194465900 \\
32 &    0.95572628769 &    0.00433005700 &   -0.0182357146 \\
      \hline\hline
    \end{tabular}
    \caption[Parts of the 1--loop coefficient $Z_{\mathrm{P}}^{(1)}$
             at $\theta=\pi/5$ and $\rho=1$]
            {\sl Parts of the 1--loop coefficient $Z_{\mathrm{P}}^{(1)}$
              at $\theta=\pi/5$ and $\rho=1$}
    \label{tab:zpparts123}
  \end{center}
\end{table}

\begin{table}[htbp]
  \begin{center}
    \begin{tabular}{|r|r|r|r|}
      \hline\hline
      $L/a$ 
      & \multicolumn{1}{c|}{$z_4$} 
      & \multicolumn{1}{c|}{$z_6$} 
      & \multicolumn{1}{c|}{$z_7$} \\
      \hline
 4 &    0.2752077673 &    1.240156503 &   -0.1954547977 \\
 6 &    0.3024134295 &    1.582932389 &   -0.1615028450 \\
 8 &    0.2903043621 &    1.870599424 &   -0.1331757885 \\
10 &    0.2677764380 &    2.136141388 &   -0.1121302285 \\
12 &    0.2446013168 &    2.391341118 &   -0.0964095495 \\
14 &    0.2234839233 &    2.641100707 &   -0.0843763356 \\
16 &    0.2049375164 &    2.887730110 &   -0.0749275370 \\
18 &    0.1888173394 &    3.132429195 &   -0.0673363702 \\
20 &    0.1748109198 &    3.375870929 &   -0.0611160764 \\
22 &    0.1625955840 &    3.618456709 &   -0.0559323228 \\
24 &    0.1518850825 &    3.860438271 &   -0.0515494198 \\
26 &    0.1424386316 &    4.101980209 &   -0.0477971213 \\
28 &    0.1340577488 &    4.343193976 &   -0.0445497094 \\
30 &    0.1265798562 &    4.584157332 &   -0.0417124957 \\
32 &    0.1198717596 &    4.824925943 &   -0.0392128935 \\
      \hline\hline
    \end{tabular}
    \caption[Parts of the 1--loop coefficient $Z_{\mathrm{P}}^{(1)}$
             at $\theta=\pi/5$ and $\rho=1$ (cont.)]
            {\sl Parts of the 1--loop coefficient $Z_{\mathrm{P}}^{(1)}$
              at $\theta=\pi/5$ and $\rho=1$ (cont.)}
    \label{tab:zpparts467}
  \end{center}
\end{table}

\begin{table}[htbp]
  \begin{center}
    \begin{tabular}{|r|r|r|r|}
      \hline\hline
      $L/a$ 
      & \multicolumn{1}{c|}{$z_1$} 
      & \multicolumn{1}{c|}{$z_2$} 
      & \multicolumn{1}{c|}{$z_3$} \\
      \hline
      \multicolumn{4}{|c|}{$\theta=0$, $\rho=1$}\\
      \hline
 4 &    0.1326808072 &    0.01528384278 &   -0.1168458223 \\
 6 &    0.1474652524 &    0.00728806063 &   -0.0779791840 \\
 8 &    0.1731329290 &    0.00439645001 &   -0.0585101791 \\
10 &    0.2035733306 &    0.00321873110 &   -0.0468218991 \\
12 &    0.2367057252 &    0.00262320959 &   -0.0390261029 \\
14 &    0.2715872938 &    0.00226224002 &   -0.0334556609 \\
16 &    0.3077062797 &    0.00201667075 &   -0.0292766717 \\
      \hline   
      \multicolumn{4}{|c|}{$\theta=0.5$, $\rho=1$}\\
      \hline
 4 &    0.0824520269 &    0.03143610800 &   -0.1238972498 \\
 6 &    0.1271965066 &    0.01394010853 &   -0.0857279226 \\
 8 &    0.1683678316 &    0.00887347640 &   -0.0652575226 \\
10 &    0.2089123772 &    0.00693665488 &   -0.0525900075 \\
12 &    0.2497370083 &    0.00597030160 &   -0.0440062052 \\
14 &    0.2910727552 &    0.00538995423 &   -0.0378156722 \\
16 &    0.3329441002 &    0.00499976060 &   -0.0331442026 \\
      \hline   
      \multicolumn{4}{|c|}{$\theta=0$, $\rho=2$}\\
      \hline
 4 &    0.6521364999 &    0.02993994420 &   -0.1209100243 \\
 6 &    0.9221596868 &    0.01125513464 &   -0.0810128062 \\
 8 &    1.1964703064 &    0.00632783134 &   -0.0608921086 \\
10 &    1.4728928412 &    0.00424833271 &   -0.0487675434 \\
12 &    1.7507932870 &    0.00311330602 &   -0.0406652864 \\
      \hline   
      \multicolumn{4}{|c|}{$\theta=0.5$, $\rho=2$}\\
      \hline
 4 &    0.0665300894 &    0.16011586673 &   -0.0811127180 \\
 6 &    0.1791766558 &    0.07550278985 &   -0.0548862358 \\
 8 &    0.2766417147 &    0.04877505559 &   -0.0413683000 \\
10 &    0.3693268329 &    0.03915526808 &   -0.0331678255 \\
12 &    0.4605318775 &    0.03486305919 &   -0.0276723609 \\
      \hline\hline
    \end{tabular}
    \caption[Parts of the 1--loop coefficient $Z_{\mathrm{P}}^{(1)}$
             at various values of $\theta$ and $\rho$]
            {\sl Parts of the 1--loop coefficient $Z_{\mathrm{P}}^{(1)}$
              at various values of $\theta$ and $\rho$
    \label{tab:zpparts123var}}
  \end{center}
\end{table}

\begin{table}[htbp]
  \begin{center}
    \begin{tabular}{|r|r|r|r|}
      \hline\hline
      $L/a$ 
      & \multicolumn{1}{c|}{$z_4$} 
      & \multicolumn{1}{c|}{$z_6$} 
      & \multicolumn{1}{c|}{$z_7$} \\
      \hline
      \multicolumn{4}{|c|}{$\theta=0$, $\rho=1$}\\
      \hline
 4 &    0.2448721196 &    1.297122470 &   -0.1526509731 \\
 6 &    0.1942219137 &    1.476291125 &   -0.1197969999 \\
 8 &    0.1640794499 &    1.643098574 &   -0.0965623402 \\
10 &    0.1420781526 &    1.806546577 &   -0.0803607200 \\
12 &    0.1251135646 &    1.968711620 &   -0.0686367894 \\
14 &    0.1116498521 &    2.130300176 &   -0.0598239325 \\
16 &    0.1007298970 &    2.291603872 &   -0.0529813561 \\
      \hline   
      \multicolumn{4}{|c|}{$\theta=0.5$, $\rho=1$}\\
      \hline
 4 &    0.2334171022 &    1.189005361 &   -0.1834591369 \\
 6 &    0.2443514694 &    1.448427447 &   -0.1510297236 \\
 8 &    0.2314225501 &    1.659729841 &   -0.1241302945 \\
10 &    0.2123099369 &    1.853796129 &   -0.1042991779 \\
12 &    0.1934250942 &    2.040359892 &   -0.0895616671 \\
14 &    0.1764740188 &    2.223182810 &   -0.0783182317 \\
16 &    0.1616952383 &    2.403947754 &   -0.0695088918 \\
      \hline   
      \multicolumn{4}{|c|}{$\theta=0$, $\rho=2$}\\
      \hline
 4 &    0.5124378929 &    2.975968192 &   -0.1883730541 \\
 6 &    0.4114552065 &    4.088045527 &   -0.1353705625 \\
 8 &    0.3378163059 &    5.172972867 &   -0.1049844154 \\
10 &    0.2849616955 &    6.247086041 &   -0.0855561121 \\
12 &    0.2458461197 &    7.315968086 &   -0.0721305763 \\
      \hline   
      \multicolumn{4}{|c|}{$\theta=0.5$, $\rho=2$}\\
      \hline
 4 &    0.2819915539 &    1.490852756 &   -0.2128630913 \\
 6 &    0.2895835015 &    1.938807978 &   -0.1612690338 \\
 8 &    0.2645149383 &    2.334410619 &   -0.1275581165 \\
10 &    0.2365500613 &    2.709957987 &   -0.1049282241 \\
12 &    0.2116794602 &    3.076362839 &   -0.0889178266 \\
      \hline\hline
    \end{tabular}
    \caption[Parts of the 1--loop coefficient $Z_{\mathrm{P}}^{(1)}$
             at various values of $\theta$ and $\rho$ (cont.)]
            {\sl Parts of the 1--loop coefficient $Z_{\mathrm{P}}^{(1)}$
              at various values of $\theta$ and $\rho$ (cont.)
    \label{tab:zpparts467var}}
  \end{center}
\end{table}

\begin{table}[htbp]
  \begin{center}
    \begin{tabular}{|r|r|r|r|}
      \hline\hline
      $L/a$ 
      & \multicolumn{1}{c|}{$\tilde{z}_1$} 
      & \multicolumn{1}{c|}{$\tilde{z}_2$} 
      & \multicolumn{1}{c|}{$\tilde{z}_3$} \\
      \hline
 4 &    0.2293123742 &    0.009573101967 &   -0.07147065652 \\
 6 &    0.3494732430 &    0.006579092370 &   -0.04825027064 \\
 8 &    0.4688244330 &    0.005555581838 &   -0.03634661389 \\
10 &    0.5891478869 &    0.005029207837 &   -0.02913523296 \\
12 &    0.7105656939 &    0.004707933201 &   -0.02430499001 \\
14 &    0.8329343379 &    0.004491803200 &   -0.02084578712 \\
16 &    0.9560901651 &    0.004336724846 &   -0.01824725568 \\
      \hline\hline
    \end{tabular}
    \caption[Parts of the 1--loop coefficient $Z_{\mathrm{P}}^{(1)}(2L)$
             with $m_1(L)=0$ at $\theta=\pi/5$ and $\rho=1$]
            {\sl Parts of the 1--loop coefficient $Z_{\mathrm{P}}^{(1)}(2L)$
             with $m_1(L)=0$ at $\theta=\pi/5$ and $\rho=1$}
    \label{tab:zptparts123}
  \end{center}
\end{table}

\begin{table}[htbp]
  \begin{center}
    \begin{tabular}{|r|r|r|r|}
      \hline\hline
      $L/a$ 
      & \multicolumn{1}{c|}{$\tilde{z}_4$} 
      & \multicolumn{1}{c|}{$\tilde{z}_6$} 
      & \multicolumn{1}{c|}{$\tilde{z}_7$} \\
      \hline
 4 &    0.2911905373 &    1.874547330 &   -0.1335040402 \\
 6 &    0.2450740029 &    2.394596702 &   -0.0965677693 \\
 8 &    0.2051995671 &    2.890423111 &   -0.0750119316 \\
10 &    0.1749691162 &    3.378169482 &   -0.0611660627 \\
12 &    0.1519872262 &    3.862442505 &   -0.0515813433 \\
14 &    0.1341272577 &    4.344969442 &   -0.0445712831 \\
16 &    0.1199210816 &    4.826518506 &   -0.0392281295 \\
      \hline\hline
    \end{tabular}
    \caption[Parts of the 1--loop coefficient $Z_{\mathrm{P}}^{(1)}(2L)$
             with $m_1(L)=0$ at $\theta=\pi/5$ and $\rho=1$ (cont.)]
            {\sl Parts of the 1--loop coefficient $Z_{\mathrm{P}}^{(1)}(2L)$
             with $m_1(L)=0$ at $\theta=\pi/5$ and $\rho=1$ (cont.)}
    \label{tab:zptparts467}
  \end{center}
\end{table}

\begin{table}[htbp]
  \begin{center}
    \begin{tabular}{|r|r|r|r|}
      \hline\hline
      $L/a$ 
      & \multicolumn{1}{c|}{$\tilde{z}_1$} 
      & \multicolumn{1}{c|}{$\tilde{z}_2$} 
      & \multicolumn{1}{c|}{$\tilde{z}_3$} \\
      \hline
      \multicolumn{4}{|c|}{$\theta=0$, $\rho=1$}\\
      \hline
 4 &    0.1844808442 &    0.004372112061 &   -0.05984829671 \\
 6 &    0.2453504776 &    0.002601676822 &   -0.03949477059 \\
 8 &    0.3142628791 &    0.002001926120 &   -0.02947930729 \\
10 &    0.3877224203 &    0.001689952794 &   -0.02352815290 \\
12 &    0.4639602752 &    0.001496341488 &   -0.01958158454 \\
14 &    0.5420480669 &    0.001364127036 &   -0.01677124455 \\
16 &    0.6214500817 &    0.001268049176 &   -0.01466752787 \\
      \hline   
      \multicolumn{4}{|c|}{$\theta=0.5$, $\rho=1$}\\
      \hline
 4 &    0.1689995368 &    0.008923199948 &   -0.06548732132 \\
 6 &    0.2505009004 &    0.005999143948 &   -0.04413591317 \\
 8 &    0.3336383383 &    0.005016999229 &   -0.03321155407 \\
10 &    0.4187398577 &    0.004517026013 &   -0.02660691792 \\
12 &    0.5054112010 &    0.004213452330 &   -0.02218859089 \\
14 &    0.5932918622 &    0.004009910129 &   -0.01902679794 \\
16 &    0.6821137770 &    0.003864208278 &   -0.01665284161 \\
      \hline   
      \multicolumn{4}{|c|}{$\theta=0$, $\rho=2$}\\
      \hline
 4 &    1.2060785837 &    0.007491917291 &   -0.06149659905 \\
 6 &    1.7559778103 &    0.003528122640 &   -0.04081983117 \\
 8 &    2.3131197146 &    0.002140383509 &   -0.03057928797 \\
10 &    2.8738200566 &    0.001463496442 &   -0.02445228153 \\
12 &    3.4366941330 &    0.001073113077 &   -0.02037239719 \\
      \hline   
      \multicolumn{4}{|c|}{$\theta=0.5$, $\rho=2$}\\
      \hline
 4 &    0.2773639094 &    0.049707671735 &   -0.04185048543 \\
 6 &    0.4609370909 &    0.035128446020 &   -0.02778302053 \\
 8 &    0.6423873199 &    0.031226685817 &   -0.02081561802 \\
10 &    0.8244441384 &    0.029527745468 &   -0.01664726569 \\
12 &    1.0074272069 &    0.028606927366 &   -0.01387127057 \\
      \hline\hline
    \end{tabular}
    \caption[Parts of the 1--loop coefficient $Z_{\mathrm{P}}^{(1)}(2L)$
             with $m_1(L)=0$ at various values of $\theta$ and $\rho$]
            {\sl Parts of the 1--loop coefficient $Z_{\mathrm{P}}^{(1)}(2L)$
             with $m_1(L)=0$ at various values of $\theta$ and $\rho$
    \label{tab:zptparts123var}}
  \end{center}
\end{table}

\begin{table}[htbp]
  \begin{center}
    \begin{tabular}{|r|r|r|r|}
      \hline\hline
      $L/a$ 
      & \multicolumn{1}{c|}{$\tilde{z}_4$} 
      & \multicolumn{1}{c|}{$\tilde{z}_6$} 
      & \multicolumn{1}{c|}{$\tilde{z}_7$} \\
      \hline
      \multicolumn{4}{|c|}{$\theta=0$, $\rho=1$}\\
      \hline
 4 &    0.1718305672 &    1.687871282 &   -0.09775258579 \\
 6 &    0.1280139314 &    2.002040662 &   -0.06907686487 \\
 8 &    0.1020232090 &    2.316619630 &   -0.05317715374 \\
10 &    0.0848215601 &    2.633590801 &   -0.04317899772 \\
12 &    0.0725826258 &    2.952117432 &   -0.03632993149 \\
14 &    0.0634269015 &    3.271619233 &   -0.03134989520 \\
16 &    0.0563194892 &    3.591757515 &   -0.02756760705 \\
      \hline   
      \multicolumn{4}{|c|}{$\theta=0.5$, $\rho=1$}\\
      \hline
 4 &    0.2321338155 &    1.663348371 &   -0.12438362939 \\
 6 &    0.1938923565 &    2.044179271 &   -0.08971537396 \\
 8 &    0.1619544256 &    2.407180440 &   -0.06959163712 \\
10 &    0.1379679076 &    2.765026025 &   -0.05670609836 \\
12 &    0.1198008298 &    3.120873229 &   -0.04780137447 \\
14 &    0.1057072432 &    3.475803721 &   -0.04129513010 \\
16 &    0.0945069004 &    3.830264175 &   -0.03633907736 \\
      \hline   
      \multicolumn{4}{|c|}{$\theta=0$, $\rho=2$}\\
      \hline
 4 &    0.3405246084 &    5.207874762 &   -0.10562217877 \\
 6 &    0.2465730856 &    7.335024524 &   -0.07229922856 \\
 8 &    0.1926801273 &    9.459097330 &   -0.05491154830 \\
10 &    0.1579528577 &   11.581344073 &   -0.04425113515 \\
12 &    0.1337701693 &   13.702563787 &   -0.03705152726 \\
      \hline   
      \multicolumn{4}{|c|}{$\theta=0.5$, $\rho=2$}\\
      \hline
 4 &    0.2661600367 &    2.343188522 &   -0.12821614899 \\
 6 &    0.2121269758 &    3.079864030 &   -0.08907791880 \\
 8 &    0.1730920116 &    3.798786095 &   -0.06801232921 \\
10 &    0.1454143632 &    4.511560279 &   -0.05494446490 \\
12 &    0.1251028064 &    5.221609660 &   -0.04606774596 \\
      \hline\hline
    \end{tabular}
    \caption[Parts of the 1--loop coefficient $Z_{\mathrm{P}}^{(1)}(2L)$
             with $m_1(L)=0$ at various values of $\theta$ and $\rho$ (cont.)]
            {\sl Parts of the 1--loop coefficient $Z_{\mathrm{P}}^{(1)}(2L)$
             with $m_1(L)=0$ at various values of $\theta$ and $\rho$ (cont.)
    \label{tab:zptparts467var}}
  \end{center}
\end{table}
\end{appendix}
\cleardoublepage
\addcontentsline{toc}{chapter}{Bibliography}
\bibliography{diss}
\bibliographystyle{h-elsevier}

\cleardoublepage
\chapter*{Lebenslauf}
\chaptermark{}

\begin{tabular}{ll}

Name: & \dcauthorname  \dcauthorsurname \\

geboren am & 10.04.1971 in Berlin \\

Nationalit{\"a}t: & Deutsch \vspace{1cm}\\

6/1991 & Abitur an der Lessing-Oberschule (Gymnasium)\\
 & in Berlin-Wedding \\

10/1991 - 01/1997 & Studium an der Technischen Universit\"at Berlin \\

 & in der Fachrichtung Physik\\

05/1997 - 06/1998 & Wissenschaftlicher Mitarbeiter beim \\

 & Deutschen Elektronensynchrotron DESY, Zeuthen\\

seit 07/1998 & Wissenschaftlicher Mitarbeiter am Institut \\
 & f\"{u}r Physik der Humboldt-Universit\"{a}t zu Berlin\\

\end{tabular}

\section*{Publikationen}

\noindent
[a] Johannes Bl\"{u}mlein and Stefan Kurth, {\sl On the Mellin transform of the  
coefficient functions of $F_L(x,Q^2)$}, hep-ph/9708388, 1998.\\[3mm]
[b] Johannes Bl\"{u}mlein and Stefan Kurth, {\sl Harmonic sums and Mellin 
transforms up to two-loop order}, Phys.~Rev.~D60:014--018, 1999.\\[3mm]
[c] ALPHA Collaboration (Achim Bode et al.), {\sl First results on the 
running coupling in QCD with two massless flavors}, 
Phys.~Lett.~B515:49--56, 2001.\\[3mm]
[d]  Bernd Gehrmann, Stefan Kurth, Juri Rolf, Ulli Wolff,
{\sl  Schr\"{o}dinger functional at negative flavour number},
Nucl.~Phys.~B612:3--24, 2001.\\[3mm]
[e] Stefan Kurth, {\sl Properties of the renormalized quark mass in the
Schr\"{o}dinger  functional with a non-vanishing background field},
Nucl.~Phys.~Proc.~Suppl.~106: 850--852, 2002.\\[3mm]
[f] Bernd Gehrmann, Stefan Kurth, Juri Rolf, Ulli Wolff,
{\sl Schr\"{o}dinger functional at $\Nf = -2$},
Nucl.~Phys.~Proc.~Suppl.~106:793--795, 2002.


\cleardoublepage
\chapter*{Selbst\"andigkeitserkl\"arung}

Hiermit erkl\"are ich, die vorliegende Arbeit selbst\"andig ohne fremde 
Hilfe verfasst zu haben und nur die angegebene Literatur und
Hilfsmittel verwendet zu haben.\\
\vspace{5cm}
\begin{flushleft}
\dcauthorname  \dcauthorsurname \\
\dcdatesubmitted
\end{flushleft}

\end{document}